\documentclass[longauth]{aa} 

\usepackage{natbib}
\bibpunct{(}{)}{;}{a}{}{,} 

\usepackage{mathptmx}
\usepackage{amsmath,amsfonts,amssymb}
\usepackage[colorlinks=True]{hyperref}
\usepackage{color}
\usepackage{float} 

\definecolor{inon}{rgb}{1.00,0.27,0.00}

\DeclareMathAlphabet\mathbfcal{OMS}{cmsy}{b}{n}
\DeclareMathAlphabet\mathnfcal{OMS}{cmsy}{m}{n}

\newcommand{\appropto}{\mathrel{\vcenter{
  \offinterlineskip\halign{\hfil$##$\cr
    \propto\cr\noalign{\kern2pt}\sim\cr\noalign{\kern-2pt}}}}}

\newcommand{\lcdm}{\ensuremath{\Lambda\textrm{CDM}}}

\newcommand{\omegam}{\ensuremath{\Omega_{\mathrm{m}}}}

\newcommand{\omegab}{\ensuremath{\Omega_{\mathrm{b}}}}

\newcommand{\Hnow}{\ensuremath{H_{0}}}
\newcommand{\hnow}{\ensuremath{h}}
\newcommand{\h}{\ensuremath{h}}
\newcommand{\sigmaeight}{\ensuremath{\sigma_{8}}}

\newcommand{\ns}{\ensuremath{n_{\mathrm{s}}}}

\newcommand{\percent}{\ensuremath{\%}}
\newcommand{\Msun}{\ensuremath{\mathrm{M}_{\odot}}}
\newcommand{\Msunh}{\ensuremath{h^{-1}\mathrm{M}_{\odot}}}
\newcommand{\Mpc}{\ensuremath{\mathrm{Mpc}}}
\newcommand{\Mpch}{\ensuremath{h^{-1}\mathrm{Mpc}}}

\newcommand{\Rfiveoo}{\ensuremath{R_{500\mathrm{c}}}}
\newcommand{\Mfiveoo}{\ensuremath{M_{500\mathrm{c}}}}

\newcommand{\redshift}{\ensuremath{z}}
\newcommand{\mass}{\ensuremath{M}}
\newcommand{\dif}{\ensuremath{\mathrm{d}}}

\newcommand{\rhocrit}{\ensuremath{\rho_{\mathrm{c}}}}

\newcommand{\erosita}{\emph{eROSITA}}

\newcommand{\wise}{\emph{WISE}}

\newcommand{\erass}{eRASS}
\newcommand{\mbias}{\ensuremath{m_{\gamma}}}

\newcommand{\zcl}{\ensuremath{z_{\mathrm{cl}}}}

\newcommand{\rshear}{\ensuremath{\gamma_{+}}}
\newcommand{\rshearmod}{\ensuremath{\gamma_{+}^{\mathrm{mod}}}}
\newcommand{\gshear}{\ensuremath{g_{+}}}
\newcommand{\gshearmod}{\ensuremath{g_{+}^{\mathrm{mod}}}}

\newcommand{\convergencemod}{\ensuremath{\kappa^{\mathrm{mod}}}}

\newcommand{\deltaSigma}{\ensuremath{\Delta{\Sigma}_{\mathrm{m}}}}
\newcommand{\deltaSigmamod}{\ensuremath{\Delta{\Sigma}_{\mathrm{m}}^{\mathrm{mod}}}}
\newcommand{\Sigmam}{\ensuremath{{\Sigma}_{\mathrm{m}}}}
\newcommand{\Sigmammod}{\ensuremath{{\Sigma}_{\mathrm{m}}^{\mathrm{mod}}}}
\newcommand{\Sigmamcen}{\ensuremath{{\Sigma}_{\mathrm{m}}^{\mathrm{cen}}}}
\newcommand{\Sigmammis}{\ensuremath{{\Sigma}_{\mathrm{m}}^{\mathrm{mis}}}}

\newcommand{\bwl}{\ensuremath{b_{\mathrm{WL}}}}
\newcommand{\mwl}{\ensuremath{M_{\mathrm{WL}}}}
\newcommand{\sigmacrit}{\ensuremath{\Sigma_{\mathrm{crit}}}}
\newcommand{\lensingeff}{\ensuremath{\beta}}

\newcommand{\rdd}{\ensuremath{R}}

\newcommand{\mpiv}{\ensuremath{M_{\mathrm{piv}}}}
\newcommand{\zpiv}{\ensuremath{z_{\mathrm{piv}}}}

\newcommand{\lext}{\ensuremath{L_{\mathrm{ext}}}}
\newcommand{\ldet}{\ensuremath{L_{\mathrm{det}}}}
\newcommand{\rate}{\ensuremath{C_{\mathrm{R}}}}
\newcommand{\ratehat}{\ensuremath{\hat{C}_{\mathrm{R}}}}
\newcommand{\Arate}{\ensuremath{A_{\mathrm{X}}}}
\newcommand{\Brate}{\ensuremath{B_{\mathrm{X}}}}
\newcommand{\deltarate}{\ensuremath{\delta_{\mathrm{X}}}}
\newcommand{\gammarate}{\ensuremath{\gamma_{\mathrm{X}}}}
\newcommand{\sigmarate}{\ensuremath{\sigma_{\mathrm{X}}}}
\newcommand{\wone}{\ensuremath{W_{1}}}
\newcommand{\wtwo}{\ensuremath{W_{2}}}

\newcommand{\Bwl}{\ensuremath{B_{\mathrm{WL}}}}
\newcommand{\deltawl}{\ensuremath{\delta_{\mathrm{WL}}}}
\newcommand{\gammawl}{\ensuremath{\gamma_{\mathrm{WL}}}}
\newcommand{\sigmawl}{\ensuremath{\sigma_{\mathrm{WL}}}}
\newcommand{\rich}{\ensuremath{\lambda}}

\newcommand{\mbcg}{\ensuremath{M_{\star,\mathrm{BCG}}}}
\newcommand{\mbcghat}{\ensuremath{\hat{M}_{\star,\mathrm{BCG}}}}
\newcommand{\Abcg}{\ensuremath{A_{\mathrm{BCG}}}}
\newcommand{\Bbcg}{\ensuremath{B_{\mathrm{BCG}}}}
\newcommand{\deltabcg}{\ensuremath{\delta_{\mathrm{BCG}}}}
\newcommand{\gammabcg}{\ensuremath{\gamma_{\mathrm{BCG}}}}
\newcommand{\sigmabcg}{\ensuremath{\sigma_{\mathrm{BCG}}}}
\newcommand{\rhowlxray}{\ensuremath{\rho_{\mathrm{WL},\mathrm{X}}}}
\newcommand{\rhowlmbcg}{\ensuremath{\rho_{\mathrm{WL},M_{\star,\mathrm{BCG}}}}}
\newcommand{\rhombcgxray}{\ensuremath{\rho_{\mathrm{X},M_{\star,\mathrm{BCG}}}}}
\newcommand{\asobs}{\ensuremath{  0.093^{+0.029}_{-0.036} }}
\newcommand{\bsobs}{\ensuremath{  1.50^{+0.20}_{-0.30} }}
\newcommand{\deltasobs}{\ensuremath{  0.7\pm 1.5 }}
\newcommand{\gammasobs}{\ensuremath{  -0.8^{+1.5}_{-1.1} }}
\newcommand{\sigmasobs}{\ensuremath{  0.56^{+0.17}_{-0.23} }}
\newcommand{\rhowlsobswlonly}{\ensuremath{  -0.05^{+0.45}_{-0.63} }}
\newcommand{\axobs}{\ensuremath{  11.547^{+0.083}_{-0.074} }}
\newcommand{\bxobs}{\ensuremath{  0.184^{+0.068}_{-0.16} }}
\newcommand{\deltaxobs}{\ensuremath{  0.22\pm 0.29 }}
\newcommand{\gammaxobs}{\ensuremath{  1.49^{+0.88}_{-0.75} }}
\newcommand{\sigmaxobs}{\ensuremath{  0.554^{+0.040}_{-0.049} }}
\newcommand{\rhowlsobswlxo}{\ensuremath{  -0.04\pm 0.43 }}
\newcommand{\rhowlxobswlxo}{\ensuremath{  -0.32^{+0.23}_{-0.49} }}
\newcommand{\rhoxobssobswlxo}{\ensuremath{  -0.10^{+0.24}_{-0.27} }}
\newcommand{\axobsprioronz}{\ensuremath{  11.433^{+0.066}_{-0.055} }}
\newcommand{\bxobsprioronz}{\ensuremath{  0.38\pm 0.11 }}
\newcommand{\deltaxobsprioronz}{\ensuremath{  0.62\pm 0.25 }}
\newcommand{\gammaxobsprioronz}{\ensuremath{  -0.36\pm 0.13 }}
\newcommand{\sigmaxobsprioronz}{\ensuremath{  0.560^{+0.044}_{-0.055} }}
\newcommand{\axobsfulllib}{\ensuremath{  11.498\pm 0.085 }}
\newcommand{\bxobsfulllib}{\ensuremath{  0.192^{+0.057}_{-0.18} }}
\newcommand{\deltaxobsfulllib}{\ensuremath{  0.37^{+0.29}_{-0.32} }}
\newcommand{\gammaxobsfulllib}{\ensuremath{  0.91^{+0.97}_{-0.77} }}
\newcommand{\sigmaxobsfulllib}{\ensuremath{  0.566^{+0.043}_{-0.051} }}
\hypersetup{linkcolor=blue,filecolor=magenta,citecolor=[RGB]{39,174,96}}
\begin{document} 

%
%

\title{
The SRG/eROSITA All-Sky Survey
}

\subtitle{
The Weak-Lensing Mass Calibration and the Stellar Mass-to-Halo Mass Relation 
from the Hyper Suprime-Cam Subaru Strategic Program
}

%
%

\author{
I-Non~Chiu\inst{1} 
\and Vittorio~Ghirardini\inst{2,3}
\and Sebastian~Grandis\inst{4}
\and Nobuhiro~Okabe\inst{5,6,7}
\and Emmanuel~Artis\inst{2}
\and Esra~Bulbul\inst{2}
\and Emre~Y.~Bahar\inst{2}
\and Fabian~Balzer\inst{2}
\and Nicolas~Clerc\inst{8}
\and Johan~Comparat\inst{2}
\and Bau-Ching~Hsieh\inst{9}
\and Florian~Kleinebreil\inst{4}
\and Matthias~Kluge\inst{2}
\and Ang~Liu\inst{2}
\and Rog{\'e}rio~Monteiro-Oliveira\inst{9}
\and Masamune~Oguri\inst{10,11}
\and Florian~Pacaud\inst{12}
\and Miriam~Ramos~Ceja\inst{2}
\and Thomas~H.~Reiprich\inst{12}
\and Jeremy~Sanders\inst{2}
\and Tim~Schrabback\inst{4}
\and Riccardo~Seppi\inst{2}
\and Martin~Sommer\inst{12}
\and Sut-Ieng Tam\inst{13}
\and Keiichi~Umetsu\inst{9}
\and Xiaoyuan~Zhang\inst{2}
}

\institute{
Department of Physics, National Cheng Kung University, No.1, University Road, Tainan City 70101, Taiwan
\and Max Planck Institute for Extraterrestrial Physics, Giessenbachstrasse 1, 85748 Garching, Germany
\and INAF, Osservatorio di Astrofisica e Scienza dello Spazio, via Piero Gobetti 93/3, I-40129 Bologna, Italy
\and Universit\"{a}t Innsbruck, Institut f\"{u}r Astro- und Teilchenphysik, Technikerstr. 25/8, 6020 Innsbruck, Austria
\and Physics Program, Graduate School of Advanced Science and Engineering, Hiroshima University, 1-3-1 Kagamiyama, Higashi-Hiroshima, Hiroshima 739-8526, Japan 
\and Hiroshima Astrophysical Science Center, Hiroshima University, 1-3-1 Kagamiyama, Higashi-Hiroshima, Hiroshima 739-8526, Japan
\and Core Research for Energetic Universe, Hiroshima University, 1-3-1, Kagamiyama, Higashi-Hiroshima, Hiroshima 739-8526, Japan
\and IRAP, Universit\'{e} de Toulouse, CNRS, UPS, CNES, Toulouse, France
\and Institute of Astronomy and Astrophysics, Academia Sinica, P.O. Box 23-141, Taipei 10617, Taiwan
\and Center for Frontier Science, Chiba University, 1-33 Yayoi-cho, Inage-ku, Chiba 263-8522, Japan 
\and Department of Physics, Graduate School of Science, Chiba University, 1-33 Yayoi-Cho, Inage-Ku, Chiba 263-8522, Japan
\and Argelander-Institut f\"{u}r Astronomie (AIfA), Universit\"{a}t Bonn, Auf dem H\"{u}gel 71, 53121 Bonn, Germany
\and Institute of Physics, National Yang Ming Chiao Tung University, 1001 University Road, Hsinchu 30010, Taiwan
}

\date{Received April 1, 2025; accepted September 17, 2005}

%
%

\abstract{
We present the weak-lensing mass calibration and constrain the BCG (brightest cluster galaxy) stellar-mass-to-halo-mass-and-redshift (\mbcg--\mass--\redshift) relation for a sample of $124$ galaxy clusters and groups at redshift $0.1<\redshift<0.8$ from the first Data Release of the \erosita\ All-Sky Survey (\erass1), using data from the Hyper Suprime-Cam (HSC) Subaru Strategic Program.
The cluster survey is conducted by the \erosita\ X-ray telescope aboard the Spectrum-Roentgen-Gamma (SRG) space observatory.
The cluster sample is X-ray-selected and optically confirmed with a negligibly low contamination rate ($\approx5\percent$).
On a basis of individual clusters, the shear profiles \gshear\ of $96$ clusters are derived using the HSC Three-Year (HSC-Y3) weak-lensing data, while the BCG stellar masses \mbcg\ of $101$ clusters are estimated using the SED template fitting to the HSC five-band ($grizY$) photometry.
The observed X-ray photon count rate \rate\ is used as the mass proxy, based on which individual halo masses \mass\ are obtained at the given \rate\ in a population modelling while accounting for systematic uncertainties in the weak-lensing modelling through a simulation-calibrated weak-lensing mass-to-halo-mass (\mwl--\mass--\redshift) relation.
The count rate (\rate--\mass--\redshift) and BCG stellar mass (\mbcg--\mass--\redshift) relations are simultaneously constrained in a forward and population modelling.
In agreement with the results based on the weak-lensing data from the DES and KiDS surveys, we obtain a \rate--\mass--\redshift\ relation with a self-similar redshift scaling and a mass trend 
that is steeper than the self-similar prediction.
We cannot simultaneously place stringent constraints on the power-law indices of the mass (\Bbcg) and redshift (\gammabcg) trends due to the parameter degeneracy arising from the sample selection and the limited sample size.
By adopting an informative prior on \gammabcg\ to break the \Bbcg--\gammabcg\ degeneracy, we obtain a 
\mbcg--\mass--\redshift\ relation with the mass slope increasing to $\Bbcg = \bxobsprioronz$.
Informed by the prior,
our results suggest that the BCG stellar mass at a fixed halo mass has remained stable with a moderate increase at a level of $\left(20\pm8\right)\percent$ since redshift $\redshift\approx0.8$.
This finding supports the picture of the ``rapid-then-slow'' BCG formation, where the majority of the stellar mass must have been assembled at much earlier cosmic time.
}

%
%


%
%

\titlerunning{The HSC analyses of eRASS1 clusters}
\authorrunning{Chiu et al.}

\maketitle

%
%

\section{Introduction}
\label{sec:introduction}

The \erosita\ (extended ROentgen Survey with an Imaging Telescope Array) X-ray telescope \citep{predehl21}, the primary soft-X-ray instrument of the Russian-German ``Spectrum-Roentgen-Gamma'' (SRG) space observatory  \citep{sunyaev21} launched in 2019, has revolutionized the study of galaxy clusters with its groundbreaking achievements in carrying out the \erosita\ All-Sky Survey \citep[hereafter \erass;][]{merloni12,merloni24}, which provides the deepest all-sky imaging in X-rays to date.
The main goal of the \erosita\ survey is to advance cosmological studies by constructing the largest sample of galaxy clusters and investigating the growth of their populations over cosmic time.
The population growth of galaxy clusters is closely related to the cosmic structure formation \citep[][]{bardeen86,allen11,kravtsov12} and places stringent constraints on cosmological parameters, such as the mean matter fraction \omegam, the degree of density-field inhomogeniety \sigmaeight, and the equation of state of dark energy \citep{weinberg13,huterer15}.

Prior to the \erosita\ era, cluster cosmology has played a central role in cosmological analyses across multiple wavelengths, including X-rays \citep{reiprich02,vikhlinin09b,mantz15,schellenberger17,garrel22}, the optical \citep{costanzi19b,to21,sunayama24}, and the millimeter wavelength \citep{salvati21,bocquet24b} via the Sunyaev–Zeldovich effect \citep[SZE;][]{sunyaev70b,sunyaev72}.
Moreover, samples selected by weak gravitational lensing (hereafter weak lensing or WL) have gained increasing attention in recent years as a promising probe \citep{chiu24,chen25}.
Cluster cosmology heavily relies on the accurate determination of the cluster mass \citep{pratt19}, for which significant progress has been made in utilizing weak lensing as a reliable method \citep{okabe10a,umetsu14,vonderlinden14a,applegate14,okabe16,schrabback18,dietrich19}.
In the past decade, the deployments of Stage-III weak-lensing surveys, such as the Dark Energy Survey \cite[DES;][]{des05}, Kilo-Degree Survey \citep[KiDS;][]{kuijken15}, and the Hyper Suprime-Cam (HSC) Subaru Strategic Program \citep[][]{aihara18a}, have enabled the weak-lensing mass calibration of sizable samples using large and homogeneous data sets with a consistent methodology \citep{simet17,mcclintock19,murata19,bellagamba19,miyatake19,umetsu20,chiu22,shirasaki24,bocquet24a}.
In particular, \cite{chiu23} conducted the first cosmological study using \erosita\ clusters selected during the performance-verification phase of the \erass\ survey in a synergy with the HSC weak-lensing mass calibration, demonstrating the potential of \erosita-based cluster cosmology when it is combined with the Stage-III weak-lensing surveys.

In 2024, the first \erosita\ Data Release \citep[DR1;][]{merloni24} from the German \erosita\ Consortium (hereafter \erosita-DE) delivered the largest sample of X-ray-selected clusters to date \citep{bulbul24,kluge24} based on the first all-sky scan (\erass1).
This milestone also led to by far the tightest cosmological constraints obtained from cluster abundance \citep{ghirardini24}, enabled by joint weak-lensing mass calibrations from the surveys of HSC (this work), DES \citep{grandis24}, and KiDS \citep{kleinebreil24}.
Leveraging the study of \cite{ghirardini24}, constraints on various non-concordance cosmological models were also obtained \citep{artis24a,artis24b}.
Importantly, the X-ray selection function of \erass1 clusters has been precisely characterized \citep{clerc24} using \erass\ digital-twin simulations \citep{seppi22}.
The combination of the largest cluster sample selected via the X-ray emission of intracluster medium (ICM), the accurate mass calibration with the state-of-the-art weak-lensing data sets, and the precise characterization of the selection function has enabled unprecedented precision in cluster studies using the \erosita\ sample: for example, the feedback of active galactic nucleus in cluster cores \citep{bahar24}, the halo assembly bias of superclusters \citep{liu24}, the properties of halo clustering \citep{seppi24}, and the halo concentration and morphology of massive clusters \citep{okabe25}.

We stress that \erosita\ clusters are 
primarily selected in X-rays with minimal dependence on the characteristics of their galaxy populations,
thus making them an ideal sample for studying the formation and evolution of galaxies hosted by massive halos.
X-ray-selected samples also offer two advantages over other selection methods: 
First, they probe a mass regime generally much lower than that of SZE-selected samples, thanks to the deep and high-resolution imaging in X-rays.
Second, by tracing the distribution of hot ICM, X-ray-selected samples are not prone to the projection effect, which currently poses severe challenges for the modelling of optically selected clusters \citep{costanzi19a,sunayama20}.

By taking full advantage of exquisite imaging from the HSC survey, this study aims to (1) perform the weak-lensing mass calibration, (2) measure the stellar mass \mbcg\ of their brightest cluster galaxies (BCGs), and (3) derive the stellar-mass-to-mass-and-redshift (\mbcg--\mass--\redshift) relation for \erosita\ clusters in the first-year (\erass1) sample.
Given the mass range of the \erass1 clusters, the \mbcg--\mass--\redshift\ relation corresponds to the high-mass end of the stellar-mass-to-halo-mass relation, a terminology widely used in the literature.
Similar to \cite{leauthaud12a} and \cite{coupon15}, a key feature of this work is that we simultaneously model the weak-lensing mass and the stellar-mass-to-halo-mass relation on a basis of individual clusters, allowing the direct determination of the mass and redshift trends of the \mbcg--\mass--\redshift\ relation without relying on the abundance-matching method \citep{vale04,kravtsov04,conroy06}, which depends on numerical simulations.
The major strengths of this work are (1) the uniform X-ray selection of \erass1 clusters extending to redshift $\redshift \approx 1$ without the selection bias toward galaxy properties and (2) the well-quantified selection function and the halo mass determination, ensuring that the resulting cosmological constraints are in 
acceptable
agreement with those from other independent probes, thereby forming a cosmology-verified cluster sample suitable for astrophysical studies.
In this study, we have identified $124$ \erass1 clusters within the common footprint of the \erass\ and HSC surveys, covering an area of $\approx500$~deg$^2$.
 
This paper is organized as follows.
In Section~\ref{sec:sample_and_data}, we describe the \erass1 clusters and the HSC data sets used in this work.
The measurements of weak-lensing observables and the BCG stellar mass are presented in Section~\ref{sec:measurements}.
The modelling is described in Section~\ref{sec:modelling}.
We present and discuss the results in Section~\ref{sec:results}, and draw conclusions in Section~\ref{sec:conclusions}.
Unless stated otherwise, the halo mass adopts the definition of \Mfiveoo, which is the mass enclosed by a sphere with a radius \Rfiveoo\ wherein the average mass density is $500$ times the cosmic critical density $\rhocrit\left(\redshift\right)$ at the cluster redshift \redshift.
In this work, we also make use of the symbol \mass\ for the halo mass interchangeably with \Mfiveoo.
We define the notation $\mathcal{N}\left(x, y^2\right)$ $\left(\mathcal{U}\left(a, b\right)\right)$ as a Gaussian distribution with the mean $x$ and variance of $y^2$ (flat distribution between $a$ and $b$).
We adopt a concordance flat \lcdm\ cosmological model with the standard cosmological parameters ($\omegam = 0.3$, $\sigmaeight = 0.8$, and $\Hnow = 70~\mathrm{km}/\mathrm{sec}/\Mpc$), which we allow to vary within reasonable ranges in the modelliing.

%
%

%
\begin{figure*}
\resizebox{\textwidth}{!}{
\includegraphics[scale=1]{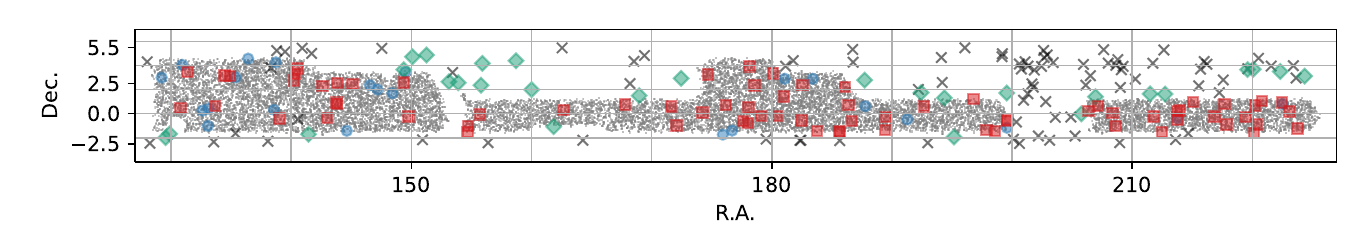}
}
\caption{
The footprint of the HSC-Y3 weak-lensing data set (grey dots) and the sky distributions of the \erass1 clusters are shown, with those categorized based on the availability of weak-lensing shear profile \gshear, the BCG stellar mass estimate \mbcg, or both.
The \erass1 clusters with available \gshear\ are marked as blue circles (23 clusters), those with \mbcg\ as green diamonds (28 clusters), and those with both measurements as red squares (73 clusters).
The other \erass1 clusters with neither \gshear\ nor \mbcg\ are shown as the grey crosses and therefore excluded from this study. 
}
\label{fig:footprint}
\end{figure*}
\begin{figure}
\resizebox{0.45\textwidth}{!}{
\includegraphics[scale=1]{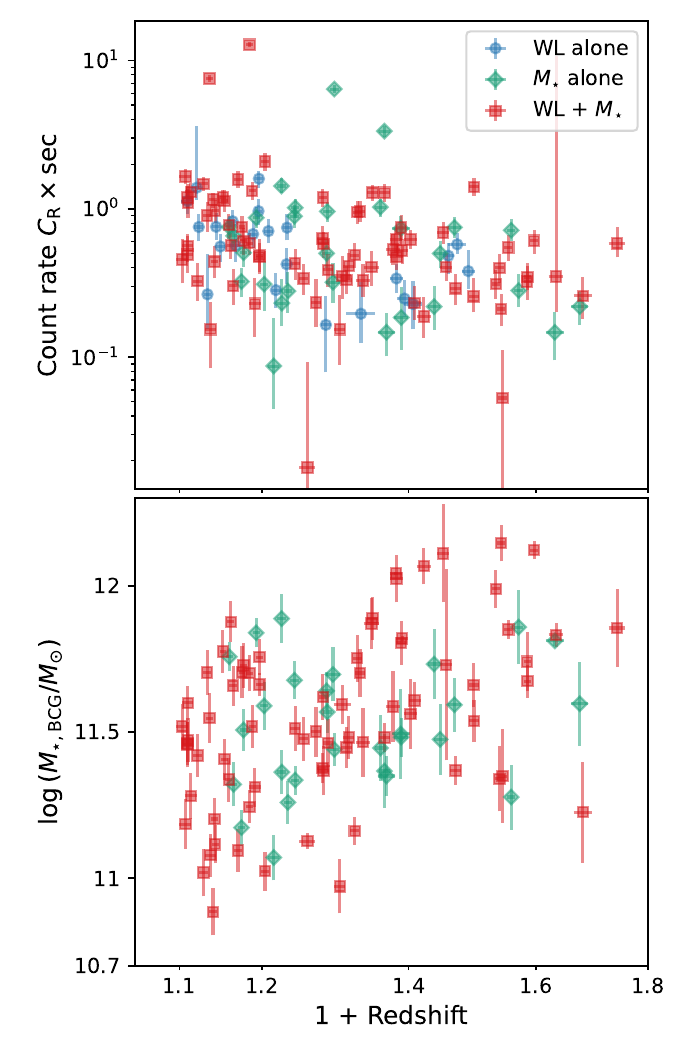}
}
\caption{
The distributions of the observed count rate \rate\ (top panel) and the BCG stellar mass \mbcg\ (left panel) as functions of the cluster redshift for the \erass1 clusters studied in this work.
The clusters are color-coded as identical as in Figure~\ref{fig:footprint}.
Note that a clear dependence of the BCG stellar mass on the cluster redshift is revealed in the bottom panel, which arises from the X-ray selection primarily favoring massive clusters at high redshift.
}
\label{fig:sample}
\end{figure}

\section{The cluster sample and data}
\label{sec:sample_and_data}

We describe the \erass1 sample of galaxy clusters in Section~\ref{sec:the_sample}.
The HSC data sets used in this work are summarized in Section~\ref{sec:the_hsc_data}.

\subsection{The cluster sample}
\label{sec:the_sample}

We use the sample of galaxy clusters overlapping the footprint of the HSC survey from the cluster catalog released in the \erosita-DE DR1 \citep{bulbul24}.
Specifically, the selection of the clusters is identical to that of the ``cosmological subsample'' used in the eRASS1 cosmological analysis \citep{ghirardini24} with (1) the extent likelihood \lext\ of $\lext>6$, (2) the galaxy richness \rich\ of $\rich>3$, and (3) the redshift range of $0.1 < \redshift < 0.8$ determined from the optical confirmation \citep{kluge24}.
The extent likelihood \lext, a quantity returned by the \erosita\ pipeline, describes the normalized logarithmic likelihood of a detected source being a spatially extended object rather than a point source.
In the interest of uniformity, the photometric redshifts of the galaxy clusters are consistently adopted with the accuracy of $\delta\redshift/\left(1 + \redshift\right)<0.005$ calibrated using spectroscopic samples \citep{kluge24}.
With such high accuracy, the uncertainty of the cluster redshift is negligible for the purpose of this work.
As a result, there are $124$ eRASS1 clusters in the common footprint between the eRASS and HSC surveys, as the final sample studied in this work.
The sky distribution of the sample is shown in Figure~\ref{fig:footprint}.

Leveraging the realistic mock simulations \citep[][see also \citealt{comparat20}]{seppi22}, the initial contamination of the cluster sample is estimated to be at a level of 
$\approx6\percent$
with the X-ray selection of $\lext>6$ alone.
After the optical confirmation, the resulting contamination is reduced to 
$\approx5\percent$
\citep{kluge24,bulbul24}, which is subdominant compared to the Poissonian noise of our sample (at a level of $\approx9\percent$) and therefore negligible in this work.
An independent modelling of the same weak-lensing shear profiles while accounting for the sample contamination in \cite{kleinebreil24} returned a result that is statistically consistent with ours (see Section~\ref{sec:rate_relation_results}), reinforcing the picture that the contamination in our sample is subdominant in this work.

The observed and corrected X-ray photon count rate (hereafter count rate \rate) in the $0.2$--$2.3$~$k$eV band is used as the mass proxy \citep{bulbul24}.
The X-ray count rate \rate\ has been shown as a reliable mass proxy for \erosita-detected clusters with well-characterized scaling with the cluster mass \citep{chiu22,grandis24,kleinebreil24}, despite with large scatter \citep{ghirardini24} that deserves being further studied in a future work.
For each cluster, the selection function---the probability $\mathnfcal{I}\left(\ratehat | \zcl, \mathnfcal{H}\right)$ of the cluster being detected---is modelled as a function of the 
intrinsic count rate \ratehat\
and evaluated at the cluster redshift \zcl\ and the sky location $\mathnfcal{H}$ \citep[][see also equation~1 in \citealt{ghirardini24}]{clerc24}.
The intrinsic count rate \ratehat\ refers to the observed one \rate\ before it is affected by observational noise.
The selection function is included in the modelling of the scaling relation in Section~\ref{sec:modelling_of_scaling_relation}.

The distribution of the clusters in the observable space of the X-ray count rate \rate\ and redshift is shown in the upper panel of Figure~\ref{fig:sample}, where the colors represent the availability of the measurements described in the following subsection.

\subsection{The HSC data sets}
\label{sec:the_hsc_data}

We use the data from the HSC survey to (1) measure the weak-lensing shear profiles and hence the weak-lensing mass, and to (2) extract the stellar mass of the BCGs. 
In what follows, a brief description about the data sets is provided.

In terms of the weak-lensing measurements, we use the latest HSC Three-Year (Y3) shape catalogs \citep{li22} that were constructed using the $i$-band imaging acquired during 2014 and 2019 with a mean seeing of $0.59$~arcsec.
The catalog has a limiting $i$-band magnitude cut of $i<24.5$~mag, resulting in an average density at a level of $\approx20$~galaxies/arcmin$^2$.
The shape measurement was rigorously calibrated against image simulations \citep[see e.g.,][]{mandelbaum18b}, delivering the multiplicative bias \mbias\ at a level of $|\mbias| < 9\times10^{-3}$ independently of redshift up to $\redshift\approx3$.
Various null tests were examined to be consistent with zero or within the requirements for cosmic-shear analyses, ensuring sufficiently accurate weak-lensing measurements for cluster studies.
The HSC-Y3 shape catalog covers a footprint area of $\approx500$~deg$^2$ and is divided into six fields, namely GAMA09H, GAMA15H, XMM, VVDS, HECTOMAP, and WIDE12H.
The clusters selected in \erass1 are only located in the fields of GAMA09H, GAMA15H, and WIDE12H, of which the data sets are used in this work.

The photometric redshift (photo-\redshift) of the weak-lensing source sample is needed to interpret the weak-lensing measurements.
Specifically, we use the photo-\redshift\ estimated by the machine-learning-based code \texttt{DEmP} \citep[Direct Empirical Photometric method;][]{hsieh14}, whose performances were fully quantified in \cite{nishizawa20}.
In short, the bias, scatter, and the outlier fraction of the photo-\redshift\ estimates for the source galaxies are estimated to be at levels of 
$|\Delta\redshift| \approx 0.003\left(1 + \redshift\right)$,
$\sigma_{\Delta\redshift} \approx 0.019\left(1 + \redshift\right)$, and 
$\approx5.4\percent$, respectively, where $\Delta\redshift$ is defined as the difference between the photometric and spectroscopic redshifts.
The full distribution $P\left(\redshift\right)$ of the photo-\redshift\ estimates is used to model the weak-lensing signals.

In \cite{kluge24}, the BCGs of \erass1 clusters were identified as the brightest passive member galaxies within a characteristic radius.
In addition, the authors estimated that approximately $85\percent$ of \erass1 clusters' BCGs coincide with the optical centers.
To ensure a homogeneous identification of BCGs, we visually select each cluster's brightest galaxy as the BCG using the HSC image.
Specifically, we select the BCG of each \erass1 cluster as the brightest galaxy whose color is similar to that of the galaxy overdensity near the X-ray center.
As a result, we find that $25$ (out of $101$) \erass1 clusters in our sample have their BCGs that differ from those automatically selected in \cite{kluge24}.
As a validation, we compare the photo-\redshift\ estimates (\texttt{photoz\_best}) of the BCGs from the HSC survey with the cluster redshifts (\texttt{Z\_LAMBDA}) and calculate the BCG photo-\redshift\ bias defined as $\Delta {\redshift}_{\mathrm{BCG}} \equiv \left(\mathtt{photoz\_best} - \mathtt{Z\_LAMBDA}\right) / \left(1 + \mathtt{Z\_LAMBDA}\right)$.
We find that the bias $\Delta {\redshift}_{\mathrm{BCG}}$ has a median value of $0.0028$ and a standard deviation of $0.015$\footnote{Among the $101$ \erass1 clusters, we have only $1$ system with the maximum value of $|\Delta {\redshift}_{\mathrm{BCG}}|\approx0.066$ in our sample.}, indicating that the BCG identification is robust in this work.

To estimate the stellar mass \mbcg\ of BCGs, we use the forced \texttt{cmodel} photometry at the $grizY$ broadband queried from the HSC Public Data Release 3 (PDR3).
No attempts to include the intracluster light (ICL) in \mbcg\ are made.
We exclude the BCGs with pixels (in one of the $grizY$ images) that are either contaminated, saturated, or masked due to bright stars, or have failures in the \texttt{cmodel} photometric fitting.
This leads to a sample of BCGs with clean \texttt{cmodel} photometry.
Aiming for the high signal-to-noise ratio and uniform HSC photometry, we only include the BCGs with at least two exposures in all $grizY$ bands.
Appendix~\ref{app:sql} contains the flags used in querying the photometry of the BCGs.
Appendix~\ref{app:imaging} shows the cutout images of these BCGs studied in this work.
The identified sky locations of the BCGs are tabulated in Table~\ref{tab:measurements}.

It is important to  note that the \texttt{cmodel} algorithm adopts a composite model, where the bulge and disk components are described by \citet[][]{devaucouleurs1948} and exponential profiles, respectively.
Consequently, the \texttt{cmodel} photometry does not fully capture the stellar distribution of the extended, diffuse ICL at large radii around BCGs.
As a result, the \texttt{cmodel} photometry is expected to underestimate the ``total'' flux of a BCG$+$ICL system \citep[see also][]{akino22}.
In fact, the exact definition of the ``total'' stellar mass of BCGs becomes nuanced, as they are embedded within the surrounding ICL.
A common approach to obtain the ``total'' stellar mass of a BCG$+$ICL system is to measure the flux within a large aperture, followed by a conversion from the observed light to the mass with an assumed mass-to-light ratio. 
In \cite{huang18c}, they estimated the stellar mass profiles of BCGs using a series of well defined apertures and found that the stellar masses estimated from the \texttt{cmodel} photometry were systematically lower than those enclosed within $100~k\mathrm{pc}$, precisely because the \texttt{cmodel} photometry does not account for the extended ICL component at large radii.
By stacking the light profiles of $\approx3000$ clusters in \cite{chen22}, the authors found that the stellar mass of a BCG alone is well approximated by the stellar mass of the bulge/disk component enclosed within $\approx50~k\mathrm{pc}$, beyond which the BCG component slowly transitions to the ICL-dominated regime at $\gtrsim100~k\mathrm{pc}$.
In this study, where the \texttt{cmodel} photometry is used, we effectively estimate only the stellar mass of the bulge/disk components of a BCG$+$ICL system (within $\lesssim50~k\mathrm{pc}$).

We note that the photometric catalog of the BCGs is constructed independently of the HSC-Y3 shape catalog, such that the availability of the optical photometry for the BCGs is not subject to the selection of the weak-lensing sources.
Consequently, some clusters have available HSC $grizY$ photometry to estimate the BCG stellar mass but no data to obtain the weak-lensing mass, and vice versa.
The final sample of $124$ eRASS1 clusters includes $23$ and $28$ systems with only weak-lensing data and BCG photometry, respectively; the remaining $73$ clusters have both data sets.

The distributions of the X-ray count rate \rate, the BCG stellar mass \mbcg\ (which we derive in Section~\ref{sec:stellar_mass_estimation}), and the cluster redshift \zcl\ are shown in Figure~\ref{fig:sample} with colors indicating the availability of the HSC data sets.

%
%

\section{Measurements}
\label{sec:measurements}

For individual clusters with the available data sets, we extract two kinds of measurements, the weak-lensing shear profile (Sections~\ref{sec:weaklensing_measurements}) and the BCG stellar mass (Section~\ref{sec:stellar_mass_estimation}).

\subsection{Weak-lensing measurements}
\label{sec:weaklensing_measurements}

To extract the weak-lensing measurements, we follow the identical procedure as detailed in \cite{chiu22}, in which they used the same HSC-Y3 data sets to derive the tangential reduced shear profiles $\gshear\left(\theta\right)$ as a function of the angular radius $\theta$ to calibrate the halo mass of clusters selected in the \erosita\ Final Equatorial-Depth Survey \citep[eFEDS;][]{liu21}.
Later, the same weak-lensing analysis was used to derive cosmological constraints using the eFEDS cluster abundance \citep{chiu23}.
Here, the weak-lensing data products produced in this work have been closely examined against other stage-III surveys \citep{kleinebreil24} and used in the eRASS1 cosmological analyses \citep{ghirardini24}.
In what follows, we provide a summary of the weak-lensing measurements.

On a basis of individual clusters, the weak-lensing measurements are composed of the tangential reduced shear profile $\gshear\left(\theta\right)$ (with the measurement uncertainty) and the corresponding redshift distribution of the source galaxies.
A galaxy is selected as a weak-lensing source of a cluster at redshift \zcl\ based on its full redshift distribution $P\left(\redshift\right)$, requiring it to be securely located at the background with $\int_{\zcl + 0.2} P\left(\redshift\right)\mathrm{d}\redshift > 0.98$.
With this ``P-cut'' selection \citep{oguri14,medezinski18a}, the resulting source sample is extremely clean without cluster member contamination out to redshift $\redshift\approx1.2$ \citep[see Figure~3 in][]{chiu22}.
Following equation~(8) in \cite{chiu22}, the shear profile $\gshear\left(\theta\right)$ is extracted around the \erosita-defined X-ray center with ten logarithmic radial bins between $0.2\Mpch$ and $3.5\Mpch$ calculated in a fixed flat \lcdm\ cosmology with $\omegam=0.3$ and $h=0.7$.
The three innermost bins are discarded in the weak-lensing mass modelling to avoid systematics, resulting in total seven logarithmic bins at a radial range of $0.5\Mpch\lesssim\rdd<3.5\Mpch$.
Note that we convert the unit of the radial binning to the angular space, i.e., in arcmin, and self-consistently fit the model of the cluster mass profile at a correct physical scale inferred from the redshift-distance relation while the cosmological parameters are varying in the forward modelling.
The uncertainty of the tangential shear profile is expressed by the weak-lensing covariance matrix \citep[see equation~(14) in][]{chiu22}; however, we only include the shape noise, as opposed to that in \cite{chiu22} containing the scatter due to uncorrelated large-scale structures \citep{hoekstra03}.
The scatter raised from uncorrelated large-scale structures is accounted for in calibrating the weak-lensing mass bias \citep[see Section~\ref{sec:wl_modelling} and][]{grandis24}. 
The stacked and lensing-weight-weighted $P\left(\redshift\right)$ of the source galaxies for each cluster is used to infer the lensing efficiency \citep[see equation~34 in][]{chiu22}.

We describe the modelling of these shear profiles to calibrate the cluster halo mass in Section~\ref{sec:wl_modelling}.

\subsection{Measurements of BCG stellar masses}
\label{sec:stellar_mass_estimation}

Similarly to the procedure in \citet[][see also \citealt{erfanianfar19}]{chiu18a}, we estimate the BCG stellar mass \mbcg\ using the SED (spectral energy distribution) fitting technique implemented in the $\chi^2$ template-fitting code, \texttt{Le~Phare} \citep{arnouts99,ilbert06}\footnote{\url{http://www.cfht.hawaii.edu/~arnouts/lephare.html}}.
We describe the analysis steps, as follows.

We use the five broadband photometry $grizY$ from the PDR3 database of the HSC survey.
For each BCG, the template fitting is fixed to the cluster redshift \zcl.
The SED fitting has two passes.
The first is to account for the zero-point offsets in the photometry. 
Specifically, we build the galaxy templates from the COSMOS field \citep{ilbert09} with the \citet{prevot84} and \cite{calzetti00} dust-extinction laws\footnote{
The extinction laws are applied to the galaxy templates according to the prescription provided within the \texttt{Le~Phare} package.
This gives the same set of the galaxy templates used to determine the photo-\redshift\ in the COSMOS field \citep{ilbert09}.
}
with $E\left(B-V\right)$ in a range between $0$ and $0.5$.
Once the best-fit template is determined for each BCG, a systematic zero-point offset of each band is computed to minimize the differences between the models and observed magnitudes among all the BCGs.
This process is iterated until a convergence is reached.
The second pass of the SED fitting is to obtain the BCG stellar mass by fitting the galaxy templates predicted by the Stellar Population Synthesis (SPS).
We build the SPS templates using the \cite{bruzual03} model with the \cite{chabrier03} initial mass function (IMF).
Three metallicities ($Z=0.004,0.008,0.02$) are used and configured with the exponentially decaying star-formation rate characterized by an $e$-folding timescale $\tau$.
We use six values of $\tau$ up to $5$~Gyr (i.e., $\frac{\tau}{\mathrm{Gyr}} = 0.1, 0.3, 1, 2, 3, 5$) because BCGs are expected to be dominated by old stellar populations without recent star-forming activities.
This is supported by our data, as extending the timescale $\tau$ to $30$~Gyr to include star-forming galaxy templates does not change the final results (see Section~\ref{sec:results}).
Moreover, the galaxy templates are configured with the \cite{calzetti00} dust-extinction law of $E\left(B-V\right)=0, 0.1, 0.2, 0.3$.
The ages of these galaxy templates can have a maximum value of $14$~Gyr.
When fitting the \cite{bruzual03} templates, we apply the zero-point offsets obtained from the first pass to the individual bands and add the same amounts of systematic uncertainties to the measurement uncertainties in quadrature.
Note that this effectively weakens the constraining power of the bands with a larger offset.
Last, we additionally increase the flux uncertainties (after adding the systematic uncertainties) by a factor of $6.5$, which is suggested by our data such that the median of the resulting reduced $\chi_{\mathrm{red}}^2$ of the BCGs is one.
This does not change the best-fit stellar mass but only increases the measurement uncertainty of the final mass estimate.
We note that the statistical uncertainties of the photometric measurements of the BCGs in the $grizY$ bands are at a level of $10^{-3}$~mag, and that the SED templates predicted by SPS are not perfect.
As a result, even small discrepancies between the model and observation can lead to significantly large $\chi_{\mathrm{red}}^2$.
By increasing the flux uncertainties while requiring the median $\chi_{\mathrm{red}}^2$ of the sample to be unity, we effectively account for the uncertainty in the derived stellar mass arising from the limitations of the SED templates. 
The measurement uncertainty of \mbcg\ is obtained by bootstrapping the \cite{bruzual03} template fitting by $1000$ times, assuming that the flux uncertainties are Gaussian.

We estimate the systematic uncertainty in our stellar mass measurements as follows.
We construct a catalog of galaxies observed in the COSMOS field from the HSC data base, estimate their stellar masses at known redshifts using the same method mentioned above, and compare the resulting stellar masses $M_{\star,\mathrm{HSC}}$ to those  in the latest \texttt{COSMOS2020} catalog ($M_{\star,\mathrm{COSMOS}}$) estimated using $30$-band photometry \citep{weaver22}.
In the stellar mass and redshift range occupied by the BCGs of the \erass1 clusters ($M_{\star}\gtrsim10^{10.7}\Msun$ and $\redshift\lesssim1$), we find a systematic offset in the stellar mass at a level of $\approx0.1~\mathrm{dex}$, i.e., $\left\langle \log\left( M_{\star,\mathrm{HSC}} / \Msun \right) - \log\left( M_{\star,\mathrm{COSMOS}} / \Msun\right) \right\rangle \approx 0.1~\mathrm{dex}$.
Moreover, this systematic offset does not depend on the redshift.
Therefore, we expect that (1) there exists a systematic uncertainty at a level of $\approx0.1$ dex in our \mbcg\ estimates with respect to those measured with the full configuration of SED templates on high-quality data such as in the COSMOS field,
and 
that (2) the systematic uncertainty only affects the normalization but not the mass or redshift trends of the scaling relation.

The resulting BCG stellar mass estimates \mbcg\ are plotted as a function of the cluster redshift in Figure~\ref{fig:sample}.
We present the results of the SED fitting in Appendix~\ref{app:imaging}.
The measurements of the BCG setllar mass \mbcg\ of individual \erass1 clusters are contained in Table~\ref{tab:measurements}.

%
%

\section{Modelling}
\label{sec:modelling}

Our goal is to constrain the scaling relations of the observables, including the X-ray count rate \rate\ and the BCG stellar mass \mbcg.
We first provide an overall framework of the modelling in Section~\ref{sec:modelling_framework}, and refer readers to \citet[][see also \citealt{bocquet19} and \citealt{chiu23}]{chiu22} for more details.
Next, we describe the modelling of the scaling relations in Section~\ref{sec:modelling_of_scaling_relation}, followed by the description of the weak-lensing modelling in Section~\ref{sec:wl_modelling}.
Finally, we give the statistical inference to obtain the parameter posteriors in Section~\ref{sec:statistical_inference}.

\subsection{Modelling framework}
\label{sec:modelling_framework}

The cluster sample is constructed based on the X-ray selection ($\lext>6$) with the aid of the optical confirmation to remove the contamination with unphysically low richness ($\rich\leq3$), together with the selection of the cluster redshift at $0.1<\redshift<0.8$ \citep[see Section~\ref{sec:the_sample} and Section~2.1 in][]{ghirardini24}.
The extent likelihood \lext\ is mainly subject to the parameters tuned to optimize in the cluster-finding algorithm and lacks a straightforward association with the physical properties of clusters.
Therefore, it is not an ideal mass proxy.
Instead, we use the observed X-ray count rate \rate\ as the mass proxy.
Meanwhile, the intrinsic X-ray count rate \ratehat\ acts as an X-ray ``selection'' observable.
The observed \rate\ and intrinsic \ratehat\ are related through the measurement uncertainty, and both can serve as selection observables.

In short, there are three selection observables $\left(\ratehat, \rich, \redshift\right)$, where the uncertainty of the cluster redshift \redshift\ is negligible, and the richness selection mainly removes obvious contamination with an insignificant impact on the resulting sample \citep{ghirardini24}.
That is, the cluster selection is primarily determined in X-rays.

The follow-up observations of individual clusters are then conducted to obtain the so-called ``follow-up'' observables, which are the weak-lensing shear profile \gshear\ and the BCG stellar mass \mbcg\ in this study.
It is extremely important to note that the follow-up observables are uniformly obtained for every cluster and that the lack of these follow-up observables for some clusters does not rely on any preferences for X-ray properties.
In other words, no additional selections are made on the weak-lensing shear profile \gshear\ and the BCG stellar mass \mbcg.

We denote the selection and follow-up observables by the symbols of $\mathbfcal{S}$ and $\mathbfcal{O}$, where 
$\mathbfcal{S} \equiv \left\lbrace \rate,\rich,\redshift \right\rbrace$ and 
$\mathbfcal{O} \equiv \left\lbrace \gshear, \mbcg \right\rbrace$, respectively.
For the scaling relations parameterized by a parameter vector $\mathbf{p}$, we maximize the probability of observing the follow-up observable $\mathbfcal{O}$ for a cluster selected with the selection observable $\mathbfcal{S}$ as
\begin{equation}
\label{eq:mass_calibration}
P\left(\mathbfcal{O} | \mathbfcal{S}, \mathbf{p} \right) = 
\frac{
N\left( \mathbfcal{O}, \mathbfcal{S} | \mathbf{p} \right)
}{
N\left( \mathbfcal{S} | \mathbf{p} \right)
}
\, ,
\end{equation}
where the label $N$ represents the number of clusters with the observables of $\mathbfcal{S}$ and/or $\mathbfcal{O}$ evaluated at the parameter vector $\mathbf{p}$.

Equation~(\ref{eq:mass_calibration}) is referred to as the ``mass calibration likelihood'' widely used in the population modelling of galaxy clusters \citep[e.g.,][]{chiu22,chiu23,ghirardini24}.
Consider the selection function as a function of only the observed quantity $\mathbfcal{S}$ rather than the intrinsic one $\hat{\mathbfcal{S}}$.
Because the selection of the clusters purely depends on $\mathbfcal{S}$ instead of $\mathbfcal{O}$, the selection-function factors of both the numerator and denominator in equation~(\ref{eq:mass_calibration}) are cancelled\footnote{To the first order, this effectively assumes that the correlated scatter between $\mathbfcal{S}$ and $\mathbfcal{O}$ is negligible. This is a valid assumption, as the primary selection observable is the X-ray count rate \rate, which is not expected to strongly correlate with the optical properties (e.g., the BCG stellar mass \mbcg\ or the weak-lensing shear profile \gshear) at a fixed halo mass.
} and have no effect \citep[see equation~(3) in][]{chiu23}.
That is, the selection function of the cluster sample is not important for the mass calibration likelihood, i.e., equation~(\ref{eq:mass_calibration}), as long as no additional selection is made on the follow-up observable $\mathbfcal{O}$, as in this study.

In this study and other \erass1 work, however, the selection function $\mathnfcal{I}\left(\ratehat | \redshift, \mathnfcal{H}\right)$ is calibrated at the sky location $\mathnfcal{H}$ and redshift \redshift\ of each cluster in terms of the ``intrinsic'' observable, namely the intrinsic count rate \ratehat\ \citep{clerc24}.
Therefore, the selection function is implicitly used in calculating both the numerator and denominator of equation~(\ref{eq:mass_calibration}) and cannot be canceled.
Note that the selection function $\mathnfcal{I}\left(\ratehat | \redshift, \mathnfcal{H}\right)$ does not depend on the parameter vector $\mathbf{p}$, which is distinct from 
the self-calibrated selection function used in \cite{chiu23}.

Because (1) we ignore the uncertainty of the cluster redshift and (2) the richness selection of $\rich>3$ has no impact on the resulting X-ray selected sample \citep{ghirardini24}, equation~(\ref{eq:mass_calibration})
can be approximated with $\mathbfcal{S} = \left\lbrace \rate, \redshift \right\rbrace$ as \citep[e.g., equation~(56) in][]{chiu22}
\begin{multline}
\label{eq:mass_calibration_approximate}
P\left(\mathbfcal{O} | \mathbfcal{S}, \mathbf{p} \right) \approx P\left( \mbcg, \gshear | \rate, \redshift, \mathbf{p} \right) \\
= \frac{
\int
 P\left(\mbcg, \gshear, \rate | \mass, \redshift, \mathbf{p}\right) 
n\left(\mass | \redshift, \mathbf{p}\right)  \dif\mass
}{
\int
P\left(\rate | \mass, \redshift, \mathbf{p}\right) 
n\left(\mass | \redshift, \mathbf{p}\right)  \dif\mass
}
\, ,
\end{multline}
where $n\left(\mass | \redshift, \mathbf{p}\right)$ is the halo mass function evaluated at the cluster redshift $\redshift$ and the parameter vector $\mathbf{p}$.

In equation~(\ref{eq:mass_calibration_approximate}), the factor $P\left(\rate | \mass, \redshift, \mathbf{p}\right)$ describes the probability of observing the X-ray count rate \rate\ given the halo mass \mass\ at the redshift \redshift\ evaluated with the parameter vector $\mathbf{p}$.
It includes two components making the observed count rate \rate\ scatter around the mean value predicted at the fixed cluster mass and redshift.
The first is the intrinsic scatter in the count rate, and the second is the measurement uncertainty due to the observational noise.
Additionally, we include the selection function in terms of the intrinsic count rate \ratehat\ at the redshift $\redshift$ and the sky location $\mathnfcal{H}$, i.e., $\mathnfcal{I}_{\redshift}\left(\ratehat\right)\equiv \mathnfcal{I}\left(\ratehat | \redshift, \mathnfcal{H}\right)$.
Therefore, we further deduce the probability $P\left(\rate | \mass, \redshift, \mathbf{p}\right)$ as
\begin{equation}
\label{eq:mass_calibration_denominator}
P\left(\rate | \mass, \redshift, \mathbf{p}\right) = \int P\left(\rate | \ratehat  \right) 
\mathnfcal{I}_{\redshift}\left(\ratehat\right) 
P\left(\ratehat  | \mass, \redshift, \mathbf{p} \right) \dif \ratehat 
\, ,
\end{equation}
where $P\left(\ratehat | \mass, \redshift, \mathbf{p} \right)$ describes the intrinsic count rate \ratehat\ scattering around the mean value predicted at the fixed cluster mass \mass\ and redshift \redshift, and the $P\left(\rate | \ratehat  \right)$ accounts for the scatter due to the measurement uncertainty.
The count rate-to-mass-and-redshift (\rate--\mass--\redshift) relation is modelled in Section~\ref{sec:modelling_of_scaling_relation}.
We assume a log-normal and constant intrinsic scatter \sigmarate\ in the X-ray count rate, as no significant variation of \sigmarate\ as a function of \mass\ and \redshift\ is suggested \citep{chiu22,chiu23,ghirardini24}.

Similarly, the factor $P\left(\mbcg, \gshear, \rate | \mass, \redshift, \mathbf{p}\right)$ in the numerator of equation~(\ref{eq:mass_calibration_approximate}) describes the probability of observing the observables $\left\lbrace\mbcg, \gshear, \rate\right\rbrace$ at the fixed cluster mass \mass\ and redshift \redshift\ given the parameter vector $\mathbf{p}$, which reads
\begin{multline}
\label{eq:mass_calibration_numerator}
P\left(\mbcg, \gshear, \rate | \mass, \redshift, \mathbf{p}\right) = \\
\int \int \int \Big[ 
P\left(\mbcg | \mbcghat \right) 
P\left(\gshear | \mwl, \redshift, \mathbf{p} \right) 
P\left(\rate | \ratehat \right) \times
\Big. \\
\Big.
\mathnfcal{I}_{\redshift}\left(\ratehat\right) 
P\left(\mbcghat, \mwl, \ratehat | \mass, \redshift, \mathbf{p} \right) 
\Big]
\dif \mbcghat~
\dif \mwl~
\dif \ratehat~
\, .
\end{multline}
In equation~(\ref{eq:mass_calibration_numerator}), the probability $P\left(\mbcghat, \mwl, \ratehat | \mass, \redshift, \mathbf{p} \right)$ describes the joint intrinsic distribution of the observables $\left\lbrace\mbcghat, \mwl, \ratehat\right\rbrace$ at the fixed cluster mass \mass\ and redshift \redshift, the factor $P\left(\mbcg | \mbcghat \right)$ takes the measurement uncertainty of the BCG stellar mass into account, and $P\left(\gshear | \mwl, \redshift, \mathbf{p} \right)$ accounts for the measurement uncertainty of the shear profile, expressed by the weak-lensing covariance matrix (Section~\ref{sec:weaklensing_measurements}).
The selection function $\mathnfcal{I}_{\redshift}\left(\ratehat\right)$ is also included to calculate equation~(\ref{eq:mass_calibration_numerator}).

The correlations between the intrinsic scatters of the observable pairs of the X-ray count rate and the BCG stellar mass ($\rho_{\mathrm{X},\mathrm{BCG}}$), the X-ray count rate and the weak-lensing mass ($\rho_{\mathrm{X},\mathrm{WL}}$), and the BCG stellar mass and the weak-lensing mass ($\rho_{\mathrm{BCG},\mathrm{WL}}$) are included in calculating the joint intrinsic distribution, 
$P\left(\mbcghat, \mwl, \ratehat | \mass, \redshift, \mathbf{p} \right) $.

We provide the following remarks for equation~(\ref{eq:mass_calibration_numerator}).
First, $P\left(\gshear | \mwl, \redshift, \mathbf{p} \right)$ models the observed shear profile $\gshear\left(\theta\right)$ by a mass profile given the weak-lensing mass \mwl\ and the redshift \redshift.
This requires a redshift-distance relation, which has dependence on cosmological parameters, to compare the mass profile in the physical scale with the observed shear profile in the angular scale.
Moreover, the mass model includes nuisance parameters to account for, e.g., the miscentering effect.
We therefore keep the parameter vector $\mathbf{p}$ in $P\left(\gshear | \mwl, \redshift, \mathbf{p} \right)$ to account for the measurement uncertainty of the shear profile.
Second, the weak-lensing mass \mwl, the mass estimate that best describes the observed shear profile, is not equal to the cluster halo mass \mass\ that is used to parameterize the halo mass function.
This is because the halo model 
is not a perfect description of observed galaxy clusters, which have e.g., substructures or triaxiality that are not included in the model \citep[see Section~4.2 in][]{chiu22}.
This leads to a biased weak-lensing mass \mwl\ with respect to the cluster halo mass \mass.
Such weak-lensing mass bias is accounted for by the weak-lensing mass-to-halo mass relation given in Section~\ref{sec:wl_modelling}.
Third, the measurement uncertainty of the BCG stellar mass is modelled as a Gaussian noise in terms of the ten-based logarithmic BCG stellar mass, i.e., $\log\left(\mbcg\right)$.

For the clusters with only the observed shear profile \gshear, the modelling of the BCG stellar mass is omitted in equation~(\ref{eq:mass_calibration_numerator}), finally leading to
\begin{multline}
\label{eq:mass_calibration_approximate_wlonly}
P\left(\mathbfcal{O} | \mathbfcal{S}, \mathbf{p} \right) \approx P\left( \gshear | \rate, \redshift, \mathbf{p} \right) \\
= \frac{
\int
P\left(\gshear, \rate | \mass, \redshift, \mathbf{p}\right) 
n\left(\mass | \redshift, \mathbf{p}\right)  \dif\mass
}{
\int
P\left(\rate | \mass, \redshift, \mathbf{p}\right) 
n\left(\mass | \redshift, \mathbf{p}\right)  \dif\mass
}
\, .
\end{multline}
For those clusters with only the BCG stellar mass estimates, we have
\begin{multline}
\label{eq:mass_calibration_approximate_bcgonly}
P\left(\mathbfcal{O} | \mathbfcal{S}, \mathbf{p} \right) \approx P\left( \mbcg | \rate, \redshift, \mathbf{p} \right) \\
= \frac{
\int
P\left(\mbcg, \rate | \mass, \redshift, \mathbf{p}\right) 
n\left(\mass | \redshift, \mathbf{p}\right)  \dif\mass
}{
\int
P\left(\rate | \mass, \redshift, \mathbf{p}\right) 
n\left(\mass | \redshift, \mathbf{p}\right)  \dif\mass
}
\, .
\end{multline}

\subsection{Modelling of the scaling relations}
\label{sec:modelling_of_scaling_relation}

In this work, we use three scaling relations to model the cluster observables as functions of the halo mass and redshift.
They are the count rate-to-mass-and-redshift (\rate--\mass--\redshift) relation, the BCG stellar-mass-to-mass-and-redshift (\mbcg--\mass--\redshift) relation, and the weak-lensing mass bias-to-mass-and-redshift (\bwl--\mass--\redshift\ or \mwl--\mass--\redshift) relation.
These scaling relations are constrained at the pivotal mass \mpiv\ and redshift \zpiv, where we use $\mpiv = 1.4\times10^{14}\Msunh$ and $\zpiv = 0.35$ throughout this work.
The correlated intrinsic scatters among these observables are also included in the modelling (see Section~\ref{sec:modelling_framework}).

\subsubsection*{The count rate-to-mass-and-redshift relation}
\label{sec:rate_relation}

For the \rate--\mass--\redshift\ relation, we use the same functional form as in the eRASS1 cosmological analysis \citep{ghirardini24}, namely
\begin{multline}
\label{eq:count_rate_relation}
\left\langle\ln\left(\frac{\ratehat}{\mathrm{counts}/\mathrm{sec}} \bigg| \mass,\redshift \right)\right\rangle = 
\ln\left(\Arate\right) + \\
\left(\Brate + \deltarate \ln\left(\frac{1 + \redshift}{1 + \zpiv}\right) \right)\ln\left(\frac{\mass}{\mpiv}\right) +
 2~\ln\left(\frac{E\left(\redshift\right)}{E\left(\zpiv\right)}\right) \\
-2~\ln\left(\frac{D_{\mathrm{L}}\left(\redshift\right)}{D_{\mathrm{L}}\left(\zpiv\right)}\right) +
\gammarate~\ln\left(\frac{1 + \redshift}{1 + \zpiv}\right)
\, ,
\end{multline}
where \Arate\ is the normalization with a unit of $\mathrm{counts}/\mathrm{sec}$ at the pivotal mass \mpiv\ and redshift \zpiv, \Brate\ is the power-law index of the mass trend with a redshift-dependent slope parameterized by \deltarate, \gammarate\ describes the power-law scaling of the redshift trend, $D_{\mathrm{L}}$ denotes the luminosity distance, and $E\left(\redshift\right)$ accounts for the self-similar redshift evolution of the X-ray luminosity \citep{kaiser86,boehringer12}.
We adopt a log-normal and constant intrinsic scatter, $\sigmarate\equiv\sqrt{\mathrm{Var}\left(\ln\ratehat | \mass, \redshift\right)}$, around the mean value predicted by equation~(\ref{eq:count_rate_relation}) at a fixed mass \mass\ and redshift \redshift.

\subsubsection*{The BCG stellar-mass-to-mass-and-redshift relation}
\label{sec:bcg_relation}

For the \mbcg--\mass--\redshift\ relation, or the stellar-mass-to-halo-mass relation, we use the power-law functional form as
\begin{multline}
\label{eq:bcg_relation}
\left\langle\ln\left(\frac{\mbcghat}{\Msun} \bigg| \mass,\redshift \right)\right\rangle = \\
\Abcg~\ln\left(10\right) +
\left( \Bbcg + \deltabcg~\left( \frac{ \frac{\redshift}{\zpiv} }{ \frac{1 + \redshift}{1 + \zpiv} } - 1\right) \right)  \ln\left(\frac{\mass}{\mpiv}\right) + 
\\
\gammabcg~\ln\left(\frac{1 + \redshift}{1 + \zpiv}\right)
\, ,
\end{multline}
where \Abcg\ is the normalization in a unit of dex, \Bbcg\ is the power-law index of the mass trend with a redshift-dependent slope parameterized by \deltabcg, and \gammabcg\ describes the power-law slope of the redshift scaling.
We assume a log-normal and constant intrinsic scatter \sigmabcg\ around the mean value predicted by equation~(\ref{eq:bcg_relation}).
Despite \Abcg\ in the unit of dex, we express the intrinsic scatter in terms of the natural log, i.e., 
$\sigmabcg = \sqrt{\mathrm{Var} \left(\ln\mbcghat | \mass,\redshift\right) }$,
in the interest of consistency with the other scaling relations.
Note that the parameterization of the redshift-dependent mass slope in equation~(\ref{eq:bcg_relation}), i.e., $\left( \left( \frac{\redshift}{\zpiv} \right) / \left( \frac{1 + \redshift}{1 + \zpiv} \right) - 1\right)$, 
is a result of following that used in \citet[][see their equation~14]{moster13}.
We have tried a different functional form, $\ln\left( 1 + \redshift\right)/\left(1 + \zpiv\right)$, 
for the redshift-dependent mass slope and find no significant impact on the final results and our conclusions.

\subsubsection*{The weak-lensing mass bias-to-mass-and-redshift}
\label{sec:wl_relation}

For the weak-lensing mass bias $\bwl\equiv \frac{\mwl}{\mass}$, we follow the identical functional form developed in \cite{grandis24} and also applied to the cosmological analysis in \cite{ghirardini24}, as
\begin{multline}
\label{eq:wlmass_relation}
\left\langle\ln\left( \bwl \big| \mass,\redshift \right)\right\rangle = 
\ln b_{\mathrm{z}}\left(\redshift \big| \deltawl, \gammawl \right) + \Bwl \ln\left(\frac{\mass}{\mpiv}\right)
\, ,
\end{multline}
where \Bwl\ describes the mass trend of the weak-lensing mass bias, and $\ln b_{\mathrm{z}}\left(\redshift \big| \deltawl, \gammawl \right)$ is the redshift-dependent normalization of the mass bias 
parameterized by two parameters, \deltarate\ and \gammarate, as
\begin{multline}
\label{eq:wlmass_relation_interpolation}
\ln b_{\mathrm{z}}\left(\redshift \big| \deltawl, \gammawl \right) = \\
         \mathfrak{I}\left(\redshift \big| \mathbf{z_\mathrm{sim}}, \mathbf{\mu_0}\right) +
\deltawl~\mathfrak{I}\left(\redshift \big| \mathbf{z_\mathrm{sim}}, \mathbf{\mu_1}\right) +
\gammawl~\mathfrak{I}\left(\redshift \big| \mathbf{z_\mathrm{sim}}, \mathbf{\mu_2}\right)
\, ,
\end{multline}
in which the notation $\mathfrak{I}\left(x_0 \big| \mathbf{x}, \mathbf{y}\right)$ stands for the linear interpolation at $x_0$ among the function that has the values $\mathbf{y}$ at $\mathbf{x}$. 
In \cite{grandis24}, they calibrated the weak-lensing mass bias using simulated clusters extracted at four snapshots of redshift, which we label them by $\mathbf{z_\mathrm{sim}}$ in equation~(\ref{eq:wlmass_relation_interpolation}), and derived the mass bias $\mathbf{\mu_0}$ with the two principle components $\mathbf{\mu_1}$ and $\mathbf{\mu_2}$ characterizing the uncertainty.
In this way, the distribution of the weak-lensing mass bias as a function of the cluster redshift can be evaluated by directly interpolating among the redshift snapshots $\mathbf{z_\mathrm{sim}}$ with the parameters of \deltawl\ and \gammawl\ modelled by Gaussian distributions of $\mathcal{N}\left(0, 1^2\right)$.
While the mean weak-lensing mass bias at the halo mass \mass\ and redshift \redshift\ is calculated with equations~(\ref{eq:wlmass_relation}) and~(\ref{eq:wlmass_relation_interpolation}), we model the intrinsic scatter of the weak-lensing mass around the mean value by a log-normal and constant scatter as $\sigmawl\equiv\sqrt{\mathrm{Var}\left(\bwl | \mass,\redshift\right)}$.
This modelling is different from those in \cite{grandis24} and \cite{kleinebreil24}, where the intrinsic scatter of the weak-lensing mass was modelled as a function of the halo mass and redshift.
This modification is consistent with the modelling of the cluster miscentering, for which we also adopt a fixed model globally for all \erass1 clusters (see Section~\ref{sec:wl_modelling} for details).
As seen in Section~\ref{sec:rate_relation_results}, the results of our modelling with a statistical approach fully agree with those of \cite{kleinebreil24}, which used a method dependent on individual clusters.

\subsection{Weak-lensing modelling}
\label{sec:wl_modelling}

In this subsection, we describe (1) the modelling of the shear profile to obtain the weak-lensing mass \mwl, and (2) the calibration of the weak-lensing mass bias \bwl. 
For the former, we closely follow the fitting procedure detailed in \cite{chiu22} with minor modifications to be consistent with that adopted in \cite{ghirardini24}.
Meanwhile, the latter task has been 
quantified in \citet[][see also \citealt{grandis21}]{grandis24} and applied to the weak-lensing mass calibration of eRASS1 clusters \citep[e.g.,][]{kleinebreil24} as well as the cosmological analyses \citep{ghirardini24}.
We briefly summarize them, as follows.

Given a cluster at redshift \zcl\ and a background population with a stacked (and lensing-weight-weighted) redshift distribution $P\left(\redshift\right)$,
the model \gshearmod\ of the tangential reduced shear profile as a function of the projected (physical) radius \rdd\ is calculated with a parameter vector $\mathbf{p}$\footnote{For the weak-lensing mass modelling, the relevant parameters in the vector $\mathbf{p}$ are (1) the miscentering parameters and (2) the cosmological parameters, which are varied with informative priors and used here to infer the angular diameter distance.} as
\begin{multline}
\label{eq:reduced_shear_model}
\gshearmod\left(\rdd | \mathbf{p}\right) = 
\frac{\rshearmod\left(\rdd | \mathbf{p}\right)}{1 - \convergencemod\left(\rdd | \mathbf{p}\right)}
\times \\
\left(1 + \convergencemod\left(\rdd | \mathbf{p}\right) \left(
\frac{\left\langle \beta^2\left(\mathbf{p}\right) \right\rangle}{\left\langle\beta\left(\mathbf{p}\right)\right\rangle^2} - 1\right)
\right)
\, ,
\end{multline}
in which \lensingeff\ is the lensing efficiency depending on the angular diameter distance\footnote{
The notation of $D_{\mathrm{A}}\left(\redshift\right)$ is the angular diameter distance to the redshift \redshift, while
$D_{\mathrm{A}}\left(z_1, z_2\right)$ represents the angular diameter distance between the redshift pairs of $z_1$ and $z_2$ with $z_2 > z_1$.
},
\begin{equation}
\label{eq:lensing_eff_definition}
\lensingeff\left(\mathbf{p}\right) = 
\begin{cases}
\frac{D_{\mathrm{A}}\left(\zcl,\redshift | \mathbf{p} \right)}{ D_{\mathrm{A}}\left(\redshift | \mathbf{p}\right) } & \text{if \redshift~>~\zcl }  \\
0 &\text{else}
\end{cases}
\, ,
\end{equation}
and \rshearmod\ and \convergencemod\ are the models of the tangential shear and convergence, respectively, which read
\begin{eqnarray}
\label{eq:rshearmodel}
\rshearmod\left(\rdd | \mathbf{p}\right) &= &
\frac{
\deltaSigmamod\left(\rdd | \mathbf{p}\right)
}{
\sigmacrit\left(\zcl | \mathbf{p}\right)
}  \, , \\
\label{eq:kappamodel}
\convergencemod\left(\rdd | \mathbf{p}\right) &= &
\frac{
\Sigmammod\left(\rdd | \mathbf{p}\right)
}{
\sigmacrit\left(\zcl | \mathbf{p}\right)
}  \, , 
\end{eqnarray}
where $\Sigmammod\left(\rdd | \mathbf{p}\right)$ is the model of the surface mass density at the projected radius \rdd, 
\sigmacrit\ is the critical surface mass density,
\begin{equation}
\label{eq:sigmacrit}
\sigmacrit\left(\zcl | \mathbf{p}\right) = \frac{c^2}{4\pi G} \frac{1}{
D_{\mathrm{A}}\left(\zcl | \mathbf{p}\right) \left\langle \beta\left(\mathbf{p}\right)  \right\rangle
}
\, ,
\end{equation}
and 
$\deltaSigmamod\left(\rdd | \mathbf{p}\right)$ is the 
excess
surface mass density,
\begin{equation}
\label{eq:delta_sigma_model}
\deltaSigmamod\left(\rdd | \mathbf{p} \right) \equiv 
\left[ \frac{2}{\rdd^2}\int_0^{\rdd}\Sigmammod\left(x | \mathbf{p} \right)x\dif x \right] ~
-
\Sigmammod\left(\rdd | \mathbf{p} \right)
\, .
\end{equation}
In equation~(\ref{eq:reduced_shear_model}), the quantities of 
$\left\langle \lensingeff \left(\mathbf{p}\right)\right\rangle$ and 
$\left\langle \lensingeff^2 \left(\mathbf{p}\right) \right\rangle$ 
are obtained with the weights of the background redshift distribution $P\left(\redshift\right)$, 
\begin{eqnarray}
\label{eq:lensing_efficiency_1_and_2}
\left\langle \lensingeff\left(\mathbf{p}\right)    \right\rangle & = & \int \lensingeff\left(\mathbf{p}\right)   P\left(\redshift\right) \dif \redshift \\
\left\langle \lensingeff^2 \left(\mathbf{p}\right) \right\rangle & = & \int \lensingeff^2\left(\mathbf{p}\right) P\left(\redshift\right) \dif \redshift 
\, .
\end{eqnarray}
The projected and physical radius \rdd\ is determined as $\rdd = D_{\mathrm{A}}\left(\zcl | \mathbf{p}\right) \times \theta$, where the $\theta$ is the clustercentric radius in a unit of radian.
The comparison between the observed shear profile and the model is in the space of the angular radius $\theta$, therefore the redshift-distance relation must be used to correctly calculate the physical radius at a fixed $\theta$ and a given parameter vector $\mathbf{p}$.

The model of the projected surface mass density \Sigmammod\ is composed of two components, the perfectly centered profile \Sigmamcen\ and that with the miscentering \Sigmammis.
They are weighted by the fraction $f_{\mathrm{mis}}$ of the miscentered clusters, in a statistical approach, as
\begin{equation}
\label{eq:sigmamod_model_with_miscentering}
\Sigmammod\left(\rdd | \mathbf{p}\right) = 
\left(1 - f_{\mathrm{mis}}\right) \times \Sigmamcen\left(\rdd | \mathbf{p}\right)  + f_{\mathrm{mis}} \times \Sigmammis\left(\rdd | \mathbf{p}\right) 
\, ,
\end{equation}
where \Sigmamcen\ is modelled by a spherical Navarro–Frenk–White \citep[hereafter NFW;][]{navarro97} profile, and 
\Sigmammis\ is modelled by a miscentered NFW model \citep[see e.g.,][]{johnston07a} with a characteristic and dimensionless miscentering scale $\sigma_{\mathrm{mis}}$ \citep[see equation~20 in][]{chiu22}, which is a varied parameter in $\mathbf{p}$.

There are minor modifications from the previous study, \cite{chiu22}.
First, a different halo concentration-to-mass ($c$--\mass) relation is used when calculating the NFW profile.
Here, we use the mean value predicted by the $c$--\mass\ relation from \cite{ragagnin21} at a given weak-lensing mass \mwl\ and the cluster redshift as the halo concentration, while they used the $c$--\mass\ relation from \cite{diemer18}. 
Second, we only include shape noises in the weak-lensing covariance matrix, as the measurement uncertainty of the observed shear profile.
The scatter (denoted as $\mathbb{C}_{\mathrm{LSS}}$) in the shear profile due to uncorrelated large-scale structures along the line of sight is accounted for when calibrating the weak-lensing mass bias.
To be exact, the scatter $\mathbb{C}_{\mathrm{LSS}}$ calculated from the formula in \cite{hoekstra03} is included in fitting the weak-lensing shear profile of simulated clusters when calibrating \bwl\ \citep[see Section~2.1.6 in][]{grandis21}. 
Consequently, the resulting scatter of \bwl\ naturally includes the effect of uncorrelated line-of-sight structures.
In \cite{chiu22}, the scatter $\mathbb{C}_{\mathrm{LSS}}$ was calculated and initially included in the weak-lensing covariance matrix as a component of the measurement uncertainty.
These two merely differ in the philosophy of the methodology, not the final results.
Third, we update the miscentering model to that determined for the eRASS1 sample.
Specifically, the miscentering of the eRASS1 clusters was 
calibrated
using the synthetic X-ray \erosita\ images of clusters in the \erosita\ digital-twin simulations 
\citep[][see also simulations produced in \citealt{comparat19,comparat20,seppi22}]{grandis24}, including both intrinsic and observational effects.
As a result, the miscentering of clusters was expressed by the probability $f_{\mathrm{mis}}$ of being miscentered and a characteristic miscentering scale $\sigma_{\mathrm{mis}}$ that depends on the extent $\mathtt{EXT}$ and the detection likelihood \ldet, two quantities returned by the \erass\ detection pipeline.

The purpose of the first two aforementioned changes is to use the identical $c$--\mass\ relation and the measurement uncertainty as in \cite{grandis24}, where the dedicated and updated calibration of the weak-lensing mass bias was carried out.
Meanwhile, for the third modification we implement the miscentering model in a statistical approach to have a much more speedy calculation.
That is, we first calculate the distribution of the characteristic miscentering scale $\sigma_{\mathrm{mis}}$ of our eRASS1 sample based on the miscentering calibration \citep[see Section 4.2 in][]{grandis24}, model it as a log-normal distribution, and apply the result as an informative Gaussian prior on $\ln\sigma_{\mathrm{mis}}$ in equation~(\ref{eq:sigmamod_model_with_miscentering}) for all the clusters consistently.
Similarly, we also do so for the parameter of the miscentering fraction $f_{\mathrm{mis}}$ for our \erass1 sample.
The resulting Gaussian priors on $f_{\mathrm{mis}}$ and $\ln\sigma_{\mathrm{mis}}$ are 
$\mathcal{N}\left(0.32, 0.043^2\right)$ and 
$\mathcal{N}\left(-0.85, 0.18^2\right)$, respectively.
This approach is equivalent to statistically correcting for the miscentering effect at a population level.
Because we also model the scatter of the weak-lensing mass bias in the same way (see the last paragraph), we do not expect a significant impact on the final results from this modification.
Indeed, our result is in excellent agreement with \cite{grandis24} and \cite{kleinebreil24}, as seen in Section~\ref{sec:rate_relation_results}.

As fully described in \citet[][see also \citealt{grandis21}]{grandis24}, the weak-lensing mass bias, $\bwl\equiv\mwl/\mass$, is calibrated by repeating the same modelling on the synthetic shear profiles of clusters extracted from the cosmological hydrodynamical TNG300 simulations \citep{pillepich18}, of which the underlying halo mass \mass\ used to parameterize the halo mass function is known.
The bias and intrinsic scatter of the weak-lensing mass are derived, specifically tailored to include the observational effects of 
(1) the misfitting due to the halo triaxiality or shape, 
(2) the miscentering of the \erosita\ X-ray center, 
(3) the cluster member contamination\footnote{The extracted weak-lensing shear profile \gshear\ shows no signature of the cluster member contamination \citep{chiu22}. We therefore set the cluster member contamination at a level of $<6\percent$ (2$\sigma$) at the cluster core of $\rdd\approx0.2\Mpch$ with a decaying behavior following a projected NFW profile.},
(4) the photo-\redshift\ bias of the background sources, 
(5) the calibration uncertainty of the galaxy shape measurement.
Importantly, the systematic uncertainties of the resulting weak-lensing mass bias and scatter include baryonic effects.
This is achieved by quantifying the systematic difference in the results of the \bwl\ calibrations between the same suite of the dark-matter-only and hydrodynamical simulations.
In the end, the constraints on the parameters of equations~(\ref{eq:wlmass_relation}) and~(\ref{eq:wlmass_relation_interpolation}) are obtained and applied as informative priors in the final modelling.
The resulting constraints are tabulated in Table~A.3. of \cite{kleinebreil24}.

In terms of the intrinsic scatter \sigmawl\ of the weak-lensing mass bias, \cite{grandis24} parameterized \sigmawl\ as a function of the cluster mass \mass\ and redshift \redshift.
In this work, we adopt a constant and average weak-lensing mass scatter \sigmawl\ without modelling its mass and redshift dependence.
This is a reasonable approach, given that we do not observe significant mass-dependent scatter and that the scatter \sigmawl\ is statistically constant out to redshift $\redshift\approx0.8$ \citep[see Table~A.3 in][]{kleinebreil24}.
In practice, we calculate the distribution of the weak-lensing bias scatter of our eRASS1 sample based on the resulting model reported in \cite{kleinebreil24}, and model it as a Gaussian distribution that we apply as an informative prior on \sigmawl\ in the final modelling.
For our sample, we find that \sigmawl\ can be well described by $\mathcal{N}\left(0.23,0.03^2\right)$.

\subsection{Statistical inference}
\label{sec:statistical_inference}

We adopt the Bayesian framework to explore the posterior $P\left(\mathbf{p} | \mathbfcal{D}\right)$ of the parameter vector $\mathbf{p}$ given the observed data vectors $\mathbfcal{D}$,
\begin{equation}
\label{eq:posteriors}
P\left(\mathbf{p} | \mathbfcal{D}\right) \propto 
\mathnfcal{L}\left(\mathbfcal{D} | \mathbf{p}\right)
\mathnfcal{P}\left(\mathbf{p}\right)
\, ,
\end{equation}
where $\mathnfcal{L}\left(\mathbfcal{D} | \mathbf{p}\right)$ is the likelihood of observing $\mathbfcal{D}$ at a given $\mathbf{p}$, and $\mathnfcal{P}\left(\mathbf{p}\right)$ is the prior on $\mathbf{p}$.
In this work, the data vector $\mathbfcal{D}$ consists of both the selection and follow-up observables $\mathbfcal{S}$ and $\mathbfcal{O}$ of individual clusters, i.e., $\mathbfcal{D} = \left\lbrace \mathbfcal{S}_{i}, \mathbfcal{O}_i | i \in \mathrm{eRASS1~clusters} \right\rbrace$.
We use the importance nested sampler \texttt{Multinest} \citep{feroz09} implemented in the \texttt{CosmoSIS} framework \citep{zuntz15} to explore the likelihood.

With the technique of the population modelling, i.e., jointly fitting individual clusters at the same time, we calculate the log-likelihood with equations~(\ref{eq:mass_calibration_approximate}),~(\ref{eq:mass_calibration_approximate_wlonly}), and~(\ref{eq:mass_calibration_approximate_bcgonly}) as
\begin{multline}
\label{eq:mass_calibration_total}
\ln\mathnfcal{L}\left(\mathbfcal{D} \big| \mathbf{p}\right) = \\
\sum_{i}\ln P\left( {\mbcg}_{i}, {\gshear}_{i} \big| {\rate}_{i}, {\zcl}_{i}, \mathbf{p}\right) +
\sum_{j}\ln P\left( {\gshear}_{j}              \big| {\rate}_{j}, {\zcl}_{j}, \mathbf{p}\right) + \\
\sum_{k}\ln P\left( {\mbcg}_{k}                \big| {\rate}_{k}, {\zcl}_{k}, \mathbf{p}\right)
\, ,
\end{multline}
where \zcl\ is the cluster redshift (and fixed to its photometric redshift), $i$ runs over the clusters with both the measurements of the weak-lensing shear profile \gshear\ and the BCG stellar mass \mbcg, and $j$ and $k$ run over those with only \gshear\ and \mbcg, respectively.
The parameter vector $\mathbf{p}$ includes the parameters\footnote{
The parameters \deltarate, \gammarate, \deltawl\ and \gammawl\ in this work correspond to the notations of $F_{\mathrm{X}}$, $G_{\mathrm{X}}$, $A_{\mathrm{WL}}$ and $B_{\mathrm{WL}}$ in \citet[][and \cite{grandis24}]{ghirardini24}, respectively.
} of 
\begin{itemize}
\item $\left\lbrace \Arate, \Brate, \deltarate, \gammarate, \sigmarate \right\rbrace$ of the \rate--\mass--\redshift\ relation in equation~(\ref{eq:count_rate_relation}),
\item $\left\lbrace \Abcg, \Bbcg, \deltabcg, \gammabcg, \sigmabcg \right\rbrace$ of the \mbcg--\mass--\redshift\ relation in equation~(\ref{eq:bcg_relation}), 
\item $\left\lbrace \Bwl, \deltawl, \gammawl, \sigmawl \right\rbrace$ of the \mwl--\mass--\redshift\ relation in equation~(\ref{eq:wlmass_relation}),
\item $\left\lbrace f_{\mathrm{mis}}, \ln\sigma_{\mathrm{mis}} \right\rbrace$ of the weak-lensing miscentering modelling in equation~(\ref{eq:sigmamod_model_with_miscentering}), and
\item $\left\lbrace\omegam, \h,  \sigmaeight, \omegab, \ns \right\rbrace$ of the cosmological parameters in a flat \lcdm\ model.
\end{itemize}

We adopt the following priors.
For the cosmological parameters, we apply flat priors, namely
$\mathcal{U}\left(0.10,0.99\right)$,
$\mathcal{U}\left(0.03,0.07\right)$,
$\mathcal{U}\left(0.50,0.90\right)$,
$\mathcal{U}\left(0.94,1.0\right)$, and
$\mathcal{U}\left(0.45,1.15\right)$
to \omegam, \omegab, \h, \ns, and \sigmaeight, respectively.
Additionally, Gaussian informative priors of 
$\mathcal{N}\left(0.3, 0.02^2\right)$, 
$\mathcal{N}\left(0.8, 0.02^2\right)$, and 
$\mathcal{N}\left(0.7, 0.05^2\right)$
are applied to \omegam, \sigmaeight, and \h, respectively.
With these informative priors, the cosmological parameters of the flat \lcdm\ model are effectively anchored to the widely accepted concordance values with the uncertainties of $\left(\omegam, \sigmaeight, \h\right)$ at levels of those obtained from the state-of-the-art constraints.

For the parameters of the weak-lensing mass bias, we apply a Gaussian prior $\mathcal{N}\left(0.047,0.021^2\right)$ on the mass-dependent power-law index \Bwl\ calibrated in \cite{grandis24} \citep[and tabulated in Table~A.3 of ][]{kleinebreil24}.
A unit Gaussian prior $\mathcal{N}\left(0,1^2\right)$ is applied on both \gammawl\ and \deltawl, by construction.
A Gaussian prior $\mathcal{N}\left(0.23, 0.03^2\right)$, informed by the eRASS1 sample we study in this work (see Section~\ref{sec:wl_modelling}), is applied on the intrinsic scatter \sigmawl\ of the weak-lensing mass bias.

We apply Gaussian priors of 
$\mathcal{N}\left(0.32, 0.043^2\right)$ and 
$\mathcal{N}\left(-0.85, 0.18^2\right)$ to the parameters of the weak-lensing miscentering model, $f_{\mathrm{mis}}$ and $\ln\sigma_{\mathrm{mis}}$, respectively (see Section~\ref{sec:wl_modelling}).

We require that the determinant of the correlated intrinsic scatter matrix to be positive when jointly fitting the observables \rate, \gshear, and \mbcg\ of a cluster.

For other parameters, only flat priors are used.
We summarize the adopted priors in Table~\ref{tab:priors}.
These priors are similar to those used in \cite{chiu22} with a major difference:
We only apply the flat prior to the intrinsic scatter \sigmarate\ of the X-ray count rate in this work, while \citet[][and the cosmological analysis in \citealt{chiu23}]{chiu22} additionally applied a Gaussian prior $\mathcal{N}\left(0.3, 0.08^2\right)$ to \sigmarate.

\begin{table}
\centering
\caption{
The priors used in the default modelling.
}
\label{tab:priors}
\resizebox{0.48\textwidth}{!}{
\begin{tabular}{cc}
\hline\hline
Parameters &Priors \\
\hline
\multicolumn{2}{c}{The X-ray count rate-to-mass-and-redshift (\rate--\mass--\redshift) relation} \\
\multicolumn{2}{c}{equation~(\ref{eq:count_rate_relation})} \\
\hline
\Arate\         &$\mathcal{U}(0.001,3.0)$   \\[3pt]
\Brate\         &$\mathcal{U}(0.0  ,5.0)$   \\[3pt]
\deltarate\     &$\mathcal{U}(-5.0,5.0)$    \\[3pt]
\gammarate\     &$\mathcal{U}(-5.0,5.0)$    \\[3pt]
\sigmarate\     &$\mathcal{U}(0.05,2.0)$    \\[3pt]
\hline
\multicolumn{2}{c}{The BCG stellar-mass-to-mass-and-redshift (\mbcg--\mass--\redshift) relation} \\
\multicolumn{2}{c}{equation~(\ref{eq:bcg_relation})} \\
\hline
\Abcg\         &$\mathcal{U}(8.0,13.0)$   \\[3pt]
\Bbcg\         &$\mathcal{U}(0.0 ,5.0)$   \\[3pt]
\deltabcg\     &$\mathcal{U}(-5.0,5.0)$   \\[3pt]
\gammabcg\     &$\mathcal{U}(-5.0,5.0)$   \\[3pt]
\sigmabcg\     &$\mathcal{U}(0.05,2.0)$   \\[3pt]
\hline
\multicolumn{2}{c}{The weak-lensing mass bias-to-mass-and-redshift (\bwl--\mass--\redshift) relation} \\
\multicolumn{2}{c}{equation~(\ref{eq:wlmass_relation})} \\
\hline
\Bwl\         &$\mathcal{N}\left(0.047,0.021^2\right)$~and~$\mathcal{U}(-1.0,1.0)$ \\[3pt]
\deltawl\     &$\mathcal{N}\left(0,1^2\right)$~and~$\mathcal{U}(-5.0,5.0)$         \\[3pt]
\gammawl\     &$\mathcal{N}\left(0,1^2\right)$~and~$\mathcal{U}(-5.0,5.0)$         \\[3pt]
\sigmawl\     &$\mathcal{N}\left(0.23,0.03^2\right)$~and~$\mathcal{U}(0.05,2.0)$   \\[3pt]
\hline
\multicolumn{2}{c}{Miscentering} \\
\hline
$f_{\mathrm{mis}}$              &$\mathcal{N}(0.32,0.043^2)$~and~$\mathcal{U}\left(0.0,1.0\right)$    \\[3pt]
$\ln \sigma_{\mathrm{mis}}$     &$\mathcal{N}(-0.85,0.18^2)$~and~$\mathcal{U}\left(-2.0,1.0\right)$   \\[3pt]
\hline
\multicolumn{2}{c}{Correlated intrinsic scatter} \\
\hline
$\rho_{\mathrm{X},\mathrm{WL}}$        &$\mathcal{U}(-0.9,0.9)$  \\[3pt]
$\rho_{\mathrm{X},\mathrm{BCG}}$       &$\mathcal{U}(-0.9,0.9)$  \\[3pt]
$\rho_{\mathrm{BCG},\mathrm{WL}}$      &$\mathcal{U}(-0.9,0.9)$  \\[3pt]
\hline
\multicolumn{2}{c}{Flat \lcdm\ Cosmology} \\
\hline
\omegam      &$\mathcal{N}\left(0.30,0.02^2\right)$~and~$\mathcal{U}(0.1,0.99)$            \\[3pt]
\sigmaeight  &$\mathcal{N}\left(0.80,0.02^2\right)$~and~$\mathcal{U}(0.45,1.15)$           \\[3pt]
\hnow        &$\mathcal{N}\left(0.7,0.05^2\right)$~and~$\mathcal{U}(0.5,0.9)$              \\[3pt]
\omegab      &$\mathcal{U}(0.03,0.07)$          										   \\[3pt]
\ns          &$\mathcal{U}(0.94,1.0)$            										   \\[3pt]
\hline
\end{tabular}
}
\tablefoot{
The first column contains the names of the parameters, while the second columns present the priors.
For a parameter with both flat and Gaussian priors, we practically allow it to vary following the Gaussian prior within a range allowed by the flat prior.
}
\end{table}
%

%
%

%
\begin{table*}
\centering
\caption{
The parameter constraints of the \rate--\mass--\redshift\ and \mbcg--\mass--\redshift\ relations, as in equation~(\ref{eq:count_rate_relation}) and equation~(\ref{eq:bcg_relation}), respectively.
}
\label{tab:parameter_constraints}
\resizebox{\textwidth}{!}{
\begin{tabular}{ccccccccccc}
Modelling
& \Arate & \Brate & \deltarate & \gammarate & \sigmarate
& \Abcg  & \Bbcg  & \deltabcg  & \gammabcg  & \sigmabcg \\[3pt]
\hline
 WL only & $  0.093^{+0.029}_{-0.036} $ & $  1.50^{+0.20}_{-0.30} $ & $  0.7\pm 1.5 $ & $  -0.8^{+1.5}_{-1.1} $ & $  0.56^{+0.17}_{-0.23} $ & $ \cdots $ & $ \cdots $ & $ \cdots $ & $ \cdots $ & $ \cdots $ \\[3pt]
 WL only + $M_{\star,\mathrm{BCG}}$ & $  0.093^{+0.029}_{-0.035} $ & $  1.53^{+0.22}_{-0.30} $ & $  0.8\pm 1.5 $ & $  -0.9^{+1.5}_{-1.1} $ & $  0.56^{+0.16}_{-0.22} $ & $  11.547^{+0.083}_{-0.074} $ & $  0.184^{+0.068}_{-0.16} $ & $  0.22\pm 0.29 $ & $  1.49^{+0.88}_{-0.75} $ & $  0.554^{+0.040}_{-0.049} $ \\[3pt]
 WL only + $M_{\star,\mathrm{BCG}}$ (priors on $\gamma_{\mathrm{BCG}}$) & $  0.084^{+0.025}_{-0.032} $ & $  1.66^{+0.25}_{-0.29} $ & $  1.4^{+1.7}_{-1.5} $ & $  -1.5^{+1.5}_{-1.2} $ & $  0.54^{+0.15}_{-0.20} $ & $  11.433^{+0.066}_{-0.055} $ & $  0.38\pm 0.11 $ & $  0.62\pm 0.25 $ & $  -0.36\pm 0.13 $ & $  0.560^{+0.044}_{-0.055} $ \\[3pt]
\hline
\end{tabular}
}
\end{table*}
\begin{figure}
\centering
\resizebox{0.5\textwidth}{!}{
\includegraphics[scale=1]{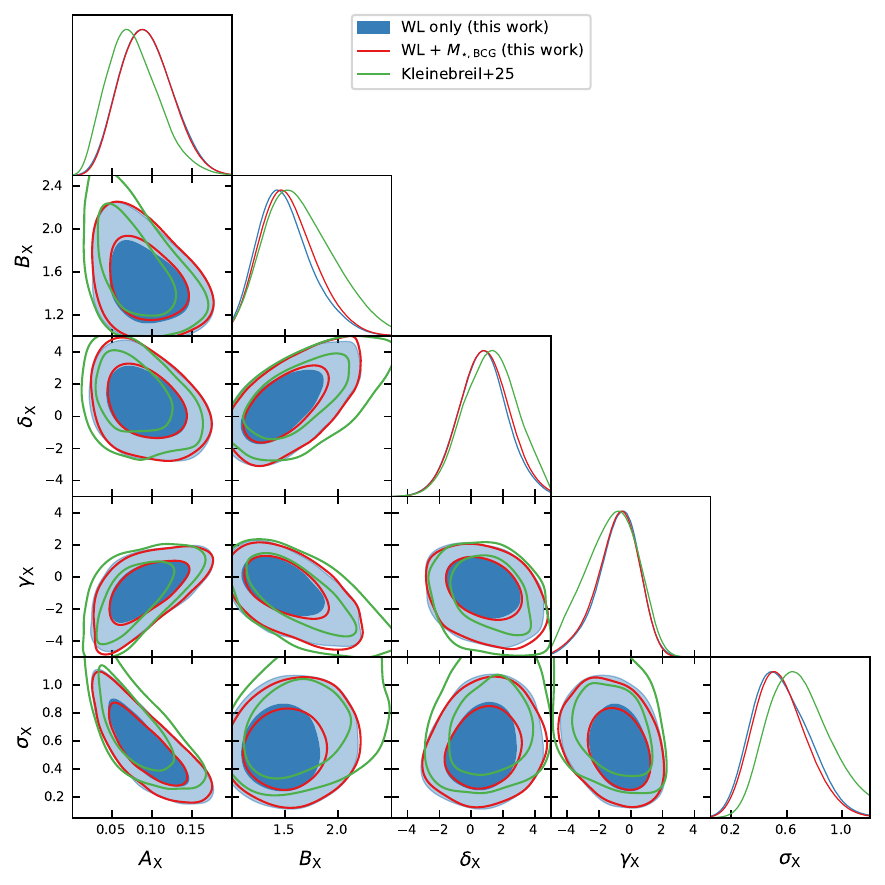}
}
\caption{
The parameter marginalized posteriors (diagonal) and the two-dimensional joint posteriors (off-diagonal) from different modelling are presented.
The results of the weak-lensing mass calibration alone are shown as the blue contours, while the inclusions of the modelling of the \mbcg--\mass--\redshift\ relation are indicated by the red contours.
We additionally indicate the results from \cite{kleinebreil24}, based on the identical cluster sample and the weak-lensing data \gshear, as the green contours. 
}
\label{fig:gtc_join}
\end{figure}
\begin{figure*}
\centering
\resizebox{\textwidth}{!}{
\includegraphics[scale=1]{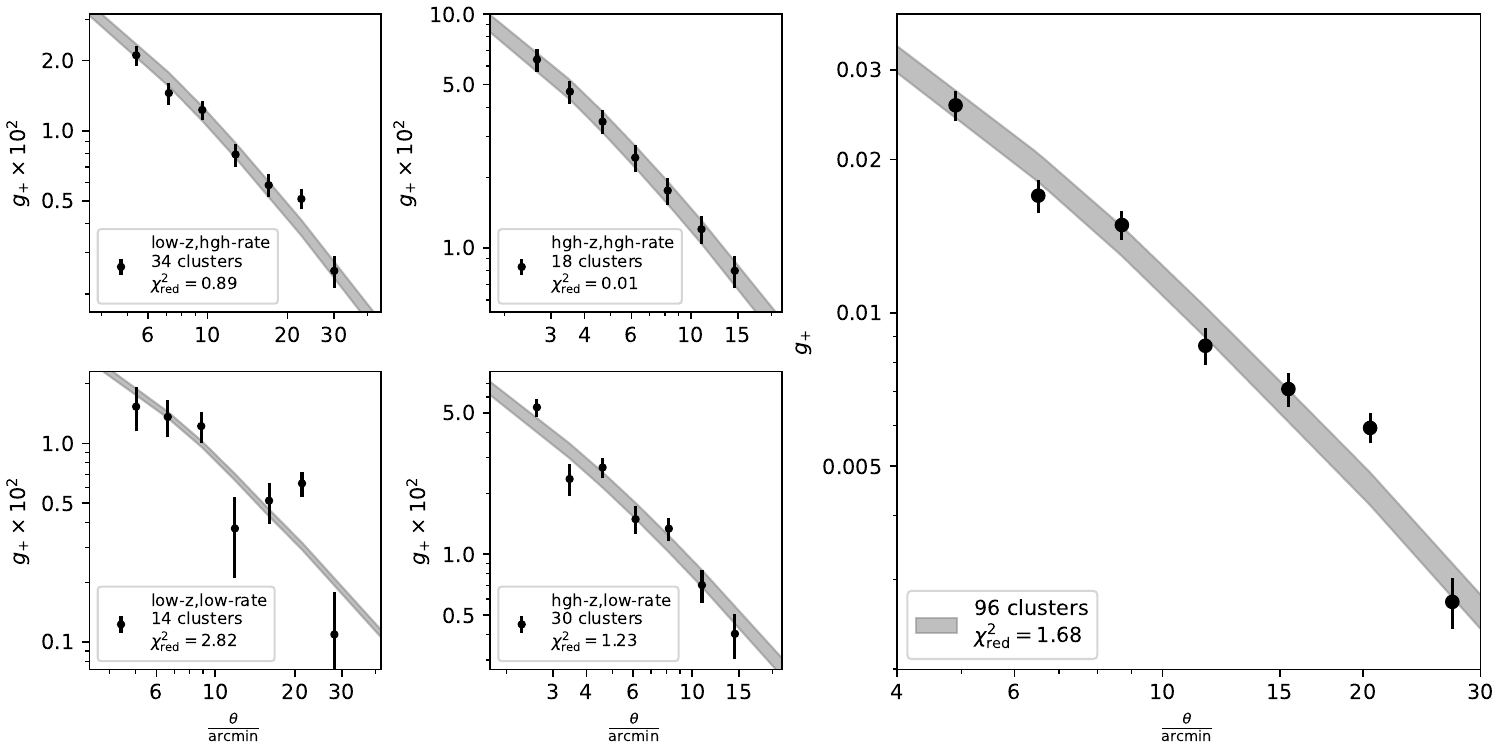}
}
\caption{
The observed shear profiles \gshear\ (black data points) and the best-fit models with fully marginalized uncertainties (grey shaded regions) are shown.
The four subplots in the left panel present the stacked profiles of the sub-samples defined with respect to the median values of the observed count rate ($\rate = 0.487$) and the redshift ($\redshift = 0.27$).
The stacked profile of all $96$ \erass1 clusters is contained in the right panel. 
The numbers of the clusters in the (sub-)samples and the reduced $\chi^2_{\mathrm{red}}$ are shown in the lower-left corners.
In both left and right panels, the radial values (the $x$ axis) of the measurements are obtained with the inverse-variance weights of the clusters in the (sub-)samples.
}
\label{fig:shear_profiles}
\end{figure*}
\begin{figure*}
\centering
\resizebox{\textwidth}{!}{
\includegraphics[scale=1]{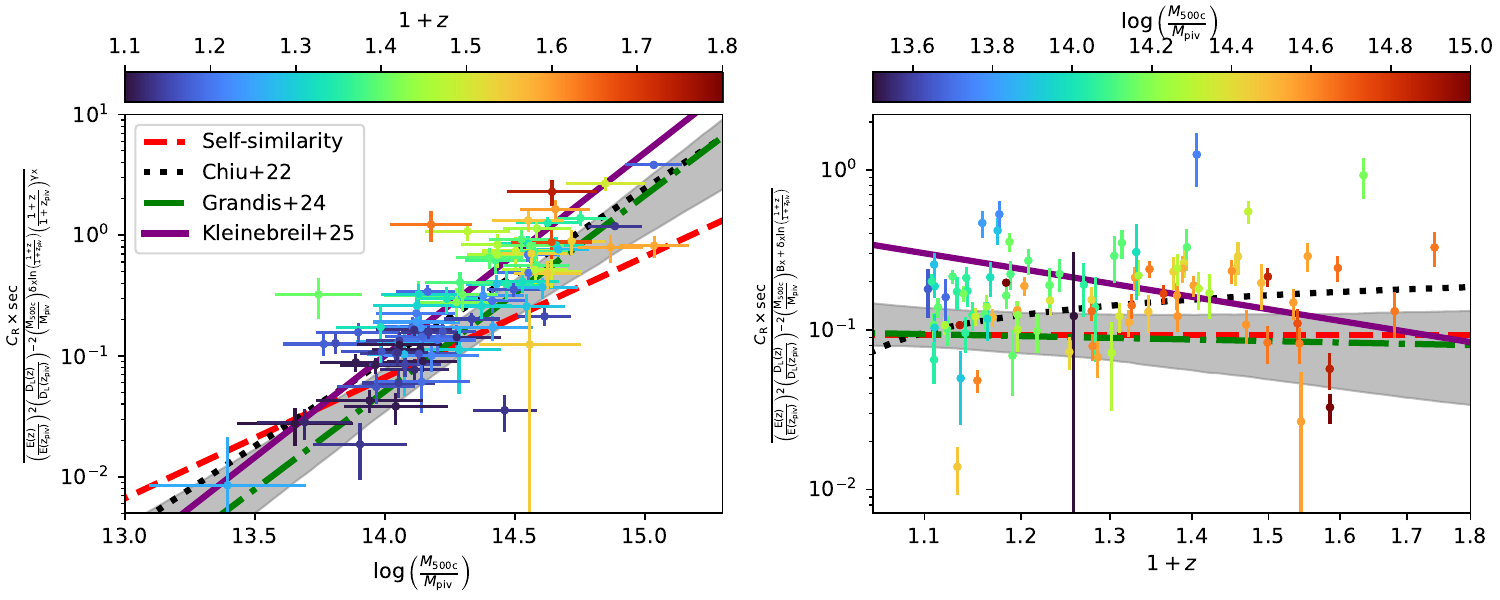}
}
\caption{
The \rate--\mass--\redshift\ relation of the \erass1 clusters and the best-fit model in equation~(\ref{eq:count_rate_relation_wlonly_result}) with its fully marginalized uncertainties (grey shaded regions).
The left panel shows the mass trend normalized at the pivotal redshift $\zpiv = 0.35$ after accounting for the redshift-dependent factors, while the right panel similarly presents the redshift scaling normalized at the pivotal mass $\mpiv = 1.4\times10^{14}\Msunh$.
In both panels, the \erass1 clusters are color-coded based on the redshift (left panel) and posterior-sampled halo mass (right panel).
For comparison, the results from \citet[][green dash-dotted lines]{grandis24} and \citet[][purple lines]{kleinebreil24}, which are based on the weak-lensing mass calibration from the DES and KiDS surveys, respectively, are also plotted.
The observed \rate--\mass--\redshift\ relation reveals a mass trend that is steeper than the self-similar prediction (the red dashed line), while its redshift trend remains statistically consistent with the self-similar scaling.
The normalization of the self-similar predictions (the red dashed lines) is fixed to the best-fit \Arate.
}
\label{fig:rate_m_z}
\end{figure*}

\section{Results and discussions}
\label{sec:results}

In a forward-modelling framework, we constrain both the \rate--\mass--\redshift\ and \mbcg--\mass--\redshift\ relations 
while accounting for the weak-lensing mass bias and scatter via the \bwl--\mass--\redshift\ relation.
In Section~\ref{sec:rate_relation_results}, we first present the weak-lensing calibrated \rate--\mass--\redshift\ relation without the modelling of \mbcg--\mass--\redshift\ relation, as an analysis of purely weak-lensing mass calibration.
Finally, We present the \mbcg--\mass--\redshift\ relation and discuss its implication in Section~\ref{sec:bcg_relation_results}.

\begin{figure*}
\centering
\resizebox{0.8\textwidth}{!}{
\includegraphics[scale=1]{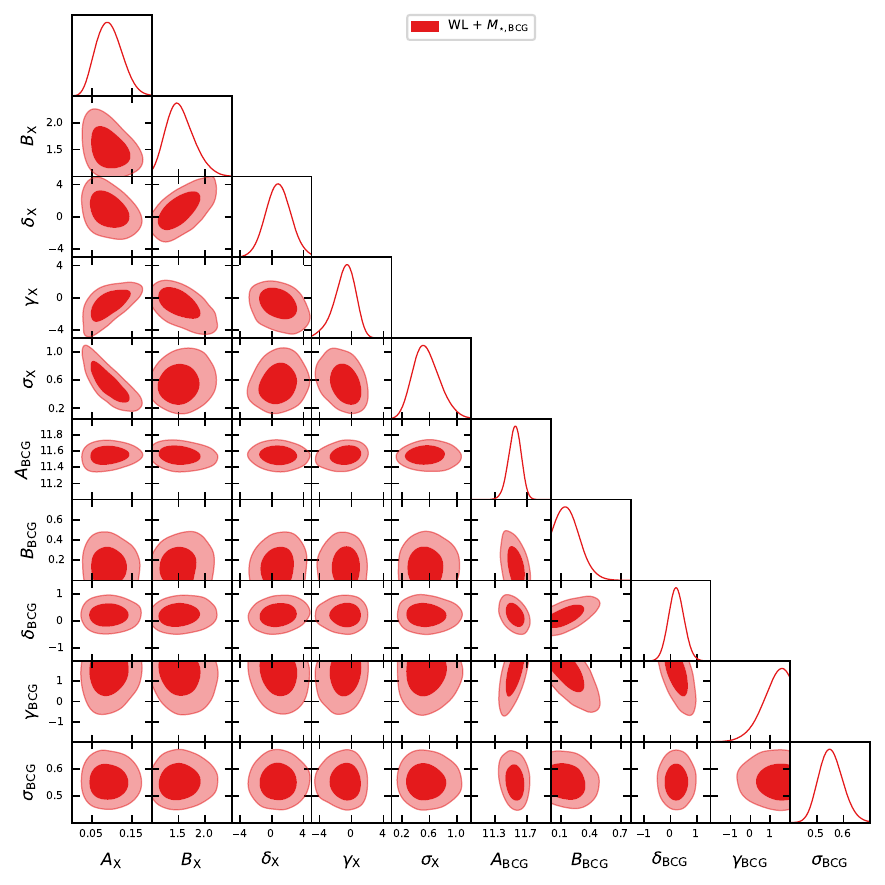}
}
\caption{
The marginalized posteriors and two-dimensional joint posteriors of the \rate--\mass--\redshift\ and \mbcg--\mass--\redshift\ relation parameters in the joint modelling are presented in a similar manner as in Figure~\ref{fig:gtc_join}.
}
\label{fig:gtc_trunc}
\end{figure*}

\subsection{The \rate--\mass--\redshift\ relation}
\label{sec:rate_relation_results}

Regardless of the BCG stellar mass estimates, there are 96 \erass1 clusters with available weak-lensing shear profiles.
Sampling the joint likelihood of those 96 clusters, i.e., equation~(\ref{eq:mass_calibration_approximate_wlonly}), without the modelling of the \mbcg--\mass--\redshift\ relation delivers the constraints on the parameters of the \rate--\mass--\redshift\ relation in equation~(\ref{eq:count_rate_relation}) as
\begin{equation}
\label{eq:count_rate_relation_wlonly_result}
\centering
\begin{array}{ccc}
\Arate         &=     &\asobs    			\, \\
\Brate         &=     &\bsobs    			\, \\
\deltarate     &=     &\deltasobs    		\, \\
\gammarate     &=     &\gammasobs    		\, \\
\sigmarate     &=     &\sigmasobs    		\, 
\end{array} \, .
\end{equation}
These numbers are tabulated in Table~\ref{tab:parameter_constraints}.
The marginalized posteriors of and the two-dimensional joint posteriors among these parameters are shown as the blue contours in Figure~\ref{fig:gtc_join}, while we present the full results of the weak-lensing-only modelling in Appendix~\ref{app:auxiliary}.
The correlated intrinsic scatter between the X-ray count rate and the weak-lensing mass is constrained as $\rhowlxray = \rhowlsobswlonly$, suggesting that the correlation between the X-ray and WL properties at a fixed halo mass is consistent with zero or ill-constrained in this work.

In Figure~\ref{fig:gtc_join}, we find excellent agreement between this work and \citet[][the green contours]{kleinebreil24}.
We stress that \cite{kleinebreil24} performed the weak-lensing mass calibration of the same sample of the 96 \erosita\ clusters overlapping the HSC-Y3 footprint based on the identical weak-lensing shear profiles extracted in this work.
The main difference is that they modelled (1) the miscentering on a per-cluster basis depending on the X-ray detection likelihood \ldet\ and extent $\mathtt{EXT}$ and (2) the redshift-dependent intrinsic scatter of the weak-lensing mass. 
Meanwhile, we derive the Gaussian priors for the sample and applied them to the parameters at a population level in the modelling (see Section~\ref{sec:wl_modelling}).
As a result, these two results obtained from independent codes are fully consistent, again ensuring the robustness of the weak-lensing modelling.

In Figure~\ref{fig:shear_profiles}, we present the stacked shear profiles and best-fit models in four sub-samples defined by the observed count rates and redshift in the left panel.
We use the median count rate (${\rate}_{\mathrm{med}}\approx0.487$) and redshift (${\redshift}_{\mathrm{med}}\approx0.27$) of the \erass1 clusters to split the sample into four bins.
The stacked shear profile for each sub-sample is calculated as the weighted average of the individual shear profiles \gshear, where the weights are defined as the inverse square of the radially dependent shape noise.
The best-fit models, with their uncertainties fully marginalized over all parameters, are shown as the grey shaded regions. 
Similarly, we produce the stacked profile of all the 96 \erass1 clusters in the right panel.
As seen, we find good agreement between the stacked profiles and our best-fit models, suggesting that the weak-lensing modelling provides excellent description of the data.

In Figure~\ref{fig:rate_m_z}, we present the mass and redshift trend of the \rate--\mass--\redshift\ relation re-scaled at the pivotal redshift $\zpiv = 0.35$ and mass $ \mpiv = 1.4\times 10^{14} \Msunh$ in the left and right panels, respectively.
The data points are \erass1 clusters with their halo mass \mass\ randomly sampled from the mass posterior $P\left(\mass | \rate, \gshear, \mathbf{p}\right)$ with the uncertainties fully marginalized over the posteriors of the parameters $\mathbf{p}$ \citep[see equation~(66) in][]{chiu22}.
To show the mass trend (left panel), on the y-axis we remove the redshift dependence by dividing the observed count rate \rate\ by the redshift-dependent factors evaluated at the best-fit parameters; similarly, we remove the mass dependence of \rate\ to show the redshift trend in the right panel.
As seen in the left panel, our results suggest a mass trend with a power-law index of \bsobs, which is significantly steeper than the self-similar prediction at the soft X-ray band ($\Brate = 1$; the red dashed line).
Meanwhile, no significant departure from the self-similar prediction of the redshift trend is revealed in the right panel ($\gammarate = \gammasobs$), despite the large errorbar.
Additionally, we find no clear evidence for the cross scaling between the halo mass and redshift ($\deltarate = \deltasobs$).
The mass trend of the X-ray count rate steeper than the self-similar prediction implies that non-gravitational process (e.g., feedback from cluster cores) plays a key role in the X-ray emission.
Moreover, this mass scaling has neither dependence on redshift nor deviation from the self-similar redshift evolution, implying that the non-gravitational mechanisms in massive clusters are already in place at high redshift ($\redshift\approx0.8$ for our cluster sample).
This picture is in excellent agreement with previous studies of \erosita-selected samples \citep{chiu22} and SZE-selected clusters \citep{bulbul19} which probed an even higher redshift to $\redshift\approx1.3$.

In Figure~\ref{fig:rate_m_z}, the black dotted lines represent the result from \cite{chiu22}, which were obtained based on $\approx300$ \erosita-selected clusters and groups in the eFEDS survey using the same HSC-Y3 WL data.
We find satisfactory agreement between \cite{chiu22} and this work with the following remarks.
First, \cite{chiu22} used a slightly different functional form to model the \rate--\mass--\redshift\ relation. Specifically, they included a simulation-calibrated correction factor for the observed count rate to probe the underlying true count rate relation in eFEDS \citep[See Section~4.1 in][]{chiu22}.
Consequently, their resulting mass and redshift trends are scaled as $\propto\mass^{\Brate -0.16}$ and $\propto{\redshift}^{0.42}\left(1 + \redshift\right)^{\gammarate}$, respectively, where they contained $\Brate = 1.58^{+0.17}_{-0.14}$ and $\deltarate = -0.44^{+0.81}_{-0.85}$.
In this work, we directly relate the observed count rate \rate\ to the halo mass without the correction factor.
In Figure~\ref{fig:rate_m_z}, we plot the overall mass and redshift trends of \cite{chiu22} as $\propto\mass^{\Brate -0.16}\appropto\mass^{1.42}$ and $\propto{\redshift}^{0.42}\left(1 + \redshift\right)^{\gammarate}\appropto{\redshift}^{0.42}\left(1 + \redshift\right)^{-0.44}$, respectively.
Second, as the major difference between \cite{chiu22} and the updated analysis of \erass\ clusters in this study \citep[also][]{grandis24,kleinebreil24}, the former applied a Gaussian prior $\mathcal{N}\left(0.3, 0.08^2\right)$ to the intrinsic scatter \sigmarate, as opposed to this work in which only an uninformative prior $\mathcal{U}\left(0.05, 2.0\right)$ is adopted.
Removing the Gaussian prior reveals a significantly large intrinsic scatter $\sigmarate\gtrsim0.55$ as preferred by the count rate observed by \erosita\ \citep[see also][]{grandis24,kleinebreil24}.
Meanwhile, there exists a strong degeneracy between the normalization \Arate\ and \sigmarate\ (as seen in Figure~\ref{fig:gtc_join}), leading to a lower \Arate\ with increasing \sigmarate\ in this work.
As seen in Figure~\ref{fig:rate_m_z}, we obtain a normalization \Arate\ mildly lower than \citet[][$\Arate\approx0.148^{+0.026}_{-0.023}$]{chiu22} at a level of $\approx1.6\sigma$.

It is worth mentioning that the large intrinsic scatter \sigmarate\ results in significant \citet{eddington13} bias, as evident from the offset between the sampled masses $P\left(\mass | \rate, \gshear, \mathbf{p}\right)$ and the best-fit \rate--\mass--\redshift\ relation in Figure~\ref{fig:rate_m_z}.
This is explained by the posterior $P\left(\mass | \rate, \gshear, \mathbf{p}\right)$ evaluated as 
$P\left(\mass | \rate, \gshear, \mathbf{p}\right) \propto P\left( \rate, \gshear | \mass , \mathbf{p}\right) P\left( \mass | \mathbf{p}\right)$, where $P\left( \mass | \mathbf{p}\right)$ is the normalized halo mass function.
At a fixed halo mass, the large scatter \sigmarate\ of the observed count rate leads to a wide distribution $P\left( \rate, \gshear | \mass , \mathbf{p}\right)$ so that the up-scatter contribution from the low-mass end of the halo mass function $P\left( \mass | \mathbf{p}\right)$ becomes significant.
Consequently, the sampled masses (data points) are expected to be lower than those following the best-fit \rate--\mass--\redshift\ relation (grey region) at a given fixed count rate in the left panel.
The Eddington bias is accounted for in this work and other \erosita\ analyses, thus the resulting scaling relations are unbiased. 
An even larger intrinsic scatter of $\sigmarate\approx1$ was found in \cite{ghirardini24}, where the authors combined the weak-lensing mass calibration and the abundance of \erass1\ clusters to constrain cosmological parameters.
It is surprising that the intrinsic scatter of the \erosita\ observed count rate, obtained from either weak lensing alone or a combination with the cluster abundance, is significantly larger than that ($\approx0.3$) of X-ray luminosity in previous studies \citep[][]{pratt09,vikhlinin09a,lovisari15,mantz16,bulbul19,chiu22,bahar22,akino22}.
This implies that significantly large scatter is introduced in measuring the X-ray count rate in the \erosita\ data pipeline, increasing the intrinsic scatter at a fixed halo mass from a scale of $30\percent$ in the X-ray luminosity to a factor of $\exp\left(1\right)\approx2.7$ in the count rate. 
A future study investigating this topic is warranted.

We show the comparisons with the weak-lensing-calibrated results of \erass1 clusters using data from the DES \citep[][the green lines]{grandis24} and KiDS \citep[][the purple lines]{kleinebreil24} surveys in Figure~\ref{fig:rate_m_z}.
We observe excellent agreement with the DES result as they constrained
$\Arate = 0.088\pm0.02$,
$\Brate = 1.62\pm0.14$,
$\deltarate = -0.85\pm0.93$,
$\gammarate = -0.32\pm0.69$, and
$\sigmarate = 0.61\pm0.19$.
Meanwhile, mild discrepancy at $\lesssim1.5\sigma$ is seen between this work the KiDS result with their errorbars generally larger than ours, as they obtained 
$\Arate = 0.17\pm0.042$,
$\Brate = 1.68\pm0.27$,
$\deltarate = 1.57\pm1.74$,
$\gammarate = -2.60\pm1.45$, and only the $1\sigma$ upper limit on the intrinsic scatter as
$\sigmarate < 0.85$.

It is worth noting that \cite{kleinebreil24} performed a direct comparison of the weak-lensing measurements between the KiDS and HSC surveys using $48$ common clusters.
They found excellent agreement in the observed shear profile \gshear\ but a significant discrepancy in the inferred excess surface mass density \deltaSigma.
Specifically, \deltaSigma\ inferred from HSC is statistically higher than that from KiDS by $\approx30\percent$ at a level of $\approx4\sigma$.
Meanwhile, only marginal difference (at a level of $\approx2.3\sigma$) was seen in \gshear\ and \deltaSigma\ between DES and KiDS.
Because the source redshift is required to derive \deltaSigma\ from the observed \gshear, the agreement in \gshear\ but not in \deltaSigma\ between the KiDS and HSC surveys implies a potential systematic uncertainty existing in the source photo-\redshift\ calibration.
Recall that $\gshear \approx \rshear = \deltaSigma / \sigmacrit \propto \deltaSigma\times\lensingeff$, where \lensingeff\ is the lensing efficiency, which monotonically increases with source redshift.
The higher \deltaSigma\ in the HSC survey compared to KiDS suggests that the photo-\redshift\ of the source sample is either underestimated in HSC, overestimated in KiDS, or a combination of both.
However, the photo-\redshift\ in the HSC survey is rigorously calibrated against spectroscopic samples and shows no bias at redshift $\redshift\lesssim1.2$ \citep{rau23}. 
In addition, the agreement between the DES and KiDS surveys suggests no significant bias in the photo-\redshift\ calibration of their source samples, which are primarily at relatively low redshift ($\redshift\lesssim1$).
A possible cause is that the photo-\redshift\ of the HSC source sample at high redshift $\redshift\gtrsim1.2$ is biased low.
This aligns well with the finding of the HSC-Y3 cosmic-shear analyses \citep{li23,dalal23,sugiyama23}, which suggest that the photo-\redshift\ of the HSC sources at $\redshift\gtrsim1.2$ is biased low by $\approx0.2$ at a level of $\approx2\sigma$.
However, the photo-\redshift\ bias at $\redshift\gtrsim1.2$ at a level of $\approx0.2$ alone cannot explain the $\approx30\percent$ difference in \Sigmam\ for the \erass1\ clusters, which are primarily at redshift $\redshift\approx0.35$.
A detailed investigation of this systematic uncertainty is beyond the scope of this paper and is clearly warranty in future work, with a focus on the photo-\redshift\ calibration at high redshift $\redshift\gtrsim1.2$.

We provide the mean of the mass posterior, $\left\langle\log\left(\frac{\Mfiveoo}{\Msunh}\right)\right\rangle$, and its sampled mass in Table~\ref{tab:measurements}.
The mass posterior is calculated as $P\left(\mass | \rate, \mathbfcal{O}, \mathbf{p}\right)$, where $\mathbfcal{O}=\left\lbrace\gshear,\mbcg\right\rbrace$, $\left\lbrace\mbcg\right\rbrace$, or $\left\lbrace\gshear\right\rbrace$, depending on the data availability.

\begin{figure*}
\centering
\resizebox{\textwidth}{!}{
\includegraphics[scale=1]{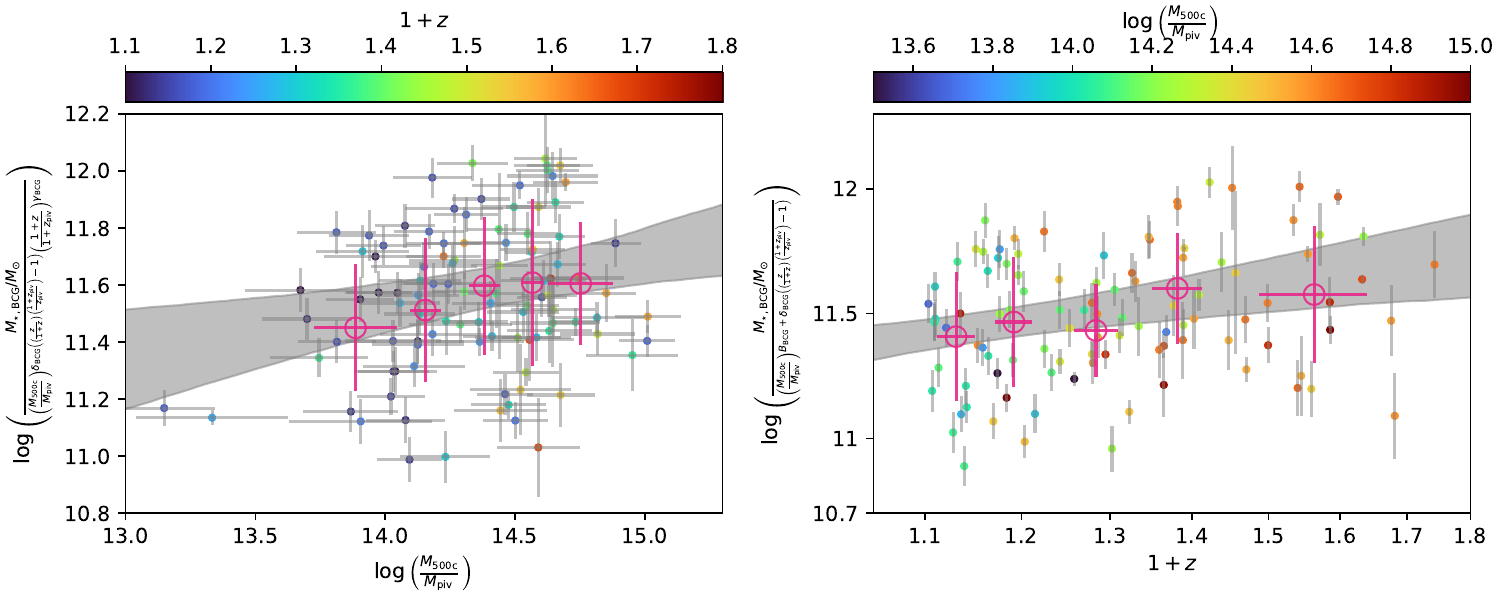}
}
\caption{
The \mbcg--\mass--\redshift\ relation of the \erass1 clusters and the best-fit model in equation~(\ref{eq:bcg_relation_result}) with its fully marginalized uncertainties (grey shaded regions).
This figure is generated following the same manner as in Figure~\ref{fig:rate_m_z}.
We additionally plot the running inverse-variance-weighted means and standard deviations as the pink open circles, using five bins of an equal number of clusters.
Errorbars of individual \erass1 clusters are omitted for clarity.
}
\label{fig:mbcg_m_z}
\end{figure*}

\subsection{The \mbcg--\mass--\redshift\ relation}
\label{sec:bcg_relation_results}

To begin with, we verify that further including the modelling of the BCG stellar mass relation, as in equation~(\ref{eq:mass_calibration_total}), yields a consistent \rate--\mass--\redshift\ relation.
This is clearly evident from the excellent agreement between the blue (WL only) and red (WL$+$BCG) contours in Figure~\ref{fig:gtc_join}, suggesting that the cluster mass estimates, informed by the X-ray count rate with an absolute weak-lensing calibration, are robust to constrain the BCG stellar mass relation.

As tabulated in Table~\ref{tab:parameter_constraints}, we obtain the constraints on the \mbcg--\mass--\redshift\ relation, i.e., equation~(\ref{eq:bcg_relation}), as 
\begin{equation}
\label{eq:bcg_relation_result}
\centering
\begin{array}{ccc}
\Abcg         &=     &\axobs    			\, \\
\Bbcg         &=     &\bxobs    			\, \\
\deltabcg     &=     &\deltaxobs    		\, \\
\gammabcg     &=     &\gammaxobs    		\, \\
\sigmabcg     &=     &\sigmaxobs    		\, 
\end{array} \, .
\end{equation}
The parameter marginalized posteriors and two-dimensional joint posteriors are presented in Figure~\ref{fig:gtc_trunc} (see the full results in Appendix~\ref{app:auxiliary}). 
Meanwhile, we constrain the correlated intrinsic scatter as 
$\rhowlxray = \rhowlsobswlxo$, 
$\rhowlmbcg = \rhowlxobswlxo$, and 
$\rhombcgxray = \rhoxobssobswlxo$, suggesting no significant intrinsic correlation among the BCG stellar mass, X-ray count rate, and the weak-lensing mass at a fixed halo mass.

It is noticeable that there exists large intrinsic scatter in the BCG stellar mass \mbcg\ at a fixed halo mass, constrained as $\sigmabcg =\sigmaxobs$ in this work.
This corresponds to an intrinsic scatter of $\approx0.24\pm0.02$~dex in a unit of $\log\left(\mbcg/\Msun\right)$, which is consistent with previous studies \citep[e.g.,][]{leauthaud12a,kravtsov18,erfanianfar19,akino22}.
Unlike the \rate--\mass--\redshift\ relation, the \mbcg--\mass--\redshift\ relation has a shallow mass slope ($\Bbcg = \bxobs$), so that the scatter $\sqrt{\mathrm{Var}{\left(\ln\mass | \mbcg \right)}}$ in the logarithmic of the halo mass at a fixed BCG stellar mass is at a level of $\approx \sigmabcg / \Bbcg  \approx 3$, corresponding to nearly a factor of $\exp\left(3\right)\approx20$ in the halo mass \mass.
That is, it is not feasible to adopt the BCG stellar mass as a precise mass proxy on a per-cluster basis.

Interestingly, \cite{goldenmarx22} found that the intrinsic scatter \sigmabcg\ of the BCG stellar mass at a fixed halo mass could be largely reduced by accounting for their local environments, as indicated by the magnitude gap between the BCG and the fourth brightest cluster galaxy \citep[see also][]{goldenmarx18,goldenmarx19}.
Moreover, \cite{goldenmarx24} measured a non-zero correlation between the magnitude gap and the central stellar mass (BCG+ICL), supporting a scenario in which BCGs primarily assemble their masses at late times through ex situ processes, such as merging.
A similar correlation between the central stellar mass and the halo concentration was also found in \cite{zu22}.
Therefore, it is possible to adopt the BCG stellar mass \mbcg\ as a reliable mass proxy for \erosita\ clusters if the formation history of BCGs is properly taken into account \citep[e.g.,][]{huang22}.

In Figure~\ref{fig:mbcg_m_z}, we present the mass and redshift trends of the \mbcg--\mass--\redshift\ relation in the left and right panels, respectively, with the re-normalization as similarly done in Figure~\ref{fig:rate_m_z}.
Additionally, we plot the running inverse-variance-weighted means and the standard deviations of the measurements as the pink circles, using five sub-sample bins of equal cluster numbers. 
As seen, the best-fit model provides a good description of the data with the noticeably large intrinsic scatter at a fixed halo mass.
Our results suggest no or mildly increasing redshift trend of the BCG stellar mass at a fixed halo mass ($\gammabcg = \gammaxobs $) at a level of $\approx2\sigma$.
However, it is in contrast to the physical picture of the hierarchical structure formation, where galaxies at late time are expected to be more massive that those at earlier time, on average \citep[see, e.g.,][]{moster13}.
This is primarily caused by the strong degeneracy between the parameters \Bbcg\ and \gammabcg\ (see Figure~\ref{fig:gtc_trunc}), which arises from a combination of the large intrinsic scatter \sigmabcg\ and the X-ray selection of the cluster sample.
While the X-ray selection favors high-mass clusters at high redshift, the excessively large scatter \sigmabcg\ (corresponding to an $\approx80\percent$ variation in \mbcg\ at a fixed halo mass \mass) 
prevents us from distinguishing between two scenarios:
(1) BCGs at a fixed halo mass are intrinsically increasing with redshift or 
(2) more massive BCGs at high redshift appear in the sample purely due to the selection bias \citep[i.e., the Malmquist bias;][]{malmquist1922}.
In the absence of low-mass clusters at high redshift, it is challenging to simultaneously constrain both the mass and redshift trends given the size of the current sample.

In light of the strong degeneracy between \Bbcg\ and \gammabcg, we additionally constrain the \mbcg--\mass--\redshift\ relation with an informative prior applied to the parameter \gammabcg.
The informative prior is derived as follows.
Based on \cite{moster13}, where the authors used an abundance-match method and accounted for the Eddington bias in the halo mass estimates, they constrained the \mbcg--\mass--\redshift\ relation across wide ranges of halo mass and redshift (see their Table 1).
We take an \erass1 cluster with 
the mean of its
mass posterior\footnote{The mass posterior is calculated in Section~\ref{sec:rate_relation_results} as $\left\langle\log\left( \Mfiveoo / \left(\Msunh\right) \right)\right\rangle$.} $\overline{\mass}$, place it at a series of equal-step redshift bins ${\redshift}_{j}$ between $0$ and $1$,
and calculate its corresponding BCG stellar mass $\left\langle\mbcg | \overline{\mass},{\redshift}_{j} \right\rangle_{\mathrm{Moster+13}}$ (without the intrinsic scatter and measurement uncertainty) following the \citet{moster13} relation assuming that it is at each redshift ${\redshift}_{j}$.
We then re-fit the functional form $A\left( \frac{1 + \redshift}{1 + \zpiv} \right)^{\gamma}$, which is the redshift scaling of the \mbcg--\mass--\redshift\ relation, to the BCG stellar masses $\left\langle\mbcg | \overline{\mass},{\redshift}_{j} \right\rangle_{\mathrm{Moster+13}}$ at different redshifts ${\redshift}_{j}$ using a $\chi^2$ minimization.
This gives the best-fit power-law index $\gamma$ of the redshift scaling for this \erass1 cluster with the mass $\overline{\mass}$.
We repeat the aforementioned procedure for all \erass1 clusters, resulting a distribution of the best-fit $\gamma$ which represents the distribution of the power-law indices for our \erass1 sample at their mass scales. 
The distribution is obtained as a nearly Gaussian $\mathcal{N}\left(-0.4, 0.14^2\right)$.
This result is free from the Malmquist bias because we do not include the intrinsic scatter and measurement uncertainty in calculating $\left\langle\mbcg | \overline{\mass},{\redshift}_{j} \right\rangle_{\mathrm{Moster+13}}$.
In the end, we apply a Gaussian prior $\mathcal{N}\left(-0.4, 0.14^2\right)$ to the parameter \gammabcg\ and obtain an improved constraint on the \mbcg--\mass--\redshift\ relation as 
\begin{equation}
\label{eq:bcg_relation_result_prioronz}
\centering
\begin{array}{ccc}
\Abcg         &=     &\axobsprioronz    			\, \\
\Bbcg         &=     &\bxobsprioronz    			\, \\
\deltabcg     &=     &\deltaxobsprioronz    		\, \\
\gammabcg     &=     &\gammaxobsprioronz    		\, \\
\sigmabcg     &=     &\sigmaxobsprioronz    		\, 
\end{array} \, .
\end{equation}
These results are also tabulated in Table~\ref{tab:parameter_constraints}.
The mass and redshift trends of the \mbcg--\mass--\redshift\ relation with the Gaussian prior on \gammabcg\ are presented in Figure~\ref{fig:mbcg_m_z_prioronz}.

\begin{figure}
\centering
\resizebox{0.45\textwidth}{!}{
\includegraphics[scale=1]{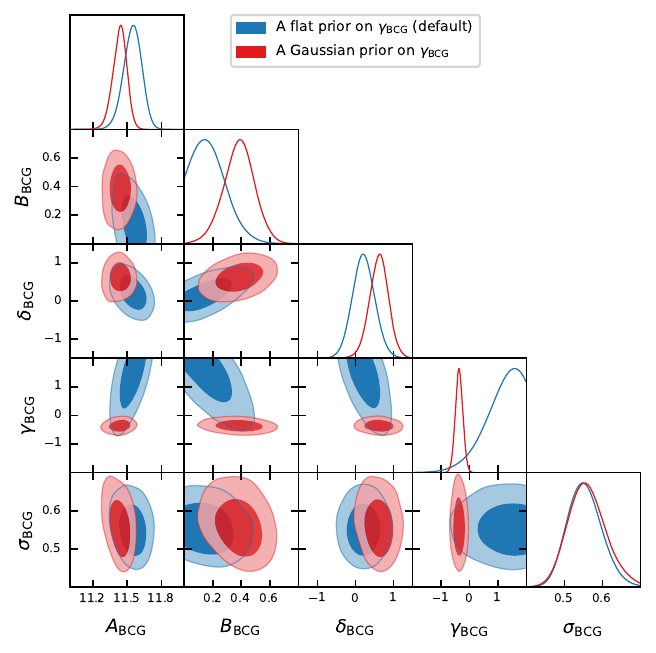}
}
\caption{
The comparison between the \mbcg--\mass--\redshift\ relations obtained with (blue contours) and without (red contours) the Gaussian prior applied on \gammabcg.
The Gaussian prior is obtained based on the overall mass and redshift distributions of the \erass1 clusters studied in this work, following the assumed relation from \cite{moster13} (see the text for details).
}
\label{fig:gtc_mbcg_gamma_prior}
\end{figure}

In Figure~\ref{fig:gtc_mbcg_gamma_prior}, we present the parameter marginalized posteriors and 
two-dimensional joint posteriors
in comparison with those without the Gaussian prior on \gammabcg.
As seen, the constraints with (blue contours) and without (red contours) the Gaussian prior are statistically consistent with each other at a level of $\approx1\sigma$.
However, the degeneracy between \Bbcg\ and \gammabcg\ is broken by including the Gaussian prior, leading to a $\approx1\sigma$ steeper mass trend ($\Bbcg = \bxobsprioronz$) with smaller errorbars.
The steeper mass trend ($\Bbcg = \bxobsprioronz$) is also in better agreement with previous studies \citep[e.g.,][]{gonzalez13,burg14,chiu16a,kravtsov18,akino22}, suggesting that our constraints based on an X-ray-selected sample could be prone to the \Bbcg--\gammabcg\ degeneracy. 
Interestingly, once the \Bbcg--\gammabcg\ degeneracy is broken, there exists a cross scaling between the mass and redshift ($\deltabcg = \deltaxobsprioronz$) at a level of $\approx2.5\sigma$.
This parameter was constrained as $-0.085\pm0.045$ in \cite{moster13}\footnote{
With the variables defined in \cite{moster13}, the parameter \gammabcg\ can be approximated as $-\gamma_{11} \left(\frac{\zpiv}{1 + \zpiv}\right)$, where we use $\zpiv = 0.35$ and \cite{moster13} obtained a constraint on $\gamma_{11}$ as $\gamma_{11} = 0.329\pm0.173$ in their Table~1.
}, which has an opposite trend that is inconsistent with ours at a mildly significant level ($\approx2.8\sigma$). 
A larger sample with more low-mass clusters at high redshift will shed light on this.
The interpretations of the other parameters \Abcg\ and \sigmabcg\ remain unaffected by the Gaussian prior.

In what follows, we discuss potential systematic uncertainties arising from the SED fitting.
In our fiducial analysis, we make an assumption that the BCGs are dominated by passively evolving galaxy populations, so we only fit the SPS model with the star-formation timescale $\tau$ up to $5$~Gyrs and low dust extinction $E\left(B-V\right)=0, 0.1, 0.2, 0.3$.
We examine the systematics regarding this assumption.
Namely, we allow a new SPS model with $\tau$ up to $30$~Gyrs configured with the dust-extinction coefficient of $E\left(B-V\right)=0, 0.1, 0.2, 0.3, 0.4, 0.5$ and repeat the whole analysis.
We find the constraints of 
$\Abcg = \axobsfulllib$, 
$\Bbcg = \bxobsfulllib$, 
$\deltabcg = \deltaxobsfulllib$,
$\gammabcg = \gammaxobsfulllib$, and
$\sigmabcg = \sigmaxobsfulllib$,
which are in excellent agreement with the fiducial results.
This reinforces the picture that the BCGs are dominated by passively evolving populations.
On the other hand, we also examine the systematics due to the inclusion of the mid-infrared photometry.
Specifically, we perform the SED fitting with the broadband photometry at $3.4\mu\mathrm{m}$ and $4.6\mu\mathrm{m}$ from \wise, and repeat the subsequent analysis.
We obtain a result that is in agreement with our fiducial results at a level of $\lesssim 1\sigma$, further ensuring the robustness of our SED fitting.
We provide a more detailed description about the SED fitting including the \wise\ photometry in Appendix~\ref{app:wise_data}.

\begin{figure*}
\centering
\resizebox{\textwidth}{!}{
\includegraphics[scale=1]{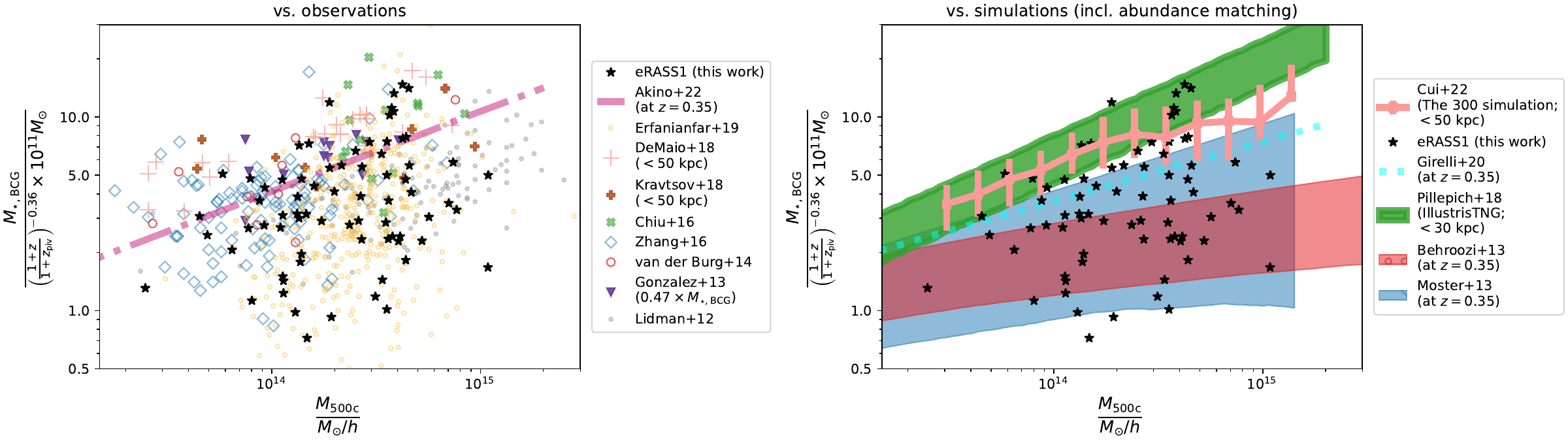}
}
\caption{
The comparisons of the mass trend of the BCG stellar mass between the \erass1 clusters and previous observational (left panel) and simulation-based (right panel) studies.
In both panels, the mass trend is re-normalized at the pivotal redshift ($\zpiv = 0.35$) using the best-fit \mbcg--\mass--\redshift\ relation as in equation~(\ref{eq:bcg_relation_result_prioronz}).
The \erass1 clusters are indicated by black stars, while previous studies are shown according to the lists on the right side of each panel.
}
\label{fig:mbcg_mcomparison}
\end{figure*}
\begin{figure*}
\centering
\resizebox{\textwidth}{!}{
\includegraphics[scale=1]{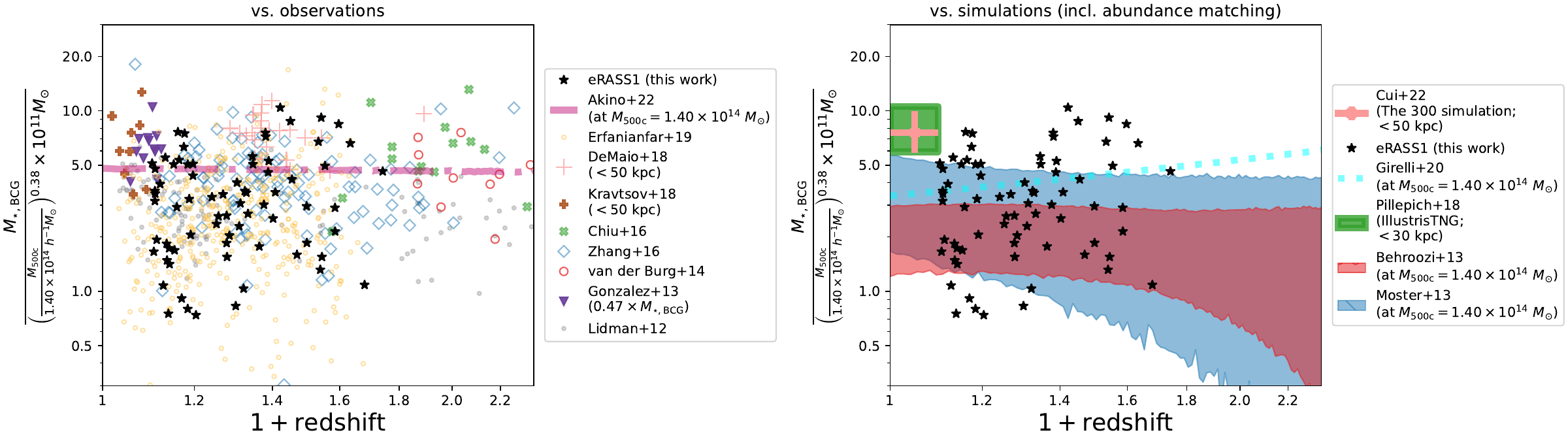}
}
\caption{
The comparisons of the redshift trend of the BCG stellar mass between the \erass1 clusters and previous observational (left panel) and simulation-based (right panel) studies.
The plots are produced in the similar way as in Figure~\ref{fig:mbcg_mcomparison} but are re-normalized at the pivotal mass ($\mpiv = 1.4\times10^{14}\Msunh$).
}
\label{fig:mbcg_zcomparison}
\end{figure*}

\subsubsection{Comparisons with previous studies}
\label{sec:mbcg_literature}

We now turn into the comparisons of the \mbcg--\mass--\redshift\ relation with previous studies.
These comparisons are illustrated in Figures~\ref{fig:mbcg_mcomparison}~and~\ref{fig:mbcg_zcomparison}, which represent the trends in mass and redshift, respectively, at the pivotal values of redshift ($\zpiv = 0.35$) and mass ($\mpiv = 1.4\times 10^{14}\Msunh$). 
For the re-normalization, we adopt the best-fit parameters in equation~(\ref{eq:bcg_relation_result_prioronz}), which are obtained with the Gaussian prior on \gammabcg.
We note that the best-fit parameters in equation~(\ref{eq:bcg_relation_result_prioronz}) incorporates the external information of the Gaussian prior, which is derived based on the overall mass and redshift distributions of \erass1 clusters informed by the previous work of \cite{moster13}.
As such, our focus in this subsection is on the consistency between the \erass1 measurements and the literature. 

In Figures~\ref{fig:mbcg_mcomparison}~and~\ref{fig:mbcg_zcomparison}, the left panels compare the results with previous observational studies, while the right panels compare them with simulation-based results.
The \erass1 cluster measurements from this work, with posterior-sampled halo masses, are indicated by black stars.

For previous observational results, we include the following:
\begin{itemize}
\item \citet[][grey points]{lidman12} compiled a heterogeneous sample of clusters spanning at $0.1\lesssim\redshift\lesssim1.6$ with the halo mass estimated from either X-ray mass proxies or the line-of-sight velocity dispersion, and derived the stellar mass from the $K$-band luminosity.
\item \citet[][purple triangles]{gonzalez13} studied $12$ X-ray-selected clusters at local Universe ($\redshift\approx0.1$) and estimated their BCG stellar masses including the ICL.
\item \citet[][red open circles]{burg14} measured the BCG stellar mass using $11$-band SED fitting and the halo mass from the line-of-sight velocity dispersion for a sample of high-redshift clusters ($0.9\lesssim\redshift\lesssim1.4$) selected in near infrared.
\item \citet[][green crosses]{chiu16a} studied $14$ SZE-selected massive clusters at redshift $0.6\lesssim\redshift\lesssim1.3$ and derived the BCG stellar mass using SED fitting to data with wavelength from the optical to mid-infrared.
\item \citet[][brown pluses]{kravtsov18} derived the BCG stellar mass (using $r$-band luminosity) and the halo mass (using X-ray mass proxies) of $9$ nearby clusters at $\redshift\lesssim0.1$.
\item \citet[][pink pluses]{demaio18} studied the stellar mass of BCGs and ICL for $23$ clusters at redshift $0.3\lesssim\redshift\lesssim0.9$ using the imaging from \textit{Hubble Space Telescope}.
\item \citet[][yellow open circles]{erfanianfar19} studied the \mbcg--\mass--\redshift\ relation of $438$ X-ray-selected clusters out to $\redshift\approx0.65$ based on the BCG stellar mass estimated from SED fitting to SDSS and \wise\ photometry.
\item \citet[][red dash-dotted lines]{akino22} measured the BCG stellar mass for $136$ X-ray-selected clusters drawn from the $XXL$ survey \citep{adami18} using the $5$-band $grizY$ photometry from HSC.
\end{itemize}
As suggested by \citet[][see also their detailed profile of the central stellar mass]{chen22}, we use the stellar mass measurements within $<50~\mathrm{kpc}$ for both \cite{kravtsov18} and \cite{demaio18} to roughly separate the BCG stellar mass \mbcg\ from the ICL. 
For these comparison samples, we consistently correct for the systematics in the halo and stellar mass estimates following the prescription in \cite{chiu18a}.
Specifically, the halo mass \mass\ is multiplied by $0.96$ ($1.12$) if it is estimated from the line-of-sight velocity dispersion (X-ray mass proxies), while the stellar mass is all corrected to that inferred by the \cite{chabrier03} IMF.

In the comparisons, we consider the following previous simulation-based results, including those based on the abundance-matching technique which assigns mass \mass\ from $N$-body simulations to observed halos statistically:
\begin{itemize}
\item Both \citet[][blue shaded regions]{moster13} and \citet[][red shaded regions]{behroozi13} determined the stellar-mass-to-halo-mass relation out to high redshift  $\redshift\gtrsim4$ using the observed stellar mass functions in the SDSS field \citep[e.g.,][]{baldry08,li09,bouwens12,moustakas13}, each with a slightly different functional form.
\item \citet[][the cyan dotted lines]{girelli20} studied the stellar-mass-to-halo-mass relation in the COSMOS field using a modified functional form from \cite{moster13} with a focus on its redshift evolution.
\end{itemize}
All the above-mentioned studies with the abundance-matching technique assumed a fixed intrinsic scatter of the central stellar mass given a halo mass at a level ranging from $\approx0.15~\mathrm{dex}$ to $\approx0.20~\mathrm{dex}$.
The $1\sigma$ confidence regions of \cite{moster13} and \cite{behroozi13} in Figures~\ref{fig:mbcg_mcomparison}~and~\ref{fig:mbcg_zcomparison} are obtained from a full marginalization of all parameters.
Meanwhile, we directly take the data points from cosmological hydrodynamic simulations (at redshift $\redshift\lesssim0.1$) of 
\begin{itemize}
\item the IllustrisTNG simulation \citep[][green shaded regions]{pillepich18}, and
\item the Three Hundred Project \citep[][pink points with errorbars]{cui22}.
\end{itemize}
No other corrections are applied to the halo mass \mass\ and the BCG stellar mass \mbcg\ in the IllustrisTNG and Three Hundred projects in Figures~\ref{fig:mbcg_mcomparison}~and~\ref{fig:mbcg_zcomparison}.
We use the BCG stellar mass within $30~\mathrm{kpc}$ for IllustrisTNG \citep{pillepich18} due to the lack of that within $50~\mathrm{kpc}$.

As seen in the figures, the \erass1 clusters (black stars) are in satisfactory agreement with observational results, which exhibit large scatter in \mbcg\ at a fixed halo mass (constrained as $\sigmabcg\approx0.6$ or $\approx0.25~\mathrm{dex}$ in this work).
After removing the redshift dependence ($\gammabcg\approx-0.4$) in the left panel of Figure~\ref{fig:mbcg_mcomparison}, the BCG stellar mass of the previous studies reveal a mass slope that is consistent with that of the \erass1 sample.
It is noteworthy that the \erass1 clusters span the widest halo mass range from $\approx10^{13}\Msun$ to $\approx10^{15}\Msun$, again highlighting the power of \erosita\ in finding galaxy clusters and groups.
Meanwhile, the agreement in the mass trend between the \erass1 clusters and the simulation-based results is seen at a similar level (the right panel in Figure~\ref{fig:mbcg_mcomparison}).
Nevertheless, we observe systematic offset in the absolute scale of the BCG stellar mass.
Specifically, the BCG stellar masses from both the hydrodynamic simulations of the IllustrisTNG and Three Hundred projects are systematically higher (by a factor of $\approx2$ to $\approx3$) than \erass1 clusters and other observational results, except for \cite{gonzalez13}, \cite{kravtsov18}, and \cite{demaio18}.
On the other hand, the results from the abundance-matching method \citep{moster13,behroozi13,cui22}, which statistically assigns the mass \mass\ from $N$-body simulations to observed halos with the stellar mass estimates, are in better agreement with the \erass1 clusters and other observational studies.
This systematic difference implies possible systematics in either the stellar mass produced in the hydrodynamic simulations, the photometric measurements in the survey pipelines \citep[see][for details in the systematics of the \texttt{cmodel} photometry]{huang18c,akino22}, or a combination of both.

After accounting for the halo-mass dependence ($\Bbcg\approx0.4$) in Figure~\ref{fig:mbcg_zcomparison}, the \erass1 measurements similarly show no significant redshift trend, in good agreement with previous studies.
This finding strongly suggests that the BCG stellar mass given a fixed halo mass (i.e., the ratio $\mbcg/\mass$) remains relatively stable at the high-mass end, with no clear evidence of evolution out to redshift $\redshift\approx1$.
The same picture of the little to no increase in the BCG stellar mass at the pivotal mass ($\mpiv = 1.4\times10^{14}\Msunh$) is also suggested by the abundance-matching results of \cite{moster13} and \cite{behroozi13}.
The increase of the BCG stellar mass at the pivotal halo mass from $\redshift=0.8$ to $\redshift=0.1$ is estimated to be at a level of $\left(\frac{1 + 0.1}{1 + 0.8}\right)^{\gammaxobsprioronz}-1\approx\left(20\pm8\right)\percent$, informed by the Gaussian prior from \cite{moster13}.
However, \cite{girelli20} reported a positive redshift trend, deviating from this pattern.
Our finding is consistent with \cite{zhang16}, where the authors obtained a mildly increasing BCG stellar mass
at a fixed halo mass for X-ray-selected clusters with a scaling of $\approx\left(1 + \redshift\right)^{-0.19\pm0.34}$, corresponding to $\approx0.13~\mathrm{dex}$ or $\approx34\percent$ from $\redshift\approx1$ to $\redshift\approx0$.
A similar picture is also suggested by \citet[][$\approx35\percent$ increase from $\redshift\approx1$ to $\redshift\approx0.3$]{lin17}.

We stress that our results quantify the scaling of the BCG stellar mass statistically as a function of the halo mass and redshift, which is complementary to the method that traces the growth of the BCG stellar mass by paring the progenitor and descendant of a cluster \citep[see][]{lin25}. 
With a sufficiently large sample of clusters covering wide ranges of the halo mass and redshift, our approach with a flexible scaling form is expected to be equivalent to the progenitor-descendant pairing strategy.
However, this is not feasible for the \erass1 sample due to the nature of the X-ray selection, which lacks low-mass groups at high redshift  that could serve as the progenitors of high-mass clusters at low redshift.

Using the merging tree in numerical simulations, \cite{lin13} found a significant growth of a factor $\approx2.3$ in the BCG stellar mass from $\redshift\approx1.5$ to $\redshift\approx0.5$ but no growth between $\redshift\approx0.5$ and $\redshift\approx0$.
Similarly, \cite{lidman12} found a significant growth in the BCG stellar mass (after accounting the halo-mass dependence) of $\approx\left(80\pm30\right)\percent$ from $\redshift\approx0.9$ to $\redshift\approx0.1$ but no growth between $\redshift\approx0.4$ and $\redshift\approx0.1$.
This suggests that BCGs must have assembled the majority of their stellar masses at much earlier epochs ($\redshift\gtrsim1$), before massive clusters formed, and have accreted only small amounts at later times.
Moreover, BCGs must grow primarily through an ex-situ process (e.g., merging) at late epochs, as no evidence of recent star formation is found.
Consequently, the BCG stellar mass fraction ($\mbcg/\mass$) exhibits a negative mass trend, i.e., $\Bbcg - 1\approx-0.6$, while the host cluster continues to grow significantly in the total mass \mass\ during the formation.
This ``rapid-then-slow'' growth of the BCG stellar mass is also suggested by the semi-analytical model in \cite{delucia07a} and further refined with an updated ICL production in \cite{contini14}.
This two-phase hierarchical formation describes that BCGs acquired their stellar masses in the core via in-situ star formation at very early epochs ($\redshift\gtrsim2$) and gradually assembled additional mass at the large radii via ex-situ mergers \citep{delucia07a,guo11}. 
On the other hand, we find that BCGs at a fixed halo mass ($\mpiv = 1.4\times10^{14}\Msunh$) assemble little stellar mass since redshift $\redshift\approx0.8$, which is in broad agreement with this picture of the ``rapid-then-slow'' mass assembly.
This picture is also in line with the recent observational result in \cite{lin25}.

%
%

\section{Conclusions}
\label{sec:conclusions}

Using the data from the HSC survey, we present the weak-lensing mass calibration of $124$ X-ray-selected galaxy clusters selected in the first scan (\erass1) of the \erosita\ All-Sky Survey \citep[\erass;][]{merloni24} and constrain their BCG stellar-mass-to-halo-mass-and-redshift (\mbcg--\mass--\redshift) relation or the so-called stellar-mass-to-halo-mass relation.
The cluster sample spans a mass range of $3\times10^{13}\Msunh\lesssim\Mfiveoo\lesssim10^{15}\Msunh$ and redshift range of $0.1<\redshift<0.8$, with the majority of low-mass systems ($\Mfiveoo\lesssim1.5\times10^{14}\Msunh$) at redshift $\redshift\lesssim0.35$.

The \erass1 clusters are X-ray-selected \citep{bulbul24} and optically confirmed \citep{kluge24}, leading to a clean ICM-based sample with a point-source contamination at a level of $\approx5\percent$.
We use the observed X-ray count rate \rate\ as the proxy for the halo mass (\Mfiveoo\ or \mass), based on which the selection function is parameterized and evaluated at the sky location and redshift of each \erass1 cluster \citep{clerc24}.

For the weak-lensing mass calibration, we measure the tangential reduced shear profile \gshear\ of each cluster using the latest HSC-Y3 data set \citep{li22}.
Following the previous studies \citep{chiu22,chiu23}, we derive the weak-lensing mass \mwl\ and infer the underlying halo mass \mass\ using the weak-lensing-mass-to-mass-and-redshift (\mwl--\mass--\redshift) relation.
The \mwl--\mass--\redshift\ relation is calibrated by hydrodynamic simulations, accounting for the modelling systematics including the halo triaxiality, the miscentering of X-ray-defined centers, the cluster member contamination to the weak-lensing observable, the bias in the photometric redshift of source galaxies, the calibration uncertainty of the galaxy shape measurement, and the baryonic feedback.

For the BCG stellar mass \mbcg, we employ the SED fitting technique on the $grizY$ five-band photometry from the third Public Data Release (PDR3) of the HSC survey.
The SED template is built from the model of Stellar Population Synthesis (SPS) assuming a passively evolving population, which is supported by our BCG data.
We additionally examine the systematics arising from the inclusion of the mid-infrared data (from \wise) in the SED fitting and find no statistically significant change to the final results.
We conclude that our BCG stellar mass estimates are homogeneous and robust.
We present the HSC cutout images of these BCGs and their SED fitting results in Appendix~\ref{app:imaging}.

Among the $124$ \erass1 clusters in this study, $73$ clusters have measurements for both the shear profile \gshear\ and the BCG stellar mass \mbcg, while $23$ have only \gshear\ measurements and $28$ have only \mbcg\ measurements.
In an approach of forward and population modelling, we simultaneously constrain both the count rate-to-mass-and-redshift (\rate--\mass--\redshift) relation and the BCG stellar-mass-to-halo-mass (\mbcg--\mass--\redshift) relation, accounting for the bias and intrinsic scatter of the weak-lensing mass via the \mwl--\mass--\redshift\ relation.
We also include the correlated intrinsic scatter among the X-ray count rate \rate, the BCG stellar mass \mbcg, and the weak-lensing mass \mwl, and find no statistically significant correlation. 

Using the weak-lensing shear profiles \gshear\ alone, we obtain a constraint on the \rate--\mass--\redshift\ relation as
\begin{multline}
\left\langle\ln\left(\frac{\ratehat}{\mathrm{counts}/\mathrm{sec}} \bigg| \mass,\redshift \right)\right\rangle = 
\ln\left(\asobs\right) +
\\
\left(\left(\bsobs\right) + \left(\deltasobs \right)\ln\left(\frac{1 + \redshift}{1 + \zpiv}\right) \right)\ln\left(\frac{\mass}{\mpiv}\right) + 
\\
2~\ln\left(\frac{E\left(\redshift\right)}{E\left(\zpiv\right)}\right) -2~\ln\left(\frac{D_{\mathrm{L}}\left(\redshift\right)}{D_{\mathrm{L}}\left(\zpiv\right)}\right) + 
\\
\left(\gammasobs\right)~\ln\left(\frac{1 + \redshift}{1 + \zpiv}\right)
\, , \nonumber
\end{multline}
with the intrinsic scatter $\sigmarate = \sigmasobs$.
The resulting \rate--\mass--\redshift\ relation suggests a mass trend significantly steeper than the self-similar prediction of the X-ray luminosity-to-mass relation but reveals no deviation from the self-similar scaling in redshift.
As also found in the companion \erosita\ studies \citep{grandis24,kleinebreil24,ghirardini24}, we confirm the excessively large intrinsic scatter \sigmarate\ compared to the typical value ($\approx0.3$) obtained for the luminosity-to-mass relation in the literature.
This implies that significant scatter is introduced in measuring the \erosita\ observed count rate converted from the physical flux in X-rays.
Our results are in good agreement with those calibrated with the weak-lensing data from the DES \citep{grandis24} and KiDS \citep{kleinebreil24} surveys.

The resulting \mbcg--\mass--\redshift\ relation is constrained as 
\begin{multline}
\label{eq:bcg_relation}
\left\langle\ln\left(\frac{\mbcghat}{\Msun} \bigg| \mass,\redshift \right)\right\rangle = \left(\axobs\right)~\ln\left(10\right) +
\\
\left( \left(\bxobs\right) + \left(\deltaxobs\right)~\left( \frac{ \frac{\redshift}{\zpiv} }{ \frac{1 + \redshift}{1 + \zpiv} } - 1\right) \right)  \ln\left(\frac{\mass}{\mpiv}\right) + 
\\
\left(\gammaxobs\right)~\ln\left(\frac{1 + \redshift}{1 + \zpiv}\right)
\, , \nonumber
\end{multline}
with the intrinsic scatter of $\sigmabcg = \sigmaxobs$.
Including the modelling of the \mbcg--\mass--\redshift\ relation has no significant impact on that of the \rate--\mass--\redshift\ relation.
The intrinsic scatter of $\sigmabcg = \sigmaxobs$ indicates significant variations in \mbcg\ at a fixed halo mass \mass.
Given the large scatter and the shallow mass slope $\Bbcg = \bxobs$, we conclude that the BCG stellar mass is 
a highly scattering
mass proxy.
Given that the X-ray selection favors high-mass clusters at high redshift, we find strong degeneracy between the power-law indices of the mass (\Bbcg) and redshift (\gammabcg) trends.
To break the degeneracy, we introduce an informative prior on \gammabcg, which results in a steeper mass trend $\Bbcg = \bxobsprioronz$.

We compare our \mbcg--\mass--\redshift\ relation with both observational and simulation-based studies in the literature.
We find satisfactory agreement in the intrinsic scatter and the scaling in the halo mass and redshift, though some studies exhibit systematic offsets in the absolute scale of the BCG stellar mass.
However, these offsets are not significant given the large scatter and systematic uncertainties in \mbcg\ among previous studies.

Our \mbcg--\mass--\redshift\ relation suggests that the BCG stellar mass fraction ($\mbcg/\mass$) at a fixed halo mass ($\mpiv = 1.4\times10^{14}\Msunh$) has remained stable since redshift $\redshift\approx0.8$, with an increase in \mbcg\ at a level of $\approx\left(20\pm8\right)\percent$.
As clusters continue to grow, the BCG stellar mass fraction significantly decreases with increasing total mass.
This result aligns well with the picture of the ``rapid-then-slow'' BCG formation scenario suggested by previous studies \citep[e.g.,][]{delucia07a,lidman12,lin13}.
A combination of the \erass1 sample and those sampling the low-mass regime at high redshift, possibly selected in the optical or with the planned next-generation X-ray missions \citep[e.g.,][]{zhang20,reynolds23,cruise25}, will shed light on the growth of the BCG mass using the statistical approach.

In summary, we present a study of the galaxy population in the context of the BCG stellar mass based on the cluster sample constructed in the first-year database released by \erosita-DE, paving a way for a future work leveraging the final-year sample with the upcoming Stage-IV experiments, such as the Rubin Observatory Legacy Survey of Space and Time \citep{lsst09}, \textit{Euclid} \citep{euclid24}, and the Nancy Grace \textit{Roman} Telescope \citep{dore19}.

%
%

\begin{acknowledgement}

The authors thank anonymous referee for providing useful comments that led to improvements of this paper.
I-Non Chiu acknowledges the support from the National Science and Technology Council in Taiwan (Grant NSTC 111-2112-M-006-037-MY3).
This work made use of the computational and storage resources in the Academic Sinica Institute of Astronomy and Astrophysics (ASIAA) and the National Center for High-Performance Computing (NCHC) in Taiwan.
I-Non Chiu thanks the hospitality of Ting-Wen Lan and Ji-Jia Tang at National Taiwan University (NTU).
This work was supported by JSPS KAKENHI Grant Number JP19KK0076.
VG acknowledges the financial contribution from the contracts Prin-MUR 2022 supported by Next Generation EU (M4.C2.1.1, n.20227RNLY3 \textit{The concordance cosmological model: stress-tests with galaxy clusters}).
E. Bulbul, V. Ghirardini, A. Liu, S. Zelmer, and X. Zhang acknowledge financial support from the European Research Council (ERC) Consolidator Grant under the European Union's Horizon 2020 research and innovation program (grant agreement CoG DarkQuest No 101002585).

This work is based on data from eROSITA, the soft X-ray instrument aboard SRG, a joint Russian-German science mission supported by the Russian Space Agency (Roskosmos), in the interests of the Russian Academy of Sciences represented by its Space Research Institute (IKI), and the Deutsches Zentrum f\"ur Luft- und Raumfahrt (DLR). The SRG spacecraft was built by Lavochkin Association (NPOL) and its subcontractors, and is operated by NPOL with support from the Max Planck Institute for Extraterrestrial Physics (MPE). The development and construction of the eROSITA X-ray instrument was led by MPE, with contributions from the Dr. Karl Remeis Observatory Bamberg \& ECAP (FAU Erlangen-Nuernberg), the University of Hamburg Observatory, the Leibniz Institute for Astrophysics Potsdam (AIP), and the Institute for Astronomy and Astrophysics of the University of Tübingen, with the support of DLR and the Max Planck Society. The Argelander Institute for Astronomy of the University of Bonn and the Ludwig Maximilians Universität Munich also participated in the science preparation for eROSITA.

The Hyper Suprime-Cam (HSC) collaboration includes the astronomical communities of Japan and Taiwan, and Princeton University. The HSC instrumentation and software were developed by the National Astronomical Observatory of Japan (NAOJ), the Kavli Institute for the Physics and Mathematics of the Universe (Kavli IPMU), the University of Tokyo, the High Energy Accelerator Research Organization (KEK), the Academia Sinica Institute for Astronomy and Astrophysics in Taiwan (ASIAA), and Princeton University. Funding was contributed by the FIRST program from the Japanese Cabinet Office, the Ministry of Education, Culture, Sports, Science and Technology (MEXT), the Japan Society for the Promotion of Science (JSPS), Japan Science and Technology Agency (JST), the Toray Science Foundation, NAOJ, Kavli IPMU, KEK, ASIAA, and Princeton University. 

This paper makes use of software developed for Vera C. Rubin Observatory. We thank the Rubin Observatory for making their code available as free software at \url{http://pipelines.lsst.io/}.

This paper is based on data collected at the Subaru Telescope and retrieved from the HSC data archive system, which is operated by the Subaru Telescope and Astronomy Data Center (ADC) at NAOJ. Data analysis was in part carried out with the cooperation of Center for Computational Astrophysics (CfCA), NAOJ. We are honored and grateful for the opportunity of observing the Universe from Maunakea, which has the cultural, historical and natural significance in Hawaii. 

The Pan-STARRS1 Surveys (PS1) and the PS1 public science archive have been made possible through contributions by the Institute for Astronomy, the University of Hawaii, the Pan-STARRS Project Office, the Max Planck Society and its participating institutes, the Max Planck Institute for Astronomy, Heidelberg, and the Max Planck Institute for Extraterrestrial Physics, Garching, The Johns Hopkins University, Durham University, the University of Edinburgh, the Queen’s University Belfast, the Harvard-Smithsonian Center for Astrophysics, the Las Cumbres Observatory Global Telescope Network Incorporated, the National Central University of Taiwan, the Space Telescope Science Institute, the National Aeronautics and Space Administration under grant No. NNX08AR22G issued through the Planetary Science Division of the NASA Science Mission Directorate, the National Science Foundation grant No. AST-1238877, the University of Maryland, Eotvos Lorand University (ELTE), the Los Alamos National Laboratory, and the Gordon and Betty Moore Foundation.

This work is possible because of the efforts in the Vera C. Rubin Observatory \citep{juric17,ivezic19} and PS1 \citep{chambers16, schlafly12, tonry12, magnier13}, and in the HSC \citep{aihara18a} developments including the deep imaging of the COSMOS field \citep{tanaka17}, the on-site quality-assurance system \citep{furusawa18}, the Hyper Suprime-Cam \citep{miyazaki15, miyazaki18, komiyama18}, the design of the filters \citep{kawanomoto18},  the data pipeline \citep{bosch18}, the design of bright-star masks \citep{coupon18}, the characterization of the photometry by the code \texttt{Synpipe} \citep{huang18a}, the photometric redshift estimation \citep{tanaka18}, the shear calibration \citep{mandelbaum18}, and the public data releases \citep{aihara18b, aihara19}.

This work made use of the following software(s): 
\texttt{IPython} \citep{ipython}, 
\texttt{SciPy} \citep{scipy}, 
\texttt{TOPCAT} \citep{topcat1,topcat2}, 
\texttt{matplotlib} \citep{matplotlib}, 
\texttt{Astropy} \citep{astropy}, 
\texttt{NumPy} \citep{van2011numpy}, 
\texttt{Pathos} \citep{pathos}, 
\texttt{pyccl} \citep{chisari19},
\texttt{getdist} \citep{getdist},
and
the \texttt{conda-forge} project \citep{conda_forge_community_2015_4774216}.

\end{acknowledgement}

%
%

\bibliographystyle{aa}
\bibliography{literature}

\begin{thebibliography}{179}
\expandafter\ifx\csname natexlab\endcsname\relax\def\natexlab#1{#1}\fi

\bibitem[{{Adami} {et~al.}(2018){Adami}, {Giles}, {Koulouridis}, {Pacaud},
  {Caretta}, {Pierre}, {Eckert}, {Ramos-Ceja}, {Gastaldello}, {Fotopoulou},
  {Guglielmo}, {Lidman}, {Sadibekova}, {Iovino}, {Maughan}, {Chiappetti},
  {Alis}, {Altieri}, {Baldry}, {Bottini}, {Birkinshaw}, {Bremer}, {Brown},
  {Cucciati}, {Driver}, {Elmer}, {Ettori}, {Evrard}, {Faccioli}, {Granett},
  {Grootes}, {Guzzo}, {Hopkins}, {Horellou}, {Lef{\`e}vre}, {Liske}, {Malek},
  {Marulli}, {Maurogordato}, {Owers}, {Paltani}, {Poggianti}, {Polletta},
  {Plionis}, {Pollo}, {Pompei}, {Ponman}, {Rapetti}, {Ricci}, {Robotham},
  {Tuffs}, {Tasca}, {Valtchanov}, {Vergani}, {Wagner}, {Willis}, \& {XXL
  Consortium}}]{adami18}
{Adami}, C., {Giles}, P., {Koulouridis}, E., {et~al.} 2018, \aap, 620, A5

\bibitem[{{Aihara} {et~al.}(2019){Aihara}, {AlSayyad}, {Ando}, {Armstrong},
  {Bosch}, {Egami}, {Furusawa}, {Furusawa}, {Goulding}, {Harikane}, {Hikage},
  {Ho}, {Hsieh}, {Huang}, {Ikeda}, {Imanishi}, {Ito}, {Iwata}, {Jaelani},
  {Kakuma}, {Kawana}, {Kikuta}, {Kobayashi}, {Koike}, {Komiyama}, {Li},
  {Liang}, {Lin}, {Luo}, {Lupton}, {Lust}, {MacArthur}, {Matsuoka}, {Mineo},
  {Miyatake}, {Miyazaki}, {More}, {Murata}, {Namiki}, {Nishizawa}, {Oguri},
  {Okabe}, {Okamoto}, {Okura}, {Ono}, {Onodera}, {Onoue}, {Osato}, {Ouchi},
  {Shibuya}, {Strauss}, {Sugiyama}, {Suto}, {Takada}, {Takagi}, {Takata},
  {Takita}, {Tanaka}, {Terai}, {Toba}, {Uchiyama}, {Utsumi}, {Wang}, {Wang}, \&
  {Yamada}}]{aihara19}
{Aihara}, H., {AlSayyad}, Y., {Ando}, M., {et~al.} 2019, \pasj, 71, 114

\bibitem[{{Aihara} {et~al.}(2018{\natexlab{a}}){Aihara}, {Arimoto},
  {Armstrong}, {Arnouts}, {Bahcall}, {Bickerton}, {Bosch}, {Bundy}, {Capak},
  {Chan}, {Chiba}, {Coupon}, {Egami}, {Enoki}, {Finet}, {Fujimori}, {Fujimoto},
  {Furusawa}, {Furusawa}, {Goto}, {Goulding}, {Greco}, {Greene}, {Gunn},
  {Hamana}, {Harikane}, {Hashimoto}, {Hattori}, {Hayashi}, {Hayashi},
  {He{\l}miniak}, {Higuchi}, {Hikage}, {Ho}, {Hsieh}, {Huang}, {Huang},
  {Ikeda}, {Imanishi}, {Inoue}, {Iwasawa}, {Iwata}, {Jaelani}, {Jian},
  {Kamata}, {Karoji}, {Kashikawa}, {Katayama}, {Kawanomoto}, {Kayo}, {Koda},
  {Koike}, {Kojima}, {Komiyama}, {Konno}, {Koshida}, {Koyama}, {Kusakabe},
  {Leauthaud}, {Lee}, {Lin}, {Lin}, {Lupton}, {Mandelbaum}, {Matsuoka},
  {Medezinski}, {Mineo}, {Miyama}, {Miyatake}, {Miyazaki}, {Momose}, {More},
  {More}, {Moritani}, {Moriya}, {Morokuma}, {Mukae}, {Murata}, {Murayama},
  {Nagao}, {Nakata}, {Niida}, {Niikura}, {Nishizawa}, {Obuchi}, {Oguri},
  {Oishi}, {Okabe}, {Okamoto}, {Okura}, {Ono}, {Onodera}, {Onoue}, {Osato},
  {Ouchi}, {Price}, {Pyo}, {Sako}, {Sawicki}, {Shibuya}, {Shimasaku},
  {Shimono}, {Shirasaki}, {Silverman}, {Simet}, {Speagle}, {Spergel},
  {Strauss}, {Sugahara}, {Sugiyama}, {Suto}, {Suyu}, {Suzuki}, {Tait},
  {Takada}, {Takata}, {Tamura}, {Tanaka}, {Tanaka}, {Tanaka}, {Tanaka},
  {Terai}, {Terashima}, {Toba}, {Tominaga}, {Toshikawa}, {Turner}, {Uchida},
  {Uchiyama}, {Umetsu}, {Uraguchi}, {Urata}, {Usuda}, {Utsumi}, {Wang}, {Wang},
  {Wong}, {Yabe}, {Yamada}, {Yamanoi}, {Yasuda}, {Yeh}, {Yonehara}, \&
  {Yuma}}]{aihara18a}
{Aihara}, H., {Arimoto}, N., {Armstrong}, R., {et~al.} 2018{\natexlab{a}},
  \pasj, 70, S4

\bibitem[{{Aihara} {et~al.}(2018{\natexlab{b}}){Aihara}, {Armstrong},
  {Bickerton}, {Bosch}, {Coupon}, {Furusawa}, {Hayashi}, {Ikeda}, {Kamata},
  {Karoji}, {Kawanomoto}, {Koike}, {Komiyama}, {Lang}, {Lupton}, {Mineo},
  {Miyatake}, {Miyazaki}, {Morokuma}, {Obuchi}, {Oishi}, {Okura}, {Price},
  {Takata}, {Tanaka}, {Tanaka}, {Tanaka}, {Uchida}, {Uraguchi}, {Utsumi},
  {Wang}, {Yamada}, {Yamanoi}, {Yasuda}, {Arimoto}, {Chiba}, {Finet},
  {Fujimori}, {Fujimoto}, {Furusawa}, {Goto}, {Goulding}, {Gunn}, {Harikane},
  {Hattori}, {Hayashi}, {He{\l}miniak}, {Higuchi}, {Hikage}, {Ho}, {Hsieh},
  {Huang}, {Huang}, {Imanishi}, {Iwata}, {Jaelani}, {Jian}, {Kashikawa},
  {Katayama}, {Kojima}, {Konno}, {Koshida}, {Kusakabe}, {Leauthaud}, {Lee},
  {Lin}, {Lin}, {Mandelbaum}, {Matsuoka}, {Medezinski}, {Miyama}, {Momose},
  {More}, {More}, {Mukae}, {Murata}, {Murayama}, {Nagao}, {Nakata}, {Niida},
  {Niikura}, {Nishizawa}, {Oguri}, {Okabe}, {Ono}, {Onodera}, {Onoue}, {Ouchi},
  {Pyo}, {Shibuya}, {Shimasaku}, {Simet}, {Speagle}, {Spergel}, {Strauss},
  {Sugahara}, {Sugiyama}, {Suto}, {Suzuki}, {Tait}, {Takada}, {Terai}, {Toba},
  {Turner}, {Uchiyama}, {Umetsu}, {Urata}, {Usuda}, {Yeh}, \&
  {Yuma}}]{aihara18b}
{Aihara}, H., {Armstrong}, R., {Bickerton}, S., {et~al.} 2018{\natexlab{b}},
  \pasj, 70, S8

\bibitem[{{Akino} {et~al.}(2022){Akino}, {Eckert}, {Okabe}, {Sereno}, {Umetsu},
  {Oguri}, {Gastaldello}, {Chiu}, {Ettori}, {Evrard}, {Farahi}, {Maughan},
  {Pierre}, {Ricci}, {Valtchanov}, {McCarthy}, {McGee}, {Miyazaki},
  {Nishizawa}, \& {Tanaka}}]{akino22}
{Akino}, D., {Eckert}, D., {Okabe}, N., {et~al.} 2022, \pasj, 74, 175

\bibitem[{{Albareti} \& et~al.(2017)}]{albareti17}
{Albareti}, F.~D. \& et~al. 2017, \apjs, 233, 25

\bibitem[{Allen {et~al.}(2011)Allen, Evrard, \& Mantz}]{allen11}
Allen, S., Evrard, A., \& Mantz, A. 2011, \araa, 49, 409

\bibitem[{Applegate {et~al.}(2014)Applegate, von~der Linden, Kelly, Allen,
  Allen, Burchat, Burke, Ebeling, Mantz, \& Morris}]{applegate14}
Applegate, D., von~der Linden, A., Kelly, P., {et~al.} 2014, \mnras, 439, 48

\bibitem[{Arnouts {et~al.}(1999)Arnouts, Cristiani, Moscardini, Matarrese,
  Lucchin, Fontana, \& Giallongo}]{arnouts99}
Arnouts, S., Cristiani, S., Moscardini, L., {et~al.} 1999, \mnras, 310, 540

\bibitem[{{Artis} {et~al.}(2024{\natexlab{a}}){Artis}, {Bulbul}, {Grandis},
  {Ghirardini}, {Clerc}, {Seppi}, {Comparat}, {Cataneo}, {von der Linden},
  {Bahar}, {Balzer}, {Chiu}, {Gruen}, {Kleinebreil}, {Kluge}, {Krippendorf},
  {Li}, {Liu}, {Malavasi}, {Merloni}, {Miyatake}, {Miyazaki}, {Nandra},
  {Okabe}, {Pacaud}, {Predehl}, {Ramos-Ceja}, {Reiprich}, {Sanders},
  {Schrabback}, {Zelmer}, \& {Zhang}}]{artis24b}
{Artis}, E., {Bulbul}, E., {Grandis}, S., {et~al.} 2024{\natexlab{a}}, arXiv
  e-prints, arXiv:2410.09499

\bibitem[{{Artis} {et~al.}(2024{\natexlab{b}}){Artis}, {Ghirardini}, {Bulbul},
  {Grandis}, {Garrel}, {Clerc}, {Seppi}, {Comparat}, {Cataneo}, {Bahar},
  {Balzer}, {Chiu}, {Gruen}, {Kleinebreil}, {Kluge}, {Krippendorf}, {Li},
  {Liu}, {Merloni}, {Miyatake}, {Miyazaki}, {Nandra}, {Okabe}, {Pacaud},
  {Predehl}, {Ramos-Ceja}, {Reiprich}, {Sanders}, {Schrabback}, {Zelmer}, \&
  {Zhang}}]{artis24a}
{Artis}, E., {Ghirardini}, V., {Bulbul}, E., {et~al.} 2024{\natexlab{b}}, \aap,
  691, A301

\bibitem[{{Astropy Collaboration} {et~al.}(2013){Astropy Collaboration},
  {Robitaille}, {Tollerud}, {Greenfield}, {Droettboom}, {Bray}, {Aldcroft},
  {Davis}, {Ginsburg}, {Price-Whelan}, {Kerzendorf}, {Conley}, {Crighton},
  {Barbary}, {Muna}, {Ferguson}, {Grollier}, {Parikh}, {Nair}, {Unther},
  {Deil}, {Woillez}, {Conseil}, {Kramer}, {Turner}, {Singer}, {Fox}, {Weaver},
  {Zabalza}, {Edwards}, {Azalee Bostroem}, {Burke}, {Casey}, {Crawford},
  {Dencheva}, {Ely}, {Jenness}, {Labrie}, {Lim}, {Pierfederici}, {Pontzen},
  {Ptak}, {Refsdal}, {Servillat}, \& {Streicher}}]{astropy}
{Astropy Collaboration}, {Robitaille}, T.~P., {Tollerud}, E.~J., {et~al.} 2013,
  \aap, 558, A33

\bibitem[{{Bahar} {et~al.}(2022){Bahar}, {Bulbul}, {Clerc}, {Ghirardini},
  {Liu}, {Nandra}, {Pacaud}, {Chiu}, {Comparat}, {Ider-Chitham}, {Klein},
  {Liu}, {Merloni}, {Migkas}, {Okabe}, {Ramos-Ceja}, {Reiprich}, {Sanders}, \&
  {Schrabback}}]{bahar22}
{Bahar}, Y.~E., {Bulbul}, E., {Clerc}, N., {et~al.} 2022, \aap, 661, A7

\bibitem[{{Bahar} {et~al.}(2024){Bahar}, {Bulbul}, {Ghirardini}, {Sanders},
  {Zhang}, {Liu}, {Clerc}, {Artis}, {Balzer}, {Biffi}, {Bose}, {Comparat},
  {Dolag}, {Garrel}, {Hadzhiyska}, {Hern{\'a}ndez-Aguayo}, {Hernquist},
  {Kluge}, {Krippendorf}, {Merloni}, {Nandra}, {Pakmor}, {Popesso},
  {Ramos-Ceja}, {Seppi}, {Springel}, {Weller}, \& {Zelmer}}]{bahar24}
{Bahar}, Y.~E., {Bulbul}, E., {Ghirardini}, V., {et~al.} 2024, \aap, 691, A188

\bibitem[{{Baldry} {et~al.}(2008){Baldry}, {Glazebrook}, \&
  {Driver}}]{baldry08}
{Baldry}, I.~K., {Glazebrook}, K., \& {Driver}, S.~P. 2008, \mnras, 388, 945

\bibitem[{{Bardeen} {et~al.}(1986){Bardeen}, {Bond}, {Kaiser}, \&
  {Szalay}}]{bardeen86}
{Bardeen}, J.~M., {Bond}, J.~R., {Kaiser}, N., \& {Szalay}, A.~S. 1986, \apj,
  304, 15

\bibitem[{{Behroozi} {et~al.}(2013){Behroozi}, {Wechsler}, {Wu}, {Busha},
  {Klypin}, \& {Primack}}]{behroozi13}
{Behroozi}, P.~S., {Wechsler}, R.~H., {Wu}, H.-Y., {et~al.} 2013, \apj, 763, 18

\bibitem[{{Bellagamba} {et~al.}(2019){Bellagamba}, {Sereno}, {Roncarelli},
  {Maturi}, {Radovich}, {Bardelli}, {Puddu}, {Moscardini}, {Getman},
  {Hildebrandt}, \& {Napolitano}}]{bellagamba19}
{Bellagamba}, F., {Sereno}, M., {Roncarelli}, M., {et~al.} 2019, \mnras, 484,
  1598

\bibitem[{{Bocquet} {et~al.}(2019){Bocquet}, {Dietrich}, {Schrabback}, {Bleem},
  {Klein}, {Allen}, {Applegate}, {Ashby}, {Bautz}, {Bayliss}, {Benson},
  {Brodwin}, {Bulbul}, {Canning}, {Capasso}, {Carlstrom}, {Chang}, {Chiu},
  {Cho}, {Clocchiatti}, {Crawford}, {Crites}, {de Haan}, {Desai}, {Dobbs},
  {Foley}, {Forman}, {Garmire}, {George}, {Gladders}, {Gonzalez}, {Grandis},
  {Gupta}, {Halverson}, {Hlavacek-Larrondo}, {Hoekstra}, {Holder}, {Holzapfel},
  {Hou}, {Hrubes}, {Huang}, {Jones}, {Khullar}, {Knox}, {Kraft}, {Lee}, {von
  der Linden}, {Luong-Van}, {Mantz}, {Marrone}, {McDonald}, {McMahon}, {Meyer},
  {Mocanu}, {Mohr}, {Morris}, {Padin}, {Patil}, {Pryke}, {Rapetti},
  {Reichardt}, {Rest}, {Ruhl}, {Saliwanchik}, {Saro}, {Sayre}, {Schaffer},
  {Shirokoff}, {Stalder}, {Stanford}, {Staniszewski}, {Stark}, {Story},
  {Strazzullo}, {Stubbs}, {Vanderlinde}, {Vieira}, {Vikhlinin}, {Williamson},
  \& {Zenteno}}]{bocquet19}
{Bocquet}, S., {Dietrich}, J.~P., {Schrabback}, T., {et~al.} 2019, \apj, 878,
  55

\bibitem[{{Bocquet} {et~al.}(2024{\natexlab{a}}){Bocquet}, {Grandis}, {Bleem},
  {Klein}, {Mohr}, {Aguena}, {Alarcon}, {Allam}, {Allen}, {Alves}, {Amon},
  {Ansarinejad}, {Bacon}, {Bayliss}, {Bechtol}, {Becker}, {Benson},
  {Bernstein}, {Brodwin}, {Brooks}, {Campos}, {Canning}, {Carlstrom}, {Carnero
  Rosell}, {Carrasco Kind}, {Carretero}, {Cawthon}, {Chang}, {Chen}, {Choi},
  {Cordero}, {Costanzi}, {da Costa}, {Pereira}, {Davis}, {DeRose}, {Desai}, {de
  Haan}, {De Vicente}, {Diehl}, {Dodelson}, {Doel}, {Doux}, {Drlica-Wagner},
  {Eckert}, {Elvin-Poole}, {Everett}, {Ferrero}, {Fert{\'e}}, {Flores},
  {Frieman}, {Garc{\'\i}a-Bellido}, {Gatti}, {Giannini}, {Gladders}, {Gruen},
  {Gruendl}, {Harrison}, {Hartley}, {Herner}, {Hinton}, {Hollowood},
  {Holzapfel}, {Honscheid}, {Huang}, {Huff}, {James}, {Jarvis}, {Khullar},
  {Kim}, {Kraft}, {Kuehn}, {Kuropatkin}, {K{\'e}ruzor{\'e}}, {Lee}, {Leget},
  {MacCrann}, {Mahler}, {Mantz}, {Marshall}, {McCullough}, {McDonald},
  {Mena-Fern{\'a}ndez}, {Miquel}, {Myles}, {Navarro-Alsina}, {Ogando},
  {Palmese}, {Pandey}, {Pieres}, {Plazas Malag{\'o}n}, {Prat}, {Raveri},
  {Reichardt}, {Roberson}, {Rollins}, {Romer}, {Romero}, {Roodman}, {Ross},
  {Rykoff}, {Salvati}, {S{\'a}nchez}, {Sanchez}, {Sanchez Cid}, {Saro},
  {Schrabback}, {Schubnell}, {Secco}, {Sevilla-Noarbe}, {Sharon}, {Sheldon},
  {Shin}, {Smith}, {Somboonpanyakul}, {Stalder}, {Stark}, {Strazzullo},
  {Suchyta}, {Swanson}, {Tarle}, {To}, {Troxel}, {Tutusaus}, {Varga}, {von der
  Linden}, {Weaverdyck}, {Weller}, {Wiseman}, {Yanny}, {Yin}, {Young}, {Zhang},
  {Zuntz}, {(The DES}, \& {SPT Collaborations)}}]{bocquet24a}
{Bocquet}, S., {Grandis}, S., {Bleem}, L.~E., {et~al.} 2024{\natexlab{a}},
  \prd, 110, 083509

\bibitem[{{Bocquet} {et~al.}(2024{\natexlab{b}}){Bocquet}, {Grandis}, {Bleem},
  {Klein}, {Mohr}, {Schrabback}, {Abbott}, {Ade}, {Aguena}, {Alarcon}, {Allam},
  {Allen}, {Alves}, {Amon}, {Anderson}, {Annis}, {Ansarinejad}, {Austermann},
  {Avila}, {Bacon}, {Bayliss}, {Beall}, {Bechtol}, {Becker}, {Bender},
  {Benson}, {Bernstein}, {Bhargava}, {Bianchini}, {Brodwin}, {Brooks},
  {Bryant}, {Campos}, {Canning}, {Carlstrom}, {Carnero Rosell}, {Carrasco
  Kind}, {Carretero}, {Castander}, {Cawthon}, {Chang}, {Chang}, {Chaubal},
  {Chen}, {Chiang}, {Choi}, {Chou}, {Citron}, {Corbett Moran}, {Cordero},
  {Costanzi}, {Crawford}, {Crites}, {da Costa}, {Pereira}, {Davis}, {Davis},
  {DeRose}, {Desai}, {de Haan}, {Diehl}, {Dobbs}, {Dodelson}, {Doux},
  {Drlica-Wagner}, {Eckert}, {Elvin-Poole}, {Everett}, {Everett}, {Ferrero},
  {Fert{\'e}}, {Flores}, {Frieman}, {Gallicchio}, {Garc{\'\i}a-Bellido},
  {Gatti}, {George}, {Giannini}, {Gladders}, {Gruen}, {Gruendl}, {Gupta},
  {Gutierrez}, {Halverson}, {Harrison}, {Hartley}, {Herner}, {Hinton},
  {Holder}, {Hollowood}, {Holzapfel}, {Honscheid}, {Hrubes}, {Huang},
  {Hubmayr}, {Huff}, {Huterer}, {Irwin}, {James}, {Jarvis}, {Khullar}, {Kim},
  {Knox}, {Kraft}, {Krause}, {Kuehn}, {Kuropatkin}, {K{\'e}ruzor{\'e}},
  {Lahav}, {Lee}, {Leget}, {Li}, {Lin}, {Lowitz}, {MacCrann}, {Mahler},
  {Mantz}, {Marshall}, {McCullough}, {McDonald}, {McMahon},
  {Mena-Fern{\'a}ndez}, {Menanteau}, {Meyer}, {Miquel}, {Montgomery}, {Myles},
  {Natoli}, {Navarro-Alsina}, {Nibarger}, {Noble}, {Novosad}, {Ogando},
  {Omori}, {Padin}, {Pandey}, {Paschos}, {Patil}, {Pieres}, {Plazas
  Malag{\'o}n}, {Porredon}, {Prat}, {Pryke}, {Raveri}, {Reichardt}, {Roberson},
  {Rollins}, {Romero}, {Roodman}, {Ruhl}, {Rykoff}, {Saliwanchik}, {Salvati},
  {S{\'a}nchez}, {Sanchez}, {Sanchez Cid}, {Saro}, {Schaffer}, {Secco},
  {Sevilla-Noarbe}, {Sharon}, {Sheldon}, {Shin}, {Sievers}, {Smecher}, {Smith},
  {Somboonpanyakul}, {Sommer}, {Stalder}, {Stark}, {Stephen}, {Strazzullo},
  {Suchyta}, {Tarle}, {To}, {Troxel}, {Tucker}, {Tutusaus}, {Varga}, {Veach},
  {Vieira}, {Vikhlinin}, {von der Linden}, {Wang}, {Weaverdyck}, {Weller},
  {Whitehorn}, {Wu}, {Yanny}, {Yefremenko}, {Yin}, {Young}, {Zebrowski},
  {Zhang}, {Zohren}, {Zuntz}, {(SPT}, \& {DES Collaborations)}}]{bocquet24b}
{Bocquet}, S., {Grandis}, S., {Bleem}, L.~E., {et~al.} 2024{\natexlab{b}},
  \prd, 110, 083510

\bibitem[{{B{\"o}hringer} {et~al.}(2012){B{\"o}hringer}, {Dolag}, \&
  {Chon}}]{boehringer12}
{B{\"o}hringer}, H., {Dolag}, K., \& {Chon}, G. 2012, \aap, 539, A120

\bibitem[{{Bosch} {et~al.}(2018){Bosch}, {Armstrong}, {Bickerton}, {Furusawa},
  {Ikeda}, {Koike}, {Lupton}, {Mineo}, {Price}, {Takata}, {Tanaka}, {Yasuda},
  {AlSayyad}, {Becker}, {Coulton}, {Coupon}, {Garmilla}, {Huang}, {Krughoff},
  {Lang}, {Leauthaud}, {Lim}, {Lust}, {MacArthur}, {Mandelbaum}, {Miyatake},
  {Miyazaki}, {Murata}, {More}, {Okura}, {Owen}, {Swinbank}, {Strauss},
  {Yamada}, \& {Yamanoi}}]{bosch18}
{Bosch}, J., {Armstrong}, R., {Bickerton}, S., {et~al.} 2018, \pasj, 70, S5

\bibitem[{{Bouwens} {et~al.}(2012){Bouwens}, {Illingworth}, {Oesch}, {Franx},
  {Labb{\'e}}, {Trenti}, {van Dokkum}, {Carollo}, {Gonz{\'a}lez}, {Smit}, \&
  {Magee}}]{bouwens12}
{Bouwens}, R.~J., {Illingworth}, G.~D., {Oesch}, P.~A., {et~al.} 2012, \apj,
  754, 83

\bibitem[{Bruzual \& Charlot(2003)}]{bruzual03}
Bruzual, G. \& Charlot, S. 2003, \mnras, 344, 1000

\bibitem[{{Bulbul} {et~al.}(2019){Bulbul}, {Chiu}, {Mohr}, {McDonald},
  {Benson}, {Bautz}, {Bayliss}, {Bleem}, {Brodwin}, {Bocquet}, {Capasso},
  {Dietrich}, {Forman}, {Hlavacek-Larrondo}, {Holzapfel}, {Khullar}, {Klein},
  {Kraft}, {Miller}, {Reichardt}, {Saro}, {Sharon}, {Stalder}, {Schrabback}, \&
  {Stanford}}]{bulbul19}
{Bulbul}, E., {Chiu}, I.~N., {Mohr}, J.~J., {et~al.} 2019, \apj, 871, 50

\bibitem[{{Bulbul} {et~al.}(2024){Bulbul}, {Liu}, {Kluge}, {Zhang}, {Sanders},
  {Bahar}, {Ghirardini}, {Artis}, {Seppi}, {Garrel}, {Ramos-Ceja}, {Comparat},
  {Balzer}, {B{\"o}ckmann}, {Br{\"u}ggen}, {Clerc}, {Dennerl}, {Dolag},
  {Freyberg}, {Grandis}, {Gruen}, {Kleinebreil}, {Krippendorf}, {Lamer},
  {Merloni}, {Migkas}, {Nandra}, {Pacaud}, {Predehl}, {Reiprich}, {Schrabback},
  {Veronica}, {Weller}, \& {Zelmer}}]{bulbul24}
{Bulbul}, E., {Liu}, A., {Kluge}, M., {et~al.} 2024, \aap, 685, A106

\bibitem[{Calzetti {et~al.}(2000)Calzetti, Armus, Bohlin, Kinney, Koornneef, \&
  Storchi-Bergmann}]{calzetti00}
Calzetti, D., Armus, L., Bohlin, R., {et~al.} 2000, \apj, 533, 682

\bibitem[{Chabrier(2003)}]{chabrier03}
Chabrier, G. 2003, \pasp, 115, 763

\bibitem[{{Chambers} {et~al.}(2016){Chambers}, {Magnier}, {Metcalfe},
  {Flewelling}, {Huber}, {Waters}, {Denneau}, {Draper}, {Farrow}, {Finkbeiner},
  {Holmberg}, {Koppenhoefer}, {Price}, {Rest}, {Saglia}, {Schlafly}, {Smartt},
  {Sweeney}, {Wainscoat}, {Burgett}, {Chastel}, {Grav}, {Heasley}, {Hodapp},
  {Jedicke}, {Kaiser}, {Kudritzki}, {Luppino}, {Lupton}, {Monet}, {Morgan},
  {Onaka}, {Shiao}, {Stubbs}, {Tonry}, {White}, {Ba{\~n}ados}, {Bell},
  {Bender}, {Bernard}, {Boegner}, {Boffi}, {Botticella}, {Calamida},
  {Casertano}, {Chen}, {Chen}, {Cole}, {Deacon}, {Frenk}, {Fitzsimmons},
  {Gezari}, {Gibbs}, {Goessl}, {Goggia}, {Gourgue}, {Goldman}, {Grant},
  {Grebel}, {Hambly}, {Hasinger}, {Heavens}, {Heckman}, {Henderson}, {Henning},
  {Holman}, {Hopp}, {Ip}, {Isani}, {Jackson}, {Keyes}, {Koekemoer}, {Kotak},
  {Le}, {Liska}, {Long}, {Lucey}, {Liu}, {Martin}, {Masci}, {McLean}, {Mindel},
  {Misra}, {Morganson}, {Murphy}, {Obaika}, {Narayan}, {Nieto-Santisteban},
  {Norberg}, {Peacock}, {Pier}, {Postman}, {Primak}, {Rae}, {Rai}, {Riess},
  {Riffeser}, {Rix}, {R{\"o}ser}, {Russel}, {Rutz}, {Schilbach}, {Schultz},
  {Scolnic}, {Strolger}, {Szalay}, {Seitz}, {Small}, {Smith}, {Soderblom},
  {Taylor}, {Thomson}, {Taylor}, {Thakar}, {Thiel}, {Thilker}, {Unger},
  {Urata}, {Valenti}, {Wagner}, {Walder}, {Walter}, {Watters}, {Werner},
  {Wood-Vasey}, \& {Wyse}}]{chambers16}
{Chambers}, K.~C., {Magnier}, E.~A., {Metcalfe}, N., {et~al.} 2016, arXiv
  e-prints, arXiv:1612.05560

\bibitem[{{Chen} {et~al.}(2025){Chen}, {Chiu}, {Oguri}, {Lin}, {Miyatake},
  {Miyazaki}, {More}, {Hamana}, {Rau}, {Sunayama}, {Sugiyama}, \&
  {Takada}}]{chen25}
{Chen}, K.-F., {Chiu}, I.~N., {Oguri}, M., {et~al.} 2025, The Open Journal of
  Astrophysics, 8, 2

\bibitem[{{Chen} {et~al.}(2022){Chen}, {Zu}, {Shao}, \& {Shan}}]{chen22}
{Chen}, X., {Zu}, Y., {Shao}, Z., \& {Shan}, H. 2022, \mnras, 514, 2692

\bibitem[{{Chisari} {et~al.}(2019){Chisari}, {Alonso}, {Krause}, {Leonard},
  {Bull}, {Neveu}, {Villarreal}, {Singh}, {McClintock}, {Ellison}, {Du},
  {Zuntz}, {Mead}, {Joudaki}, {Lorenz}, {Tr{\"o}ster}, {Sanchez}, {Lanusse},
  {Ishak}, {Hlozek}, {Blazek}, {Campagne}, {Almoubayyed}, {Eifler}, {Kirby},
  {Kirkby}, {Plaszczynski}, {Slosar}, {Vrastil}, {Wagoner}, \& {LSST Dark
  Energy Science Collaboration}}]{chisari19}
{Chisari}, N.~E., {Alonso}, D., {Krause}, E., {et~al.} 2019, \apjs, 242, 2

\bibitem[{Chiu {et~al.}(2016)Chiu, Mohr, McDonald, Bocquet, Ashby, Bayliss,
  Benson, Bleem, Brodwin, Desai, Dietrich, Forman, Gangkofner, Gonzalez,
  Hennig, Liu, Reichardt, Saro, Stalder, Stanford, Song, Schrabback,
  {\v{S}}uhada, Strazzullo, \& Zenteno}]{chiu16a}
Chiu, I., Mohr, J., McDonald, M., {et~al.} 2016, \mnras, 455, 258

\bibitem[{{Chiu} {et~al.}(2018){Chiu}, {Mohr}, {McDonald}, {Bocquet}, {Desai},
  {Klein}, {Israel}, {Ashby}, {Stanford}, {Benson}, {Brodwin}, {Abbott},
  {Abdalla}, {Allam}, {Annis}, {Bayliss}, {Benoit-L{\'e}vy}, {Bertin}, {Bleem},
  {Brooks}, {Buckley-Geer}, {Bulbul}, {Capasso}, {Carlstrom}, {Rosell},
  {Carretero}, {Castander}, {Cunha}, {D'Andrea}, {da Costa}, {Davis}, {Diehl},
  {Dietrich}, {Doel}, {Drlica-Wagner}, {Eifler}, {Evrard}, {Flaugher},
  {Garc{\'{\i}}a-Bellido}, {Garmire}, {Gaztanaga}, {Gerdes}, {Gonzalez},
  {Gruen}, {Gruendl}, {Gschwend}, {Gupta}, {Gutierrez}, {Hlavacek-L},
  {Honscheid}, {James}, {Jeltema}, {Kraft}, {Krause}, {Kuehn}, {Kuhlmann},
  {Kuropatkin}, {Lahav}, {Lima}, {Maia}, {Marshall}, {Melchior}, {Menanteau},
  {Miquel}, {Murray}, {Nord}, {Ogando}, {Plazas}, {Rapetti}, {Reichardt},
  {Romer}, {Roodman}, {Sanchez}, {Saro}, {Scarpine}, {Schindler}, {Schubnell},
  {Sharon}, {Smith}, {Smith}, {Soares-Santos}, {Sobreira}, {Stalder}, {Stern},
  {Strazzullo}, {Suchyta}, {Swanson}, {Tarle}, {Vikram}, {Walker}, {Weller}, \&
  {Zhang}}]{chiu18a}
{Chiu}, I., {Mohr}, J.~J., {McDonald}, M., {et~al.} 2018, \mnras, 478, 3072

\bibitem[{{Chiu} {et~al.}(2024){Chiu}, {Chen}, {Oguri}, {Rau}, {Hamana}, {Lin},
  {Miyatake}, {Miyazaki}, {More}, {Sunayama}, {Sugiyama}, \& {Takada}}]{chiu24}
{Chiu}, I.~N., {Chen}, K.-F., {Oguri}, M., {et~al.} 2024, The Open Journal of
  Astrophysics, 7, 90

\bibitem[{{Chiu} {et~al.}(2022){Chiu}, {Ghirardini}, {Liu}, {Grandis},
  {Bulbul}, {Bahar}, {Comparat}, {Bocquet}, {Clerc}, {Klein}, {Liu}, {Li},
  {Miyatake}, {Mohr}, {More}, {Oguri}, {Okabe}, {Pacaud}, {Ramos-Ceja},
  {Reiprich}, {Schrabback}, \& {Umetsu}}]{chiu22}
{Chiu}, I.~N., {Ghirardini}, V., {Liu}, A., {et~al.} 2022, \aap, 661, A11

\bibitem[{{Chiu} {et~al.}(2023){Chiu}, {Klein}, {Mohr}, \& {Bocquet}}]{chiu23}
{Chiu}, I.~N., {Klein}, M., {Mohr}, J., \& {Bocquet}, S. 2023, \mnras, 522,
  1601

\bibitem[{{Clerc} {et~al.}(2024){Clerc}, {Comparat}, {Seppi}, {Artis}, {Bahar},
  {Balzer}, {Bulbul}, {Dauser}, {Garrel}, {Ghirardini}, {Grandis}, {Kirsch},
  {Kluge}, {Liu}, {Pacaud}, {Ramos-Ceja}, {Reiprich}, {Sanders}, {Wilms}, \&
  {Zhang}}]{clerc24}
{Clerc}, N., {Comparat}, J., {Seppi}, R., {et~al.} 2024, \aap, 687, A238

\bibitem[{{Comparat} {et~al.}(2020){Comparat}, {Eckert}, {Finoguenov},
  {Schmidt}, {Sanders}, {Nagai}, {Lau}, {K{\"a}}, {fer}, {Pacaud}, {Clerc},
  {Reiprich}, {Bulbul}, {Chitham}, {Chiang}, {Ghirardini}, {Gonzalez-Perez},
  {Gozaliasl}, {Fitzpatrick}, {Klypin}, {Merloni}, {Nandra}, {Liu}, {Prada},
  {Ramos-Ceja}, {Salvato}, {Seppi}, {Tempel}, \& {Yepes}}]{comparat20}
{Comparat}, J., {Eckert}, D., {Finoguenov}, A., {et~al.} 2020, The Open Journal
  of Astrophysics, 3, 13

\bibitem[{{Comparat} {et~al.}(2019){Comparat}, {Merloni}, {Salvato}, {Nandra},
  {Boller}, {Georgakakis}, {Finoguenov}, {Dwelly}, {Buchner}, {Del Moro},
  {Clerc}, {Wang}, {Zhao}, {Prada}, {Yepes}, {Brusa}, {Krumpe}, \&
  {Liu}}]{comparat19}
{Comparat}, J., {Merloni}, A., {Salvato}, M., {et~al.} 2019, \mnras, 487, 2005

\bibitem[{conda-forge community(2015)}]{conda_forge_community_2015_4774216}
conda-forge community. 2015, {The conda-forge Project: Community-based Software
  Distribution Built on the conda Package Format and Ecosystem}

\bibitem[{{Conroy} {et~al.}(2006){Conroy}, {Wechsler}, \&
  {Kravtsov}}]{conroy06}
{Conroy}, C., {Wechsler}, R.~H., \& {Kravtsov}, A.~V. 2006, \apj, 647, 201

\bibitem[{{Contini} {et~al.}(2014){Contini}, {De Lucia}, {Villalobos}, \&
  {Borgani}}]{contini14}
{Contini}, E., {De Lucia}, G., {Villalobos}, {\'A}., \& {Borgani}, S. 2014,
  \mnras, 437, 3787

\bibitem[{{Costanzi} {et~al.}(2019{\natexlab{a}}){Costanzi}, {Rozo}, {Rykoff},
  {Farahi}, {Jeltema}, {Evrard}, {Mantz}, {Gruen}, {Mandelbaum}, {DeRose},
  {McClintock}, {Varga}, {Zhang}, {Weller}, {Wechsler}, \&
  {Aguena}}]{costanzi19a}
{Costanzi}, M., {Rozo}, E., {Rykoff}, E.~S., {et~al.} 2019{\natexlab{a}},
  \mnras, 482, 490

\bibitem[{{Costanzi} {et~al.}(2019{\natexlab{b}}){Costanzi}, {Rozo}, {Simet},
  {Zhang}, {Evrard}, {Mantz}, {Rykoff}, {Jeltema}, {Gruen}, {Allen},
  {McClintock}, {Romer}, {von der Linden}, {Farahi}, {DeRose}, {Varga},
  {Weller}, {Giles}, {Hollowood}, {Bhargava}, {Bermeo-Hernandez}, {Chen},
  {Abbott}, {Abdalla}, {Avila}, {Bechtol}, {Brooks}, {Buckley-Geer}, {Burke},
  {Rosell}, {Kind}, {Carretero}, {Crocce}, {Cunha}, {da Costa}, {Davis}, {De
  Vicente}, {Diehl}, {Dietrich}, {Doel}, {Eifler}, {Estrada}, {Flaugher},
  {Fosalba}, {Frieman}, {Garc{\'\i}a-Bellido}, {Gaztanaga}, {Gerdes},
  {Giannantonio}, {Gruendl}, {Gschwend}, {Gutierrez}, {Hartley}, {Honscheid},
  {Hoyle}, {James}, {Krause}, {Kuehn}, {Kuropatkin}, {Lima}, {Lin}, {Maia},
  {March}, {Marshall}, {Martini}, {Menanteau}, {Miller}, {Miquel}, {Mohr},
  {Ogando}, {Plazas}, {Roodman}, {Sanchez}, {Scarpine}, {Schindler},
  {Schubnell}, {Serrano}, {Sevilla-Noarbe}, {Sheldon}, {Smith},
  {Soares-Santos}, {Sobreira}, {Suchyta}, {Swanson}, {Tarle}, {Thomas}, \&
  {Wechsler}}]{costanzi19b}
{Costanzi}, M., {Rozo}, E., {Simet}, M., {et~al.} 2019{\natexlab{b}}, \mnras,
  488, 4779

\bibitem[{{Coupon} {et~al.}(2015){Coupon}, {Arnouts}, {van Waerbeke},
  {Moutard}, {Ilbert}, {van Uitert}, {Erben}, {Garilli}, {Guzzo}, {Heymans},
  {Hildebrandt}, {Hoekstra}, {Kilbinger}, {Kitching}, {Mellier}, {Miller},
  {Scodeggio}, {Bonnett}, {Branchini}, {Davidzon}, {De Lucia}, {Fritz}, {Fu},
  {Hudelot}, {Hudson}, {Kuijken}, {Leauthaud}, {Le F{\`e}vre}, {McCracken},
  {Moscardini}, {Rowe}, {Schrabback}, {Semboloni}, \& {Velander}}]{coupon15}
{Coupon}, J., {Arnouts}, S., {van Waerbeke}, L., {et~al.} 2015, \mnras, 449,
  1352

\bibitem[{{Coupon} {et~al.}(2018){Coupon}, {Czakon}, {Bosch}, {Komiyama},
  {Medezinski}, {Miyazaki}, \& {Oguri}}]{coupon18}
{Coupon}, J., {Czakon}, N., {Bosch}, J., {et~al.} 2018, \pasj, 70, S7

\bibitem[{{Cruise} {et~al.}(2025){Cruise}, {Guainazzi}, {Aird}, {Carrera},
  {Costantini}, {Corrales}, {Dauser}, {Eckert}, {Gastaldello}, {Matsumoto},
  {Osten}, {Petrucci}, {Porquet}, {Pratt}, {Rea}, {Reiprich}, {Simionescu},
  {Spiga}, \& {Troja}}]{cruise25}
{Cruise}, M., {Guainazzi}, M., {Aird}, J., {et~al.} 2025, Nature Astronomy, 9,
  36

\bibitem[{{Cui} {et~al.}(2022){Cui}, {Dave}, {Knebe}, {Rasia}, {Gray},
  {Pearce}, {Power}, {Yepes}, {Anbajagane}, {Ceverino}, {Contreras-Santos}, {de
  Andres}, {De Petris}, {Ettori}, {Haggar}, {Li}, {Wang}, {Yang}, {Borgani},
  {Dolag}, {Zu}, {Kuchner}, {Ca{\~n}as}, {Ferragamo}, \& {Gianfagna}}]{cui22}
{Cui}, W., {Dave}, R., {Knebe}, A., {et~al.} 2022, \mnras, 514, 977

\bibitem[{{Dalal} {et~al.}(2023){Dalal}, {Li}, {Nicola}, {Zuntz}, {Strauss},
  {Sugiyama}, {Zhang}, {Rau}, {Mandelbaum}, {Takada}, {More}, {Miyatake},
  {Kannawadi}, {Shirasaki}, {Taniguchi}, {Takahashi}, {Osato}, {Hamana},
  {Oguri}, {Nishizawa}, {Malag{\'o}n}, {Sunayama}, {Alonso}, {Slosar}, {Luo},
  {Armstrong}, {Bosch}, {Hsieh}, {Komiyama}, {Lupton}, {Lust}, {MacArthur},
  {Miyazaki}, {Murayama}, {Nishimichi}, {Okura}, {Price}, {Tait}, {Tanaka}, \&
  {Wang}}]{dalal23}
{Dalal}, R., {Li}, X., {Nicola}, A., {et~al.} 2023, \prd, 108, 123519

\bibitem[{{De Lucia} \& Blaizot(2007)}]{delucia07a}
{De Lucia}, G. \& Blaizot, J. 2007, \mnras, 375, 2

\bibitem[{{de Vaucouleurs}(1948)}]{devaucouleurs1948}
{de Vaucouleurs}, G. 1948, Annales d'Astrophysique, 11, 247

\bibitem[{{DeMaio} {et~al.}(2018){DeMaio}, {Gonzalez}, {Zabludoff}, {Zaritsky},
  {Connor}, {Donahue}, \& {Mulchaey}}]{demaio18}
{DeMaio}, T., {Gonzalez}, A.~H., {Zabludoff}, A., {et~al.} 2018, \mnras, 474,
  3009

\bibitem[{{Diemer}(2018)}]{diemer18}
{Diemer}, B. 2018, \apjs, 239, 35

\bibitem[{{Dietrich} {et~al.}(2019){Dietrich}, {Bocquet}, {Schrabback},
  {Applegate}, {Hoekstra}, {Grandis}, {Mohr}, {Allen}, {Bayliss}, {Benson},
  {Bleem}, {Brodwin}, {Bulbul}, {Capasso}, {Chiu}, {Crawford}, {Gonzalez}, {de
  Haan}, {Klein}, {von der Linden}, {Mantz}, {Marrone}, {McDonald},
  {Raghunathan}, {Rapetti}, {Reichardt}, {Saro}, {Stalder}, {Stark}, {Stern},
  \& {Stubbs}}]{dietrich19}
{Dietrich}, J.~P., {Bocquet}, S., {Schrabback}, T., {et~al.} 2019, \mnras, 483,
  2871

\bibitem[{{Dore} {et~al.}(2019){Dore}, {Hirata}, {Wang}, {Weinberg}, {Eifler},
  {Foley}, {Heinrich}, {Krause}, {Perlmutter}, {Pisani}, {Scolnic}, {Spergel},
  {Suntzeff}, {Aldering}, {Baltay}, {Capak}, {Choi}, {Dvorkin}, {Fall}, {Fang},
  {Fruchter}, {Galbany}, {Ho}, {Hounsell}, {Izard}, {Jain}, {Koekemoer},
  {Kruk}, {Leauthaud}, {Malhotra}, {Mandelbaum}, {Massara}, {Masters},
  {Miyatake}, {Plazas}, {Rhoads}, {Rhodes}, {Rose}, {Rubin}, {Sako},
  {Samushia}, {Shirasaki}, {Simet}, {Takada}, {Troxel}, {Wu}, {Yoshida}, \&
  {Zhai}}]{dore19}
{Dore}, O., {Hirata}, C., {Wang}, Y., {et~al.} 2019, \baas, 51, 341

\bibitem[{Eddington(1913)}]{eddington13}
Eddington, A. 1913, \mnras, 73, 359

\bibitem[{{Erfanianfar} {et~al.}(2019){Erfanianfar}, {Finoguenov}, {Furnell},
  {Popesso}, {Biviano}, {Wuyts}, {Collins}, {Mirkazemi}, {Comparat},
  {Khosroshahi}, {Nandra}, {Capasso}, {Rykoff}, {Wilman}, {Merloni}, {Clerc},
  {Salvato}, {Chitham}, {Kelvin}, {Gozaliasl}, {Weijmans}, {Brownstein},
  {Egami}, {Pereira}, {Schneider}, {Kirkpatrick}, {Damsted}, \&
  {Kukkola}}]{erfanianfar19}
{Erfanianfar}, G., {Finoguenov}, A., {Furnell}, K., {et~al.} 2019, \aap, 631,
  A175

\bibitem[{{Euclid Collaboration} {et~al.}(2024){Euclid Collaboration},
  {Mellier}, {Abdurro'uf}, {Acevedo Barroso}, {Ach{\'u}carro}, {Adamek},
  {Adam}, {Addison}, {Aghanim}, {Aguena}, {Ajani}, {Akrami}, {Al-Bahlawan},
  {Alavi}, {Albuquerque}, {Alestas}, {Alguero}, {Allaoui}, {Allen}, {Allevato},
  {Alonso-Tetilla}, {Altieri}, {Alvarez-Candal}, {Alvi}, {Amara}, {Amendola},
  {Amiaux}, {Andika}, {Andreon}, {Andrews}, {Angora}, {Angulo}, {Annibali},
  {Anselmi}, {Anselmi}, {Arcari}, {Archidiacono}, {Aric{\`o}}, {Arnaud},
  {Arnouts}, {Asgari}, {Asorey}, {Atayde}, {Atek}, {Atrio-Barandela}, {Aubert},
  {Aubourg}, {Auphan}, {Auricchio}, {Aussel}, {Aussel}, {Avelino},
  {Avgoustidis}, {Avila}, {Awan}, {Azzollini}, {Baccigalupi}, {Bachelet},
  {Bacon}, {Baes}, {Bagley}, {Bahr-Kalus}, {Balaguera-Antolinez}, {Balbinot},
  {Balcells}, {Baldi}, {Baldry}, {Balestra}, {Ballardini}, {Ballester},
  {Balogh}, {Ba{\~n}ados}, {Barbier}, {Bardelli}, {Baron}, {Barreiro},
  {Barrena}, {Barriere}, {Barros}, {Barthelemy}, {Bartolo}, {Basset},
  {Battaglia}, {Battisti}, {Baugh}, {Baumont}, {Bazzanini}, {Beaulieu},
  {Beckmann}, {Belikov}, {Bel}, {Bellagamba}, {Bella}, {Bellini}, {Benabed},
  {Bender}, {Benevento}, {Bennett}, {Benson}, {Bergamini}, {Bermejo-Climent},
  {Bernardeau}, {Bertacca}, {Berthe}, {Berthier}, {Bethermin}, {Beutler},
  {Bevillon}, {Bhargava}, {Bhatawdekar}, {Bianchi}, {Bisigello}, {Biviano},
  {Blake}, {Blanchard}, {Blazek}, {Blot}, {Bosco}, {Bodendorf}, {Boenke},
  {B{\"o}hringer}, {Boldrini}, {Bolzonella}, {Bonchi}, {Bonici}, {Bonino},
  {Bonino}, {Bonvin}, {Bon}, {Booth}, {Borgani}, {Borlaff}, {Borsato}, {Bosco},
  {Bose}, {Botticella}, {Boucaud}, {Bouche}, {Boucher}, {Boutigny}, {Bouvard},
  {Bouwens}, {Bouy}, {Bowler}, {Bozza}, {Bozzo}, {Branchini}, {Brando},
  {Brau-Nogue}, {Brekke}, {Bremer}, {Brescia}, {Breton}, {Brinchmann},
  {Brinckmann}, {Brockley-Blatt}, {Brodwin}, {Brouard}, {Brown}, {Bruton},
  {Bucko}, {Buddelmeijer}, {Buenadicha}, {Buitrago}, {Burger}, {Burigana},
  {Busillo}, {Busonero}, {Cabanac}, {Cabayol-Garcia}, {Cagliari}, {Caillat},
  {Caillat}, {Calabrese}, {Calabro}, {Calderone}, {Calura}, {Camacho Quevedo},
  {Camera}, {Campos}, {Canas-Herrera}, {Candini}, {Cantiello}, {Capobianco},
  {Cappellaro}, {Cappelluti}, {Cappi}, {Caputi}, {Cara}, {Carbone}, {Cardone},
  {Carella}, {Carlberg}, {Carle}, {Carminati}, {Caro}, {Carrasco}, {Carretero},
  {Carrilho}, \& {Carron Duque}}]{euclid24}
{Euclid Collaboration}, {Mellier}, Y., {Abdurro'uf}, {et~al.} 2024, arXiv
  e-prints, arXiv:2405.13491

\bibitem[{{Feroz} {et~al.}(2009){Feroz}, {Hobson}, \& {Bridges}}]{feroz09}
{Feroz}, F., {Hobson}, M.~P., \& {Bridges}, M. 2009, \mnras, 398, 1601

\bibitem[{{Furusawa} {et~al.}(2018){Furusawa}, {Koike}, {Takata}, {Okura},
  {Miyatake}, {Lupton}, {Bickerton}, {Price}, {Bosch}, {Yasuda}, {Mineo},
  {Yamada}, {Miyazaki}, {Nakata}, {Koshida}, {Komiyama}, {Utsumi},
  {Kawanomoto}, {Jeschke}, {Noumaru}, {Schubert}, {Iwata}, {Finet},
  {Fujiyoshi}, {Tajitsu}, {Terai}, \& {Lee}}]{furusawa18}
{Furusawa}, H., {Koike}, M., {Takata}, T., {et~al.} 2018, \pasj, 70, S3

\bibitem[{{Garrel} {et~al.}(2022){Garrel}, {Pierre}, {Valageas}, {Eckert},
  {Marulli}, {Veropalumbo}, {Pacaud}, {Clerc}, {Sereno}, {Umetsu},
  {Moscardini}, {Bhargava}, {Adami}, {Chiappetti}, {Gastaldello},
  {Koulouridis}, {Le Fevre}, \& {Plionis}}]{garrel22}
{Garrel}, C., {Pierre}, M., {Valageas}, P., {et~al.} 2022, \aap, 663, A3

\bibitem[{{Ghirardini} {et~al.}(2024){Ghirardini}, {Bulbul}, {Artis}, {Clerc},
  {Garrel}, {Grandis}, {Kluge}, {Liu}, {Bahar}, {Balzer}, {Chiu}, {Comparat},
  {Gruen}, {Kleinebreil}, {Krippendorf}, {Merloni}, {Nandra}, {Okabe},
  {Pacaud}, {Predehl}, {Ramos-Ceja}, {Reiprich}, {Sanders}, {Schrabback},
  {Seppi}, {Zelmer}, {Zhang}, {Bornemann}, {Brunner}, {Burwitz}, {Coutinho},
  {Dennerl}, {Freyberg}, {Friedrich}, {Gaida}, {Gueguen}, {Haberl}, {Kink},
  {Lamer}, {Li}, {Liu}, {Maitra}, {Meidinger}, {Mueller}, {Miyatake},
  {Miyazaki}, {Robrade}, {Schwope}, \& {Stewart}}]{ghirardini24}
{Ghirardini}, V., {Bulbul}, E., {Artis}, E., {et~al.} 2024, \aap, 689, A298

\bibitem[{{Girelli} {et~al.}(2020){Girelli}, {Pozzetti}, {Bolzonella},
  {Giocoli}, {Marulli}, \& {Baldi}}]{girelli20}
{Girelli}, G., {Pozzetti}, L., {Bolzonella}, M., {et~al.} 2020, \aap, 634, A135

\bibitem[{{Golden-Marx} \& {Miller}(2018)}]{goldenmarx18}
{Golden-Marx}, J.~B. \& {Miller}, C.~J. 2018, \apj, 860, 2

\bibitem[{{Golden-Marx} \& {Miller}(2019)}]{goldenmarx19}
{Golden-Marx}, J.~B. \& {Miller}, C.~J. 2019, \apj, 878, 14

\bibitem[{{Golden-Marx} {et~al.}(2022){Golden-Marx}, {Miller}, {Zhang},
  {Ogando}, {Palmese}, {Abbott}, {Aguena}, {Allam}, {Andrade-Oliveira},
  {Annis}, {Bacon}, {Bertin}, {Brooks}, {Buckley-Geer}, {Carnero Rosell},
  {Carrasco Kind}, {Castander}, {Costanzi}, {Crocce}, {da Costa}, {Pereira},
  {De Vicente}, {Desai}, {Diehl}, {Doel}, {Drlica-Wagner}, {Everett}, {Evrard},
  {Ferrero}, {Flaugher}, {Fosalba}, {Frieman}, {Garc{\'\i}a-Bellido},
  {Gaztanaga}, {Gerdes}, {Gruen}, {Gruendl}, {Gschwend}, {Gutierrez},
  {Hartley}, {Hinton}, {Hollowood}, {Honscheid}, {Hoyle}, {James}, {Jeltema},
  {Kim}, {Krause}, {Kuehn}, {Kuropatkin}, {Lahav}, {Lima}, {Maia}, {Marshall},
  {Melchior}, {Menanteau}, {Miquel}, {Mohr}, {Morgan}, {Paz-Chinch{\'o}n},
  {Petravick}, {Pieres}, {Plazas Malag{\'o}n}, {Prat}, {Romer}, {Sanchez},
  {Santiago}, {Scarpine}, {Schubnell}, {Serrano}, {Sevilla-Noarbe}, {Smith},
  {Soares-Santos}, {Suchyta}, {Tarle}, \& {Varga}}]{goldenmarx22}
{Golden-Marx}, J.~B., {Miller}, C.~J., {Zhang}, Y., {et~al.} 2022, \apj, 928,
  28

\bibitem[{{Golden-Marx} {et~al.}(2025){Golden-Marx}, {Zhang}, {Ogando},
  {Yanny}, {da Silva Pereira}, {Hilton}, {Aguena}, {Allam}, {Andrade-Oliveira},
  {Bacon}, {Brooks}, {Carnero Rosell}, {Carretero}, {Cheng}, {da Costa}, {De
  Vicente}, {Desai}, {Doel}, {Everett}, {Ferrero}, {Frieman},
  {Garc{\'\i}a-Bellido}, {Gatti}, {Giannini}, {Gruen}, {Gruendl}, {Gutierrez},
  {Hinton}, {Hollowood}, {Honscheid}, {James}, {Kuehn}, {Lee},
  {Mena-Fern{\'a}ndez}, {Menanteau}, {Miquel}, {Mohr}, {Palmese}, {Pieres},
  {Plazas Malag{\'o}n}, {Samuroff}, {Sanchez}, {Schubnell}, {Sevilla-Noarbe},
  {Smith}, {Suchyta}, {Tarle}, {Vikram}, {Walker}, {Weaverdyck}, \&
  {Wiseman}}]{goldenmarx24}
{Golden-Marx}, J.~B., {Zhang}, Y., {Ogando}, R.~L.~C., {et~al.} 2025, \mnras,
  538, 622

\bibitem[{{Gonzalez} {et~al.}(2013){Gonzalez}, {Sivanandam}, {Zabludoff}, \&
  {Zaritsky}}]{gonzalez13}
{Gonzalez}, A.~H., {Sivanandam}, S., {Zabludoff}, A.~I., \& {Zaritsky}, D.
  2013, \apj, 778, 14

\bibitem[{{Grandis} {et~al.}(2021){Grandis}, {Bocquet}, {Mohr}, {Klein}, \&
  {Dolag}}]{grandis21}
{Grandis}, S., {Bocquet}, S., {Mohr}, J.~J., {Klein}, M., \& {Dolag}, K. 2021,
  \mnras, 507, 5671

\bibitem[{{Grandis} {et~al.}(2024){Grandis}, {Ghirardini}, {Bocquet}, {Garrel},
  {Mohr}, {Liu}, {Kluge}, {Kimmig}, {Reiprich}, {Alarcon}, {Amon}, {Artis},
  {Bahar}, {Balzer}, {Bechtol}, {Becker}, {Bernstein}, {Bulbul}, {Campos},
  {Carnero Rosell}, {Carrasco Kind}, {Cawthon}, {Chang}, {Chen}, {Chiu},
  {Choi}, {Clerc}, {Comparat}, {Cordero}, {Davis}, {Derose}, {Diehl},
  {Dodelson}, {Doux}, {Drlica-Wagner}, {Eckert}, {Elvin-Poole}, {Everett},
  {Ferte}, {Gatti}, {Giannini}, {Giles}, {Gruen}, {Gruendl}, {Harrison},
  {Hartley}, {Herner}, {Huff}, {Kleinebreil}, {Kuropatkin}, {Leget},
  {Maccrann}, {Mccullough}, {Merloni}, {Myles}, {Nandra}, {Navarro-Alsina},
  {Okabe}, {Pacaud}, {Pandey}, {Prat}, {Predehl}, {Ramos}, {Raveri}, {Rollins},
  {Roodman}, {Ross}, {Rykoff}, {Sanchez}, {Sanders}, {Schrabback}, {Secco},
  {Seppi}, {Sevilla-Noarbe}, {Sheldon}, {Shin}, {Troxel}, {Tutusaus}, {Varga},
  {Wu}, {Yanny}, {Yin}, {Zhang}, {Zhang}, {Alves}, {Bhargava}, {Brooks},
  {Burke}, {Carretero}, {Costanzi}, {da Costa}, {Pereira}, {De Vicente},
  {Desai}, {Doel}, {Ferrero}, {Flaugher}, {Friedel}, {Frieman},
  {Garc{\'\i}a-Bellido}, {Gutierrez}, {Hinton}, {Hollowood}, {Honscheid},
  {James}, {Jeffrey}, {Lahav}, {Lee}, {Marshall}, {Menanteau}, {Ogando},
  {Pieres}, {Plazas Malag{\'o}n}, {Romer}, {Sanchez}, {Schubnell}, {Smith},
  {Suchyta}, {Swanson}, {Tarle}, {Weaverdyck}, \& {Weller}}]{grandis24}
{Grandis}, S., {Ghirardini}, V., {Bocquet}, S., {et~al.} 2024, \aap, 687, A178

\bibitem[{{Guo} {et~al.}(2011){Guo}, {White}, {Boylan-Kolchin}, {De Lucia},
  {Kauffmann}, {Lemson}, {Li}, {Springel}, \& {Weinmann}}]{guo11}
{Guo}, Q., {White}, S., {Boylan-Kolchin}, M., {et~al.} 2011, \mnras, 413, 101

\bibitem[{{Hoekstra}(2003)}]{hoekstra03}
{Hoekstra}, H. 2003, \mnras, 339, 1155

\bibitem[{{Hsieh} \& {Yee}(2014)}]{hsieh14}
{Hsieh}, B.~C. \& {Yee}, H.~K.~C. 2014, \apj, 792, 102

\bibitem[{{Huang} {et~al.}(2022){Huang}, {Leauthaud}, {Bradshaw}, {Hearin},
  {Behroozi}, {Lange}, {Greene}, {DeRose}, {Speagle}, \& {Xhakaj}}]{huang22}
{Huang}, S., {Leauthaud}, A., {Bradshaw}, C., {et~al.} 2022, \mnras, 515, 4722

\bibitem[{{Huang} {et~al.}(2018{\natexlab{a}}){Huang}, {Leauthaud}, {Greene},
  {Bundy}, {Lin}, {Tanaka}, {Miyazaki}, \& {Komiyama}}]{huang18c}
{Huang}, S., {Leauthaud}, A., {Greene}, J.~E., {et~al.} 2018{\natexlab{a}},
  \mnras, 475, 3348

\bibitem[{{Huang} {et~al.}(2018{\natexlab{b}}){Huang}, {Leauthaud}, {Murata},
  {Bosch}, {Price}, {Lupton}, {Mandelbaum}, {Lackner}, {Bickerton}, {Miyazaki},
  {Coupon}, \& {Tanaka}}]{huang18a}
{Huang}, S., {Leauthaud}, A., {Murata}, R., {et~al.} 2018{\natexlab{b}}, \pasj,
  70, S6

\bibitem[{Hunter(2007)}]{matplotlib}
Hunter, J.~D. 2007, Computing In Science \& Engineering, 9, 90

\bibitem[{{Huterer} {et~al.}(2015){Huterer}, {Kirkby}, {Bean}, {Connolly},
  {Dawson}, {Dodelson}, {Evrard}, {Jain}, {Jarvis}, {Linder}, {Mandelbaum},
  {May}, {Raccanelli}, {Reid}, {Rozo}, {Schmidt}, {Sehgal}, {Slosar}, {van
  Engelen}, {Wu}, \& {Zhao}}]{huterer15}
{Huterer}, D., {Kirkby}, D., {Bean}, R., {et~al.} 2015, Astroparticle Physics,
  63, 23

\bibitem[{Ilbert {et~al.}(2006)Ilbert, Arnouts, McCracken, Bolzonella, Bertin,
  {Le F{\`{e}}vre}, Mellier, Zamorani, Pell{\`{o}}, Iovino, Tresse, {Le Brun},
  Bottini, Garilli, Maccagni, Picat, Scaramella, Scodeggio, Vettolani,
  Zanichelli, Adami, Bardelli, Cappi, Charlot, Ciliegi, Contini, Cucciati,
  Foucaud, Franzetti, Gavignaud, Guzzo, Marano, Marinoni, Mazure, Meneux,
  Merighi, Paltani, Pollo, Pozzetti, Radovich, Zucca, Bondi, Bongiorno,
  Busarello, {de La Torre}, Gregorini, Lamareille, Mathez, Merluzzi, Ripepi,
  Rizzo, \& Vergani}]{ilbert06}
Ilbert, O., Arnouts, S., McCracken, H., {et~al.} 2006, \aap, 457, 841

\bibitem[{{Ilbert} {et~al.}(2009){Ilbert}, {Capak}, {Salvato}, {Aussel},
  {McCracken}, {Sanders}, {Scoville}, {Kartaltepe}, {Arnouts}, {Le Floc'h},
  {Mobasher}, {Taniguchi}, {Lamareille}, {Leauthaud}, {Sasaki}, {Thompson},
  {Zamojski}, {Zamorani}, {Bardelli}, {Bolzonella}, {Bongiorno}, {Brusa},
  {Caputi}, {Carollo}, {Contini}, {Cook}, {Coppa}, {Cucciati}, {de la Torre},
  {de Ravel}, {Franzetti}, {Garilli}, {Hasinger}, {Iovino}, {Kampczyk},
  {Kneib}, {Knobel}, {Kovac}, {Le Borgne}, {Le Brun}, {Le F{\`e}vre}, {Lilly},
  {Looper}, {Maier}, {Mainieri}, {Mellier}, {Mignoli}, {Murayama}, {Pell{\`o}},
  {Peng}, {P{\'e}rez-Montero}, {Renzini}, {Ricciardelli}, {Schiminovich},
  {Scodeggio}, {Shioya}, {Silverman}, {Surace}, {Tanaka}, {Tasca}, {Tresse},
  {Vergani}, \& {Zucca}}]{ilbert09}
{Ilbert}, O., {Capak}, P., {Salvato}, M., {et~al.} 2009, \apj, 690, 1236

\bibitem[{{Ivezic} {et~al.}(2019){Ivezic}, {Kahn}, {Tyson}, {Abel}, {Acosta},
  {Allsman}, {Alonso}, {AlSayyad}, {Anderson}, {Andrew}, {Angel}, {Angeli},
  {Ansari}, {Antilogus}, {Araujo}, {Armstrong}, {Arndt}, {Astier}, {Aubourg},
  {Auza}, {Axelrod}, {Bard}, {Barr}, {Barrau}, {Bartlett}, {Bauer}, {Bauman},
  {Baumont}, {Bechtol}, {Bechtol}, {Becker}, {Becla}, {Beldica}, {Bellavia},
  {Bianco}, {Biswas}, {Blanc}, {Blazek}, {Bland ford}, {Bloom}, {Bogart},
  {Bond}, {Booth}, {Borgland}, {Borne}, {Bosch}, {Boutigny}, {Brackett},
  {Bradshaw}, {Brand t}, {Brown}, {Bullock}, {Burchat}, {Burke}, {Cagnoli},
  {Calabrese}, {Callahan}, {Callen}, {Carlin}, {Carlson}, {Chand rasekharan},
  {Charles-Emerson}, {Chesley}, {Cheu}, {Chiang}, {Chiang}, {Chirino}, {Chow},
  {Ciardi}, {Claver}, {Cohen-Tanugi}, {Cockrum}, {Coles}, {Connolly}, {Cook},
  {Cooray}, {Covey}, {Cribbs}, {Cui}, {Cutri}, {Daly}, {Daniel}, {Daruich},
  {Daubard}, {Daues}, {Dawson}, {Delgado}, {Dellapenna}, {de Peyster}, {de
  Val-Borro}, {Digel}, {Doherty}, {Dubois}, {Dubois-Felsmann}, {Durech},
  {Economou}, {Eifler}, {Eracleous}, {Emmons}, {Fausti Neto}, {Ferguson},
  {Figueroa}, {Fisher-Levine}, {Focke}, {Foss}, {Frank}, {Freemon}, {Gangler},
  {Gawiser}, {Geary}, {Gee}, {Geha}, {Gessner}, {Gibson}, {Gilmore},
  {Glanzman}, {Glick}, {Goldina}, {Goldstein}, {Goodenow}, {Graham},
  {Gressler}, {Gris}, {Guy}, {Guyonnet}, {Haller}, {Harris}, {Hascall},
  {Haupt}, {Hernand ez}, {Herrmann}, {Hileman}, {Hoblitt}, {Hodgson}, {Hogan},
  {Howard}, {Huang}, {Huffer}, {Ingraham}, {Innes}, {Jacoby}, {Jain}, {Jammes},
  {Jee}, {Jenness}, {Jernigan}, {Jevremovi{\'c}}, {Johns}, {Johnson},
  {Johnson}, {Jones}, {Juramy-Gilles}, {Juri{\'c}}, {Kalirai}, {Kallivayalil},
  {Kalmbach}, {Kantor}, {Karst}, {Kasliwal}, {Kelly}, {Kessler}, {Kinnison},
  {Kirkby}, {Knox}, {Kotov}, {Krabbendam}, {Krughoff}, {Kub{\'a}nek},
  {Kuczewski}, {Kulkarni}, {Ku}, {Kurita}, {Lage}, {Lambert}, {Lange},
  {Langton}, {Le Guillou}, {Levine}, {Liang}, {Lim}, {Lintott}, {Long},
  {Lopez}, {Lotz}, {Lupton}, {Lust}, {MacArthur}, {Mahabal}, {Mand elbaum},
  {Markiewicz}, {Marsh}, {Marshall}, {Marshall}, {May}, {McKercher}, {McQueen},
  {Meyers}, {Migliore}, {Miller}, {Mills}, {Miraval}, {Moeyens}, {Moolekamp},
  {Monet}, {Moniez}, {Monkewitz}, {Montgomery}, {Morrison}, {Mueller},
  {Muller}, {Mu{\~n}oz Arancibia}, {Neill}, {Newbry}, {Nief}, {Nomerotski},
  {Nordby}, {O'Connor}, {Oliver}, {Olivier}, {Olsen}, {O'Mullane}, {Ortiz},
  {Osier}, {Owen}, {Pain}, {Palecek}, {Parejko}, {Parsons}, {Pease},
  {Peterson}, {Peterson}, {Petravick}, {Libby Petrick}, {Petry},
  {Pierfederici}, {Pietrowicz}, {Pike}, {Pinto}, {Plante}, {Plate}, {Plutchak},
  {Price}, {Prouza}, {Radeka}, {Rajagopal}, {Rasmussen}, {Regnault}, {Reil},
  {Reiss}, {Reuter}, {Ridgway}, {Riot}, {Ritz}, {Robinson}, {Roby}, {Roodman},
  {Rosing}, {Roucelle}, {Rumore}, {Russo}, {Saha}, {Sassolas}, {Schalk},
  {Schellart}, {Schindler}, {Schmidt}, {Schneider}, {Schneider}, {Schoening},
  {Schumacher}, {Schwamb}, {Sebag}, {Selvy}, {Sembroski}, {Seppala}, {Serio},
  {Serrano}, {Shaw}, {Shipsey}, {Sick}, {Silvestri}, {Slater}, {Smith},
  {Smith}, {Sobhani}, {Soldahl}, {Storrie-Lombardi}, {Stover}, {Strauss},
  {Street}, {Stubbs}, {Sullivan}, {Sweeney}, {Swinbank}, {Szalay}, {Takacs},
  {Tether}, {Thaler}, {Thayer}, {Thomas}, {Thornton}, {Thukral}, {Tice},
  {Trilling}, {Turri}, {Van Berg}, {Vanden Berk}, {Vetter}, {Virieux},
  {Vucina}, {Wahl}, {Walkowicz}, {Walsh}, {Walter}, {Wang}, {Wang}, {Warner},
  {Wiecha}, {Willman}, {Winters}, {Wittman}, {Wolff}, {Wood-Vasey}, {Wu},
  {Xin}, {Yoachim}, \& {Zhan}}]{ivezic19}
{Ivezic}, {\v{Z}}., {Kahn}, S.~M., {Tyson}, J.~A., {et~al.} 2019, \apj, 873,
  111

\bibitem[{{Johnston} {et~al.}(2007){Johnston}, {Sheldon}, {Tasitsiomi},
  {Frieman}, {Wechsler}, \& {McKay}}]{johnston07a}
{Johnston}, D.~E., {Sheldon}, E.~S., {Tasitsiomi}, A., {et~al.} 2007, \apj,
  656, 27

\bibitem[{{Juric} {et~al.}(2017){Juric}, {Kantor}, {Lim}, {Lupton},
  {Dubois-Felsmann}, {Jenness}, {Axelrod}, {Aleksi{\'c}}, {Allsman},
  {AlSayyad}, {Alt}, {Armstrong}, {Basney}, {Becker}, {Becla}, {Biswas},
  {Bosch}, {Boutigny}, {Kind}, {Ciardi}, {Connolly}, {Daniel}, {Daues},
  {Economou}, {Chiang}, {Fausti}, {Fisher-Levine}, {Freemon}, {Gris},
  {Hernandez}, {Hoblitt}, {Ivezi{\'c}}, {Jammes}, {Jevremovi{\'c}}, {Jones},
  {Kalmbach}, {Kasliwal}, {Krughoff}, {Lurie}, {Lust}, {MacArthur}, {Melchior},
  {Moeyens}, {Nidever}, {Owen}, {Parejko}, {Peterson}, {Petravick},
  {Pietrowicz}, {Price}, {Reiss}, {Shaw}, {Sick}, {Slater}, {Strauss},
  {Sullivan}, {Swinbank}, {Van Dyk}, {Vuj{\v{c}}i{\'c}}, {Withers}, \&
  {Yoachim}}]{juric17}
{Juric}, M., {Kantor}, J., {Lim}, K.~T., {et~al.} 2017, Astronomical Society of
  the Pacific Conference Series, Vol. 512, {The LSST Data Management System},
  ed. N.~P.~F. {Lorente}, K.~{Shortridge}, \& R.~{Wayth}, 279

\bibitem[{Kaiser \& Silk(1986)}]{kaiser86}
Kaiser, N. \& Silk, J. 1986, \nat, 324, 529

\bibitem[{{Kawanomoto} {et~al.}(2018){Kawanomoto}, {Uraguchi}, {Komiyama},
  {Miyazaki}, {Furusawa}, {Finet}, {Hattori}, {Wang}, {Yasuda}, \&
  {Suzuki}}]{kawanomoto18}
{Kawanomoto}, S., {Uraguchi}, F., {Komiyama}, Y., {et~al.} 2018, \pasj, 70, 66

\bibitem[{{Kleinebreil} {et~al.}(2024){Kleinebreil}, {Grandis}, {Schrabback},
  {Ghirardini}, {Chiu}, {Liu}, {Kluge}, {Reiprich}, {Artis}, {Bahar}, {Balzer},
  {Bulbul}, {Clerc}, {Comparat}, {Garrel}, {Gruen}, {Li}, {Miyatake},
  {Miyazaki}, {Ramos-Ceja}, {Sanders}, {Seppi}, {Okabe}, \&
  {Zhang}}]{kleinebreil24}
{Kleinebreil}, F., {Grandis}, S., {Schrabback}, T., {et~al.} 2024, arXiv
  e-prints, arXiv:2402.08456

\bibitem[{{Kluge} {et~al.}(2024){Kluge}, {Comparat}, {Liu}, {Balzer}, {Bulbul},
  {Ider Chitham}, {Ghirardini}, {Garrel}, {Bahar}, {Artis}, {Bender}, {Clerc},
  {Dwelly}, {Fabricius}, {Grandis}, {Hern{\'a}ndez-Lang}, {Hill}, {Joshi},
  {Lamer}, {Merloni}, {Nandra}, {Pacaud}, {Predehl}, {Ramos-Ceja}, {Reiprich},
  {Salvato}, {Sanders}, {Schrabback}, {Seppi}, {Zelmer}, {Zenteno}, \&
  {Zhang}}]{kluge24}
{Kluge}, M., {Comparat}, J., {Liu}, A., {et~al.} 2024, \aap, 688, A210

\bibitem[{{Komiyama} {et~al.}(2018){Komiyama}, {Obuchi}, {Nakaya}, {Kamata},
  {Kawanomoto}, {Utsumi}, {Miyazaki}, {Uraguchi}, {Furusawa}, {Morokuma},
  {Uchida}, {Miyatake}, {Mineo}, {Fujimori}, {Aihara}, {Karoji}, {Gunn}, \&
  {Wang}}]{komiyama18}
{Komiyama}, Y., {Obuchi}, Y., {Nakaya}, H., {et~al.} 2018, \pasj, 70, S2

\bibitem[{Kravtsov \& Borgani(2012)}]{kravtsov12}
Kravtsov, A. \& Borgani, S. 2012, \araa, 50, 353

\bibitem[{{Kravtsov} {et~al.}(2004){Kravtsov}, {Berlind}, {Wechsler}, {Klypin},
  {Gottl{\"o}ber}, {Allgood}, \& {Primack}}]{kravtsov04}
{Kravtsov}, A.~V., {Berlind}, A.~A., {Wechsler}, R.~H., {et~al.} 2004, \apj,
  609, 35

\bibitem[{{Kravtsov} {et~al.}(2018){Kravtsov}, {Vikhlinin}, \&
  {Meshcheryakov}}]{kravtsov18}
{Kravtsov}, A.~V., {Vikhlinin}, A.~A., \& {Meshcheryakov}, A.~V. 2018,
  Astronomy Letters, 44, 8

\bibitem[{{Kuijken} {et~al.}(2015){Kuijken}, {Heymans}, {Hildebrandt},
  {Nakajima}, {Erben}, {de Jong}, {Viola}, {Choi}, {Hoekstra}, {Miller}, {van
  Uitert}, {Amon}, {Blake}, {Brouwer}, {Buddendiek}, {Conti}, {Eriksen},
  {Grado}, {Harnois-D{\'e}raps}, {Helmich}, {Herbonnet}, {Irisarri},
  {Kitching}, {Klaes}, {La Barbera}, {Napolitano}, {Radovich}, {Schneider},
  {Sif{\'o}n}, {Sikkema}, {Simon}, {Tudorica}, {Valentijn}, {Verdoes Kleijn},
  \& {van Waerbeke}}]{kuijken15}
{Kuijken}, K., {Heymans}, C., {Hildebrandt}, H., {et~al.} 2015, \mnras, 454,
  3500

\bibitem[{{Lang}(2014)}]{lang14}
{Lang}, D. 2014, \aj, 147, 108

\bibitem[{{Lang} {et~al.}(2016){Lang}, {Hogg}, \& {Schlegel}}]{lang16}
{Lang}, D., {Hogg}, D.~W., \& {Schlegel}, D.~J. 2016, \aj, 151, 36

\bibitem[{Leauthaud {et~al.}(2012)Leauthaud, Tinker, Bundy, Behroozi, Massey,
  Rhodes, George, Kneib, Benson, Wechsler, Busha, Capak, Cort{\^{e}}s, Ilbert,
  Koekemoer, {Le F{\`{e}}vre}, Lilly, McCracken, Salvato, Schrabback, Scoville,
  Smith, \& Taylor}]{leauthaud12a}
Leauthaud, A., Tinker, J., Bundy, K., {et~al.} 2012, \apj, 744, 159

\bibitem[{{Lewis}(2019)}]{getdist}
{Lewis}, A. 2019, arXiv e-prints, arXiv:1910.13970

\bibitem[{{Li} \& {White}(2009)}]{li09}
{Li}, C. \& {White}, S. D.~M. 2009, \mnras, 398, 2177

\bibitem[{{Li} {et~al.}(2022){Li}, {Miyatake}, {Luo}, {More}, {Oguri},
  {Hamana}, {Mandelbaum}, {Shirasaki}, {Takada}, {Armstrong}, {Kannawadi},
  {Takita}, {Miyazaki}, {Nishizawa}, {Plazas Malagon}, {Strauss}, {Tanaka}, \&
  {Yoshida}}]{li22}
{Li}, X., {Miyatake}, H., {Luo}, W., {et~al.} 2022, \pasj, 74, 421

\bibitem[{{Li} {et~al.}(2023){Li}, {Zhang}, {Sugiyama}, {Dalal}, {Terasawa},
  {Rau}, {Mandelbaum}, {Takada}, {More}, {Strauss}, {Miyatake}, {Shirasaki},
  {Hamana}, {Oguri}, {Luo}, {Nishizawa}, {Takahashi}, {Nicola}, {Osato},
  {Kannawadi}, {Sunayama}, {Armstrong}, {Bosch}, {Komiyama}, {Lupton}, {Lust},
  {MacArthur}, {Miyazaki}, {Murayama}, {Nishimichi}, {Okura}, {Price}, {Tait},
  {Tanaka}, \& {Wang}}]{li23}
{Li}, X., {Zhang}, T., {Sugiyama}, S., {et~al.} 2023, \prd, 108, 123518

\bibitem[{Lidman {et~al.}(2012)Lidman, Suherli, Muzzin, Wilson, Demarco,
  Brough, Rettura, Cox, DeGroot, Yee, Gilbank, Hoekstra, Balogh, Ellingson,
  Hicks, Nantais, Noble, Lacy, Surace, \& Webb}]{lidman12}
Lidman, C., Suherli, J., Muzzin, A., {et~al.} 2012, \mnras, 427, 550

\bibitem[{{Lin} {et~al.}(2013){Lin}, {Brodwin}, {Gonzalez}, {Bode},
  {Eisenhardt}, {Stanford}, \& {Vikhlinin}}]{lin13}
{Lin}, Y.-T., {Brodwin}, M., {Gonzalez}, A.~H., {et~al.} 2013, \apj, 771, 61

\bibitem[{{Lin} {et~al.}(2025){Lin}, {Chen}, {Chen}, {Chuang}, \&
  {Oguri}}]{lin25}
{Lin}, Y.-T., {Chen}, K.-F., {Chen}, T.-C., {Chuang}, C.-Y., \& {Oguri}, M.
  2025, arXiv e-prints, arXiv:2503.13592

\bibitem[{{Lin} {et~al.}(2017){Lin}, {Hsieh}, {Lin}, {Oguri}, {Chen}, {Tanaka},
  {Chiu}, {Huang}, {Kodama}, {Leauthaud}, {More}, {Nishizawa}, {Bundy}, {Lin},
  \& {Miyazaki}}]{lin17}
{Lin}, Y.-T., {Hsieh}, B.-C., {Lin}, S.-C., {et~al.} 2017, \apj, 851, 139

\bibitem[{{Liu} {et~al.}(2022){Liu}, {Bulbul}, {Ghirardini}, {Liu}, {Klein},
  {Clerc}, {{\"O}zsoy}, {Ramos-Ceja}, {Pacaud}, {Comparat}, {Okabe}, {Bahar},
  {Biffi}, {Brunner}, {Br{\"u}ggen}, {Buchner}, {Ider Chitham}, {Chiu},
  {Dolag}, {Gatuzz}, {Gonzalez}, {Hoang}, {Lamer}, {Merloni}, {Nandra},
  {Oguri}, {Ota}, {Predehl}, {Reiprich}, {Salvato}, {Schrabback}, {Sanders},
  {Seppi}, \& {Thibaud}}]{liu21}
{Liu}, A., {Bulbul}, E., {Ghirardini}, V., {et~al.} 2022, \aap, 661, A2

\bibitem[{{Liu} {et~al.}(2024){Liu}, {Bulbul}, {Shin}, {von der Linden},
  {Ghirardini}, {Kluge}, {Sanders}, {Grandis}, {Zhang}, {Artis}, {Bahar},
  {Balzer}, {Clerc}, {Malavasi}, {Merloni}, {Nandra}, {Ramos-Ceja}, \&
  {Zelmer}}]{liu24}
{Liu}, A., {Bulbul}, E., {Shin}, T., {et~al.} 2024, \aap, 688, A186

\bibitem[{{Lovisari} {et~al.}(2015){Lovisari}, {Reiprich}, \&
  {Schellenberger}}]{lovisari15}
{Lovisari}, L., {Reiprich}, T.~H., \& {Schellenberger}, G. 2015, \aap, 573,
  A118

\bibitem[{{LSST Science Collaboration} {et~al.}(2009){LSST Science
  Collaboration}, {Abell}, {Allison}, {Anderson}, {Andrew}, {Angel}, {Armus},
  {Arnett}, {Asztalos}, {Axelrod}, {Bailey}, {Ballantyne}, {Bankert},
  {Barkhouse}, {Barr}, {Barrientos}, {Barth}, {Bartlett}, {Becker}, {Becla},
  {Beers}, {Bernstein}, {Biswas}, {Blanton}, {Bloom}, {Bochanski}, {Boeshaar},
  {Borne}, {Bradac}, {Brandt}, {Bridge}, {Brown}, {Brunner}, {Bullock},
  {Burgasser}, {Burge}, {Burke}, {Cargile}, {Chandrasekharan}, {Chartas},
  {Chesley}, {Chu}, {Cinabro}, {Claire}, {Claver}, {Clowe}, {Connolly}, {Cook},
  {Cooke}, {Cooray}, {Covey}, {Culliton}, {de Jong}, {de Vries}, {Debattista},
  {Delgado}, {Dell'Antonio}, {Dhital}, {Di Stefano}, {Dickinson}, {Dilday},
  {Djorgovski}, {Dobler}, {Donalek}, {Dubois-Felsmann}, {Durech},
  {Eliasdottir}, {Eracleous}, {Eyer}, {Falco}, {Fan}, {Fassnacht}, {Ferguson},
  {Fernandez}, {Fields}, {Finkbeiner}, {Figueroa}, {Fox}, {Francke}, {Frank},
  {Frieman}, {Fromenteau}, {Furqan}, {Galaz}, {Gal-Yam}, {Garnavich},
  {Gawiser}, {Geary}, {Gee}, {Gibson}, {Gilmore}, {Grace}, {Green}, {Gressler},
  {Grillmair}, {Habib}, {Haggerty}, {Hamuy}, {Harris}, {Hawley}, {Heavens},
  {Hebb}, {Henry}, {Hileman}, {Hilton}, {Hoadley}, {Holberg}, {Holman},
  {Howell}, {Infante}, {Ivezic}, {Jacoby}, {Jain}, {R}, {Jedicke}, {Jee},
  {Garrett Jernigan}, {Jha}, {Johnston}, {Jones}, {Juric}, {Kaasalainen},
  {Styliani}, {Kafka}, {Kahn}, {Kaib}, {Kalirai}, {Kantor}, {Kasliwal},
  {Keeton}, {Kessler}, {Knezevic}, {Kowalski}, {Krabbendam}, {Krughoff},
  {Kulkarni}, {Kuhlman}, {Lacy}, {Lepine}, {Liang}, {Lien}, {Lira}, {Long},
  {Lorenz}, {Lotz}, {Lupton}, {Lutz}, {Macri}, {Mahabal}, {Mandelbaum},
  {Marshall}, {May}, {McGehee}, {Meadows}, {Meert}, {Milani}, {Miller},
  {Miller}, {Mills}, {Minniti}, {Monet}, {Mukadam}, {Nakar}, {Neill}, {Newman},
  {Nikolaev}, {Nordby}, {O'Connor}, {Oguri}, {Oliver}, {Olivier}, {Olsen},
  {Olsen}, {Olszewski}, {Oluseyi}, {Padilla}, {Parker}, {Pepper}, {Peterson},
  {Petry}, {Pinto}, {Pizagno}, {Popescu}, {Prsa}, {Radcka}, {Raddick},
  {Rasmussen}, {Rau}, {Rho}, {Rhoads}, {Richards}, {Ridgway}, {Robertson},
  {Roskar}, {Saha}, {Sarajedini}, {Scannapieco}, {Schalk}, {Schindler}, \&
  {Schmidt}}]{lsst09}
{LSST Science Collaboration}, {Abell}, P.~A., {Allison}, J., {et~al.} 2009,
  arXiv e-prints, arXiv:0912.0201

\bibitem[{{Magnier} {et~al.}(2013){Magnier}, {Schlafly}, {Finkbeiner}, {Juric},
  {Tonry}, {Burgett}, {Chambers}, {Flewelling}, {Kaiser}, {Kudritzki},
  {Morgan}, {Price}, {Sweeney}, \& {Stubbs}}]{magnier13}
{Magnier}, E.~A., {Schlafly}, E., {Finkbeiner}, D., {et~al.} 2013, \apjs, 205,
  20

\bibitem[{{Malmquist}(1922)}]{malmquist1922}
{Malmquist}, K.~G. 1922, Meddelanden fran Lunds Astronomiska Observatorium
  Serie I, 100, 1

\bibitem[{{Mandelbaum} {et~al.}(2018{\natexlab{a}}){Mandelbaum}, {Lanusse},
  {Leauthaud}, {Armstrong}, {Simet}, {Miyatake}, {Meyers}, {Bosch}, {Murata},
  {Miyazaki}, \& {Tanaka}}]{mandelbaum18b}
{Mandelbaum}, R., {Lanusse}, F., {Leauthaud}, A., {et~al.} 2018{\natexlab{a}},
  \mnras, 481, 3170

\bibitem[{{Mandelbaum} {et~al.}(2018{\natexlab{b}}){Mandelbaum}, {Miyatake},
  {Hamana}, {Oguri}, {Simet}, {Armstrong}, {Bosch}, {Murata}, {Lanusse},
  {Leauthaud}, {Coupon}, {More}, {Takada}, {Miyazaki}, {Speagle}, {Shirasaki},
  {Sif{\'o}n}, {Huang}, {Nishizawa}, {Medezinski}, {Okura}, {Okabe}, {Czakon},
  {Takahashi}, {Coulton}, {Hikage}, {Komiyama}, {Lupton}, {Strauss}, {Tanaka},
  \& {Utsumi}}]{mandelbaum18}
{Mandelbaum}, R., {Miyatake}, H., {Hamana}, T., {et~al.} 2018{\natexlab{b}},
  \pasj, 70, S25

\bibitem[{Mantz {et~al.}(2016)Mantz, Allen, Morris, \& Schmidt}]{mantz16}
Mantz, A., Allen, S., Morris, R., \& Schmidt, R. 2016, \mnras, 456, 4020

\bibitem[{{Mantz} {et~al.}(2015){Mantz}, {von der Linden}, {Allen},
  {Applegate}, {Kelly}, {Morris}, {Rapetti}, {Schmidt}, {Adhikari}, {Allen},
  {Burchat}, {Burke}, {Cataneo}, {Donovan}, {Ebeling}, {Shandera}, \&
  {Wright}}]{mantz15}
{Mantz}, A.~B., {von der Linden}, A., {Allen}, S.~W., {et~al.} 2015, \mnras,
  446, 2205

\bibitem[{{McClintock} {et~al.}(2019){McClintock}, {Varga}, {Gruen}, {Rozo},
  {Rykoff}, {Shin}, {Melchior}, {DeRose}, {Seitz}, {Dietrich}, {Sheldon},
  {Zhang}, {von der Linden}, {Jeltema}, {Mantz}, {Romer}, {Allen}, {Becker},
  {Bermeo}, {Bhargava}, {Costanzi}, {Everett}, {Farahi}, {Hamaus}, {Hartley},
  {Hollowood}, {Hoyle}, {Israel}, {Li}, {MacCrann}, {Morris}, {Palmese},
  {Plazas}, {Pollina}, {Rau}, {Simet}, {Soares-Santos}, {Troxel}, {Vergara
  Cervantes}, {Wechsler}, {Zuntz}, {Abbott}, {Abdalla}, {Allam}, {Annis},
  {Avila}, {Bridle}, {Brooks}, {Burke}, {Carnero Rosell}, {Carrasco Kind},
  {Carretero}, {Castander}, {Crocce}, {Cunha}, {D'Andrea}, {da Costa}, {Davis},
  {De Vicente}, {Diehl}, {Doel}, {Drlica-Wagner}, {Evrard}, {Flaugher},
  {Fosalba}, {Frieman}, {Garc{\'\i}a-Bellido}, {Gaztanaga}, {Gerdes},
  {Giannantonio}, {Gruendl}, {Gutierrez}, {Honscheid}, {James}, {Kirk},
  {Krause}, {Kuehn}, {Lahav}, {Li}, {Lima}, {March}, {Marshall}, {Menanteau},
  {Miquel}, {Mohr}, {Nord}, {Ogando}, {Roodman}, {Sanchez}, {Scarpine},
  {Schindler}, {Sevilla-Noarbe}, {Smith}, {Smith}, {Sobreira}, {Suchyta},
  {Swanson}, {Tarle}, {Tucker}, {Vikram}, {Walker}, {Weller}, \& {DES
  Collaboration}}]{mcclintock19}
{McClintock}, T., {Varga}, T.~N., {Gruen}, D., {et~al.} 2019, \mnras, 482, 1352

\bibitem[{{McKerns} {et~al.}(2012){McKerns}, {Strand}, {Sullivan}, {Fang}, \&
  {Aivazis}}]{pathos}
{McKerns}, M.~M., {Strand}, L., {Sullivan}, T., {Fang}, A., \& {Aivazis}, M.
  A.~G. 2012, arXiv e-prints, arXiv:1202.1056

\bibitem[{{Medezinski} {et~al.}(2018){Medezinski}, {Oguri}, {Nishizawa},
  {Speagle}, {Miyatake}, {Umetsu}, {Leauthaud}, {Murata}, {Mandelbaum},
  {Sif{\'o}n}, {Strauss}, {Huang}, {Simet}, {Okabe}, {Tanaka}, \&
  {Komiyama}}]{medezinski18a}
{Medezinski}, E., {Oguri}, M., {Nishizawa}, A.~J., {et~al.} 2018, \pasj, 70, 30

\bibitem[{{Merloni} {et~al.}(2024){Merloni}, {Lamer}, {Liu}, {Ramos-Ceja},
  {Brunner}, {Bulbul}, {Dennerl}, {Doroshenko}, {Freyberg}, {Friedrich},
  {Gatuzz}, {Georgakakis}, {Haberl}, {Igo}, {Kreykenbohm}, {Liu}, {Maitra},
  {Malyali}, {Mayer}, {Nandra}, {Predehl}, {Robrade}, {Salvato}, {Sanders},
  {Stewart}, {Tub{\'\i}n-Arenas}, {Weber}, {Wilms}, {Arcodia}, {Artis},
  {Aschersleben}, {Avakyan}, {Aydar}, {Bahar}, {Balzer}, {Becker}, {Berger},
  {Boller}, {Bornemann}, {Br{\"u}ggen}, {Brusa}, {Buchner}, {Burwitz},
  {Camilloni}, {Clerc}, {Comparat}, {Coutinho}, {Czesla}, {Dannhauer},
  {Dauner}, {Dauser}, {Dietl}, {Dolag}, {Dwelly}, {Egg}, {Ehl}, {Freund},
  {Friedrich}, {Gaida}, {Garrel}, {Ghirardini}, {Gokus}, {Gr{\"u}nwald},
  {Grandis}, {Grotova}, {Gruen}, {Gueguen}, {H{\"a}mmerich}, {Hamaus},
  {Hasinger}, {Haubner}, {Homan}, {Ider Chitham}, {Joseph}, {Joyce},
  {K{\"o}nig}, {Kaltenbrunner}, {Khokhriakova}, {Kink}, {Kirsch}, {Kluge},
  {Knies}, {Krippendorf}, {Krumpe}, {Kurpas}, {Li}, {Liu}, {Locatelli},
  {Lorenz}, {M{\"u}ller}, {Magaudda}, {Mannes}, {McCall}, {Meidinger},
  {Michailidis}, {Migkas}, {Mu{\~n}oz-Giraldo}, {Musiimenta}, {Nguyen-Dang},
  {Ni}, {Olechowska}, {Ota}, {Pacaud}, {Pasini}, {Perinati}, {Pires},
  {Pommranz}, {Ponti}, {Poppenhaeger}, {P{\"u}hlhofer}, {Rau}, {Reh},
  {Reiprich}, {Roster}, {Saeedi}, {Santangelo}, {Sasaki}, {Schmitt},
  {Schneider}, {Schrabback}, {Schuster}, {Schwope}, {Seppi}, {Serim},
  {Shreeram}, {Sokolova-Lapa}, {Starck}, {Stelzer}, {Stierhof}, {Suleimanov},
  {Tenzer}, {Traulsen}, {Tr{\"u}mper}, {Tsuge}, {Urrutia}, {Veronica},
  {Waddell}, {Willer}, {Wolf}, {Yeung}, {Zainab}, {Zangrandi}, {Zhang},
  {Zhang}, \& {Zheng}}]{merloni24}
{Merloni}, A., {Lamer}, G., {Liu}, T., {et~al.} 2024, \aap, 682, A34

\bibitem[{{Merloni} {et~al.}(2012){Merloni}, {Predehl}, {Becker},
  {B{\"o}hringer}, {Boller}, {Brunner}, {Brusa}, {Dennerl}, {Freyberg},
  {Friedrich}, {Georgakakis}, {Haberl}, {Hasinger}, {Meidinger}, {Mohr},
  {Nandra}, {Rau}, {Reiprich}, {Robrade}, {Salvato}, {Santangelo}, {Sasaki},
  {Schwope}, {Wilms}, \& {German eROSITA Consortium}}]{merloni12}
{Merloni}, A., {Predehl}, P., {Becker}, W., {et~al.} 2012, arXiv e-prints,
  arXiv:1209.3114

\bibitem[{{Miyatake} {et~al.}(2019){Miyatake}, {Battaglia}, {Hilton},
  {Medezinski}, {Nishizawa}, {More}, {Aiola}, {Bahcall}, {Bond}, {Calabrese},
  {Choi}, {Devlin}, {Dunkley}, {Dunner}, {Fuzia}, {Gallardo}, {Gralla},
  {Hasselfield}, {Halpern}, {Hikage}, {Hill}, {Hincks}, {Hlo{\v{z}}ek},
  {Huffenberger}, {Hughes}, {Koopman}, {Kosowsky}, {Louis}, {Madhavacheril},
  {McMahon}, {Mandelbaum}, {Marriage}, {Maurin}, {Miyazaki}, {Moodley},
  {Murata}, {Naess}, {Newburgh}, {Niemack}, {Nishimichi}, {Okabe}, {Oguri},
  {Osato}, {Page}, {Partridge}, {Robertson}, {Sehgal}, {Sherwin}, {Shirasaki},
  {Sievers}, {Sif{\'o}n}, {Simon}, {Spergel}, {Staggs}, {Stein}, {Takada},
  {Trac}, {Umetsu}, {van Engelen}, \& {Wollack}}]{miyatake19}
{Miyatake}, H., {Battaglia}, N., {Hilton}, M., {et~al.} 2019, \apj, 875, 63

\bibitem[{{Miyazaki}(2015)}]{miyazaki15}
{Miyazaki}, S. 2015, IAU General Assembly, 22, 2255916

\bibitem[{{Miyazaki} {et~al.}(2018){Miyazaki}, {Komiyama}, {Kawanomoto}, {Doi},
  {Furusawa}, {Hamana}, {Hayashi}, {Ikeda}, {Kamata}, {Karoji}, {Koike},
  {Kurakami}, {Miyama}, {Morokuma}, {Nakata}, {Namikawa}, {Nakaya}, {Nariai},
  {Obuchi}, {Oishi}, {Okada}, {Okura}, {Tait}, {Takata}, {Tanaka}, {Tanaka},
  {Terai}, {Tomono}, {Uraguchi}, {Usuda}, {Utsumi}, {Yamada}, {Yamanoi},
  {Aihara}, {Fujimori}, {Mineo}, {Miyatake}, {Oguri}, {Uchida}, {Tanaka},
  {Yasuda}, {Takada}, {Murayama}, {Nishizawa}, {Sugiyama}, {Chiba}, {Futamase},
  {Wang}, {Chen}, {Ho}, {Liaw}, {Chiu}, {Ho}, {Lai}, {Lee}, {Jeng}, {Iwamura},
  {Armstrong}, {Bickerton}, {Bosch}, {Gunn}, {Lupton}, {Loomis}, {Price},
  {Smith}, {Strauss}, {Turner}, {Suzuki}, {Miyazaki}, {Muramatsu}, {Yamamoto},
  {Endo}, {Ezaki}, {Ito}, {Kawaguchi}, {Sofuku}, {Taniike}, {Akutsu}, {Dojo},
  {Kasumi}, {Matsuda}, {Imoto}, {Miwa}, {Suzuki}, {Takeshi}, \&
  {Yokota}}]{miyazaki18}
{Miyazaki}, S., {Komiyama}, Y., {Kawanomoto}, S., {et~al.} 2018, \pasj, 70, S1

\bibitem[{{Moster} {et~al.}(2013){Moster}, {Naab}, \& {White}}]{moster13}
{Moster}, B.~P., {Naab}, T., \& {White}, S. D.~M. 2013, \mnras, 428, 3121

\bibitem[{{Moustakas} {et~al.}(2013){Moustakas}, {Coil}, {Aird}, {Blanton},
  {Cool}, {Eisenstein}, {Mendez}, {Wong}, {Zhu}, \& {Arnouts}}]{moustakas13}
{Moustakas}, J., {Coil}, A.~L., {Aird}, J., {et~al.} 2013, \apj, 767, 50

\bibitem[{{Murata} {et~al.}(2019){Murata}, {Oguri}, {Nishimichi}, {Takada},
  {Mandelbaum}, {More}, {Shirasaki}, {Nishizawa}, \& {Osato}}]{murata19}
{Murata}, R., {Oguri}, M., {Nishimichi}, T., {et~al.} 2019, \pasj, 71, 107

\bibitem[{Navarro {et~al.}(1997)Navarro, Frenk, \& White}]{navarro97}
Navarro, J., Frenk, C., \& White, S. 1997, \apj, 490, 493

\bibitem[{{Nishizawa} {et~al.}(2020){Nishizawa}, {Hsieh}, {Tanaka}, \&
  {Takata}}]{nishizawa20}
{Nishizawa}, A.~J., {Hsieh}, B.-C., {Tanaka}, M., \& {Takata}, T. 2020, arXiv
  e-prints, arXiv:2003.01511

\bibitem[{{Oguri}(2014)}]{oguri14}
{Oguri}, M. 2014, \mnras, 444, 147

\bibitem[{{Okabe} {et~al.}(2025){Okabe}, {Reiprich}, {Grandis}, {Chiu},
  {Oguri}, {Umetsu}, {Bulbul}, {Bahar}, {Balzer}, {Clerc}, {Comparat},
  {Ghirardini}, {Kleinebreil}, {Kluge}, {Liu}, {Lin}, {Monteiro-Oliveira},
  {Pacaud}, {Ramos Ceja}, {Sanders}, {Schrabback}, {Seppi}, {Sommer}, \&
  {Zhang}}]{okabe25}
{Okabe}, N., {Reiprich}, T., {Grandis}, S., {et~al.} 2025, arXiv e-prints,
  arXiv:2503.09952

\bibitem[{{Okabe} \& {Smith}(2016)}]{okabe16}
{Okabe}, N. \& {Smith}, G.~P. 2016, \mnras, 461, 3794

\bibitem[{{Okabe} {et~al.}(2010){Okabe}, {Takada}, {Umetsu}, {Futamase}, \&
  {Smith}}]{okabe10a}
{Okabe}, N., {Takada}, M., {Umetsu}, K., {Futamase}, T., \& {Smith}, G.~P.
  2010, \pasj, 62, 811

\bibitem[{P\'erez \& Granger(2007)}]{ipython}
P\'erez, F. \& Granger, B.~E. 2007, Computing in Science and Engineering, 9, 21

\bibitem[{{Pillepich} {et~al.}(2018){Pillepich}, {Reiprich}, {Porciani},
  {Borm}, \& {Merloni}}]{pillepich18}
{Pillepich}, A., {Reiprich}, T.~H., {Porciani}, C., {Borm}, K., \& {Merloni},
  A. 2018, \mnras, 481, 613

\bibitem[{Pratt {et~al.}(2009)Pratt, Croston, Arnaud, \&
  B{\"{o}}hringer}]{pratt09}
Pratt, G., Croston, J., Arnaud, M., \& B{\"{o}}hringer, H. 2009, \aap, 498, 361

\bibitem[{{Pratt} {et~al.}(2019){Pratt}, {Arnaud}, {Biviano}, {Eckert},
  {Ettori}, {Nagai}, {Okabe}, \& {Reiprich}}]{pratt19}
{Pratt}, G.~W., {Arnaud}, M., {Biviano}, A., {et~al.} 2019, \ssr, 215, 25

\bibitem[{{Predehl} {et~al.}(2021){Predehl}, {Andritschke}, {Arefiev},
  {Babyshkin}, {Batanov}, {Becker}, {B{\"o}hringer}, {Bogomolov}, {Boller},
  {Borm}, {Bornemann}, {Br{\"a}uninger}, {Br{\"u}ggen}, {Brunner}, {Brusa},
  {Bulbul}, {Buntov}, {Burwitz}, {Burkert}, {Clerc}, {Churazov}, {Coutinho},
  {Dauser}, {Dennerl}, {Doroshenko}, {Eder}, {Emberger}, {Eraerds},
  {Finoguenov}, {Freyberg}, {Friedrich}, {Friedrich}, {F{\"u}rmetz},
  {Georgakakis}, {Gilfanov}, {Granato}, {Grossberger}, {Gueguen}, {Gureev},
  {Haberl}, {H{\"a}lker}, {Hartner}, {Hasinger}, {Huber}, {Ji}, {Kienlin},
  {Kink}, {Korotkov}, {Kreykenbohm}, {Lamer}, {Lomakin}, {Lapshov}, {Liu},
  {Maitra}, {Meidinger}, {Menz}, {Merloni}, {Mernik}, {Mican}, {Mohr},
  {M{\"u}ller}, {Nandra}, {Nazarov}, {Pacaud}, {Pavlinsky}, {Perinati},
  {Pfeffermann}, {Pietschner}, {Ramos-Ceja}, {Rau}, {Reiffers}, {Reiprich},
  {Robrade}, {Salvato}, {Sanders}, {Santangelo}, {Sasaki}, {Scheuerle},
  {Schmid}, {Schmitt}, {Schwope}, {Shirshakov}, {Steinmetz}, {Stewart},
  {Str{\"u}der}, {Sunyaev}, {Tenzer}, {Tiedemann}, {Tr{\"u}mper}, {Voron},
  {Weber}, {Wilms}, \& {Yaroshenko}}]{predehl21}
{Predehl}, P., {Andritschke}, R., {Arefiev}, V., {et~al.} 2021, \aap, 647, A1

\bibitem[{{Prevot} {et~al.}(1984){Prevot}, {Lequeux}, {Maurice}, {Prevot}, \&
  {Rocca-Volmerange}}]{prevot84}
{Prevot}, M.~L., {Lequeux}, J., {Maurice}, E., {Prevot}, L., \&
  {Rocca-Volmerange}, B. 1984, \aap, 132, 389

\bibitem[{{Ragagnin} {et~al.}(2021){Ragagnin}, {Saro}, {Singh}, \&
  {Dolag}}]{ragagnin21}
{Ragagnin}, A., {Saro}, A., {Singh}, P., \& {Dolag}, K. 2021, \mnras, 500, 5056

\bibitem[{{Rau} {et~al.}(2023){Rau}, {Dalal}, {Zhang}, {Li}, {Nishizawa},
  {More}, {Mandelbaum}, {Miyatake}, {Strauss}, \& {Takada}}]{rau23}
{Rau}, M.~M., {Dalal}, R., {Zhang}, T., {et~al.} 2023, \mnras, 524, 5109

\bibitem[{{Reiprich} \& {B{\"o}hringer}(2002)}]{reiprich02}
{Reiprich}, T.~H. \& {B{\"o}hringer}, H. 2002, \apj, 567, 716

\bibitem[{{Reynolds} {et~al.}(2023){Reynolds}, {Kara}, {Mushotzky}, {Ptak},
  {Koss}, {Williams}, {Allen}, {Bauer}, {Bautz}, {Bogadhee}, {Burdge},
  {Cappelluti}, {Cenko}, {Chartas}, {Chan}, {Corrales}, {Daylan}, {Falcone},
  {Foord}, {Grant}, {Habouzit}, {Haggard}, {Herrmann}, {Hodges-Kluck},
  {Kargaltsev}, {King}, {Kounkel}, {Lopez}, {Marchesi}, {McDonald}, {Meyer},
  {Miller}, {Nynka}, {Okajima}, {Pacucci}, {Russell}, {Safi-Harb}, {Strassun},
  {Trindade Falc{\~a}o}, {Walker}, {Wilms}, {Yukita}, \& {Zhang}}]{reynolds23}
{Reynolds}, C.~S., {Kara}, E.~A., {Mushotzky}, R.~F., {et~al.} 2023, in Society
  of Photo-Optical Instrumentation Engineers (SPIE) Conference Series, Vol.
  12678, UV, X-Ray, and Gamma-Ray Space Instrumentation for Astronomy XXIII,
  ed. O.~H. {Siegmund} \& K.~{Hoadley}, 126781E

\bibitem[{{Salvati} {et~al.}(2022){Salvati}, {Saro}, {Bocquet}, {Costanzi},
  {Ansarinejad}, {Benson}, {Bleem}, {Calzadilla}, {Carlstrom}, {Chang},
  {Chown}, {Crites}, {de Haan}, {Dobbs}, {Everett}, {Floyd}, {Grandis},
  {George}, {Halverson}, {Holder}, {Holzapfel}, {Hrubes}, {Lee}, {Luong-Van},
  {McDonald}, {McMahon}, {Meyer}, {Millea}, {Mocanu}, {Mohr}, {Natoli},
  {Omori}, {Padin}, {Pryke}, {Reichardt}, {Ruhl}, {Ruppin}, {Schaffer},
  {Schrabback}, {Shirokoff}, {Staniszewski}, {Stark}, {Vieira}, \&
  {Williamson}}]{salvati21}
{Salvati}, L., {Saro}, A., {Bocquet}, S., {et~al.} 2022, \apj, 934, 129

\bibitem[{{Schellenberger} \& {Reiprich}(2017)}]{schellenberger17}
{Schellenberger}, G. \& {Reiprich}, T.~H. 2017, \mnras, 471, 1370

\bibitem[{{Schlafly} {et~al.}(2012){Schlafly}, {Finkbeiner}, {Juri{\'c}},
  {Magnier}, {Burgett}, {Chambers}, {Grav}, {Hodapp}, {Kaiser}, {Kudritzki},
  {Martin}, {Morgan}, {Price}, {Rix}, {Stubbs}, {Tonry}, \&
  {Wainscoat}}]{schlafly12}
{Schlafly}, E.~F., {Finkbeiner}, D.~P., {Juri{\'c}}, M., {et~al.} 2012, \apj,
  756, 158

\bibitem[{{Schrabback} {et~al.}(2018){Schrabback}, {Applegate}, {Dietrich},
  {Hoekstra}, {Bocquet}, {Gonzalez}, {von der Linden}, {McDonald}, {Morrison},
  {Raihan}, {Allen}, {Bayliss}, {Benson}, {Bleem}, {Chiu}, {Desai}, {Foley},
  {de Haan}, {High}, {Hilbert}, {Mantz}, {Massey}, {Mohr}, {Reichardt}, {Saro},
  {Simon}, {Stern}, {Stubbs}, \& {Zenteno}}]{schrabback18}
{Schrabback}, T., {Applegate}, D., {Dietrich}, J.~P., {et~al.} 2018, \mnras,
  474, 2635

\bibitem[{{Seppi} {et~al.}(2022){Seppi}, {Comparat}, {Bulbul}, {Nandra},
  {Merloni}, {Clerc}, {Liu}, {Ghirardini}, {Liu}, {Salvato}, {Sanders},
  {Wilms}, {Dwelly}, {Dauser}, {K{\"o}nig}, {Ramos-Ceja}, {Garrel}, \&
  {Reiprich}}]{seppi22}
{Seppi}, R., {Comparat}, J., {Bulbul}, E., {et~al.} 2022, \aap, 665, A78

\bibitem[{{Seppi} {et~al.}(2024){Seppi}, {Comparat}, {Ghirardini}, {Garrel},
  {Artis}, {S{\'a}nchez}, {Liu}, {Clerc}, {Bulbul}, {Grandis}, {Kluge},
  {Reiprich}, {Merloni}, {Zhang}, {Bahar}, {Shreeram}, {Sanders}, {Ramos-Ceja},
  \& {Krumpe}}]{seppi24}
{Seppi}, R., {Comparat}, J., {Ghirardini}, V., {et~al.} 2024, \aap, 686, A196

\bibitem[{{Shirasaki} {et~al.}(2024){Shirasaki}, {Sif{\'o}n}, {Miyatake},
  {Lau}, {Zhang}, {Bahcall}, {Battaglia}, {Devlin}, {Dunkley}, {Farahi},
  {Hilton}, {Lin}, {Nagai}, {Staggs}, {Sunayama}, {Spergel}, \&
  {Wollack}}]{shirasaki24}
{Shirasaki}, M., {Sif{\'o}n}, C., {Miyatake}, H., {et~al.} 2024, \prd, 110,
  103006

\bibitem[{{Simet} {et~al.}(2017){Simet}, {McClintock}, {Mandelbaum}, {Rozo},
  {Rykoff}, {Sheldon}, \& {Wechsler}}]{simet17}
{Simet}, M., {McClintock}, T., {Mandelbaum}, R., {et~al.} 2017, \mnras, 466,
  3103

\bibitem[{{Sugiyama} {et~al.}(2023){Sugiyama}, {Miyatake}, {More}, {Li},
  {Shirasaki}, {Takada}, {Kobayashi}, {Takahashi}, {Nishimichi}, {Nishizawa},
  {Rau}, {Zhang}, {Dalal}, {Mandelbaum}, {Strauss}, {Hamana}, {Oguri}, {Osato},
  {Kannawadi}, {Hsieh}, {Luo}, {Armstrong}, {Bosch}, {Komiyama}, {Lupton},
  {Lust}, {Miyazaki}, {Murayama}, {Okura}, {Price}, {Tait}, {Tanaka}, \&
  {Wang}}]{sugiyama23}
{Sugiyama}, S., {Miyatake}, H., {More}, S., {et~al.} 2023, \prd, 108, 123521

\bibitem[{Sunayama {et~al.}(2024)Sunayama, Miyatake, Sugiyama, More, Li, Dalal,
  Rau, Shi, Chiu, Shirasaki, Zhang, \& Nishizawa}]{sunayama24}
Sunayama, T., Miyatake, H., Sugiyama, S., {et~al.} 2024, Phys. Rev. D, 110,
  083511

\bibitem[{{Sunayama} {et~al.}(2020){Sunayama}, {Park}, {Takada}, {Kobayashi},
  {Nishimichi}, {Kurita}, {More}, {Oguri}, \& {Osato}}]{sunayama20}
{Sunayama}, T., {Park}, Y., {Takada}, M., {et~al.} 2020, \mnras, 496, 4468

\bibitem[{{Sunyaev} {et~al.}(2021){Sunyaev}, {Arefiev}, {Babyshkin},
  {Bogomolov}, {Borisov}, {Buntov}, {Brunner}, {Burenin}, {Churazov},
  {Coutinho}, {Eder}, {Eismont}, {Freyberg}, {Gilfanov}, {Gureyev}, {Hasinger},
  {Khabibullin}, {Kolmykov}, {Komovkin}, {Krivonos}, {Lapshov}, {Levin},
  {Lomakin}, {Lutovinov}, {Medvedev}, {Merloni}, {Mernik}, {Mikhailov},
  {Molodtsov}, {Mzhelsky}, {M{\"u}ller}, {Nandra}, {Nazarov}, {Pavlinsky},
  {Poghodin}, {Predehl}, {Robrade}, {Sazonov}, {Scheuerle}, {Shirshakov},
  {Tkachenko}, \& {Voron}}]{sunyaev21}
{Sunyaev}, R., {Arefiev}, V., {Babyshkin}, V., {et~al.} 2021, \aap, 656, A132

\bibitem[{Sunyaev \& Zel'dovich(1972)}]{sunyaev72}
Sunyaev, R. \& Zel'dovich, Y. 1972, Comments on Astrophysics and Space Physics,
  4, 173

\bibitem[{Sunyaev \& Zel'dovich(1970)}]{sunyaev70b}
Sunyaev, R.~A. \& Zel'dovich, Y.~B. 1970, \apss, 7, 3

\bibitem[{{Tanaka} {et~al.}(2018){Tanaka}, {Coupon}, {Hsieh}, {Mineo},
  {Nishizawa}, {Speagle}, {Furusawa}, {Miyazaki}, \& {Murayama}}]{tanaka18}
{Tanaka}, M., {Coupon}, J., {Hsieh}, B.-C., {et~al.} 2018, \pasj, 70, S9

\bibitem[{{Tanaka} {et~al.}(2017){Tanaka}, {Hasinger}, {Silverman},
  {Bickerton}, {Furusawa}, {Harikane}, {Hu}, {Ikeda}, {Li}, {McCracken},
  {Price}, {Strauss}, {Koike}, {Komiyama}, {Mineo}, {Miyazaki}, {Nishizawa},
  {Takata}, {Utsumi}, {Yamada}, \& {Yasuda}}]{tanaka17}
{Tanaka}, M., {Hasinger}, G., {Silverman}, J.~D., {et~al.} 2017, arXiv
  e-prints, arXiv:1706.00566

\bibitem[{{Taylor}(2005)}]{topcat1}
{Taylor}, M.~B. 2005, in Astronomical Society of the Pacific Conference Series,
  Vol. 347, Astronomical Data Analysis Software and Systems XIV, ed.
  P.~{Shopbell}, M.~{Britton}, \& R.~{Ebert}, 29

\bibitem[{{Taylor}(2006)}]{topcat2}
{Taylor}, M.~B. 2006, in Astronomical Society of the Pacific Conference Series,
  Vol. 351, Astronomical Data Analysis Software and Systems XV, ed.
  C.~{Gabriel}, C.~{Arviset}, D.~{Ponz}, \& S.~{Enrique}, 666

\bibitem[{{The Dark Energy Survey Collaboration}(2005)}]{des05}
{The Dark Energy Survey Collaboration}. 2005, arXiv e-prints, astro:0510346

\bibitem[{{To} {et~al.}(2021){To}, {Krause}, {Rozo}, {Wu}, {Gruen}, {Wechsler},
  {Eifler}, {Rykoff}, {Costanzi}, {Becker}, {Bernstein}, {Blazek}, {Bocquet},
  {Bridle}, {Cawthon}, {Choi}, {Crocce}, {Davis}, {DeRose}, {Drlica-Wagner},
  {Elvin-Poole}, {Fang}, {Farahi}, {Friedrich}, {Gatti}, {Gaztanaga},
  {Giannantonio}, {Hartley}, {Hoyle}, {Jarvis}, {MacCrann}, {McClintock},
  {Miranda}, {Pereira}, {Park}, {Porredon}, {Prat}, {Rau}, {Ross}, {Samuroff},
  {S{\'a}nchez}, {Sevilla-Noarbe}, {Sheldon}, {Troxel}, {Varga}, {Vielzeuf},
  {Zhang}, {Zuntz}, {Abbott}, {Aguena}, {Amon}, {Annis}, {Avila}, {Bertin},
  {Bhargava}, {Brooks}, {Burke}, {Carnero Rosell}, {Carrasco Kind},
  {Carretero}, {Chang}, {Conselice}, {da Costa}, {Davis}, {Desai}, {Diehl},
  {Dietrich}, {Everett}, {Evrard}, {Ferrero}, {Flaugher}, {Fosalba}, {Frieman},
  {Garc{\'\i}a-Bellido}, {Gruendl}, {Gutierrez}, {Hinton}, {Hollowood},
  {Honscheid}, {Huterer}, {James}, {Jeltema}, {Kron}, {Kuehn}, {Kuropatkin},
  {Lima}, {Maia}, {Marshall}, {Menanteau}, {Miquel}, {Morgan}, {Muir}, {Myles},
  {Palmese}, {Paz-Chinch{\'o}n}, {Plazas}, {Romer}, {Roodman}, {Sanchez},
  {Santiago}, {Scarpine}, {Serrano}, {Smith}, {Suchyta}, {Swanson}, {Tarle},
  {Thomas}, {Tucker}, {Weller}, {Wester}, {Wilkinson}, \& {DES
  Collaboration}}]{to21}
{To}, C., {Krause}, E., {Rozo}, E., {et~al.} 2021, \prl, 126, 141301

\bibitem[{{Tonry} {et~al.}(2012){Tonry}, {Stubbs}, {Lykke}, {Doherty},
  {Shivvers}, {Burgett}, {Chambers}, {Hodapp}, {Kaiser}, {Kudritzki},
  {Magnier}, {Morgan}, {Price}, \& {Wainscoat}}]{tonry12}
{Tonry}, J.~L., {Stubbs}, C.~W., {Lykke}, K.~R., {et~al.} 2012, \apj, 750, 99

\bibitem[{Umetsu {et~al.}(2014)Umetsu, Medezinski, Nonino, Merten, Postman,
  Meneghetti, Donahue, Czakon, Molino, Seitz, Gruen, Lemze, Balestra, Benitez,
  Biviano, Broadhurst, Ford, Grillo, Koekemoer, Melchior, Mercurio, Moustakas,
  Rosati, \& Zitrin}]{umetsu14}
Umetsu, K., Medezinski, E., Nonino, M., {et~al.} 2014, \apj, 795, 163

\bibitem[{{Umetsu} {et~al.}(2020){Umetsu}, {Sereno}, {Lieu}, {Miyatake},
  {Medezinski}, {Nishizawa}, {Giles}, {Gastaldello}, {McCarthy}, {Kilbinger},
  {Birkinshaw}, {Ettori}, {Okabe}, {Chiu}, {Coupon}, {Eckert}, {Fujita},
  {Higuchi}, {Koulouridis}, {Maughan}, {Miyazaki}, {Oguri}, {Pacaud}, {Pierre},
  {Rapetti}, \& {Smith}}]{umetsu20}
{Umetsu}, K., {Sereno}, M., {Lieu}, M., {et~al.} 2020, \apj, 890, 148

\bibitem[{{Vale} \& {Ostriker}(2004)}]{vale04}
{Vale}, A. \& {Ostriker}, J.~P. 2004, \mnras, 353, 189

\bibitem[{van~der Burg {et~al.}(2014)van~der Burg, Muzzin, Hoekstra, Wilson,
  Lidman, \& Yee}]{burg14}
van~der Burg, R., Muzzin, A., Hoekstra, H., {et~al.} 2014, \aap, 561, A79

\bibitem[{Van Der~Walt {et~al.}(2011)Van Der~Walt, Colbert, \&
  Varoquaux}]{van2011numpy}
Van Der~Walt, S., Colbert, S.~C., \& Varoquaux, G. 2011, Computing in Science
  \& Engineering, 13, 22

\bibitem[{Vikhlinin {et~al.}(2009{\natexlab{a}})Vikhlinin, Burenin, Ebeling,
  Forman, Hornstrup, Jones, Kravtsov, Murray, Nagai, Quintana, \&
  Voevodkin}]{vikhlinin09a}
Vikhlinin, A., Burenin, R., Ebeling, H., {et~al.} 2009{\natexlab{a}}, \apj,
  692, 1033

\bibitem[{Vikhlinin {et~al.}(2009{\natexlab{b}})Vikhlinin, Kravtsov, Burenin,
  Ebeling, Forman, Hornstrup, Jones, Murray, Nagai, Quintana, \&
  Voevodkin}]{vikhlinin09b}
Vikhlinin, A., Kravtsov, A., Burenin, R., {et~al.} 2009{\natexlab{b}}, \apj,
  692, 1060

\bibitem[{{Virtanen} {et~al.}(2020){Virtanen}, {Gommers}, {Oliphant},
  {Haberland}, {Reddy}, {Cournapeau}, {Burovski}, {Peterson}, {Weckesser},
  {Bright}, {van der Walt}, {Brett}, {Wilson}, {Jarrod Millman}, {Mayorov},
  {Nelson}, {Jones}, {Kern}, {Larson}, {Carey}, {Polat}, {Feng}, {Moore}, {Vand
  erPlas}, {Laxalde}, {Perktold}, {Cimrman}, {Henriksen}, {Quintero}, {Harris},
  {Archibald}, {Ribeiro}, {Pedregosa}, {van Mulbregt}, \&
  {Contributors}}]{scipy}
{Virtanen}, P., {Gommers}, R., {Oliphant}, T.~E., {et~al.} 2020, Nature
  Methods, 17, 261

\bibitem[{{von der Linden} {et~al.}(2014){von der Linden}, {Allen},
  {Applegate}, {Kelly}, {Allen}, {Ebeling}, {Burchat}, {Burke}, {Donovan},
  {Morris}, {Blandford}, {Erben}, \& {Mantz}}]{vonderlinden14a}
{von der Linden}, A., {Allen}, M.~T., {Applegate}, D.~E., {et~al.} 2014,
  \mnras, 439, 2

\bibitem[{{Weaver} {et~al.}(2022){Weaver}, {Kauffmann}, {Ilbert}, {McCracken},
  {Moneti}, {Toft}, {Brammer}, {Shuntov}, {Davidzon}, {Hsieh}, {Laigle},
  {Anastasiou}, {Jespersen}, {Vinther}, {Capak}, {Casey}, {McPartland},
  {Milvang-Jensen}, {Mobasher}, {Sanders}, {Zalesky}, {Arnouts}, {Aussel},
  {Dunlop}, {Faisst}, {Franx}, {Furtak}, {Fynbo}, {Gould}, {Greve}, {Gwyn},
  {Kartaltepe}, {Kashino}, {Koekemoer}, {Kokorev}, {Le F{\`e}vre}, {Lilly},
  {Masters}, {Magdis}, {Mehta}, {Peng}, {Riechers}, {Salvato}, {Sawicki},
  {Scarlata}, {Scoville}, {Shirley}, {Silverman}, {Sneppen}, {Smolc̆i{\'c}},
  {Steinhardt}, {Stern}, {Tanaka}, {Taniguchi}, {Teplitz}, {Vaccari}, {Wang},
  \& {Zamorani}}]{weaver22}
{Weaver}, J.~R., {Kauffmann}, O.~B., {Ilbert}, O., {et~al.} 2022, \apjs, 258,
  11

\bibitem[{{Weinberg} {et~al.}(2013){Weinberg}, {Mortonson}, {Eisenstein},
  {Hirata}, {Riess}, \& {Rozo}}]{weinberg13}
{Weinberg}, D.~H., {Mortonson}, M.~J., {Eisenstein}, D.~J., {et~al.} 2013,
  \physrep, 530, 87

\bibitem[{Wright {et~al.}(2010)Wright, Eisenhardt, Mainzer, Ressler, Cutri,
  Jarrett, Kirkpatrick, Padgett, McMillan, Skrutskie, Stanford, Cohen, Walker,
  Mather, Leisawitz, {Gautier III}, McLean, Benford, Lonsdale, Blain, Mendez,
  Irace, Duval, Liu, Royer, Heinrichsen, Howard, Shannon, Kendall, Walsh,
  Larsen, Cardon, Schick, Schwalm, Abid, Fabinsky, Naes, \& Tsai}]{wright10}
Wright, E., Eisenhardt, P., Mainzer, A., {et~al.} 2010, \aj, 140, 1868

\bibitem[{{Zhang} {et~al.}(2020){Zhang}, {Ramos-Ceja}, {Pacaud}, \&
  {Reiprich}}]{zhang20}
{Zhang}, C., {Ramos-Ceja}, M.~E., {Pacaud}, F., \& {Reiprich}, T.~H. 2020,
  \aap, 642, A17

\bibitem[{{Zhang} {et~al.}(2016){Zhang}, {Miller}, {McKay}, {Rooney}, {Evrard},
  {Romer}, {Perfecto}, {Song}, {Desai}, {Mohr}, {Wilcox}, {Bermeo-Hernandez},
  {Jeltema}, {Hollowood}, {Bacon}, {Capozzi}, {Collins}, {Das}, {Gerdes},
  {Hennig}, {Hilton}, {Hoyle}, {Kay}, {Liddle}, {Mann}, {Mehrtens}, {Nichol},
  {Papovich}, {Sahl{\'e}n}, {Soares-Santos}, {Stott}, {Viana}, {Abbott},
  {Abdalla}, {Banerji}, {Bauer}, {Benoit-L{\'e}vy}, {Bertin}, {Brooks},
  {Buckley-Geer}, {Burke}, {Carnero Rosell}, {Castander}, {Diehl}, {Doel},
  {Cunha}, {Eifler}, {Fausti Neto}, {Fernandez}, {Flaugher}, {Fosalba},
  {Frieman}, {Gaztanaga}, {Gruen}, {Gruendl}, {Honscheid}, {James}, {Kuehn},
  {Kuropatkin}, {Lahav}, {Maia}, {Makler}, {Marshall}, {Martini}, {Miquel},
  {Ogando}, {Plazas}, {Roodman}, {Rykoff}, {Sako}, {Sanchez}, {Scarpine},
  {Schubnell}, {Sevilla}, {Smith}, {Sobreira}, {Suchyta}, {Swanson}, {Tarle},
  {Thaler}, {Tucker}, {Vikram}, \& {da Costa}}]{zhang16}
{Zhang}, Y., {Miller}, C., {McKay}, T., {et~al.} 2016, \apj, 816, 98

\bibitem[{{Zu} {et~al.}(2022){Zu}, {Song}, {Shao}, {Chen}, {Zheng}, {Gao},
  {Yu}, {Shan}, \& {Jing}}]{zu22}
{Zu}, Y., {Song}, Y., {Shao}, Z., {et~al.} 2022, \mnras, 511, 1789

\bibitem[{{Zuntz} {et~al.}(2015){Zuntz}, {Paterno}, {Jennings}, {Rudd},
  {Manzotti}, {Dodelson}, {Bridle}, {Sehrish}, \& {Kowalkowski}}]{zuntz15}
{Zuntz}, J., {Paterno}, M., {Jennings}, E., {et~al.} 2015, Astronomy and
  Computing, 12, 45

\end{thebibliography}

%
%

\onecolumn

\appendix

\section{The SQL query of the BCG photometry}
\label{app:sql}

In the HSC PDR3 database, we query the reliable \texttt{cmodel} photometry in $grizY$ for the BCGs using the following SQL script.
\begin{verbatim}
SELECT
photo.object_id,
photo.ra,
photo.dec,
photo.[g,r,i,z,y]_cmodel_mag - photo.a_[g,r,i,z,y] 
- offsets.[g,r,i,z,y]_mag_offset - correction.corr_[r,i]mag,
photo.[g,r,i,z,y]_cmodel_magerr
FROM pdr3_wide.forced                        as photo
LEFT JOIN pdr3_wide.masks                    as masks           USING (object_id)
LEFT JOIN pdr3_wide.mag_corr                 as correction      USING (object_id)
LEFT JOIN pdr3_wide.stellar_sequence_offsets as offsets         USING (skymap_id)
WHERE
photo.isprimary                                 is True  AND
photo.[g,r,i,z,y]_pixelflags_edge               is False AND
photo.[g,r,i,z,y]_pixelflags_interpolatedcenter is False AND
photo.[g,r,i,z,y]_pixelflags_crcenter           is False AND
photo.[g,r,i,z,y]_pixelflags_saturatedcenter    is False AND
photo.[g,r,i,z,y]_pixelflags_suspectcenter      is False AND
photo.[g,r,i,z,y]_cmodel_flag                   is False AND
masks.[g,r,i,z,y]_mask_brightstar_halo          is False AND
masks.[g,r,i,z,y]_mask_brightstar_ghost         is False AND
masks.[g,r,i,z,y]_mask_brightstar_blooming      is False AND
photo.[g,r,i,z,y]_inputcount_value              >= 2
\end{verbatim}

\section{The cutout images and the SED fitting of the BCGs}
\label{app:imaging}

\begin{figure*}[!ht]
\centering
\resizebox{0.33\textwidth}{!}{\includegraphics[scale=1]{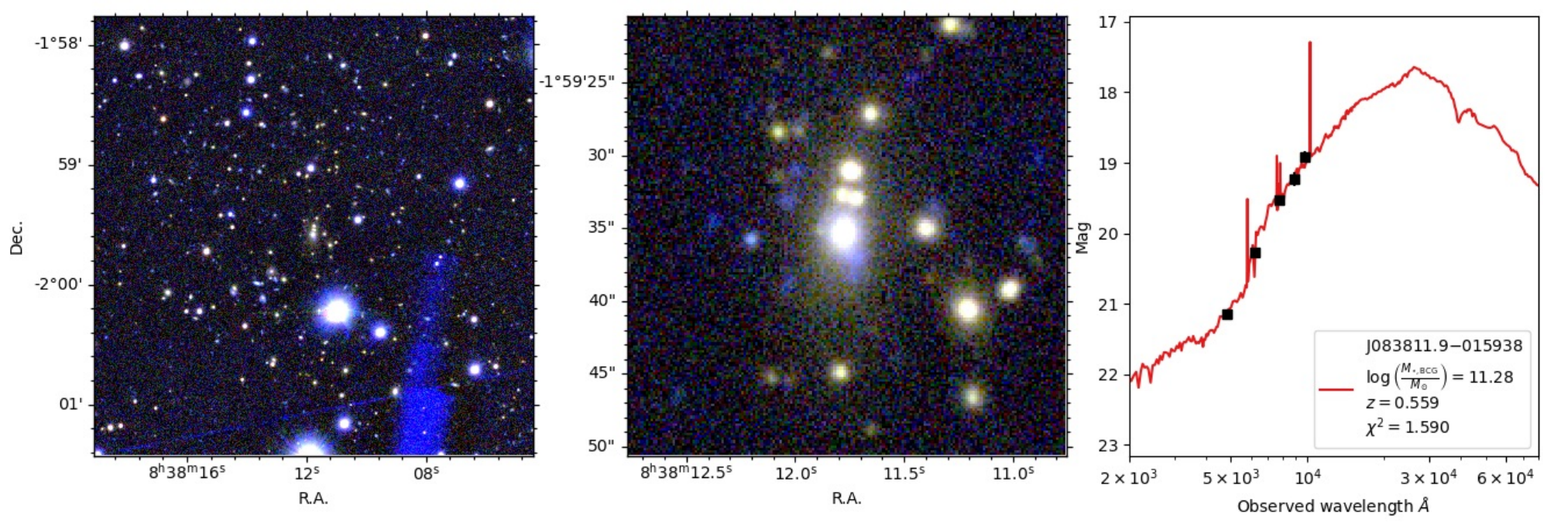}}
\resizebox{0.33\textwidth}{!}{\includegraphics[scale=1]{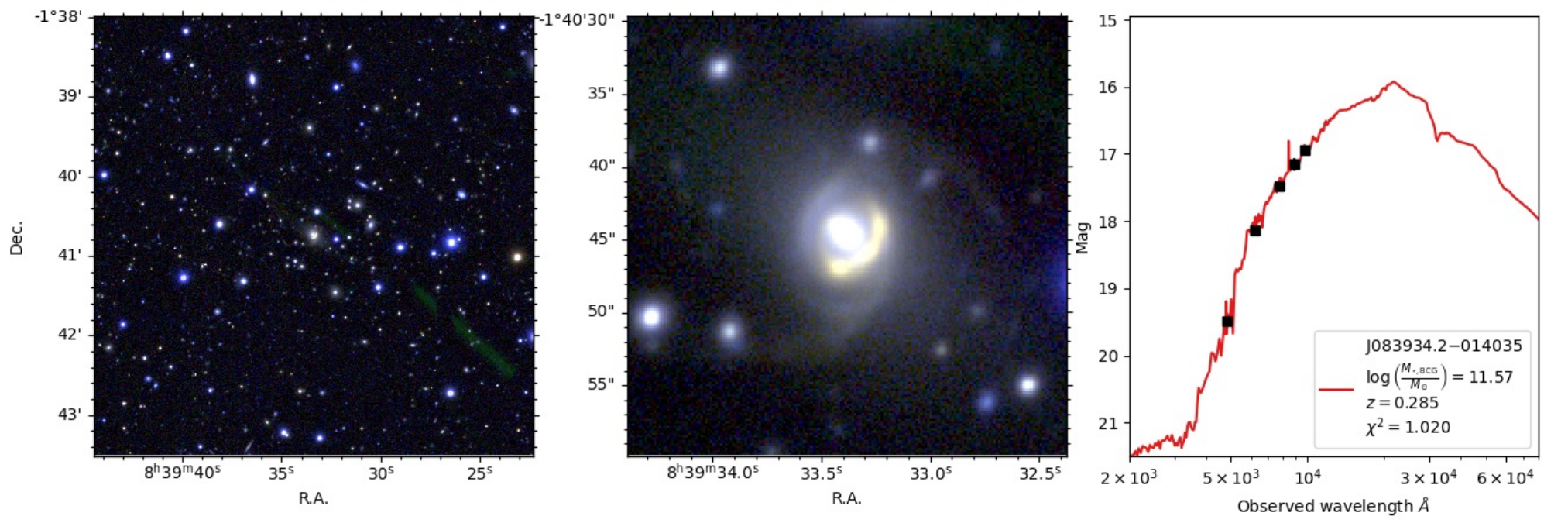}}
\resizebox{0.33\textwidth}{!}{\includegraphics[scale=1]{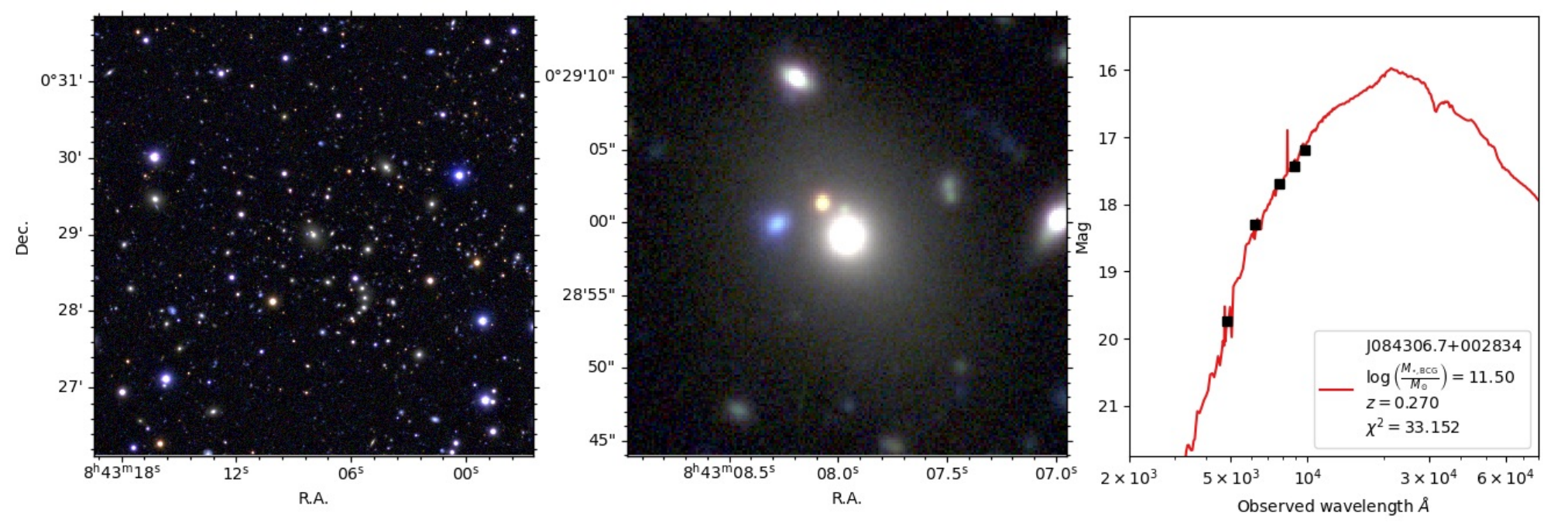}}\\
\resizebox{0.33\textwidth}{!}{\includegraphics[scale=1]{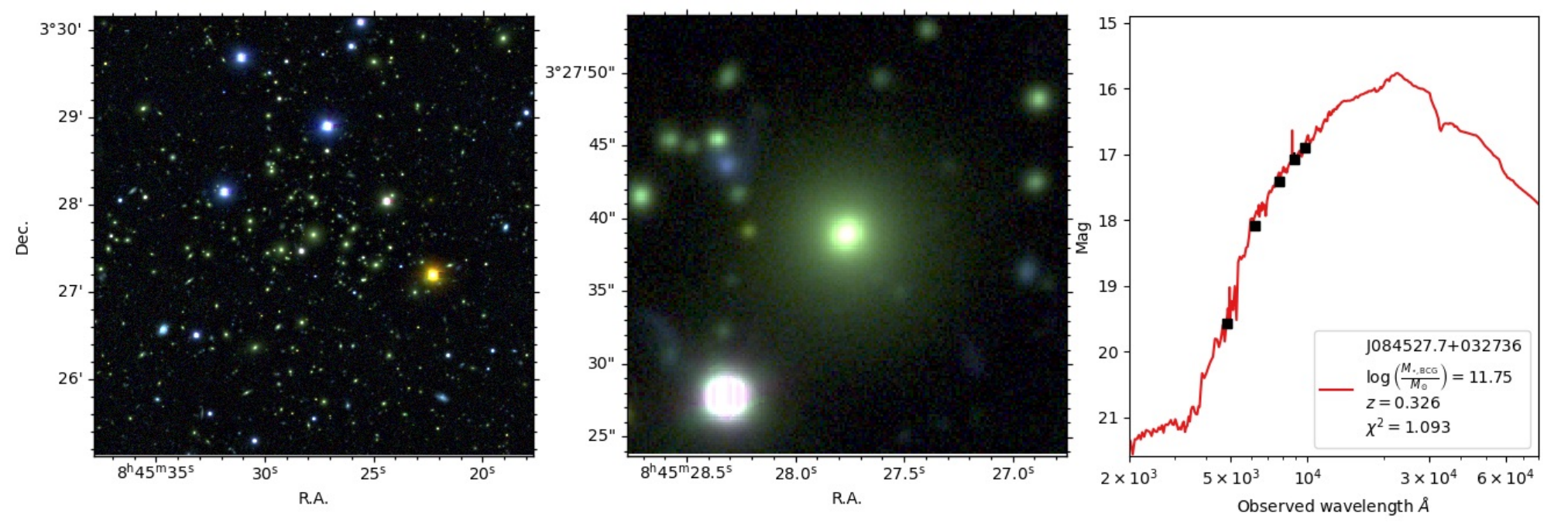}}
\resizebox{0.33\textwidth}{!}{\includegraphics[scale=1]{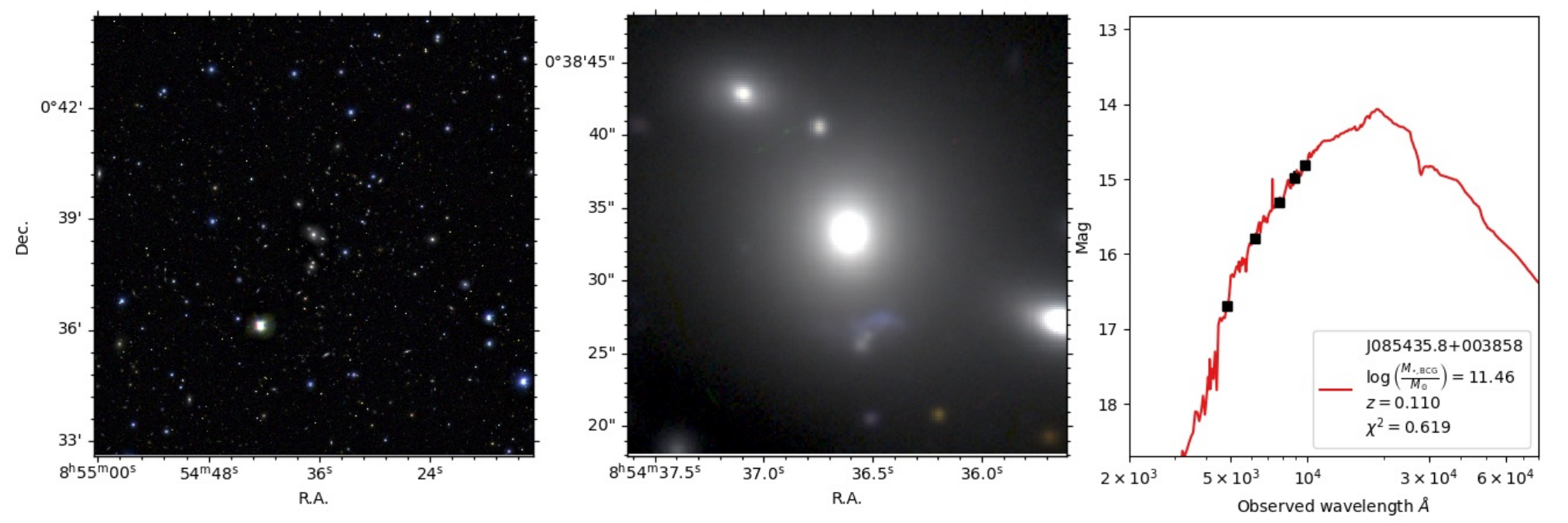}}
\resizebox{0.33\textwidth}{!}{\includegraphics[scale=1]{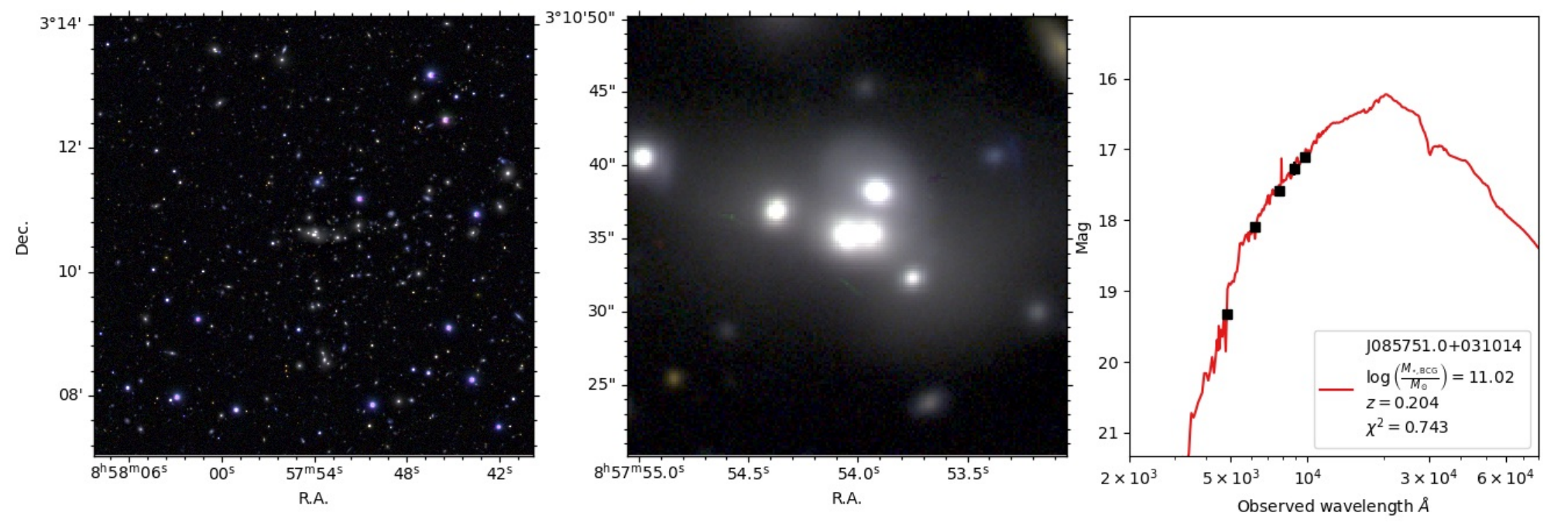}}\\
\resizebox{0.33\textwidth}{!}{\includegraphics[scale=1]{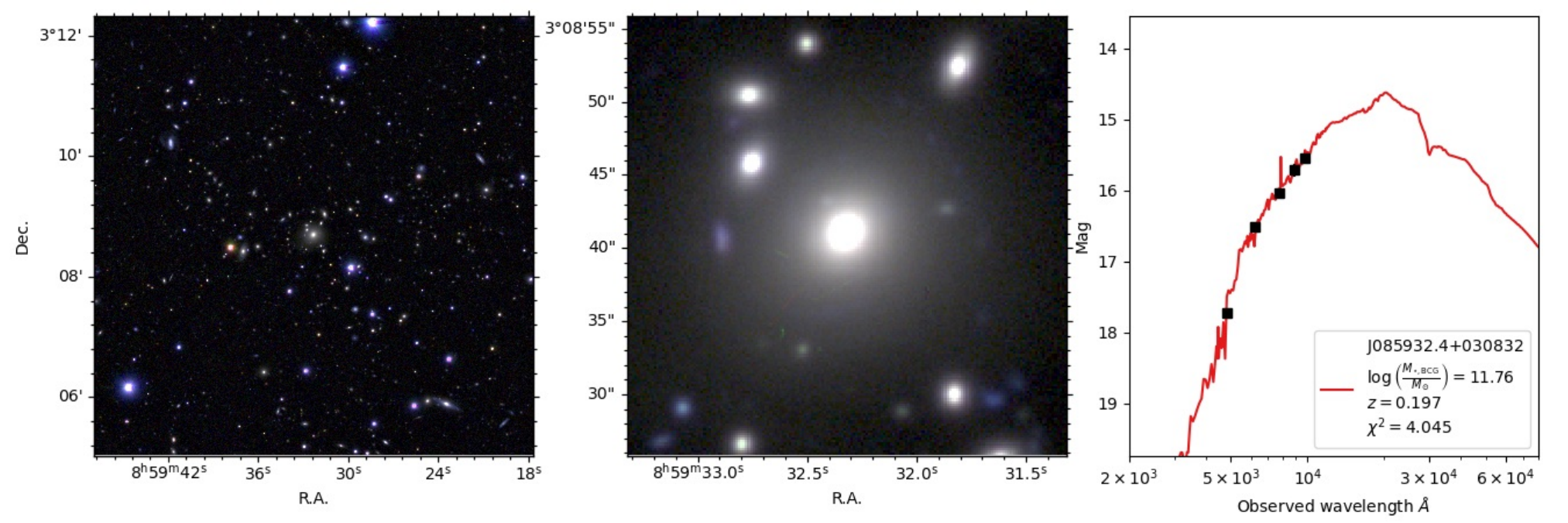}}
\resizebox{0.33\textwidth}{!}{\includegraphics[scale=1]{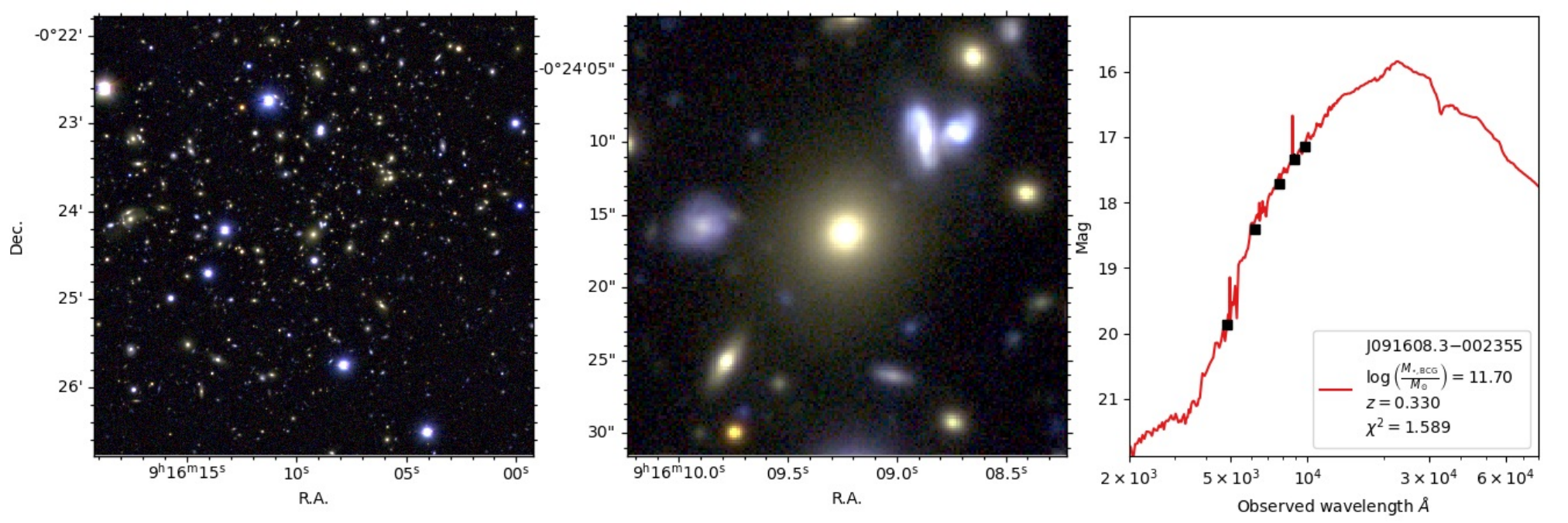}}
\resizebox{0.33\textwidth}{!}{\includegraphics[scale=1]{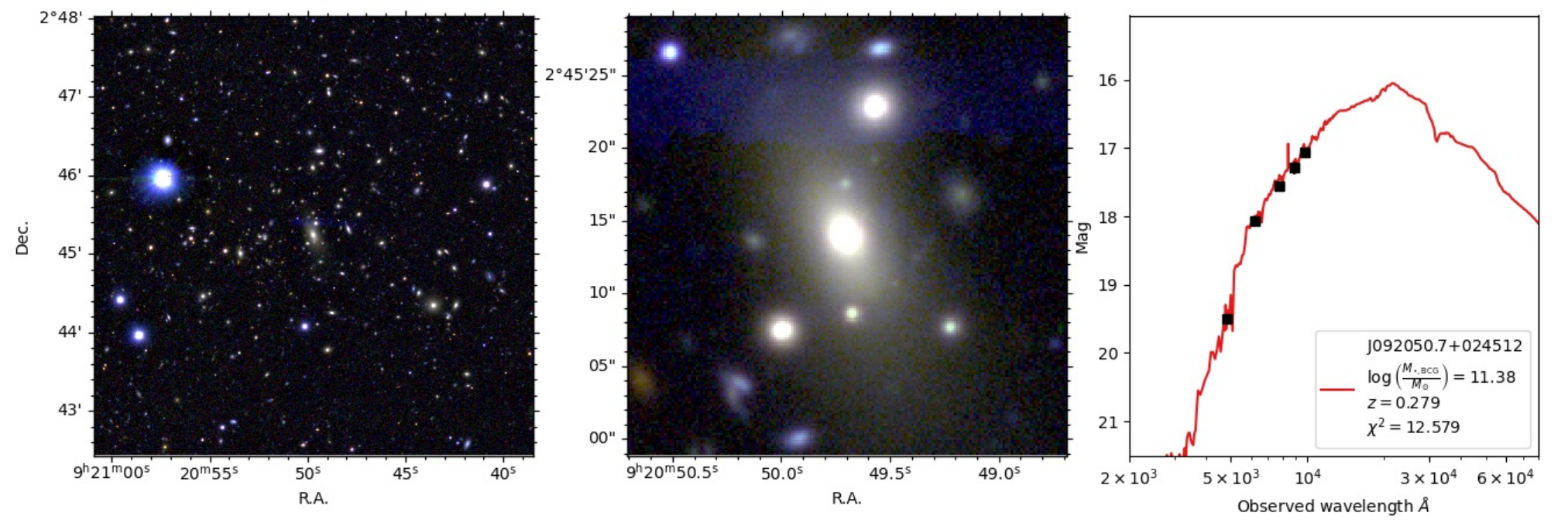}}\\
\resizebox{0.33\textwidth}{!}{\includegraphics[scale=1]{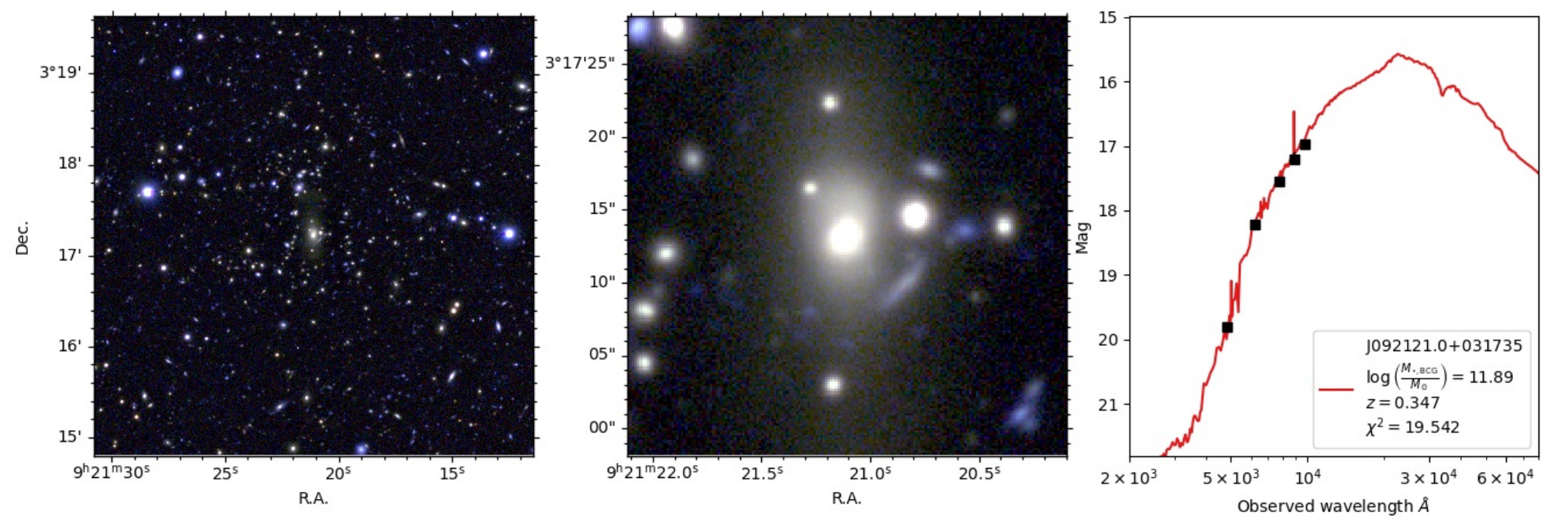}}
\resizebox{0.33\textwidth}{!}{\includegraphics[scale=1]{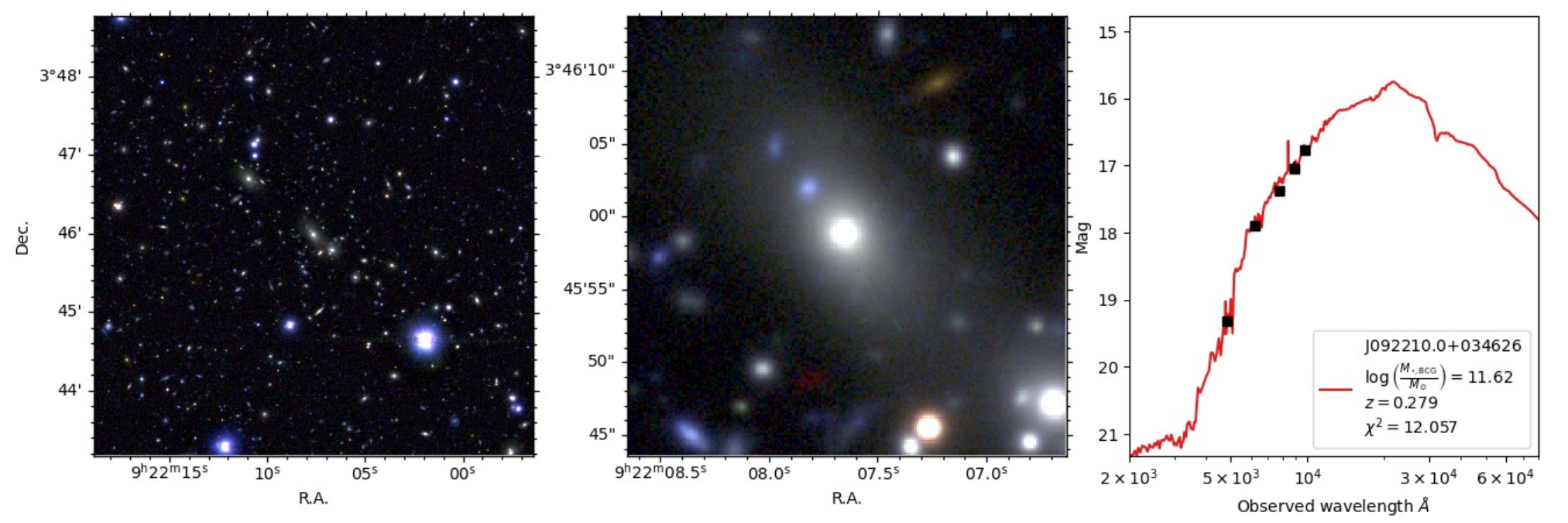}}
\resizebox{0.33\textwidth}{!}{\includegraphics[scale=1]{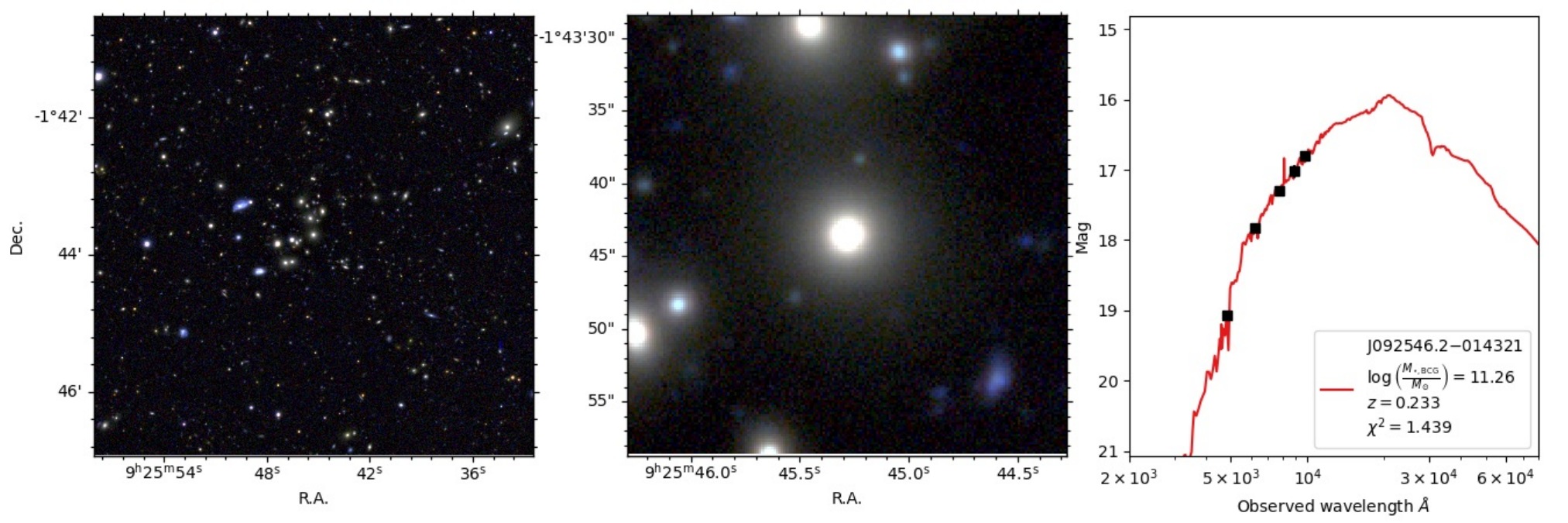}}\\
\resizebox{0.33\textwidth}{!}{\includegraphics[scale=1]{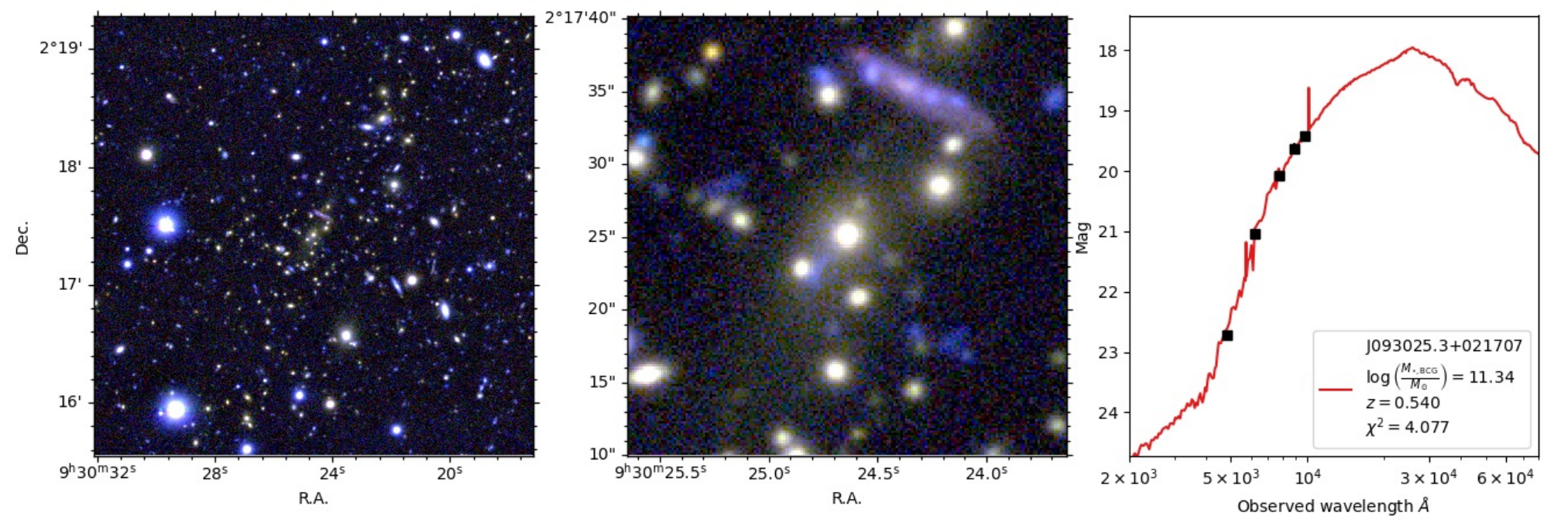}}
\resizebox{0.33\textwidth}{!}{\includegraphics[scale=1]{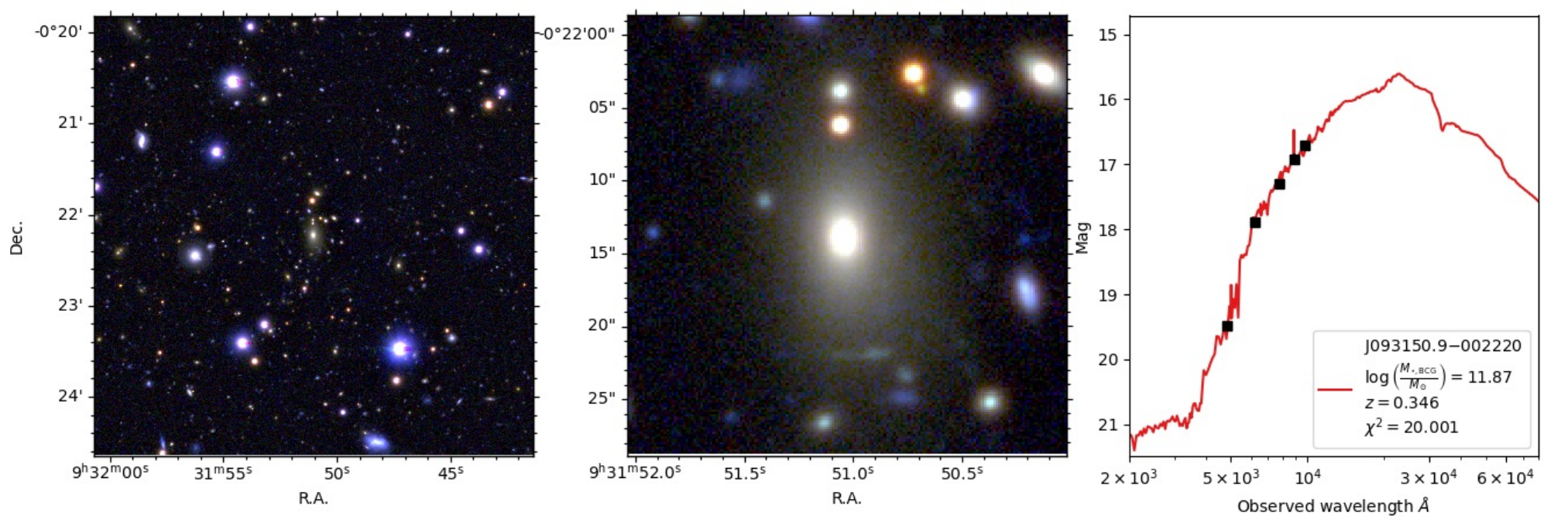}}
\resizebox{0.33\textwidth}{!}{\includegraphics[scale=1]{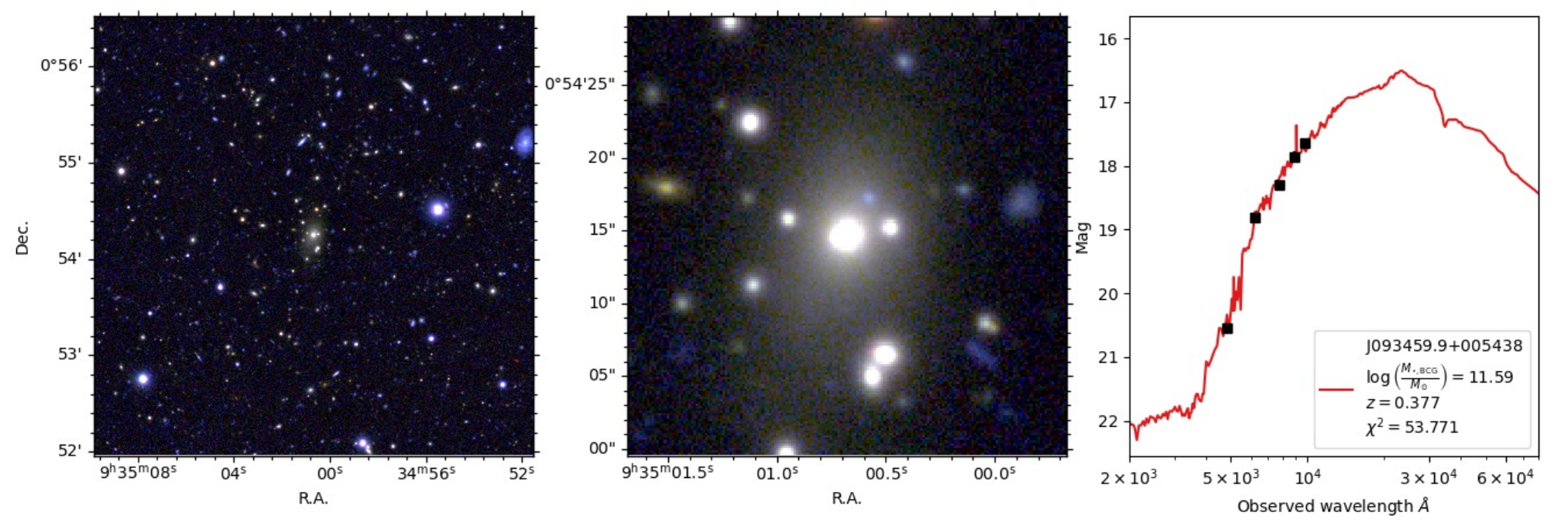}}\\
\resizebox{0.33\textwidth}{!}{\includegraphics[scale=1]{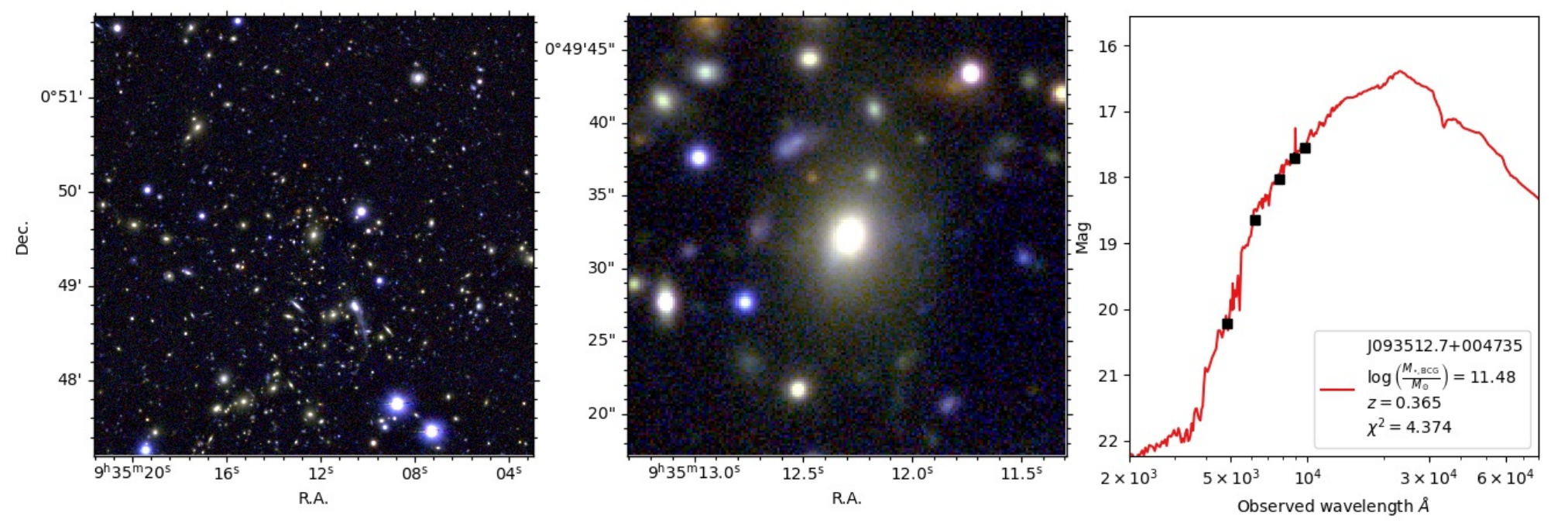}}
\resizebox{0.33\textwidth}{!}{\includegraphics[scale=1]{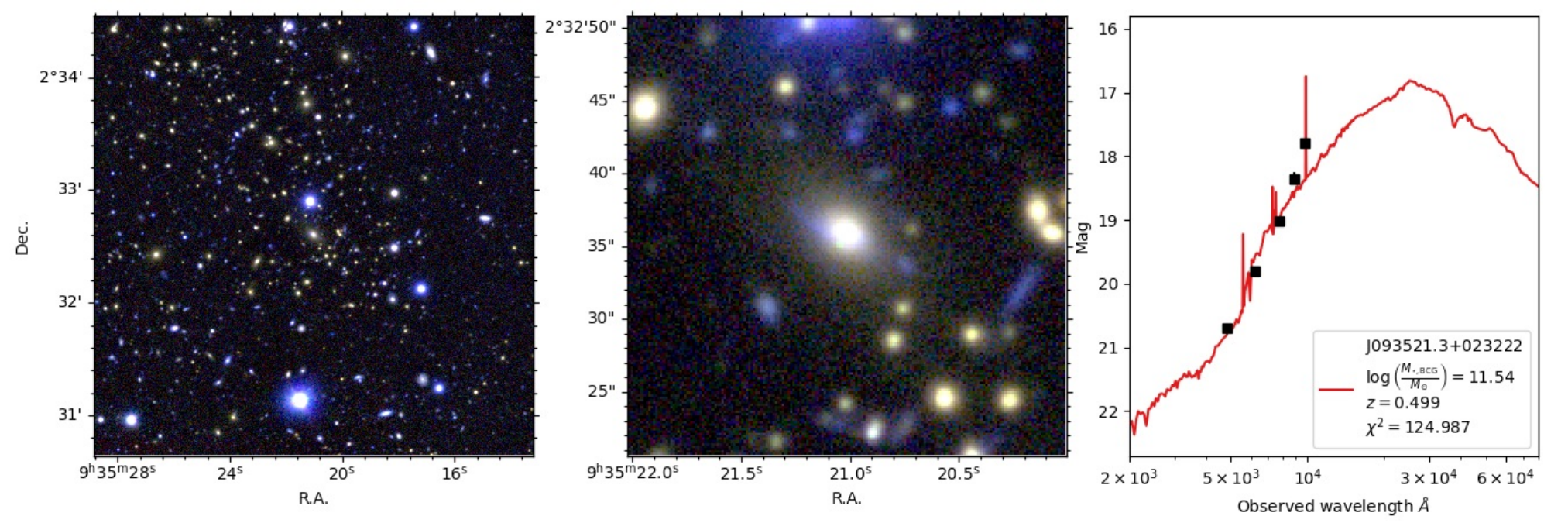}}
\resizebox{0.33\textwidth}{!}{\includegraphics[scale=1]{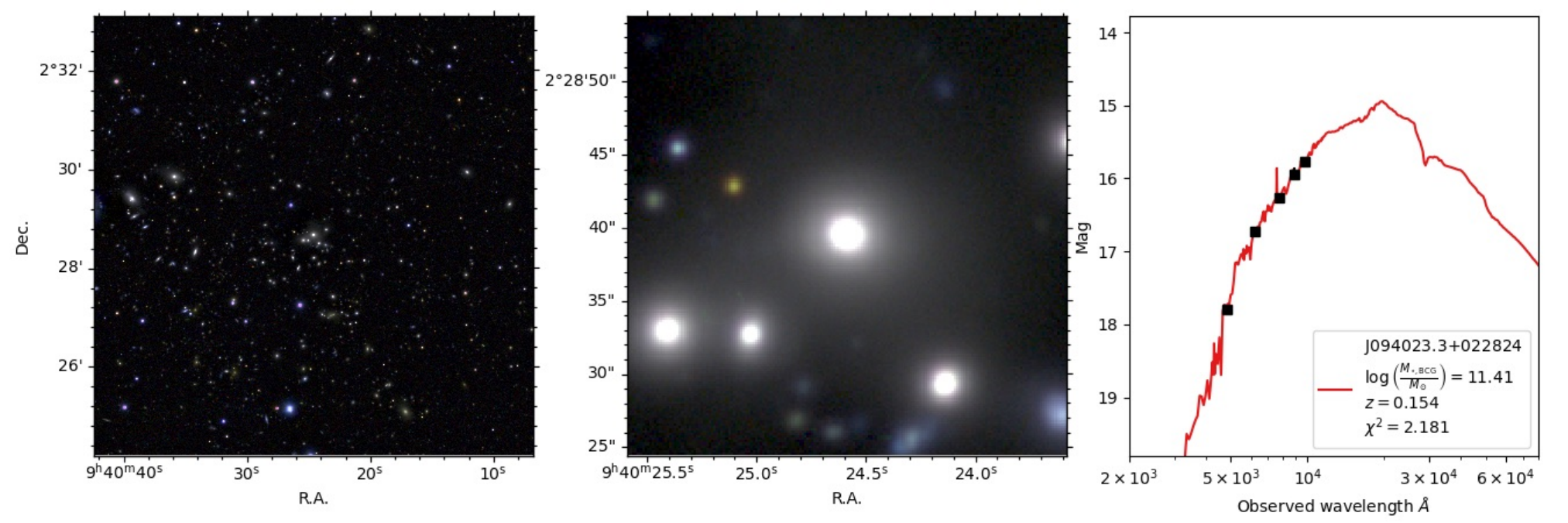}}\\
\resizebox{0.33\textwidth}{!}{\includegraphics[scale=1]{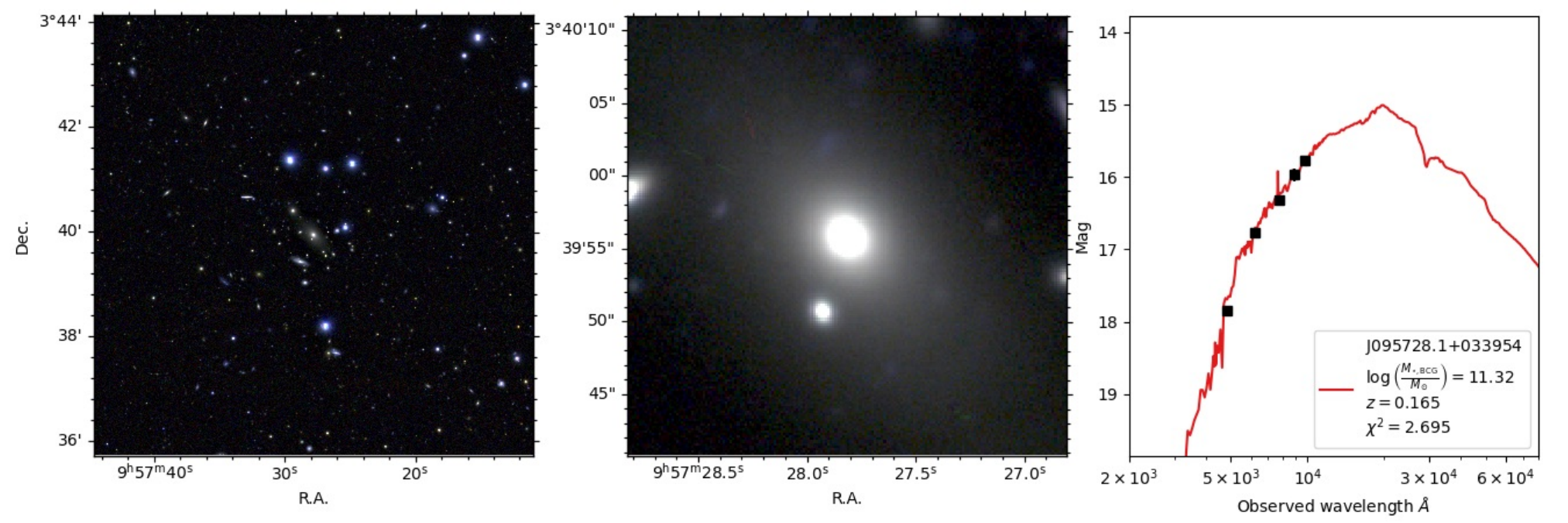}}
\resizebox{0.33\textwidth}{!}{\includegraphics[scale=1]{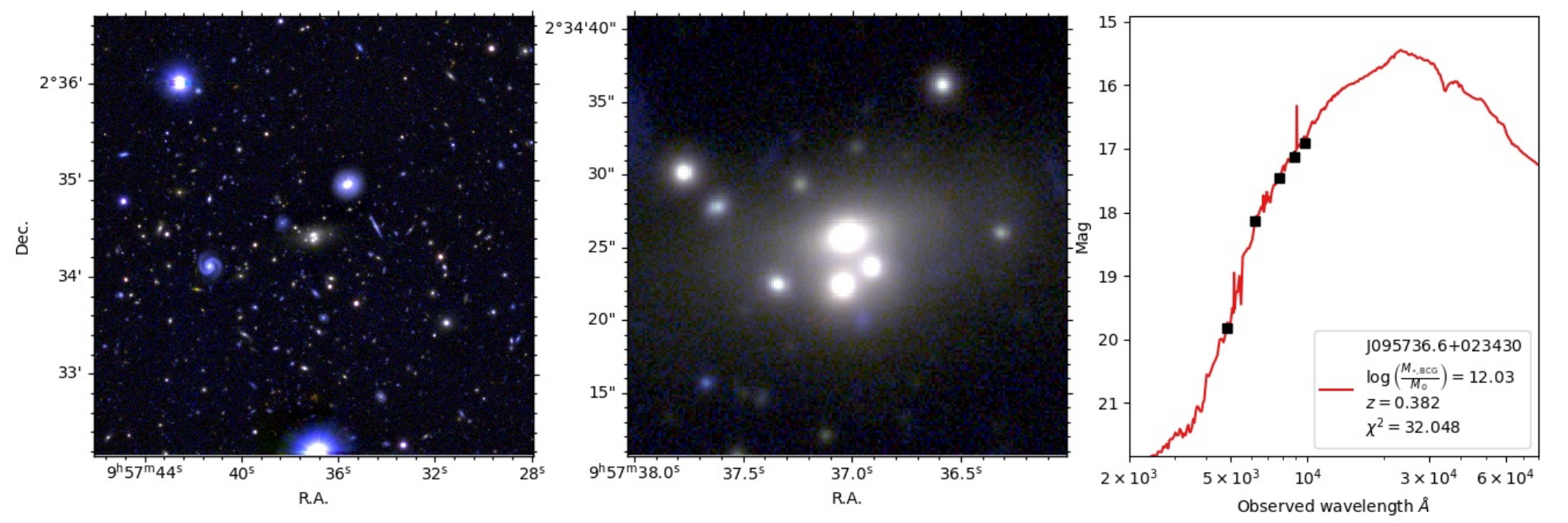}}
\resizebox{0.33\textwidth}{!}{\includegraphics[scale=1]{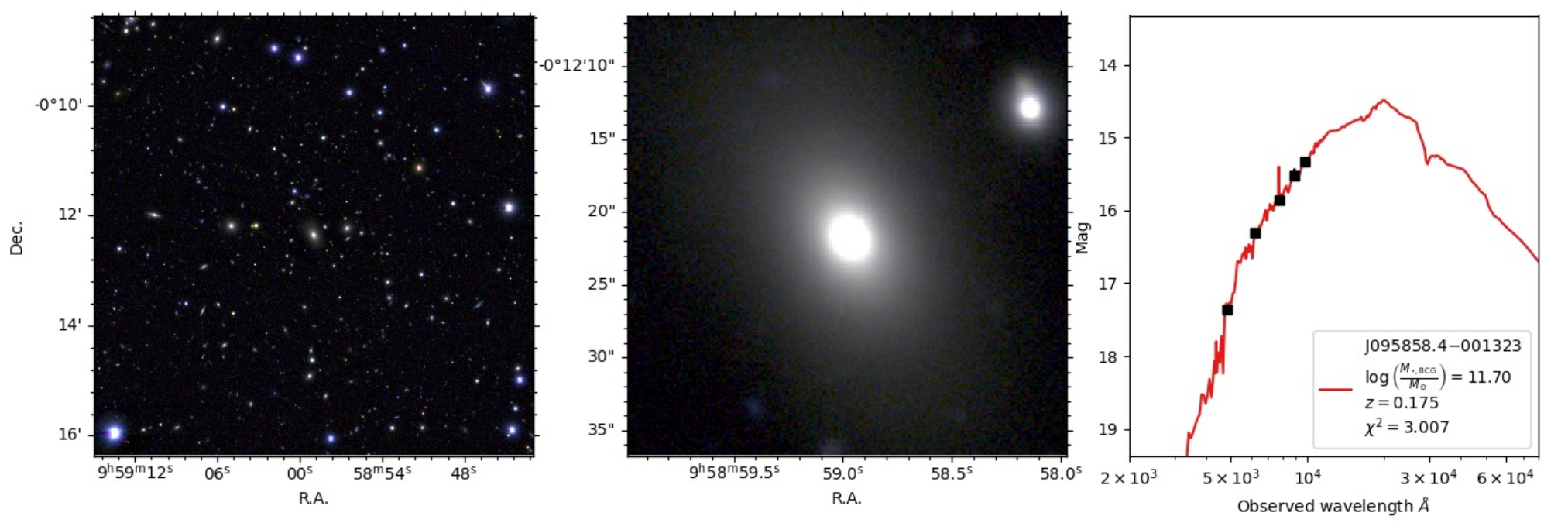}}\\
\resizebox{0.33\textwidth}{!}{\includegraphics[scale=1]{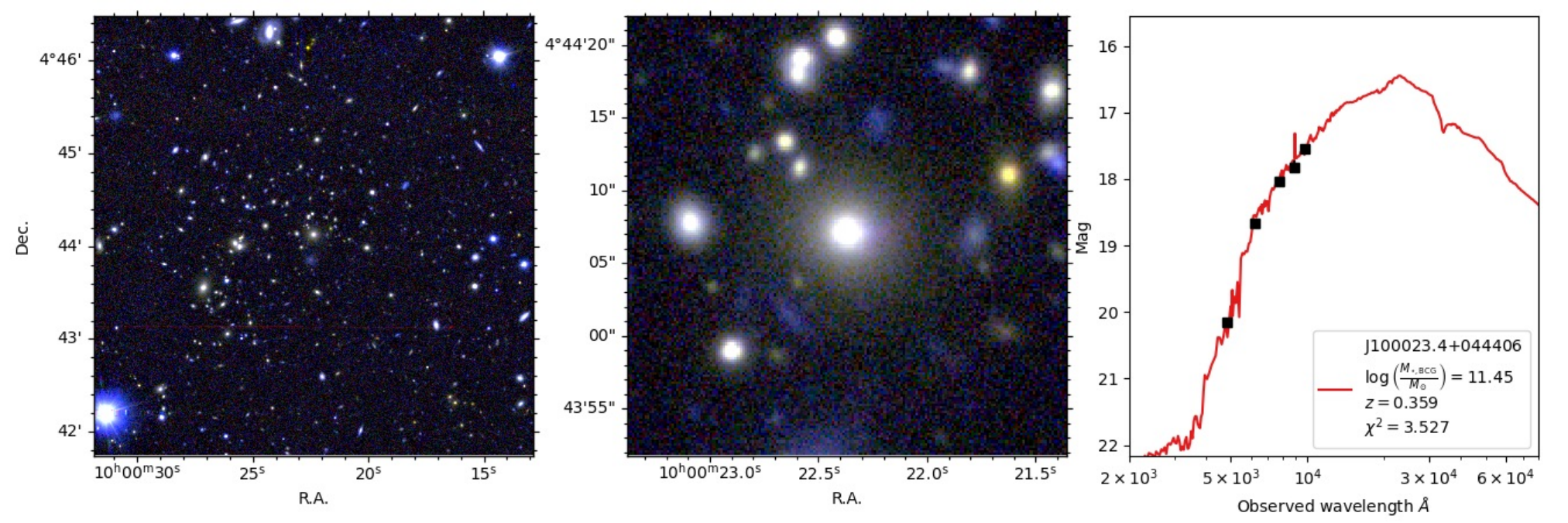}}
\resizebox{0.33\textwidth}{!}{\includegraphics[scale=1]{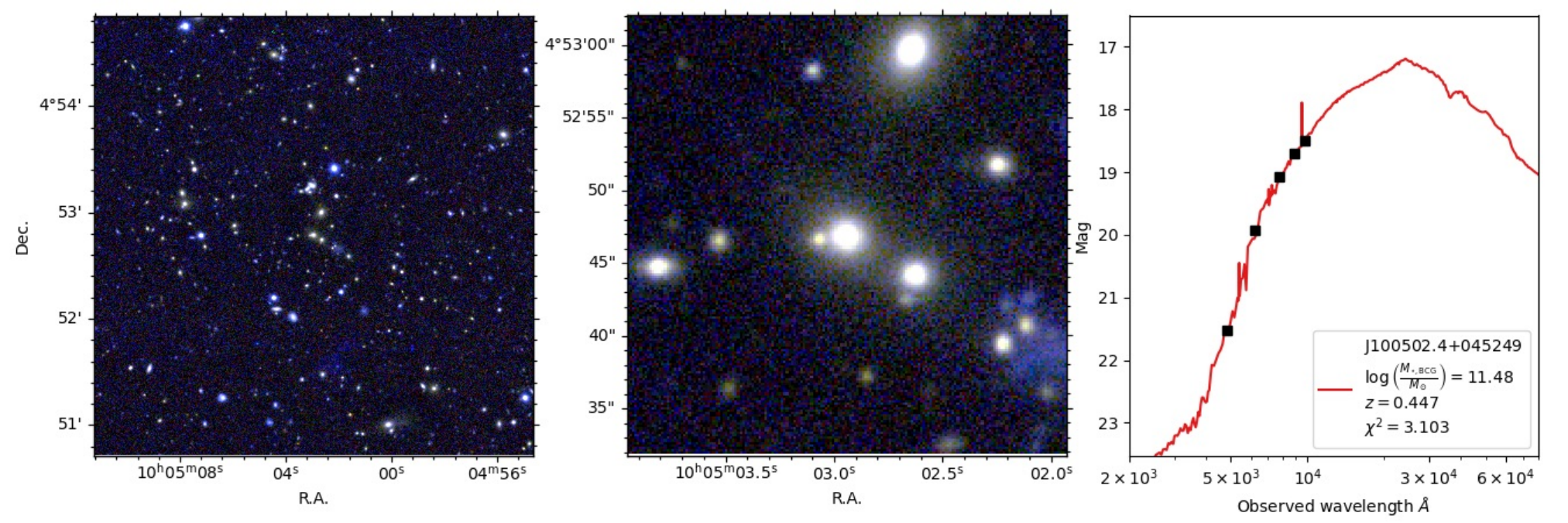}}
\resizebox{0.33\textwidth}{!}{\includegraphics[scale=1]{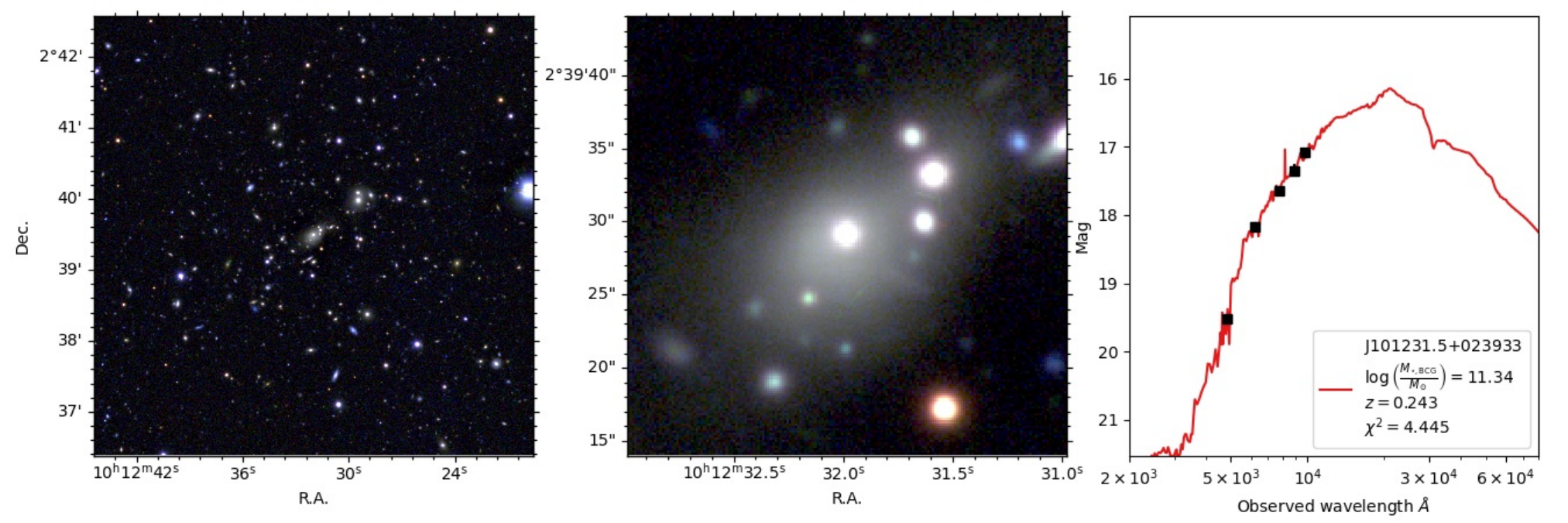}}\\
\resizebox{0.33\textwidth}{!}{\includegraphics[scale=1]{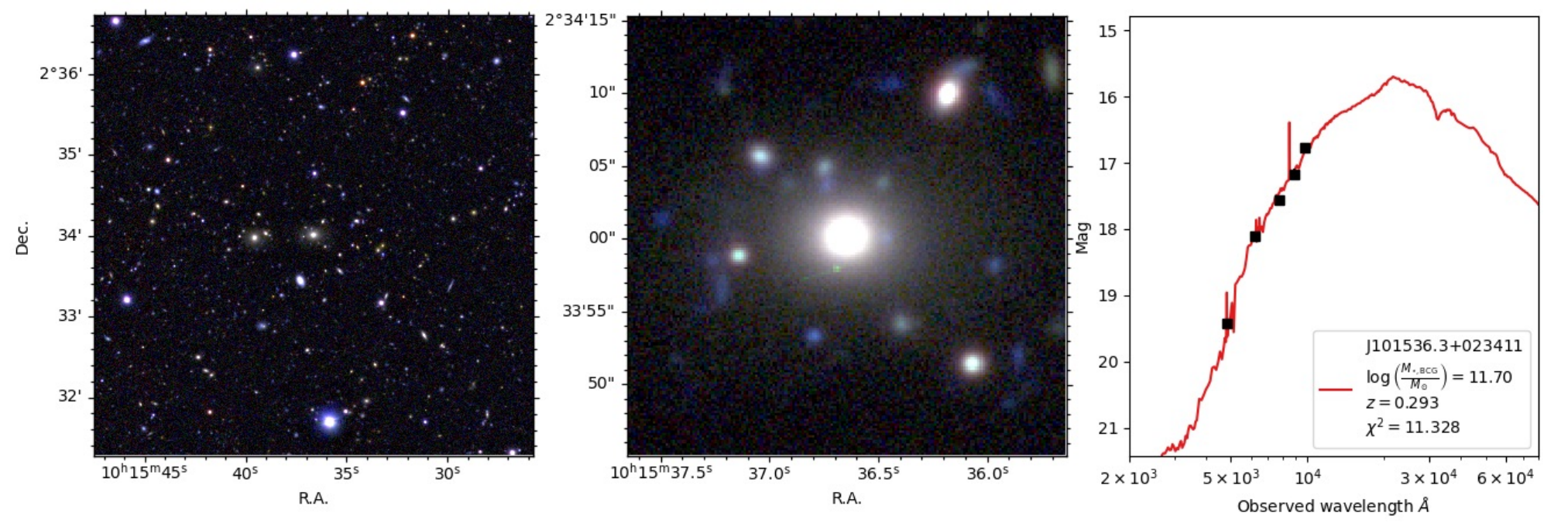}}
\resizebox{0.33\textwidth}{!}{\includegraphics[scale=1]{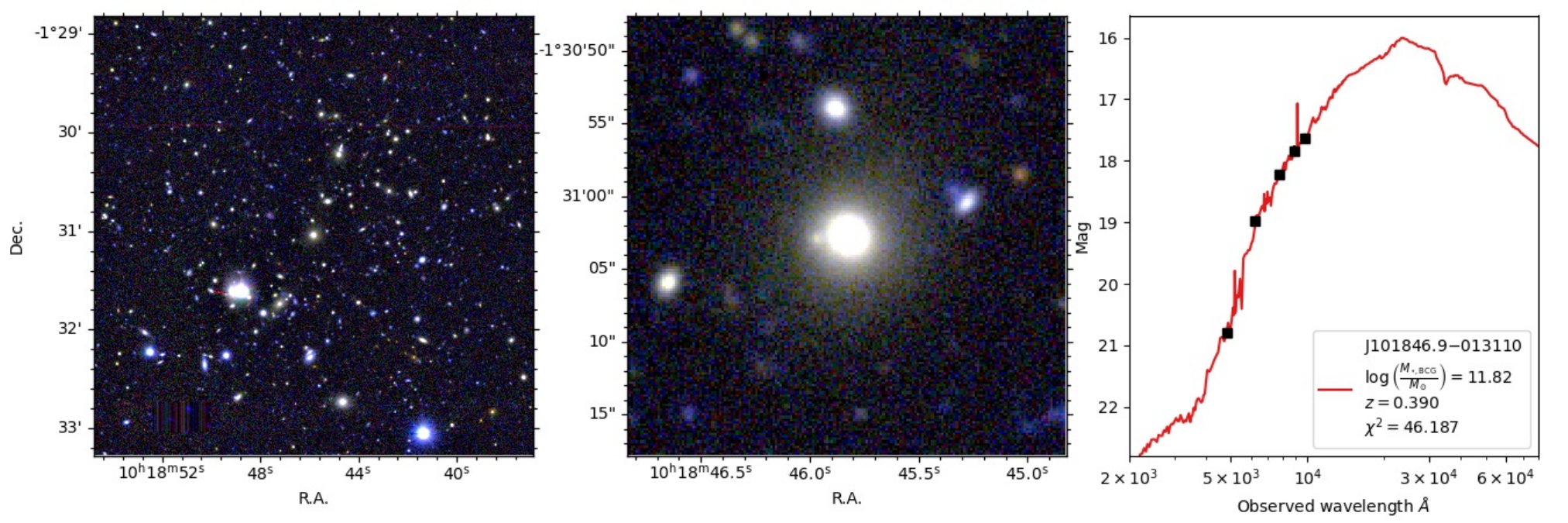}}
\resizebox{0.33\textwidth}{!}{\includegraphics[scale=1]{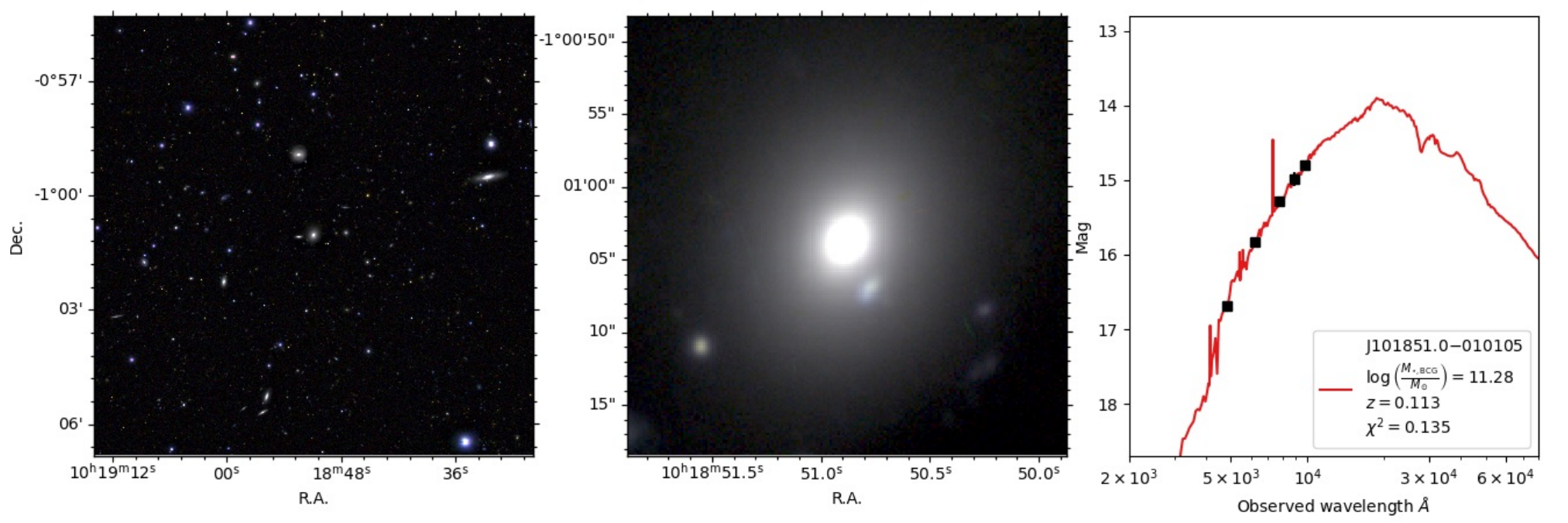}}\\
\resizebox{0.33\textwidth}{!}{\includegraphics[scale=1]{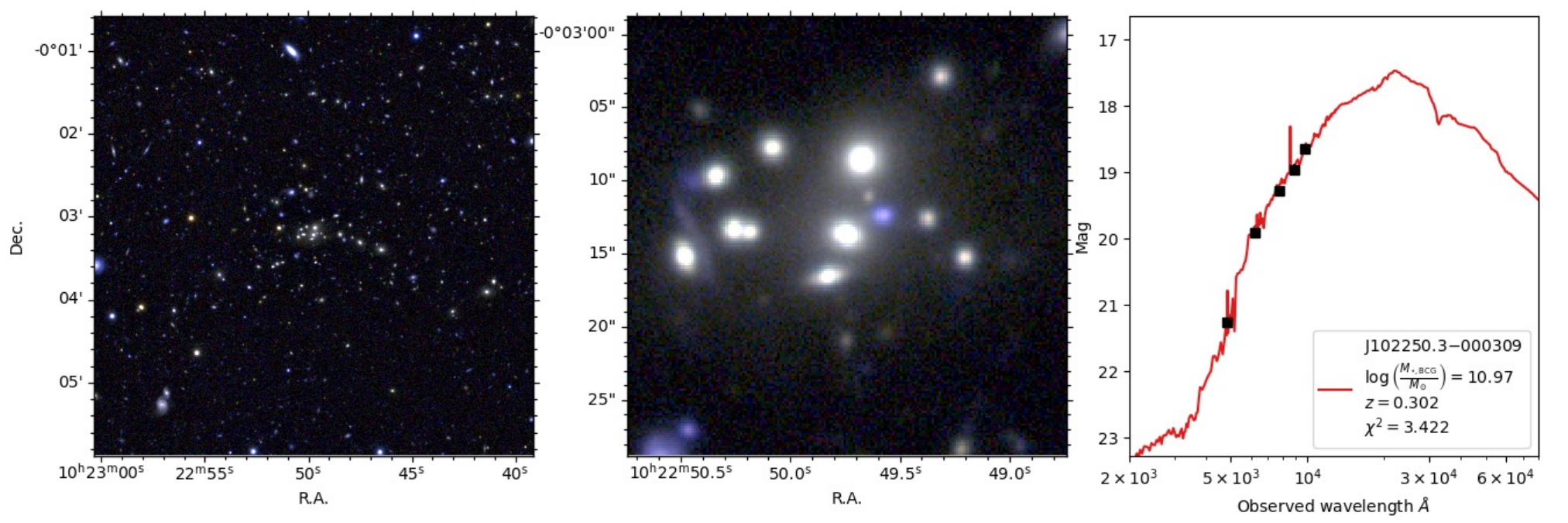}}
\resizebox{0.33\textwidth}{!}{\includegraphics[scale=1]{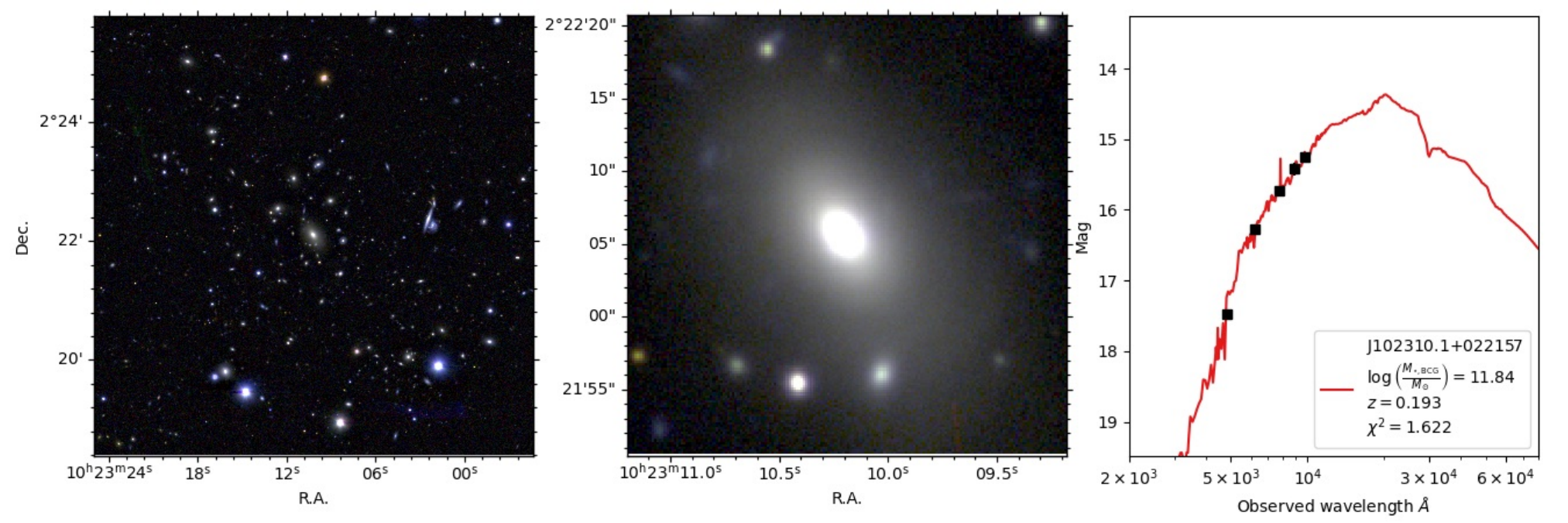}}
\resizebox{0.33\textwidth}{!}{\includegraphics[scale=1]{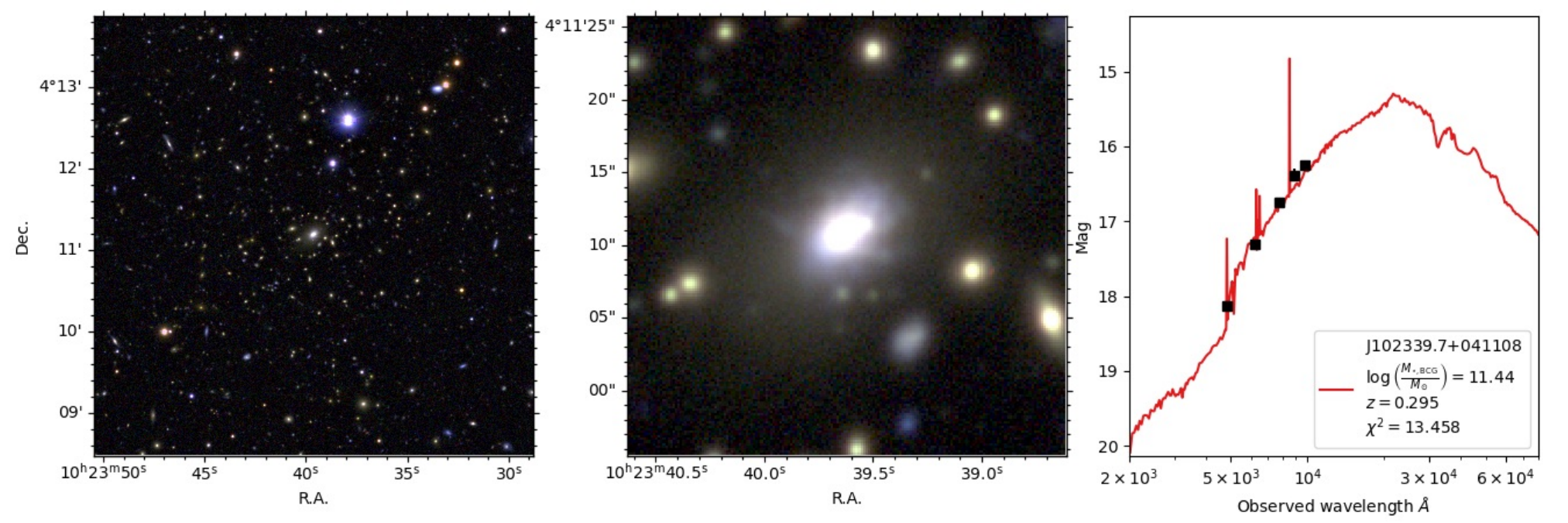}}\\
\resizebox{0.33\textwidth}{!}{\includegraphics[scale=1]{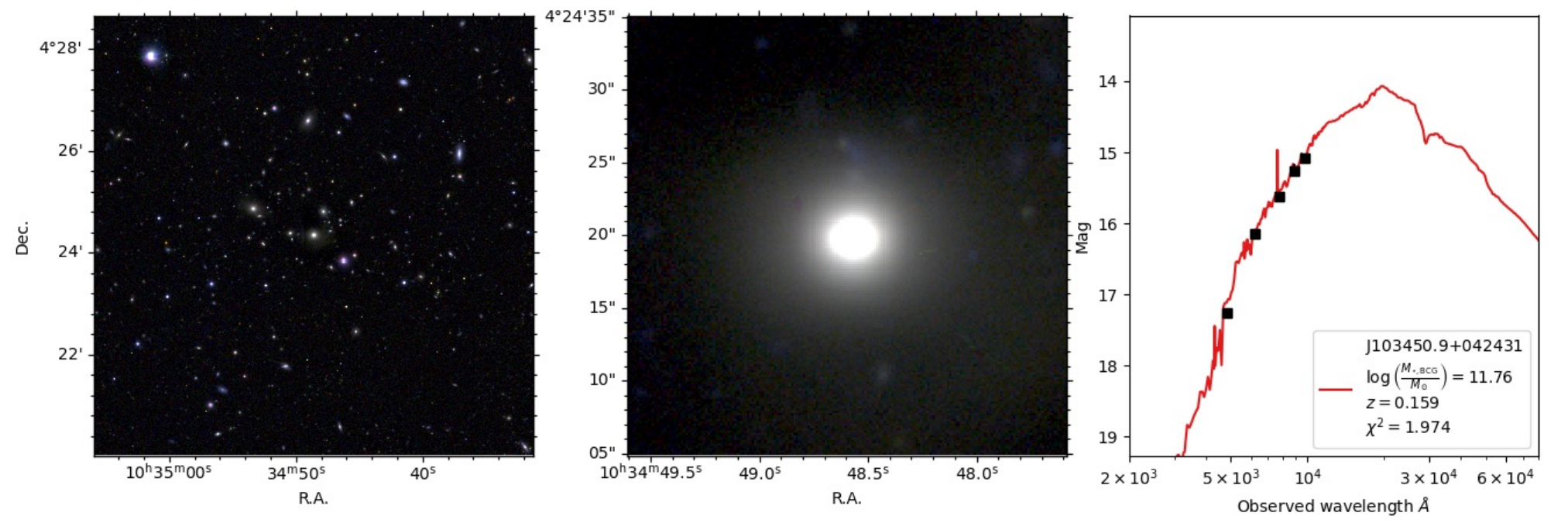}}
\resizebox{0.33\textwidth}{!}{\includegraphics[scale=1]{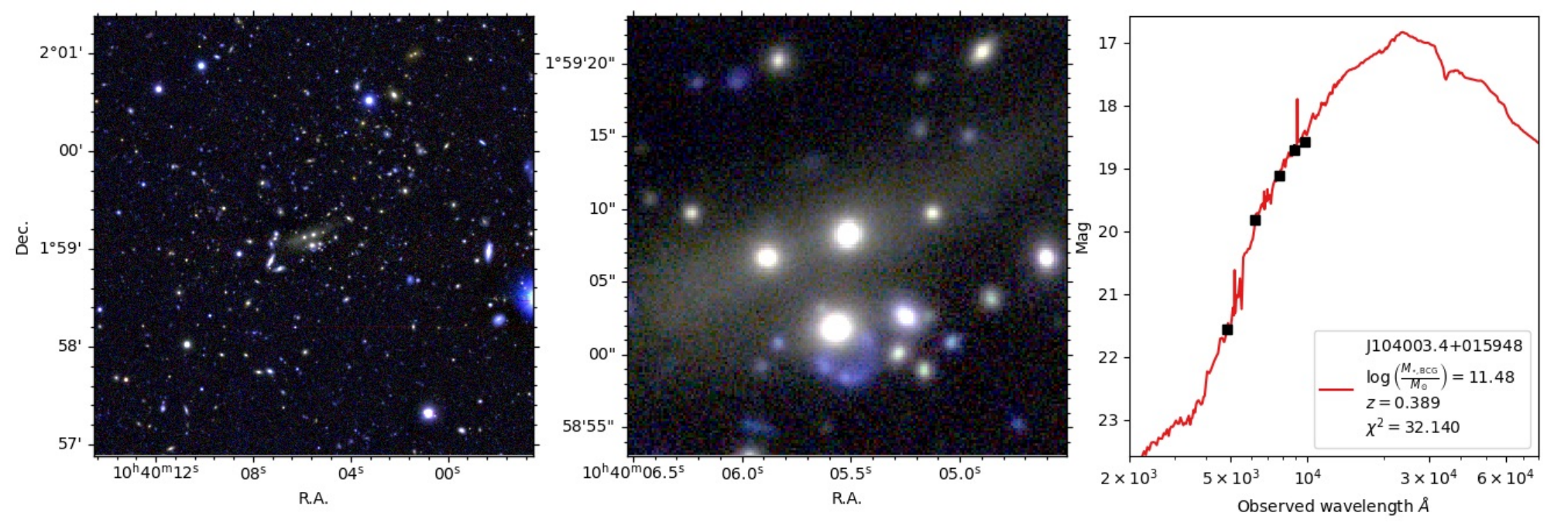}}
\resizebox{0.33\textwidth}{!}{\includegraphics[scale=1]{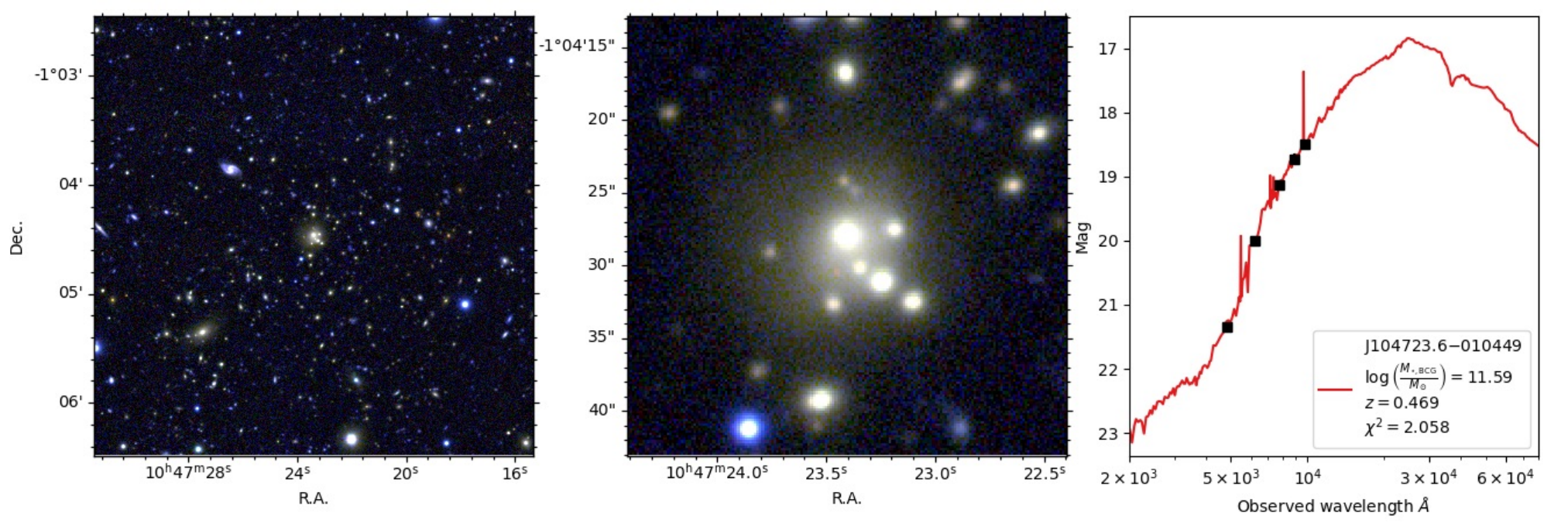}}\\
\resizebox{0.33\textwidth}{!}{\includegraphics[scale=1]{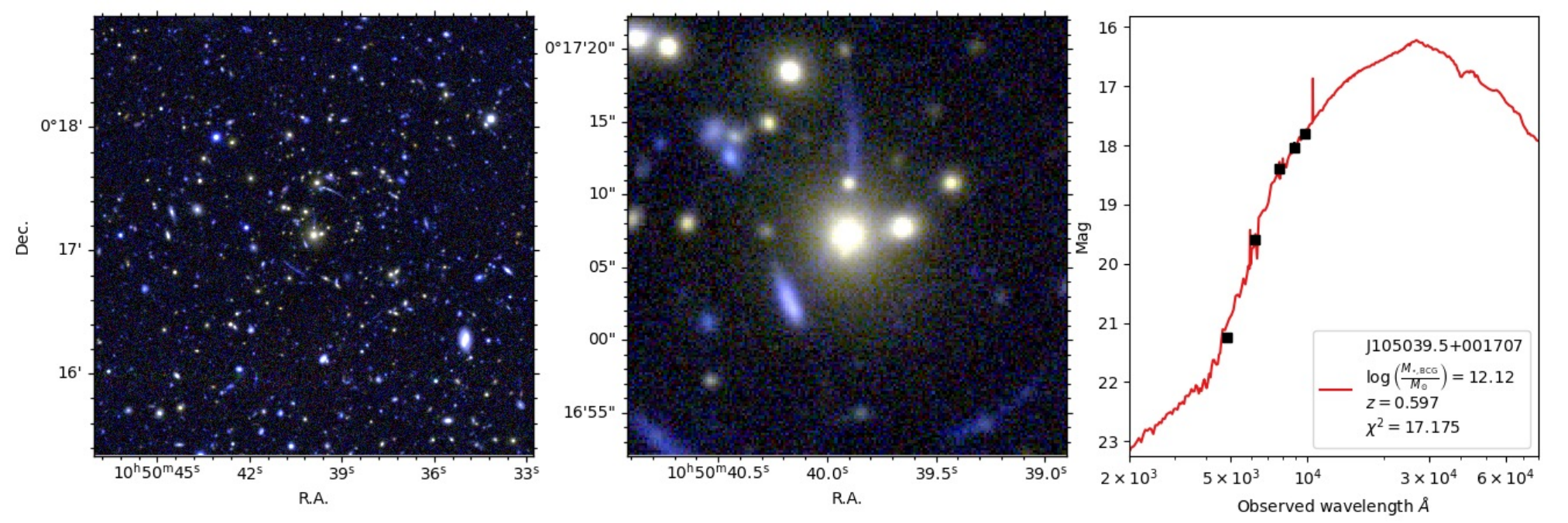}}
\resizebox{0.33\textwidth}{!}{\includegraphics[scale=1]{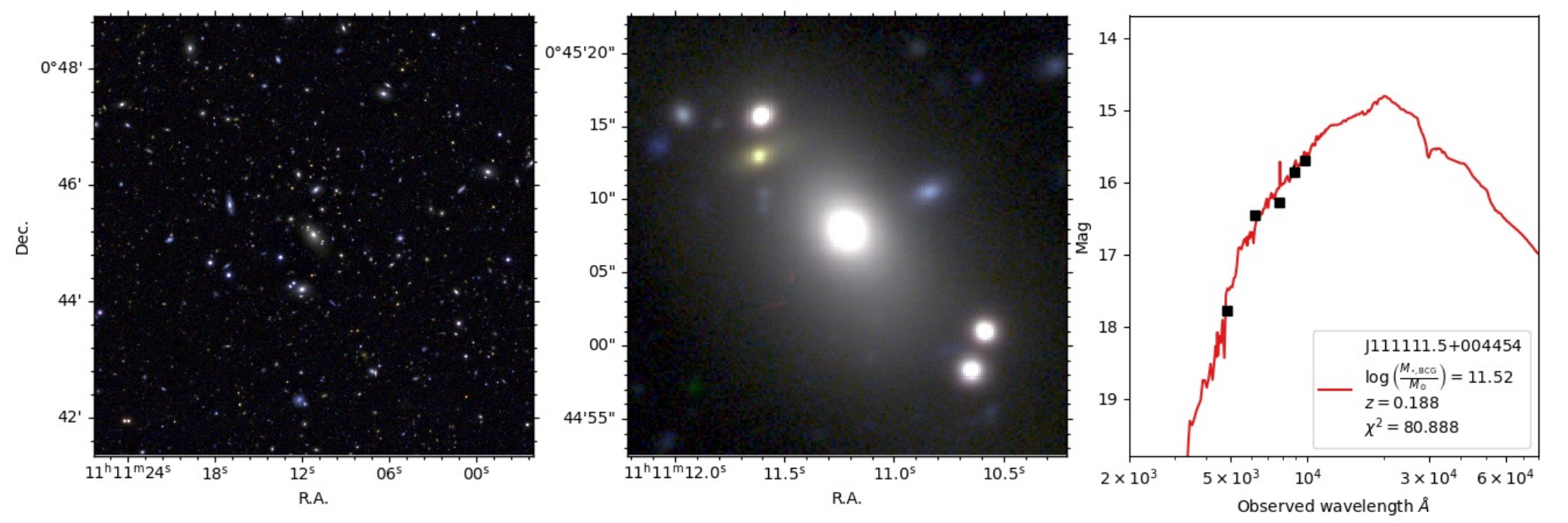}}
\resizebox{0.33\textwidth}{!}{\includegraphics[scale=1]{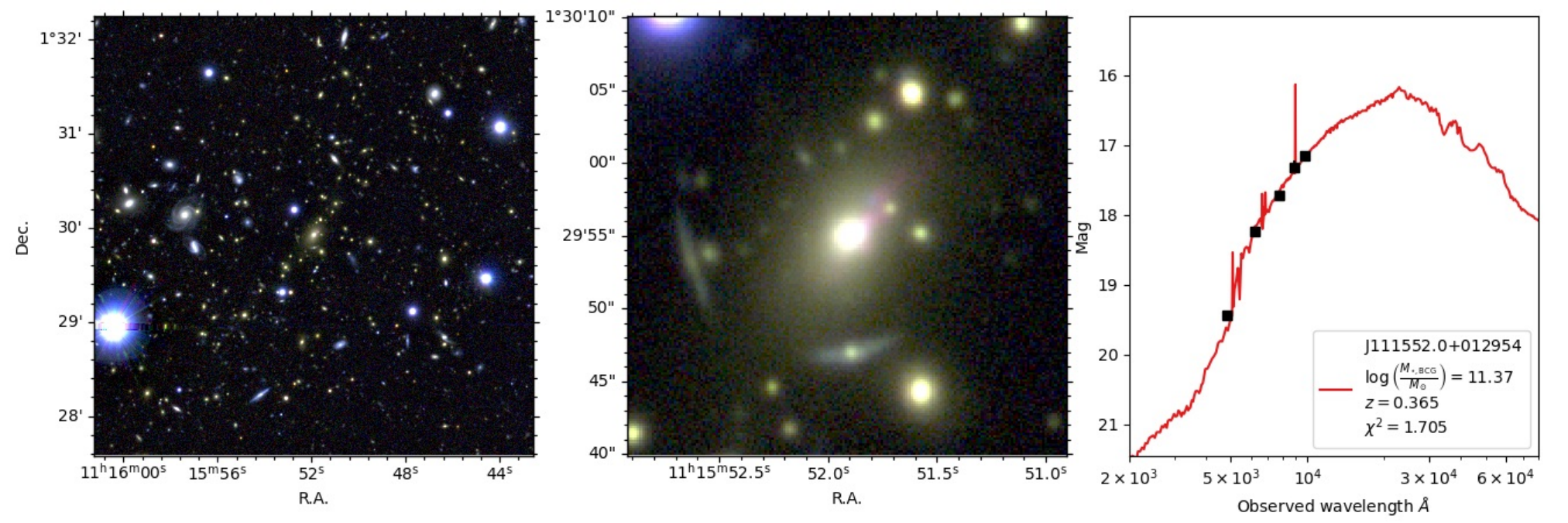}}
\caption{See Appendix~\ref{app:imaging} for details.}
\label{fig:cutout1}
\end{figure*}

\begin{figure*}[!ht]
\centering
\resizebox{0.33\textwidth}{!}{\includegraphics[scale=1]{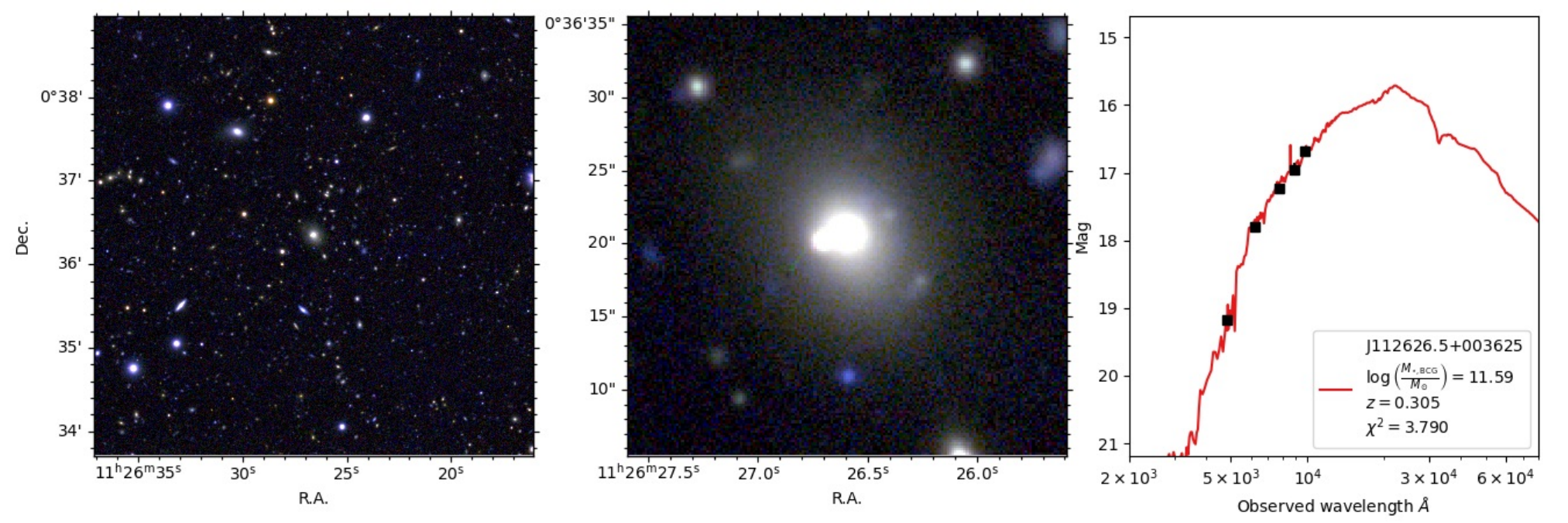}}
\resizebox{0.33\textwidth}{!}{\includegraphics[scale=1]{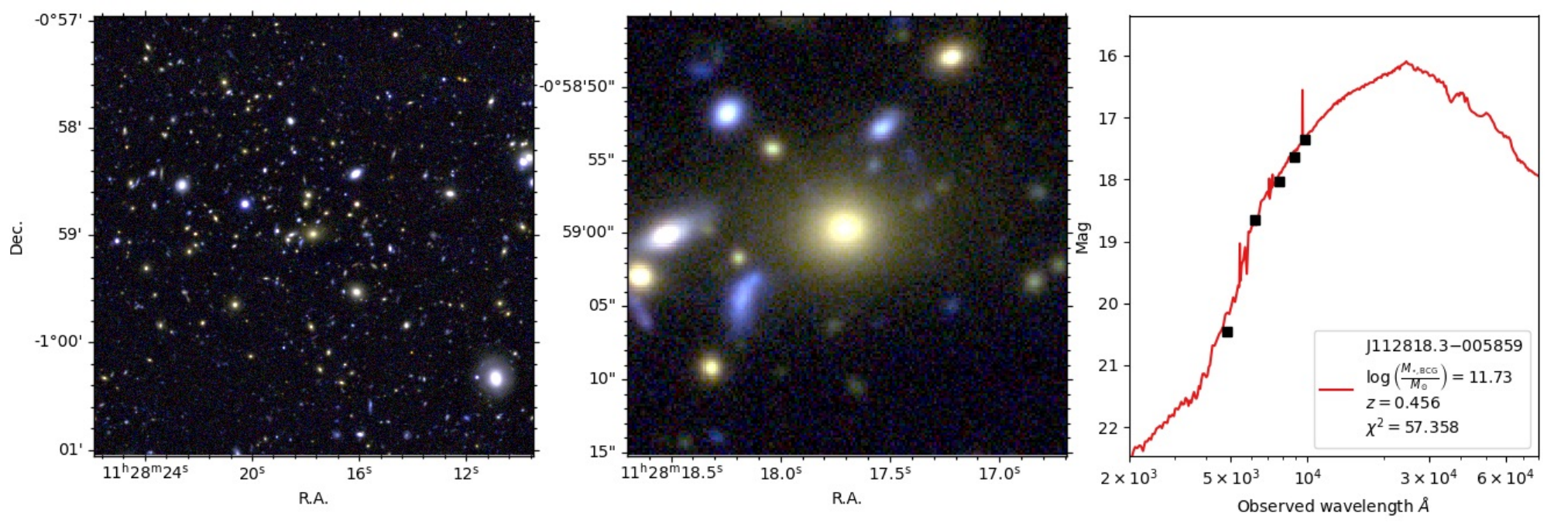}}
\resizebox{0.33\textwidth}{!}{\includegraphics[scale=1]{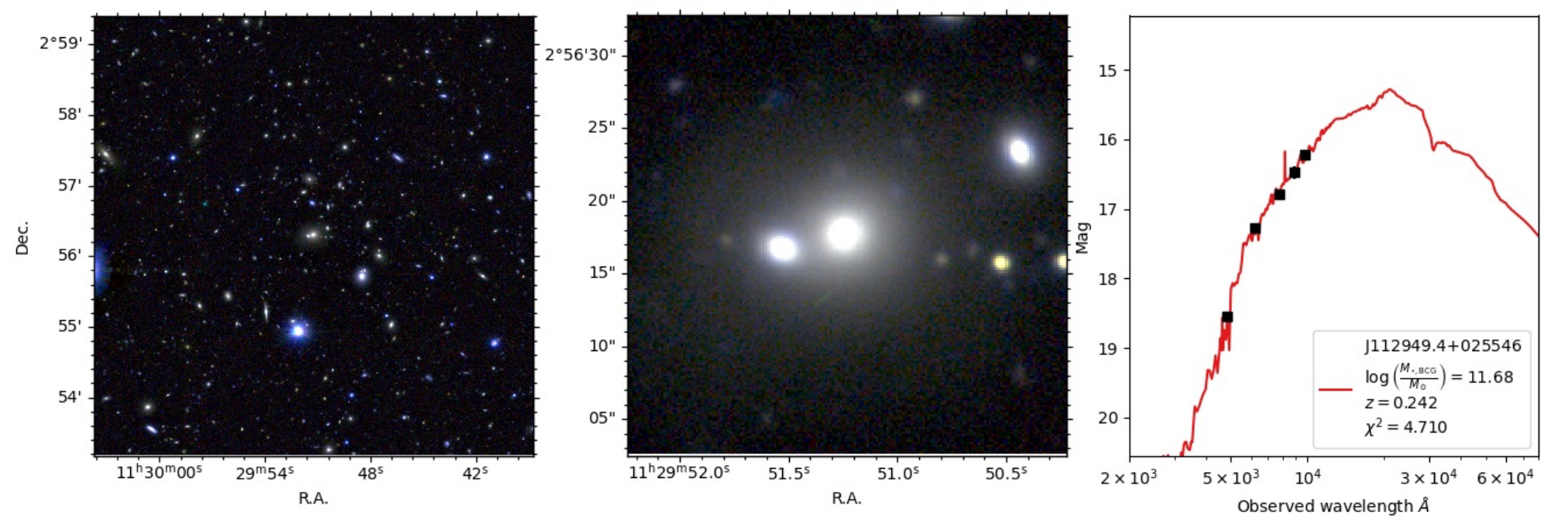}}\\
\resizebox{0.33\textwidth}{!}{\includegraphics[scale=1]{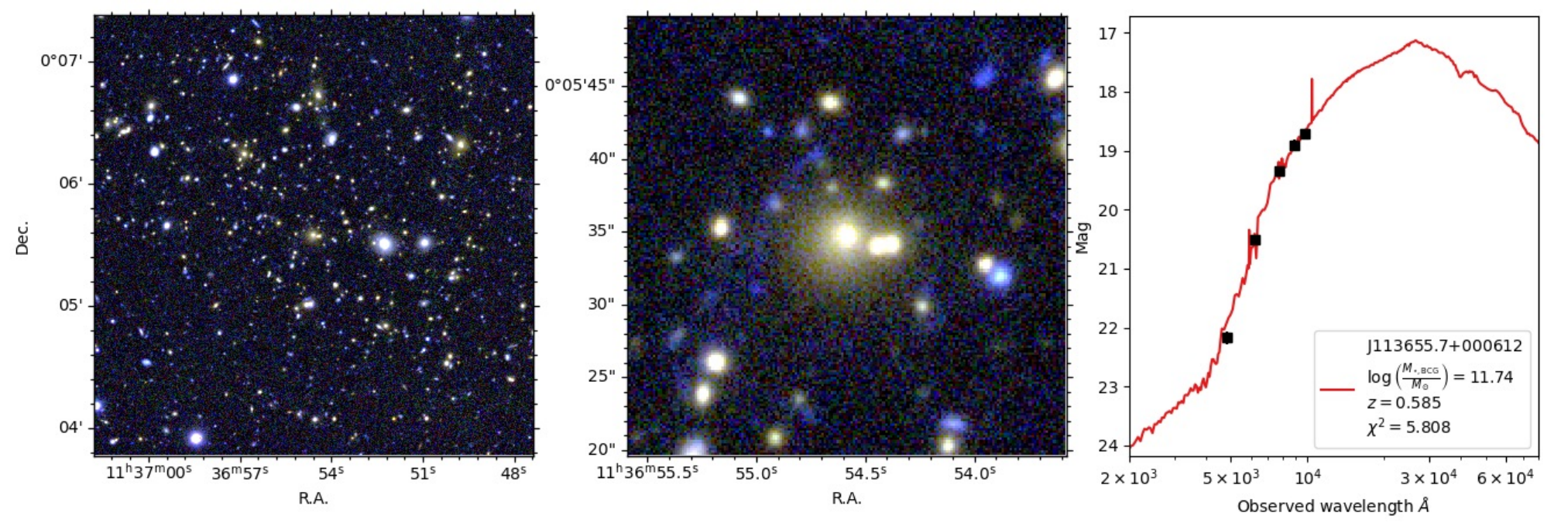}}
\resizebox{0.33\textwidth}{!}{\includegraphics[scale=1]{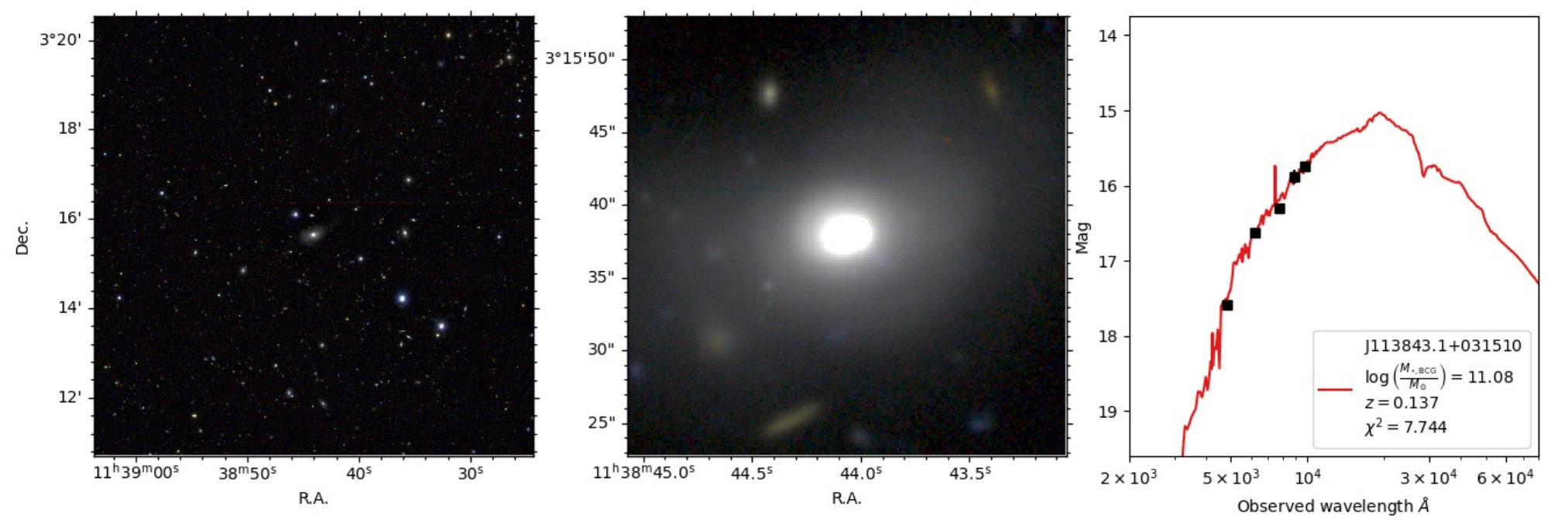}}
\resizebox{0.33\textwidth}{!}{\includegraphics[scale=1]{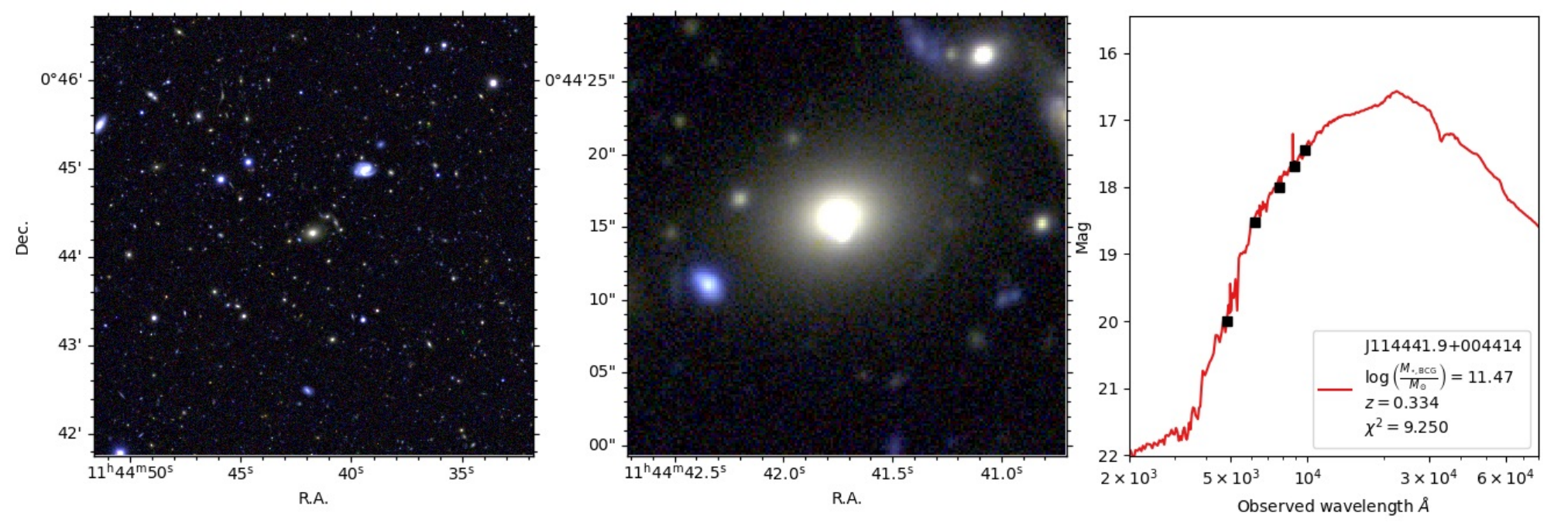}}\\
\resizebox{0.33\textwidth}{!}{\includegraphics[scale=1]{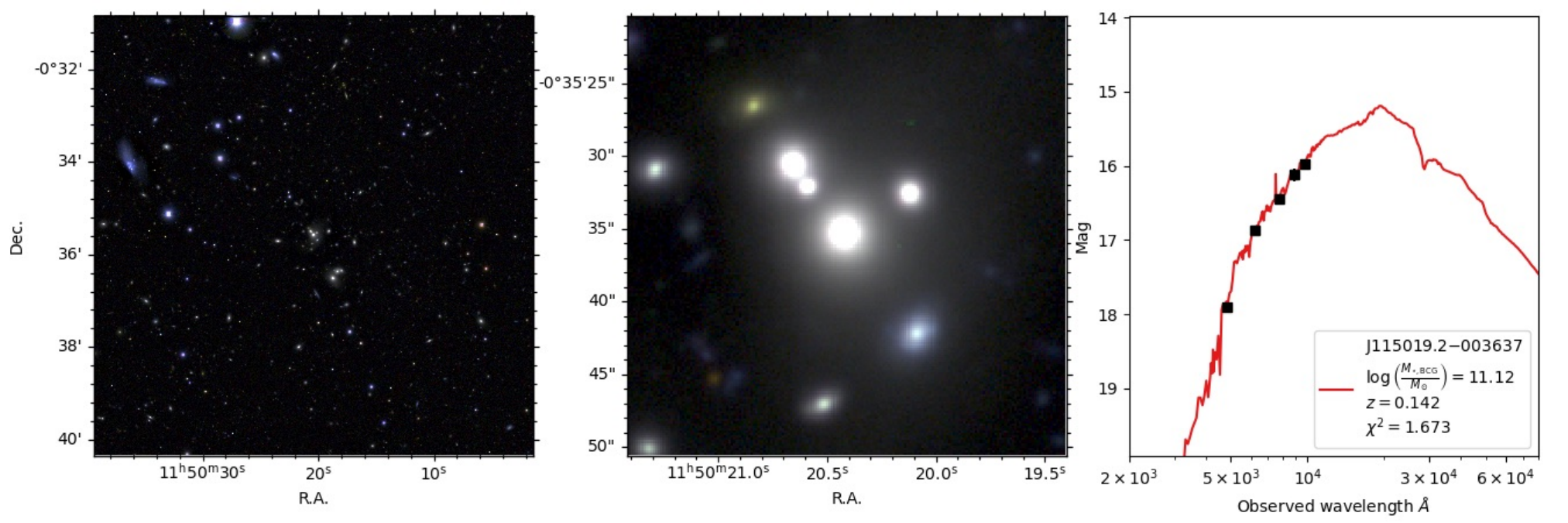}}
\resizebox{0.33\textwidth}{!}{\includegraphics[scale=1]{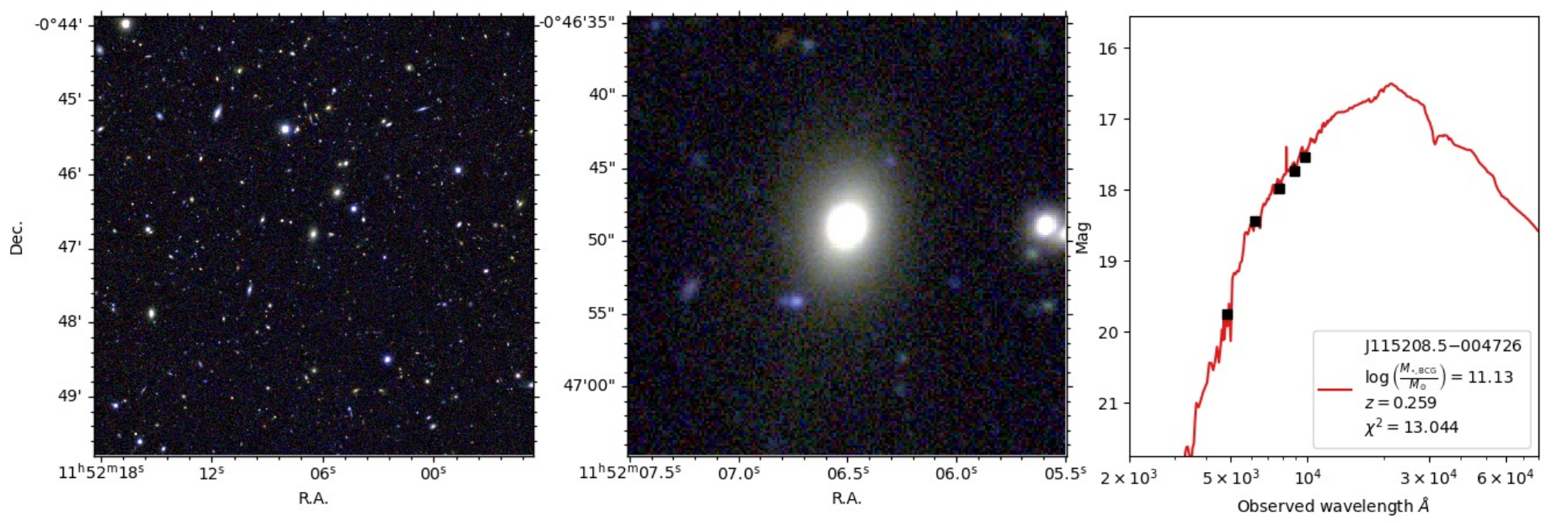}}
\resizebox{0.33\textwidth}{!}{\includegraphics[scale=1]{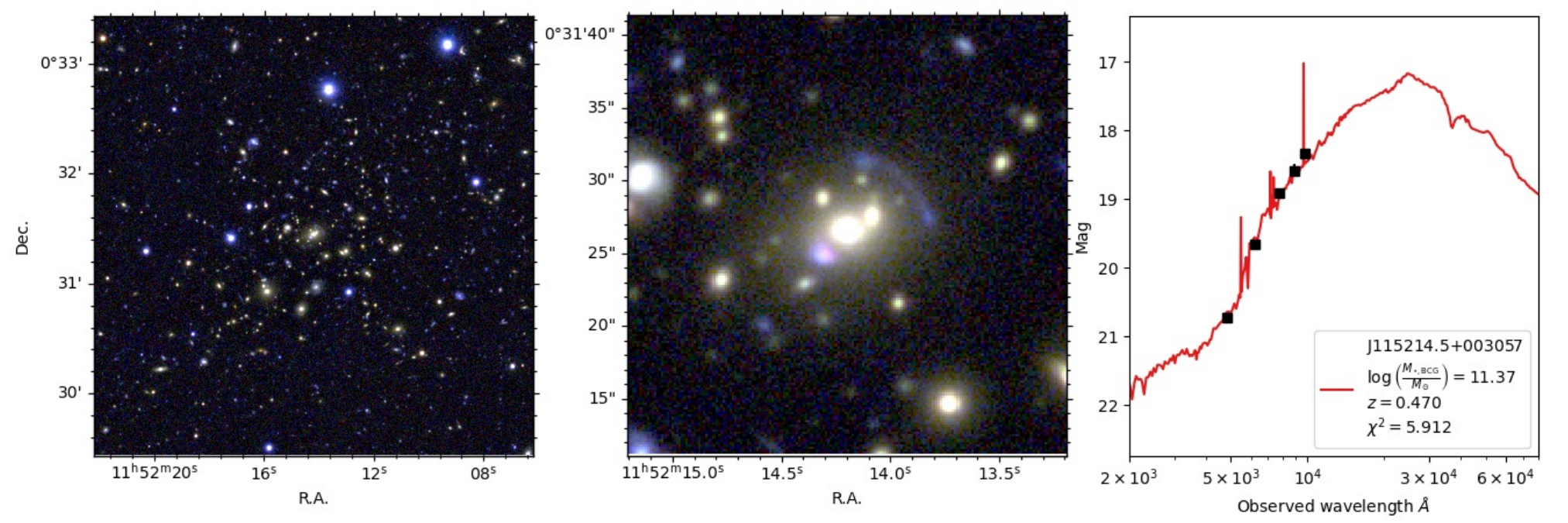}}\\
\resizebox{0.33\textwidth}{!}{\includegraphics[scale=1]{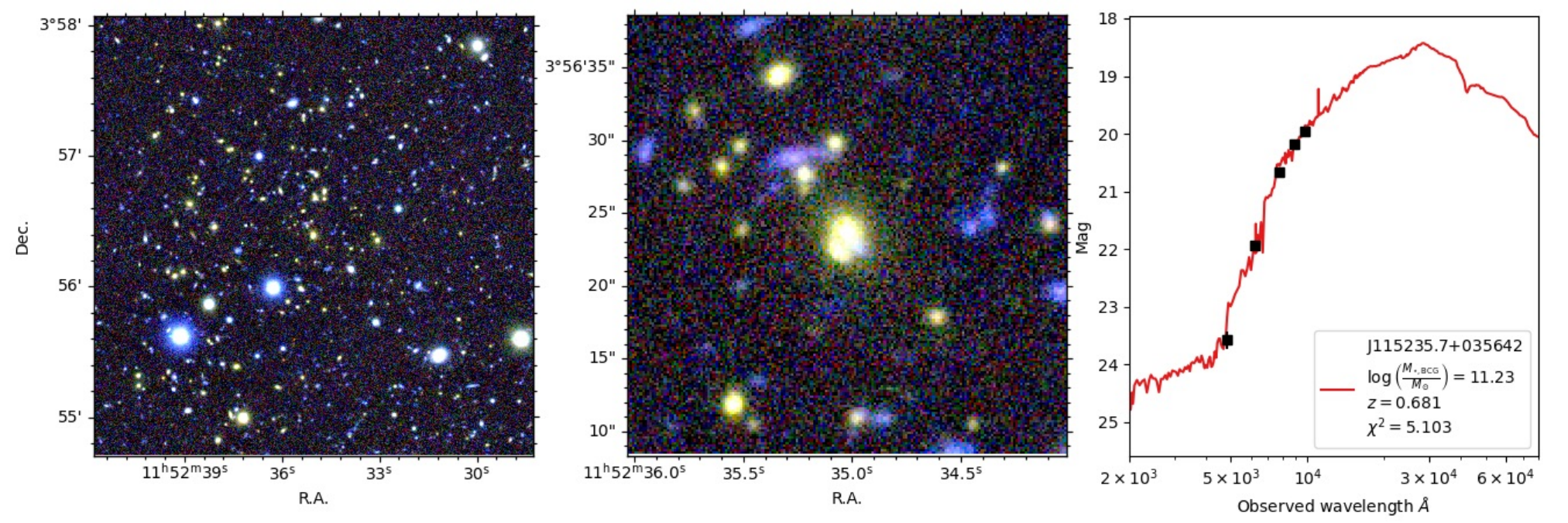}}
\resizebox{0.33\textwidth}{!}{\includegraphics[scale=1]{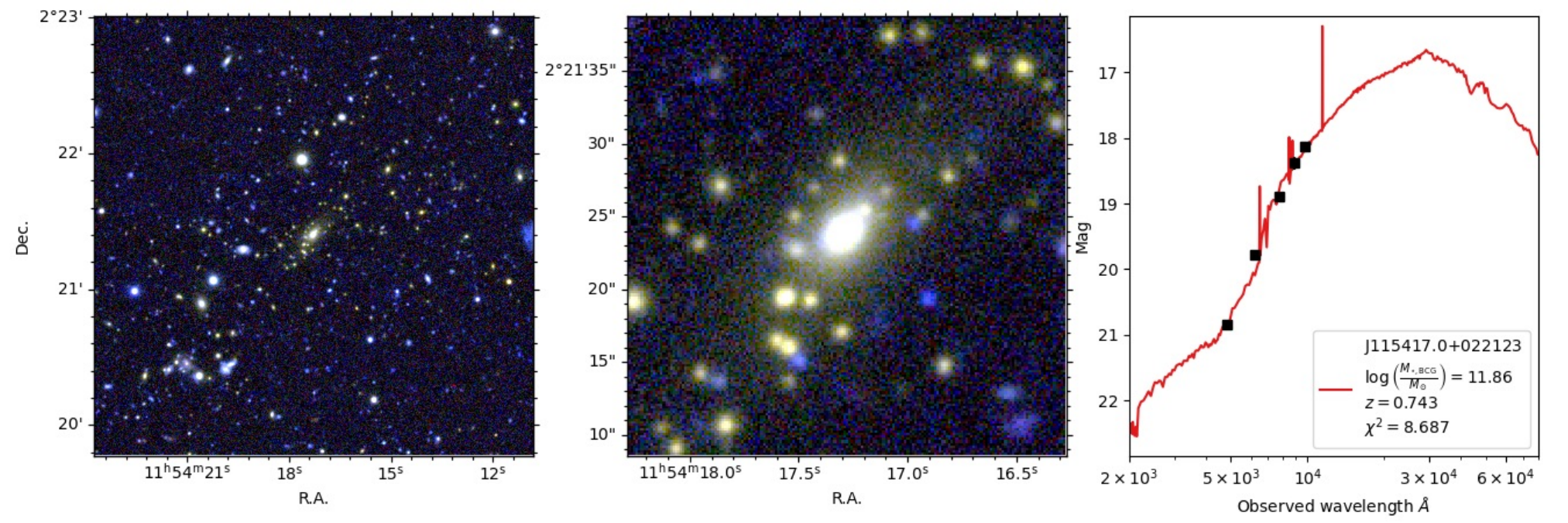}}
\resizebox{0.33\textwidth}{!}{\includegraphics[scale=1]{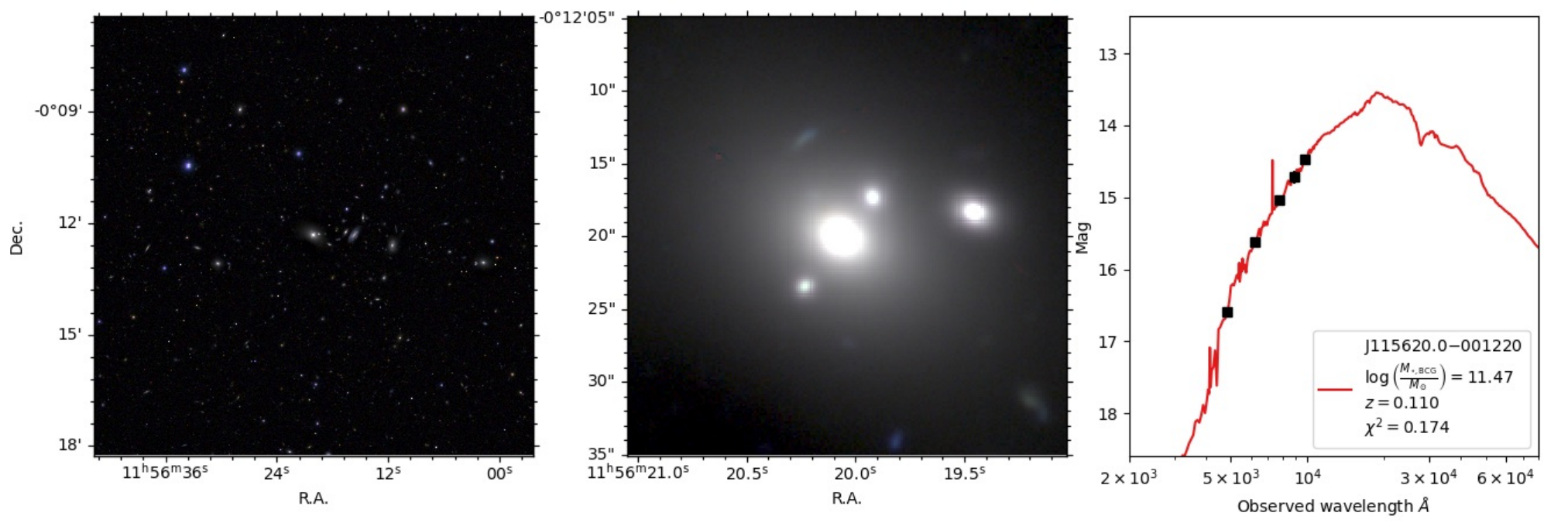}}\\
\resizebox{0.33\textwidth}{!}{\includegraphics[scale=1]{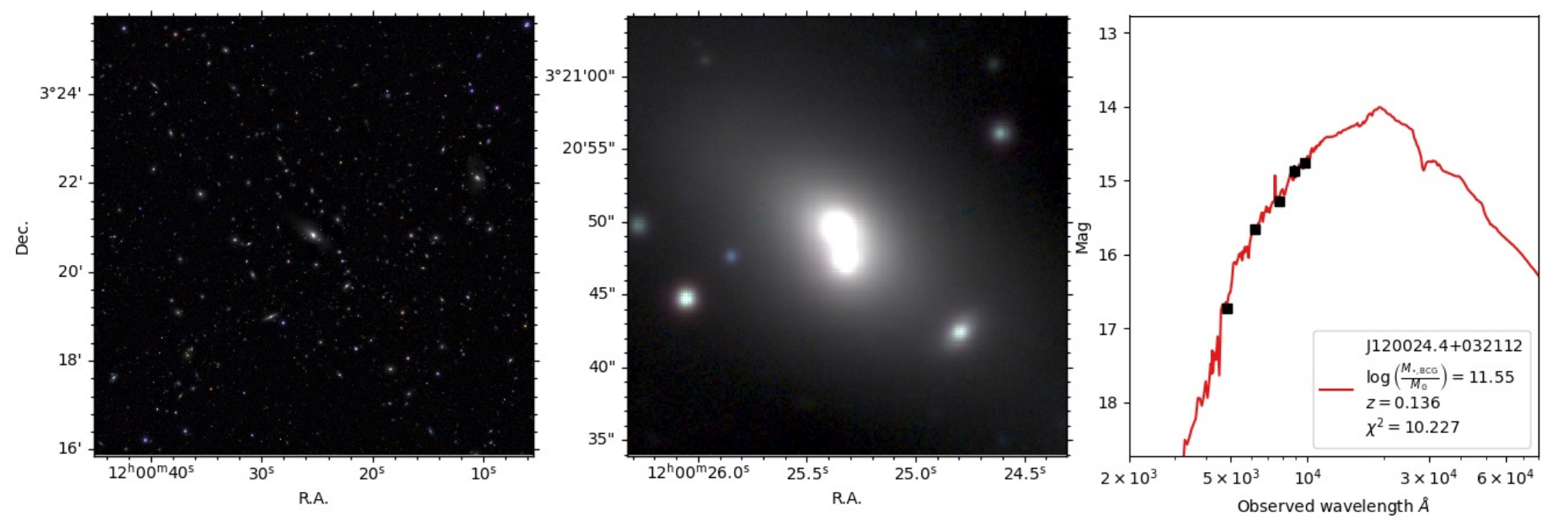}}
\resizebox{0.33\textwidth}{!}{\includegraphics[scale=1]{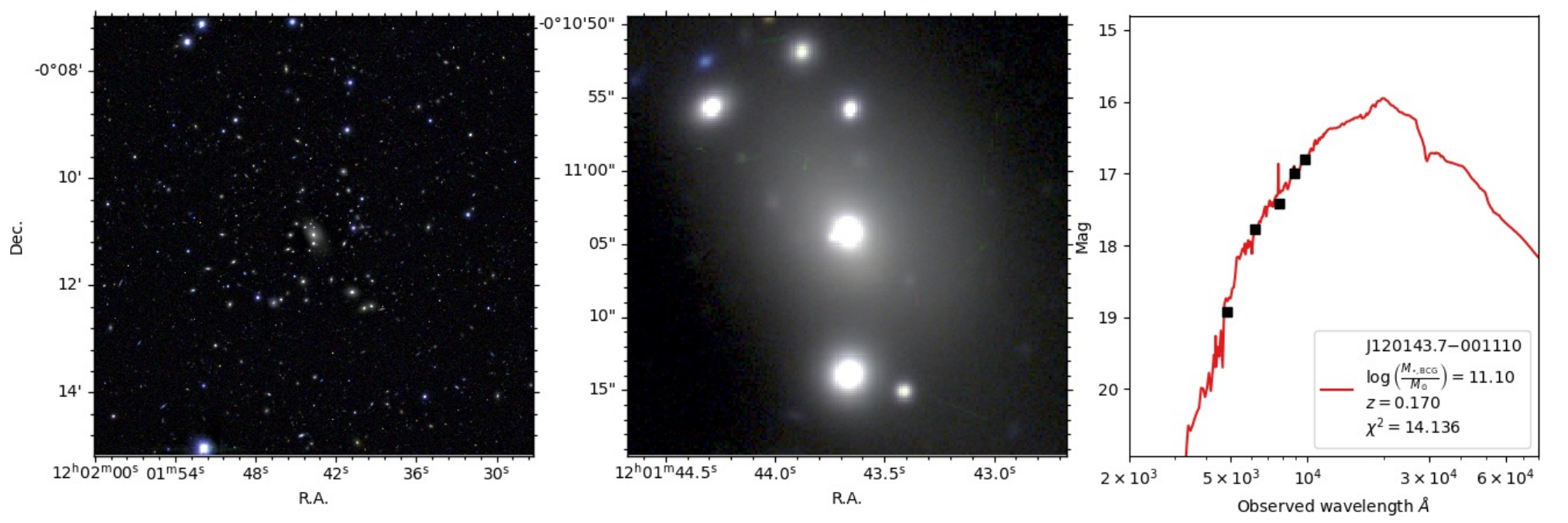}}
\resizebox{0.33\textwidth}{!}{\includegraphics[scale=1]{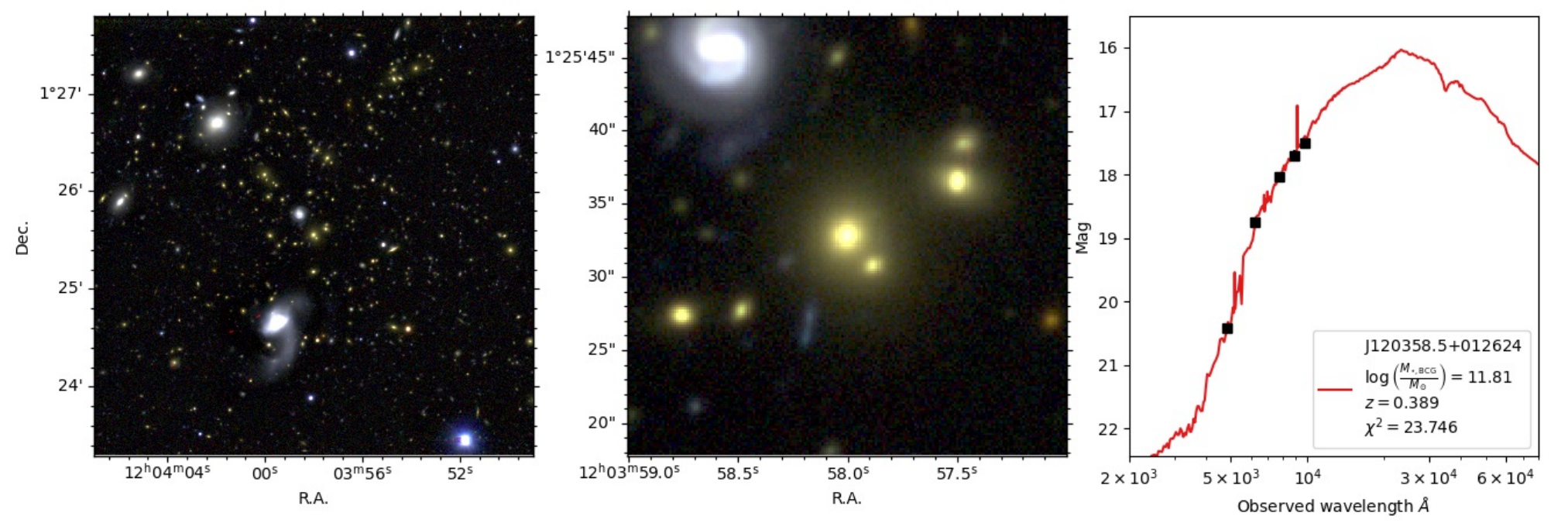}}\\
\resizebox{0.33\textwidth}{!}{\includegraphics[scale=1]{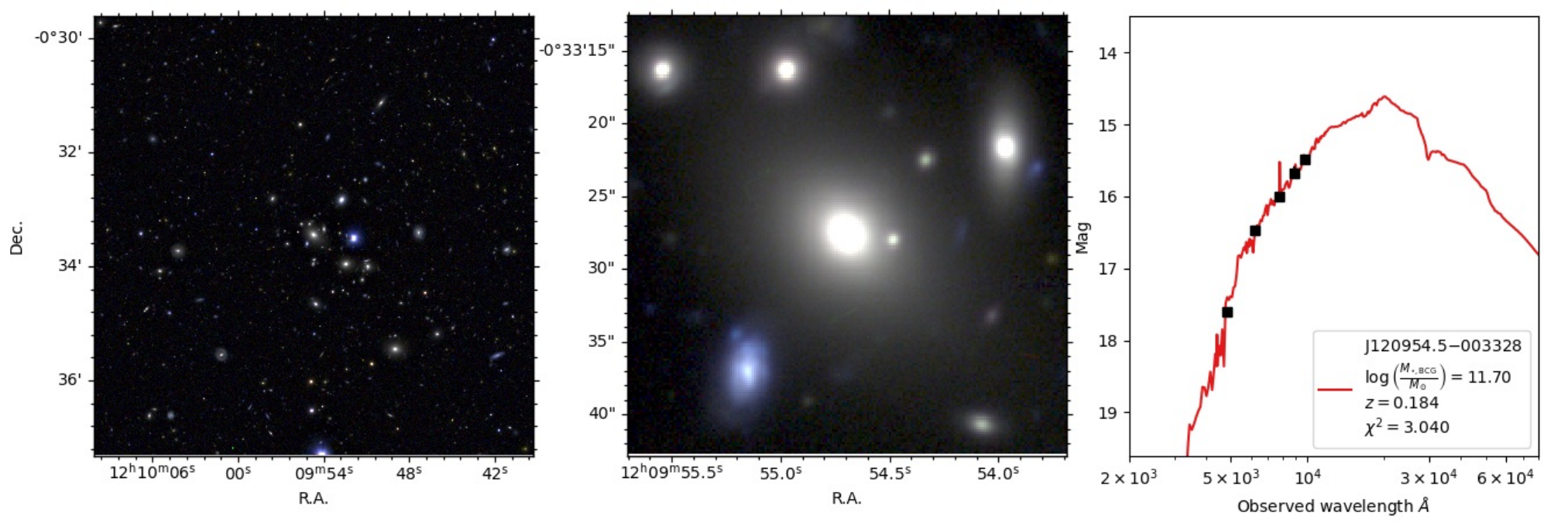}}
\resizebox{0.33\textwidth}{!}{\includegraphics[scale=1]{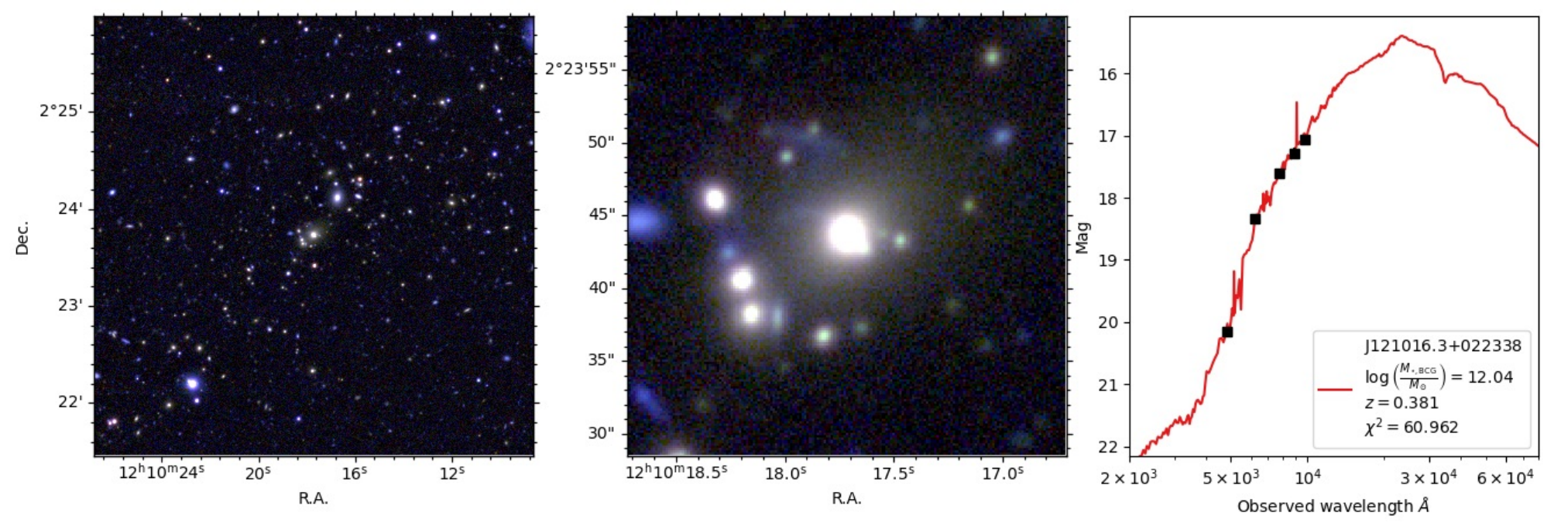}}
\resizebox{0.33\textwidth}{!}{\includegraphics[scale=1]{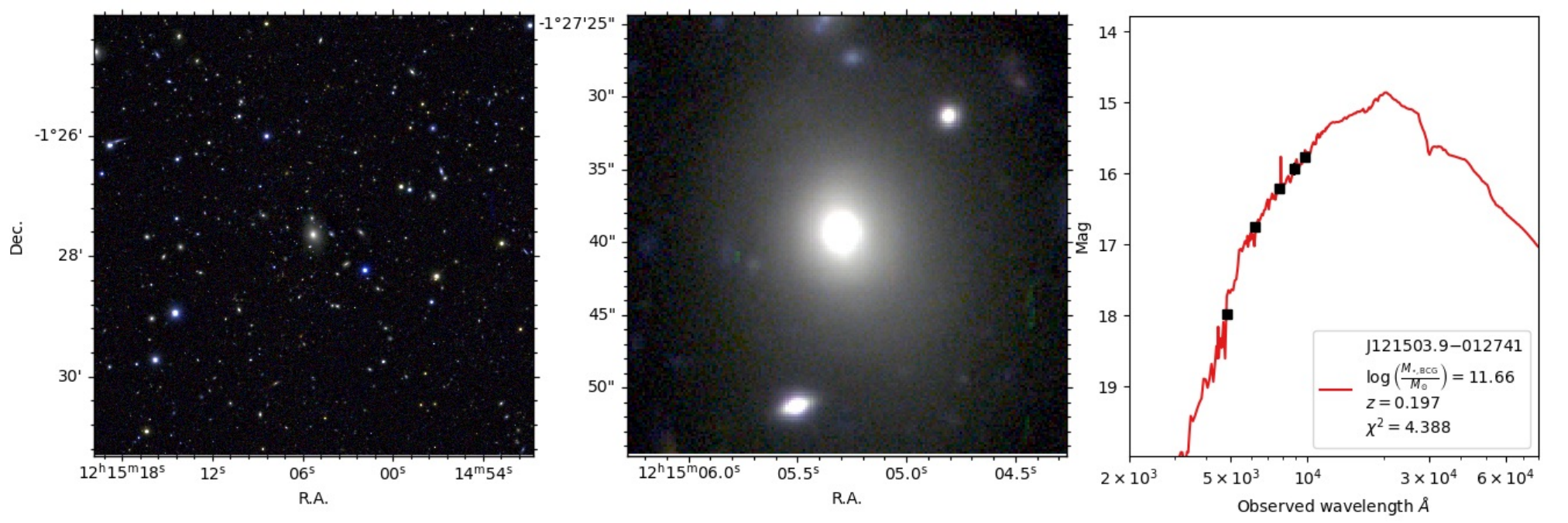}}\\
\resizebox{0.33\textwidth}{!}{\includegraphics[scale=1]{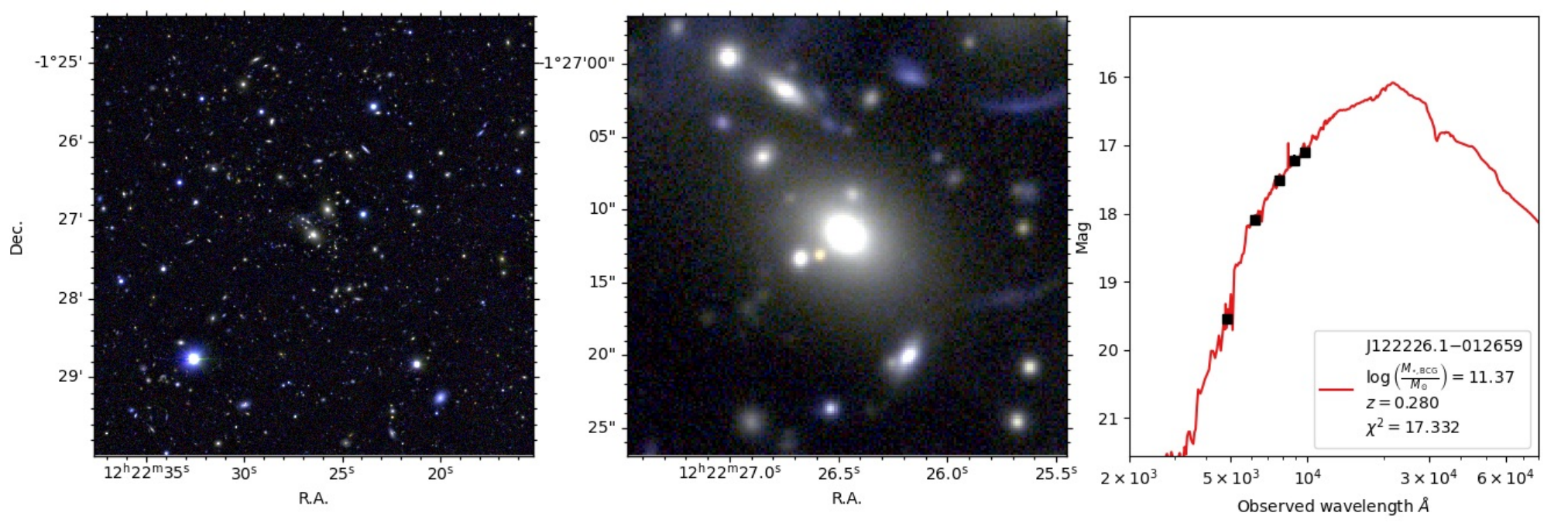}}
\resizebox{0.33\textwidth}{!}{\includegraphics[scale=1]{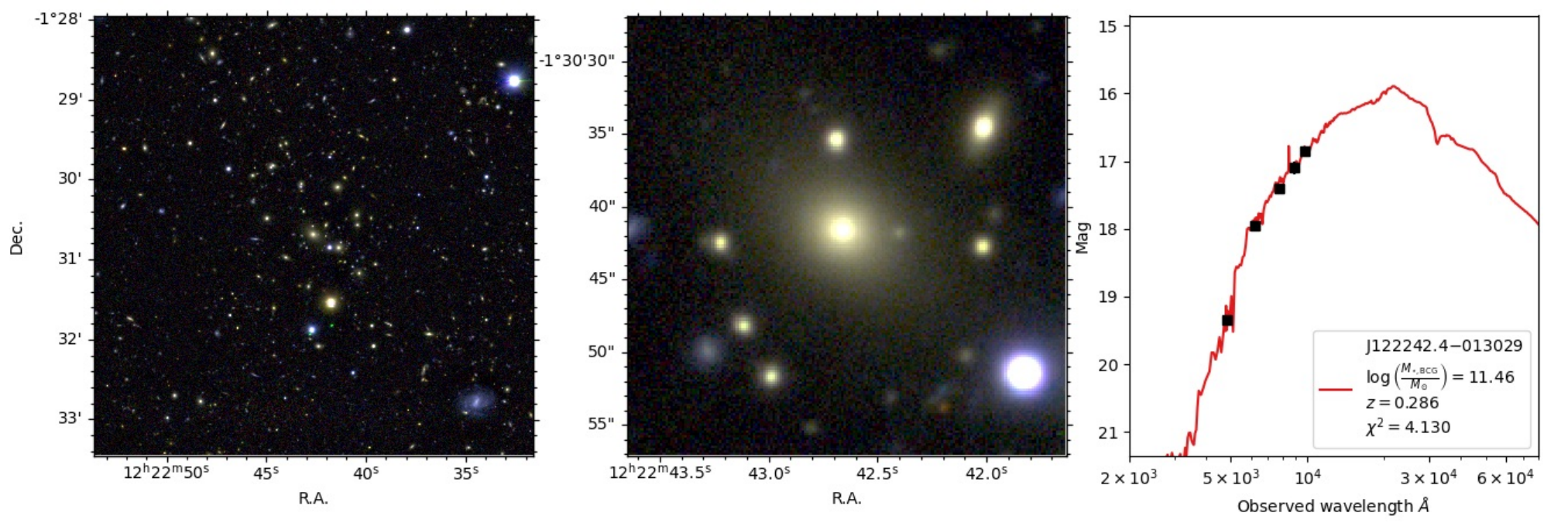}}
\resizebox{0.33\textwidth}{!}{\includegraphics[scale=1]{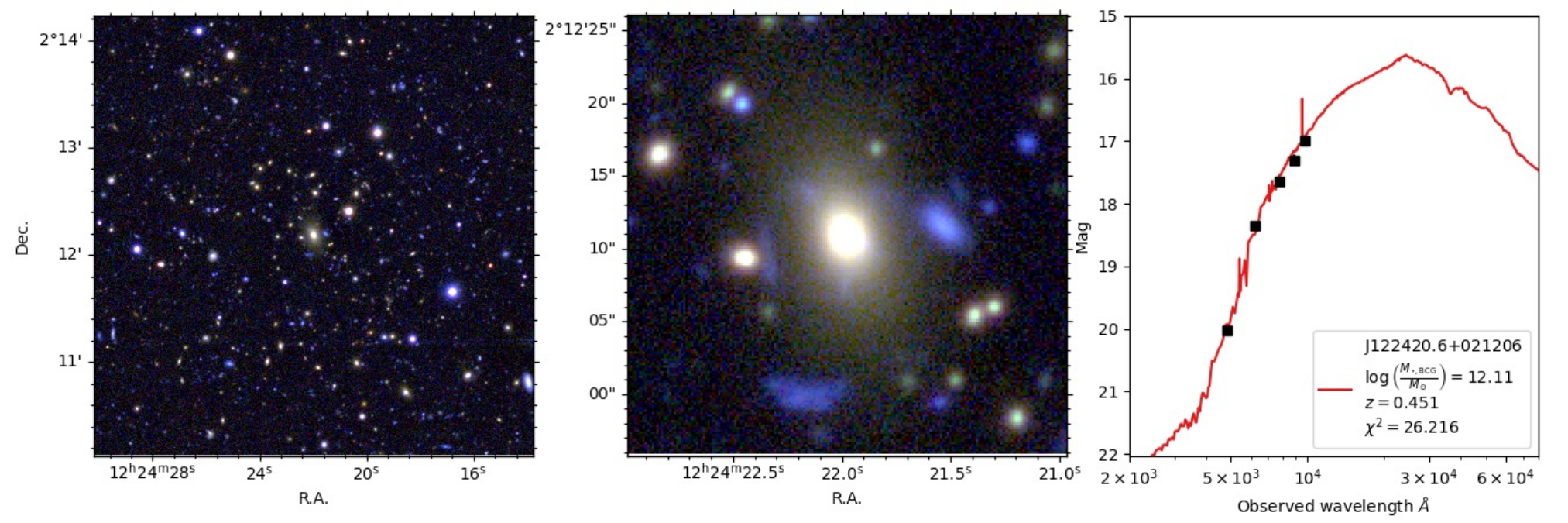}}\\
\resizebox{0.33\textwidth}{!}{\includegraphics[scale=1]{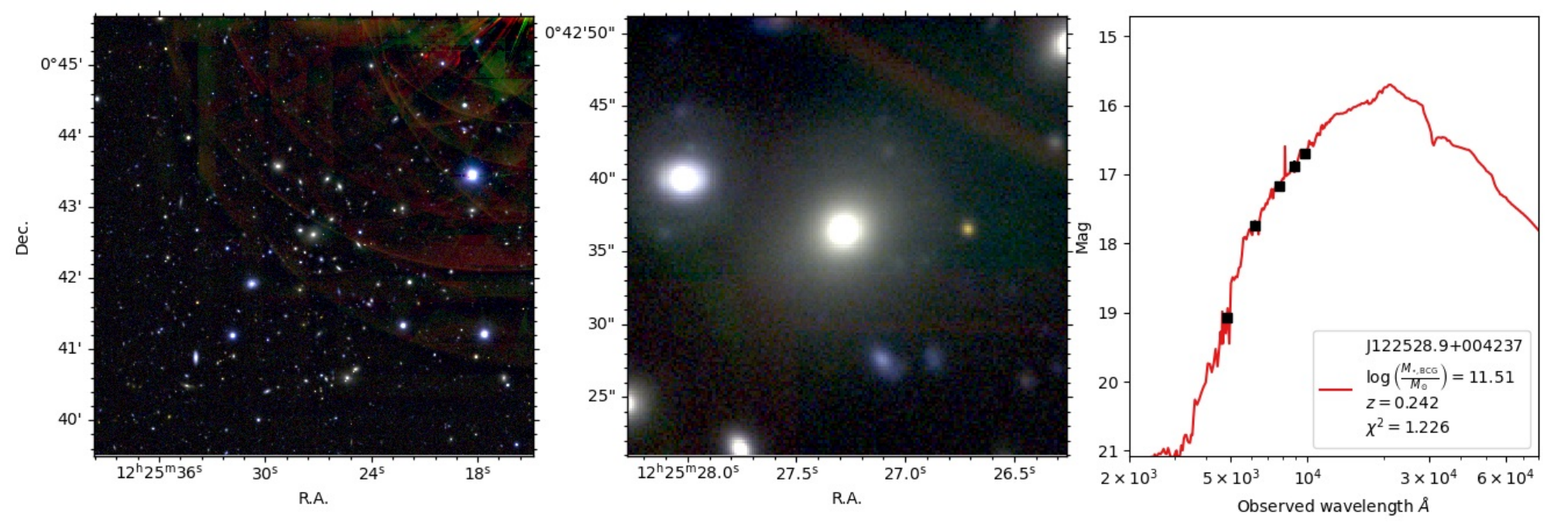}}
\resizebox{0.33\textwidth}{!}{\includegraphics[scale=1]{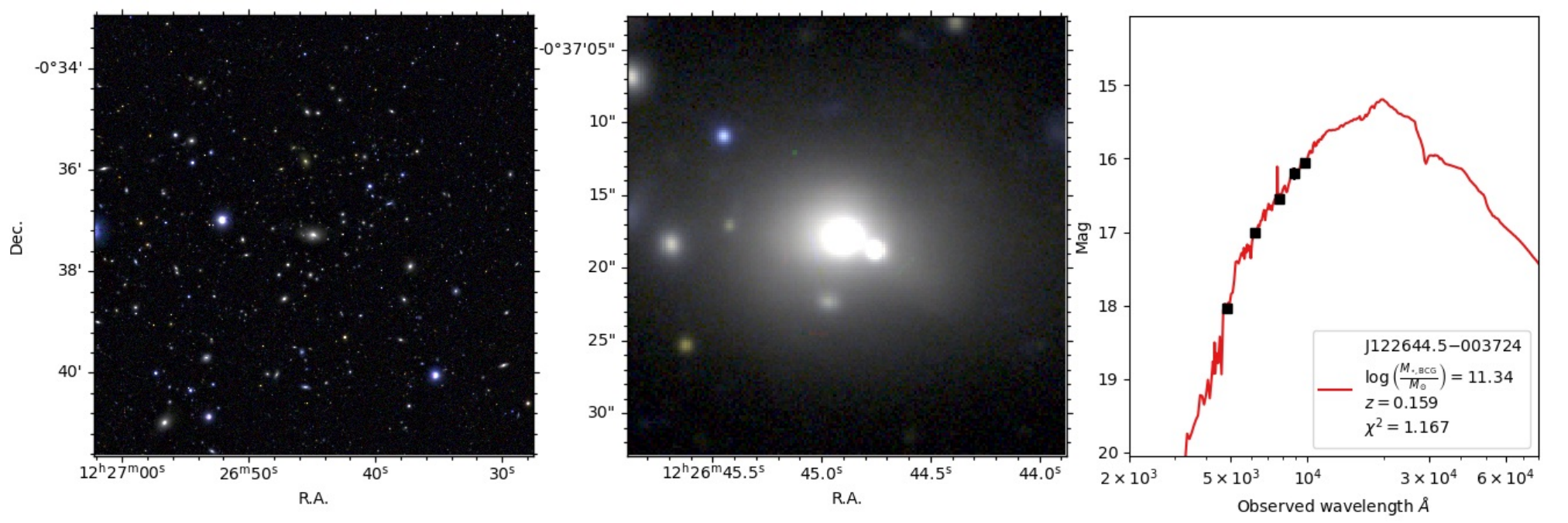}}
\resizebox{0.33\textwidth}{!}{\includegraphics[scale=1]{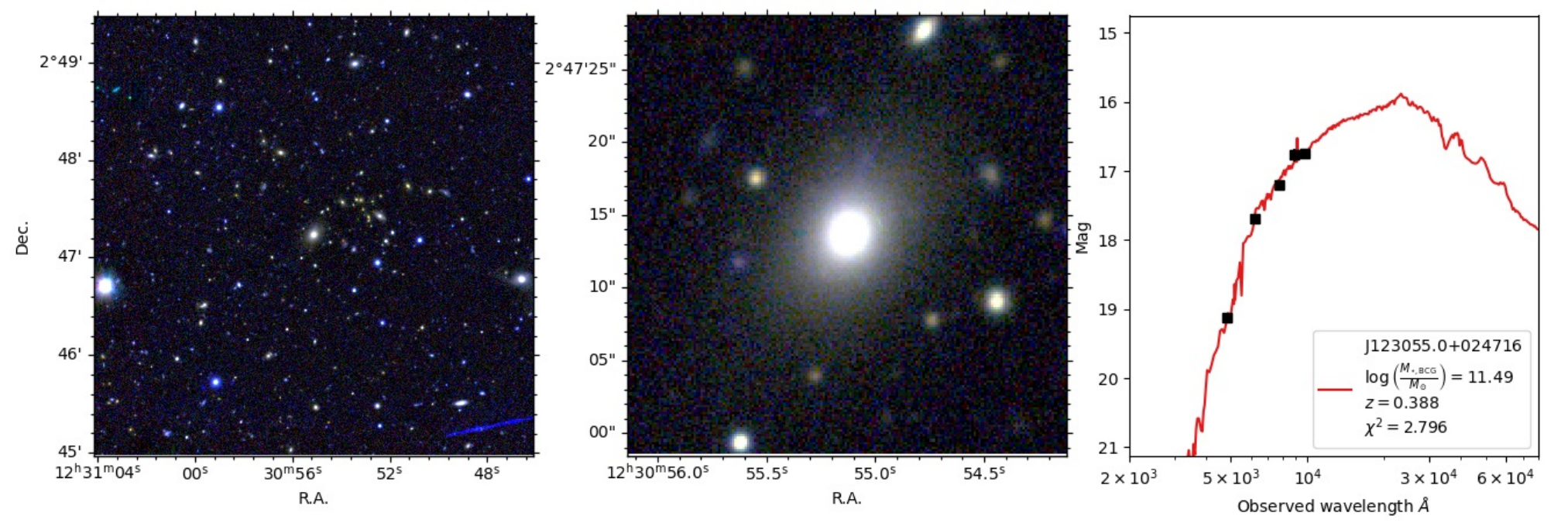}}\\
\resizebox{0.33\textwidth}{!}{\includegraphics[scale=1]{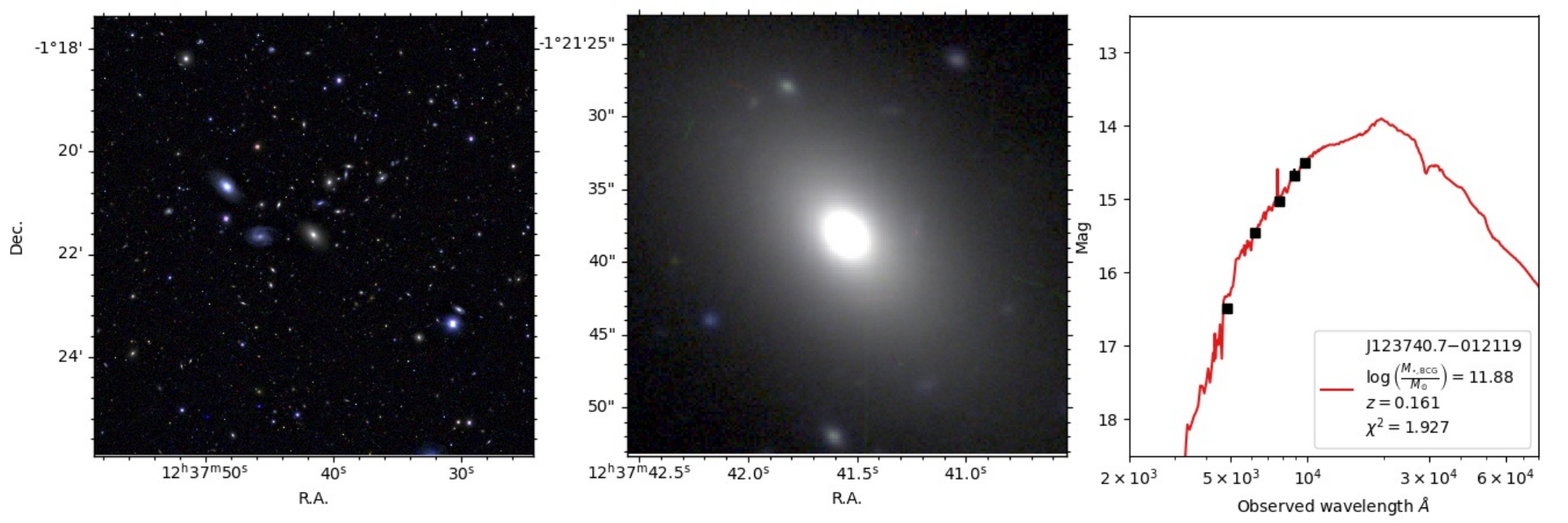}}
\resizebox{0.33\textwidth}{!}{\includegraphics[scale=1]{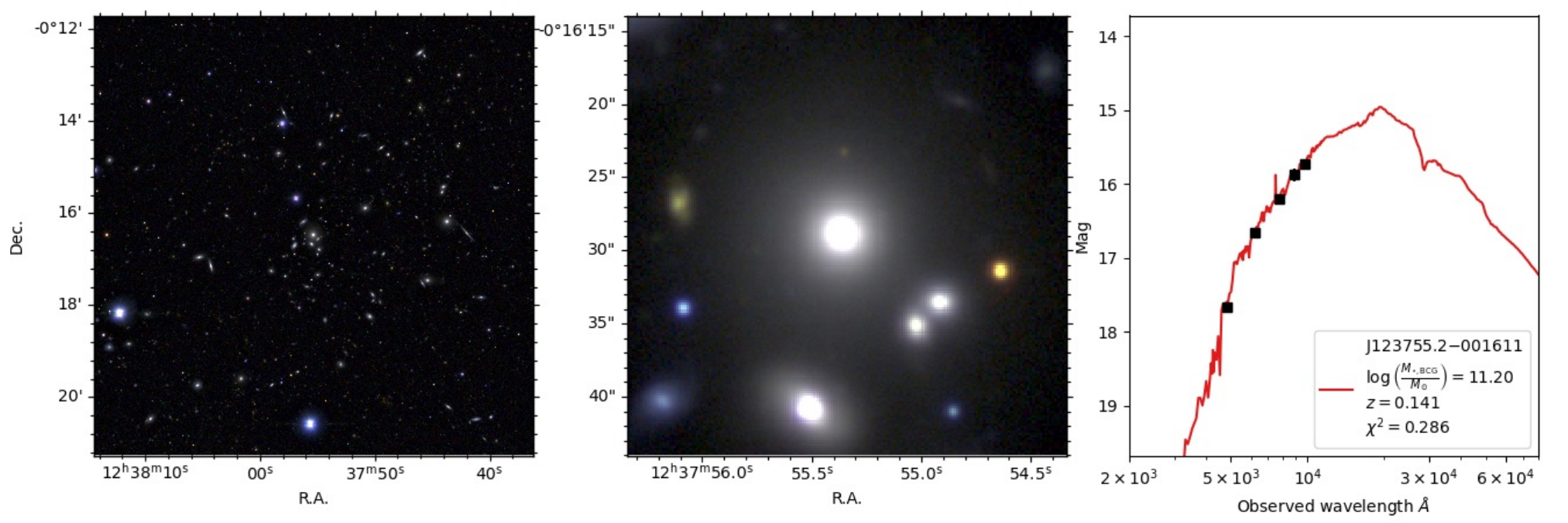}}
\resizebox{0.33\textwidth}{!}{\includegraphics[scale=1]{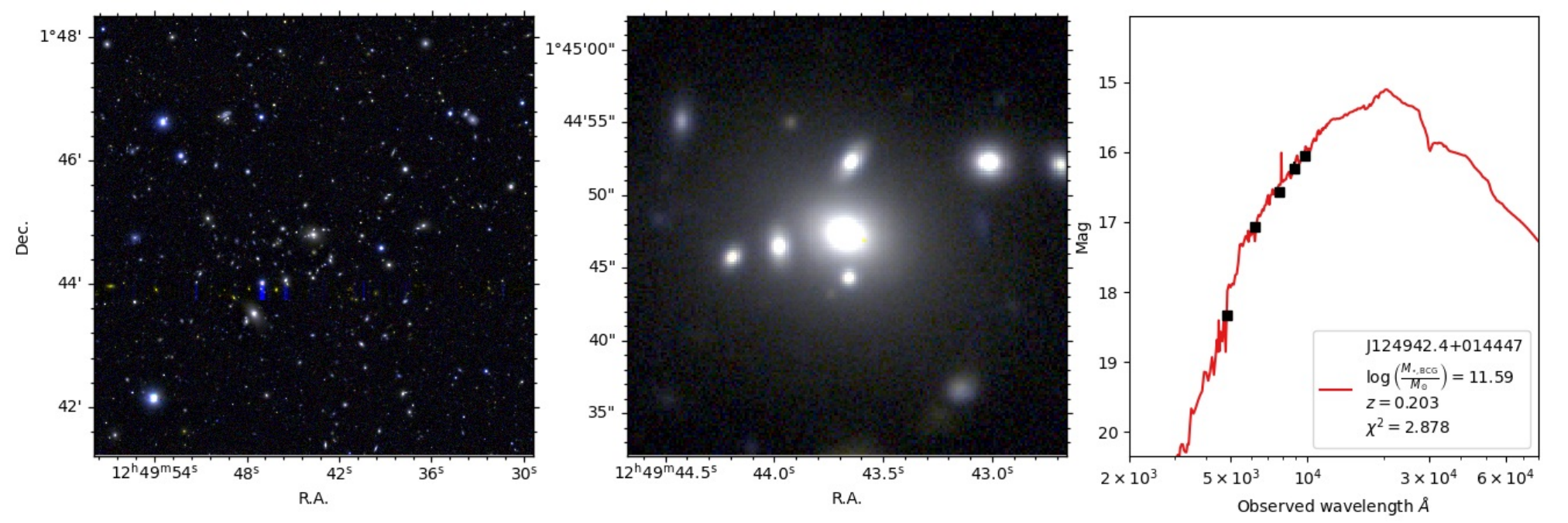}}\\
\resizebox{0.33\textwidth}{!}{\includegraphics[scale=1]{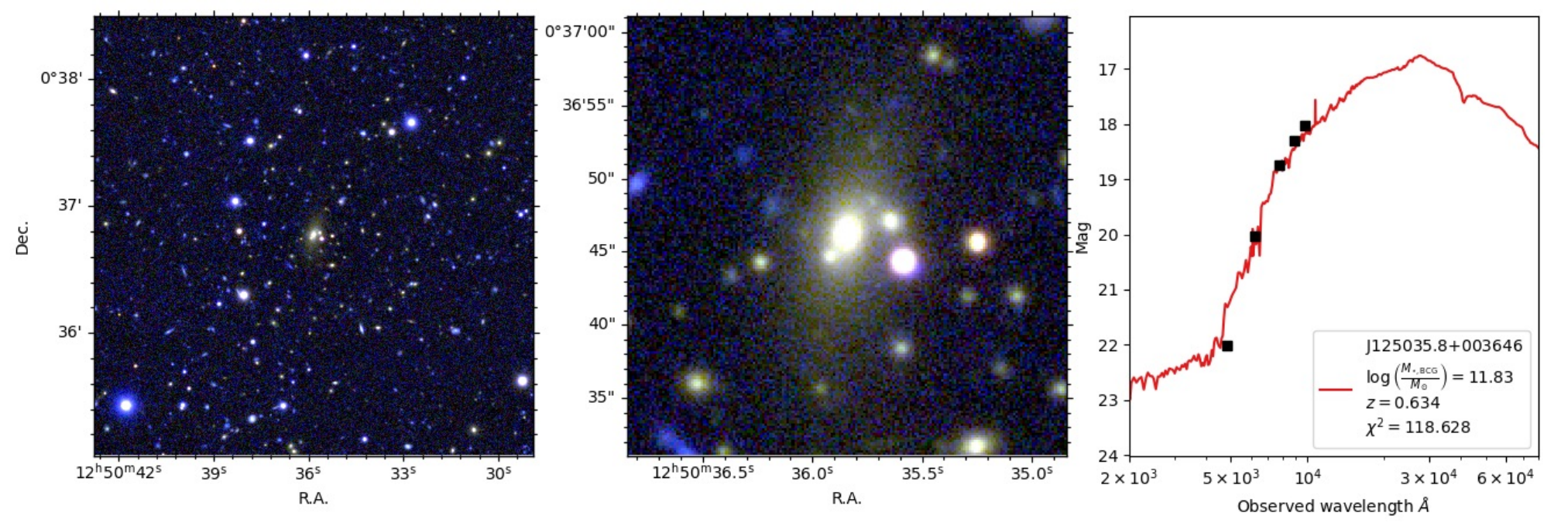}}
\resizebox{0.33\textwidth}{!}{\includegraphics[scale=1]{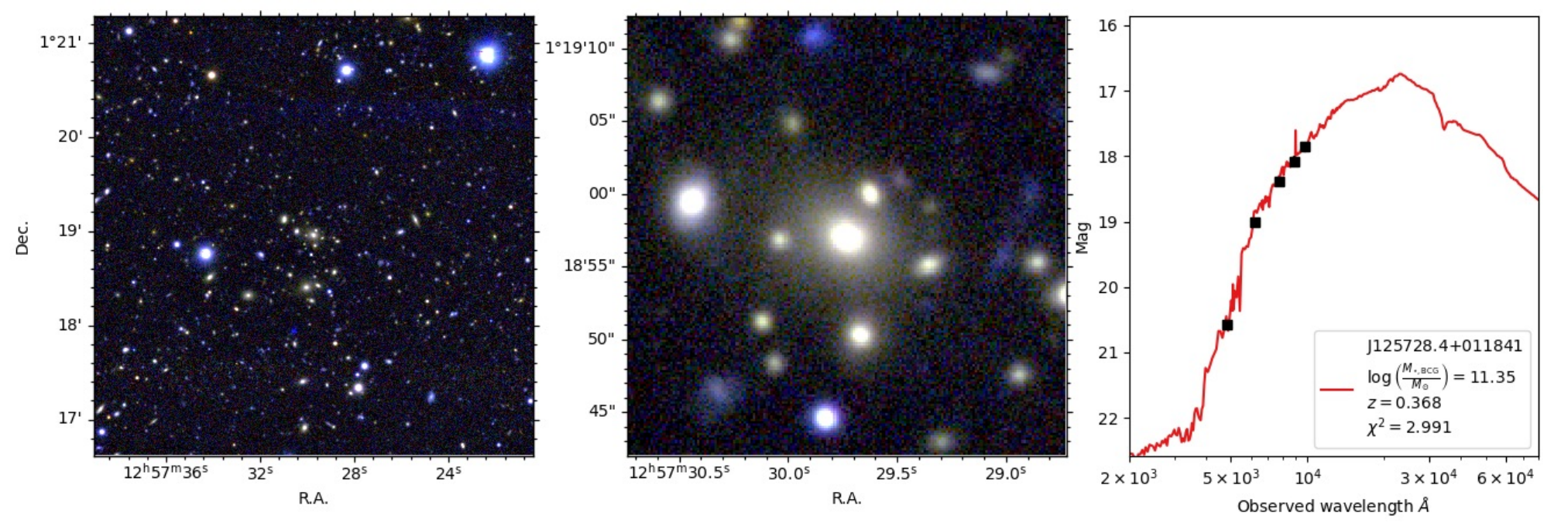}}
\resizebox{0.33\textwidth}{!}{\includegraphics[scale=1]{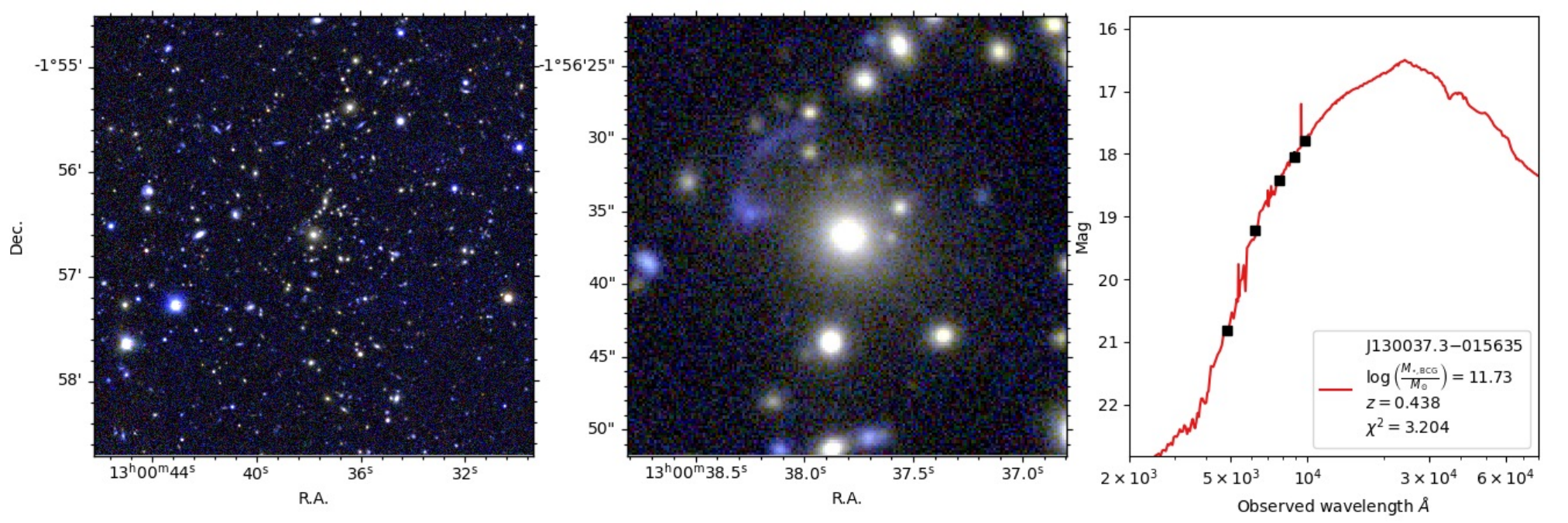}}\\
\resizebox{0.33\textwidth}{!}{\includegraphics[scale=1]{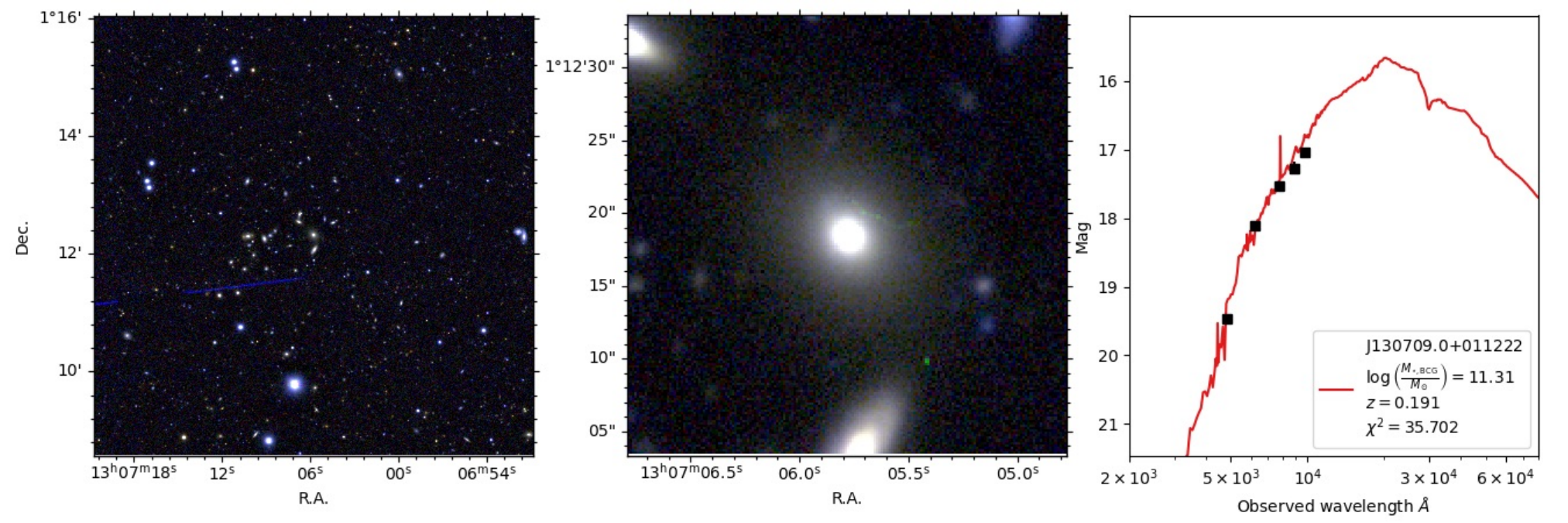}}
\resizebox{0.33\textwidth}{!}{\includegraphics[scale=1]{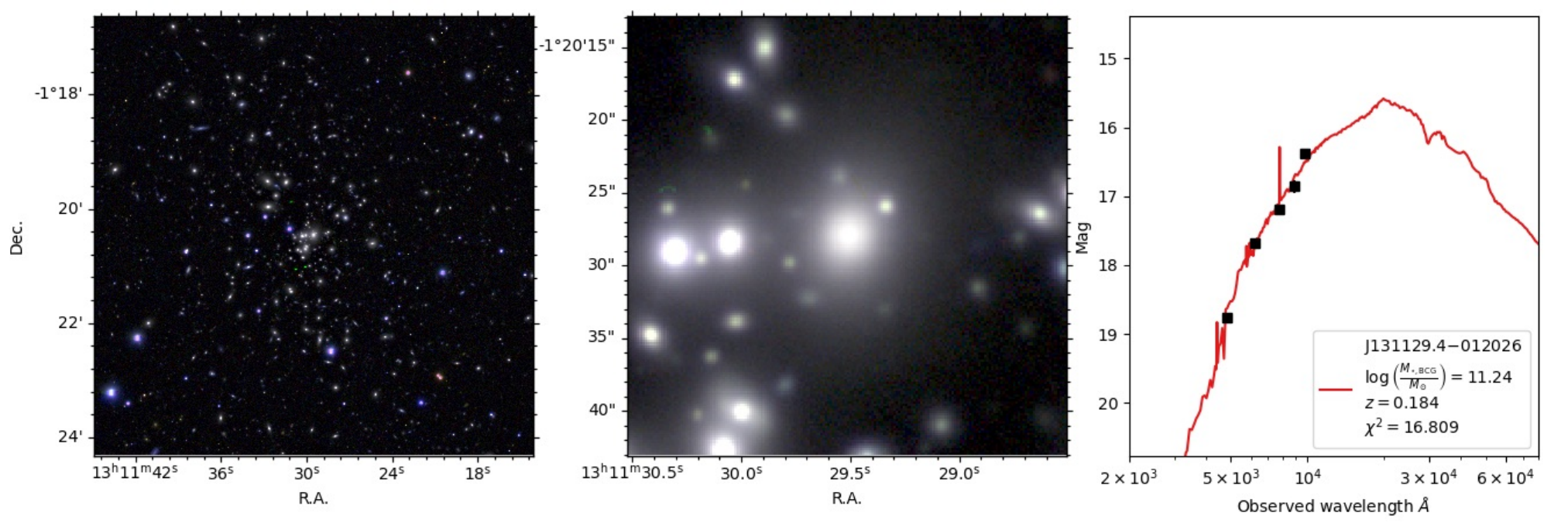}}
\resizebox{0.33\textwidth}{!}{\includegraphics[scale=1]{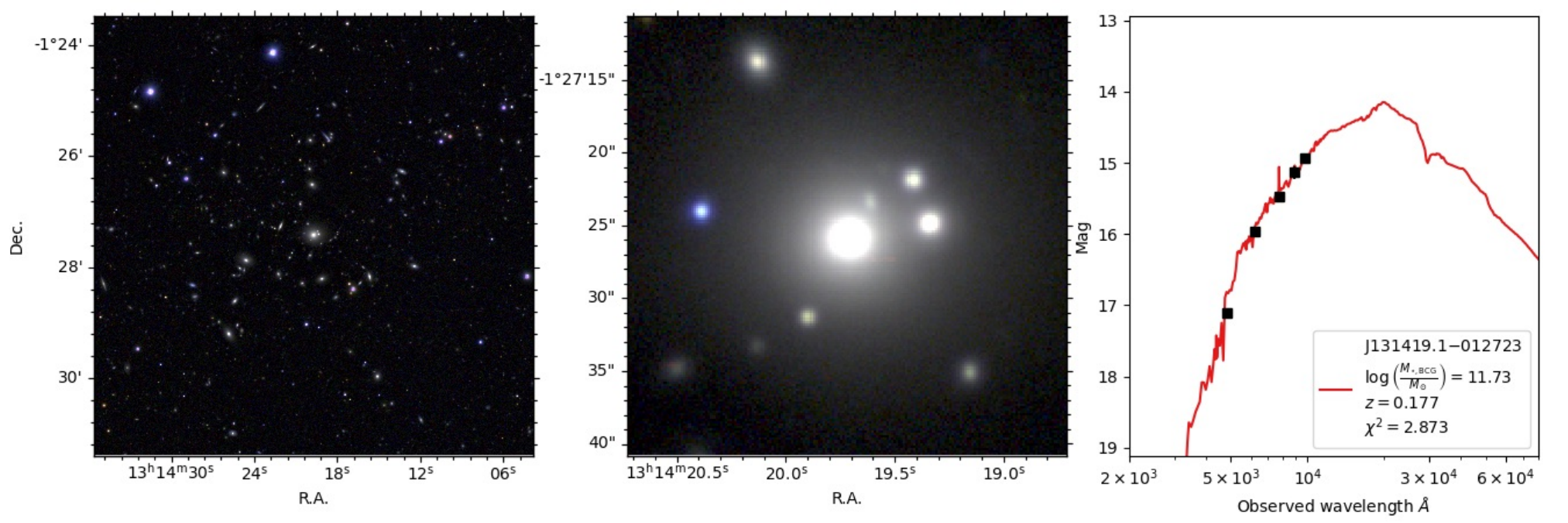}}\\
\resizebox{0.33\textwidth}{!}{\includegraphics[scale=1]{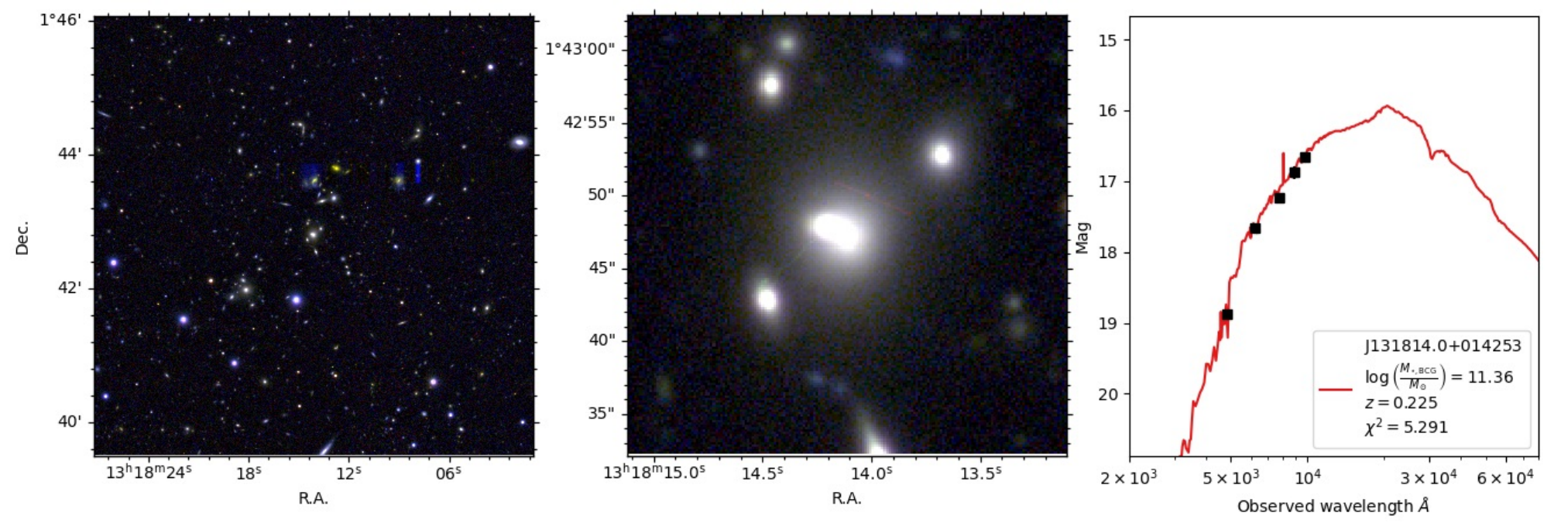}}
\resizebox{0.33\textwidth}{!}{\includegraphics[scale=1]{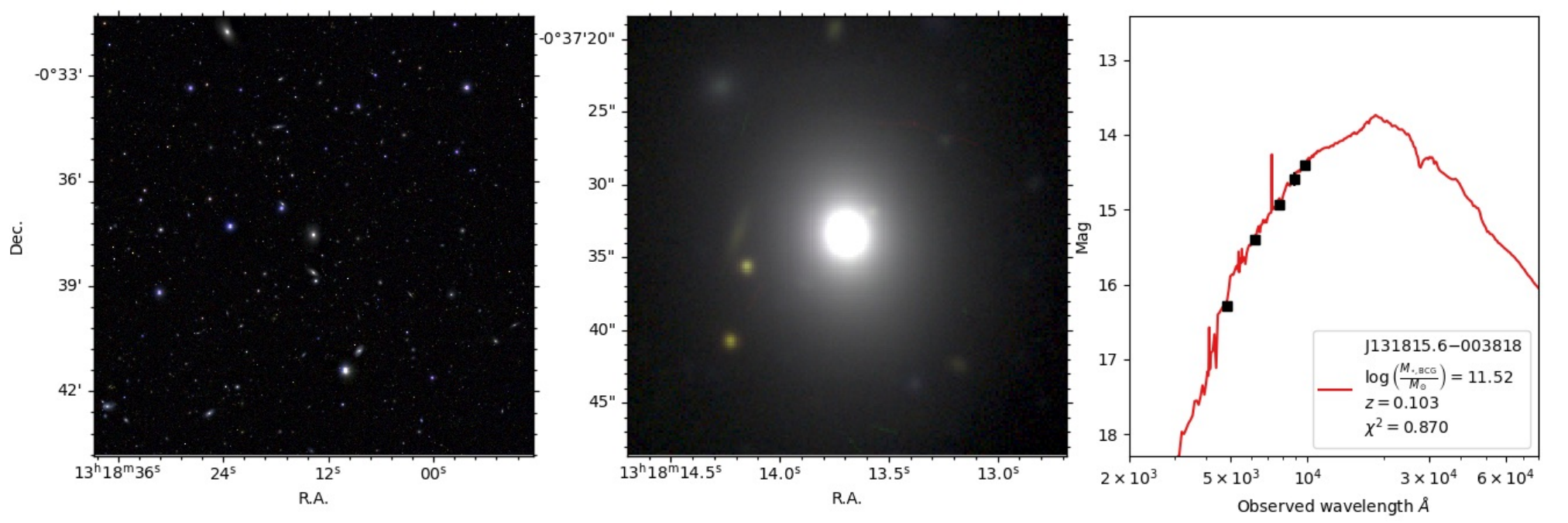}}
\resizebox{0.33\textwidth}{!}{\includegraphics[scale=1]{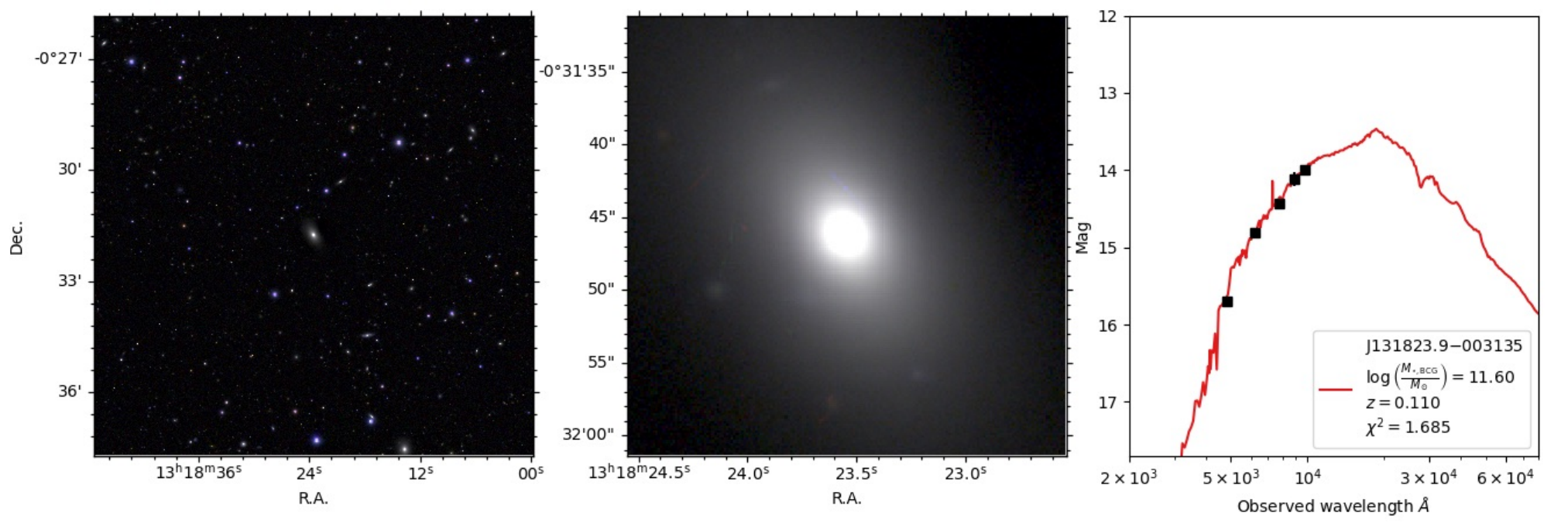}}
\caption{See Appendix~\ref{app:imaging} for details.}
\label{fig:cutout2}
\end{figure*}

\begin{figure*}[!ht]
\centering
\resizebox{0.33\textwidth}{!}{\includegraphics[scale=1]{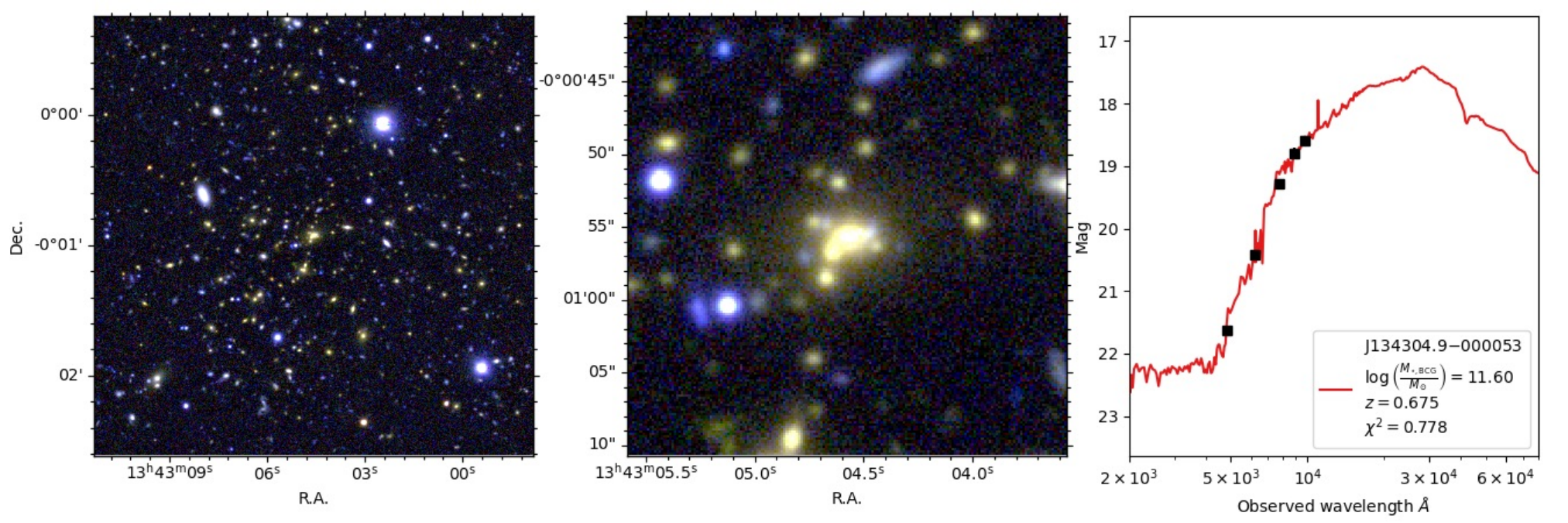}}
\resizebox{0.33\textwidth}{!}{\includegraphics[scale=1]{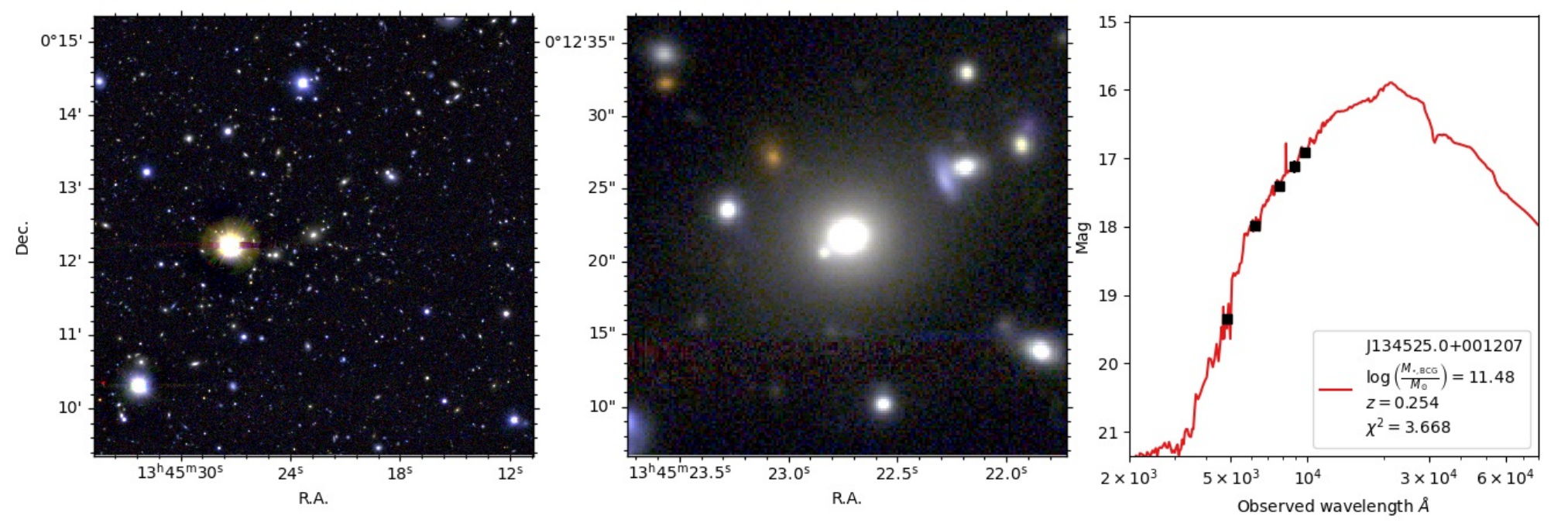}}
\resizebox{0.33\textwidth}{!}{\includegraphics[scale=1]{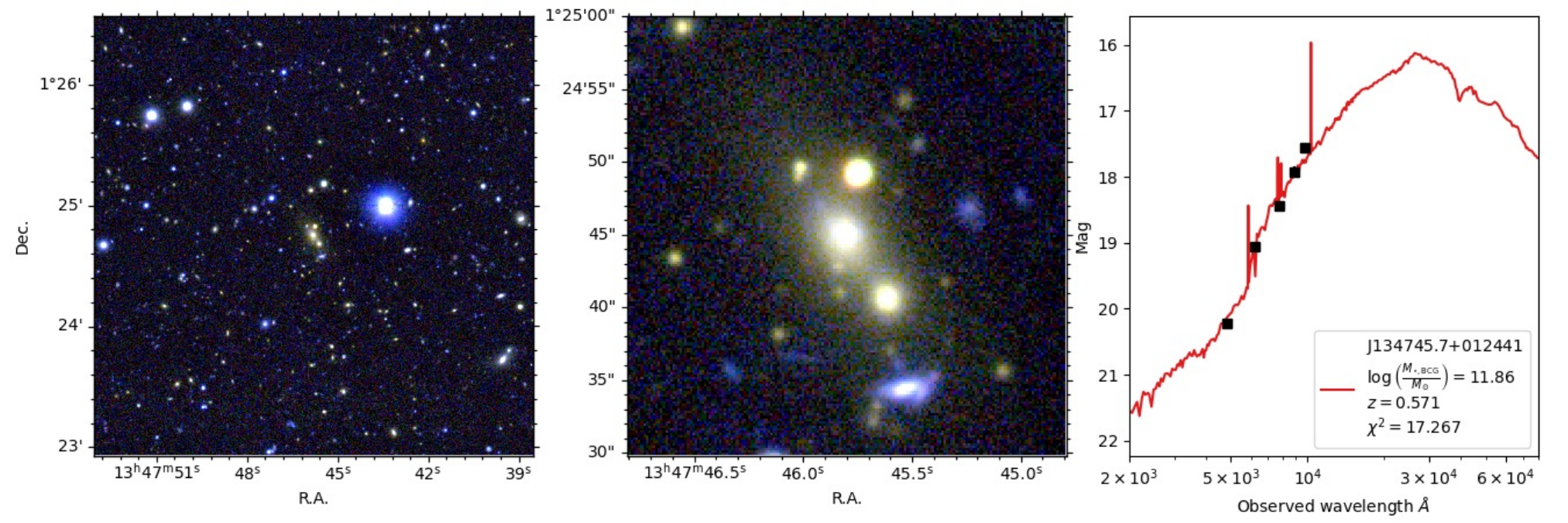}}\\
\resizebox{0.33\textwidth}{!}{\includegraphics[scale=1]{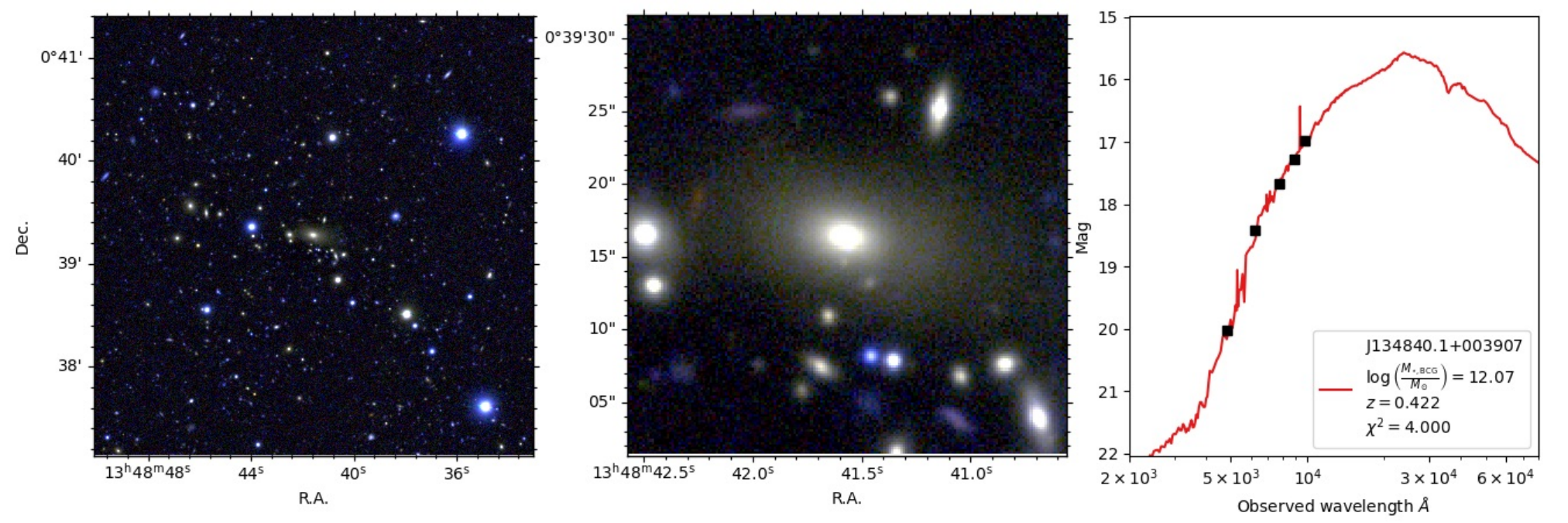}}
\resizebox{0.33\textwidth}{!}{\includegraphics[scale=1]{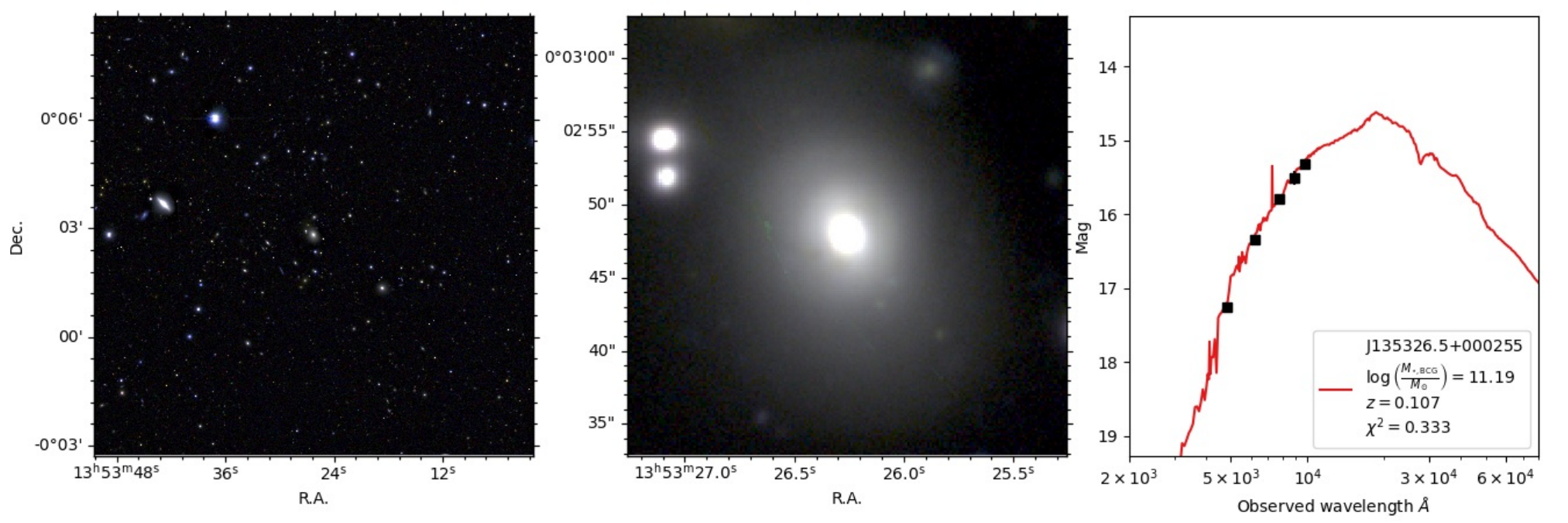}}
\resizebox{0.33\textwidth}{!}{\includegraphics[scale=1]{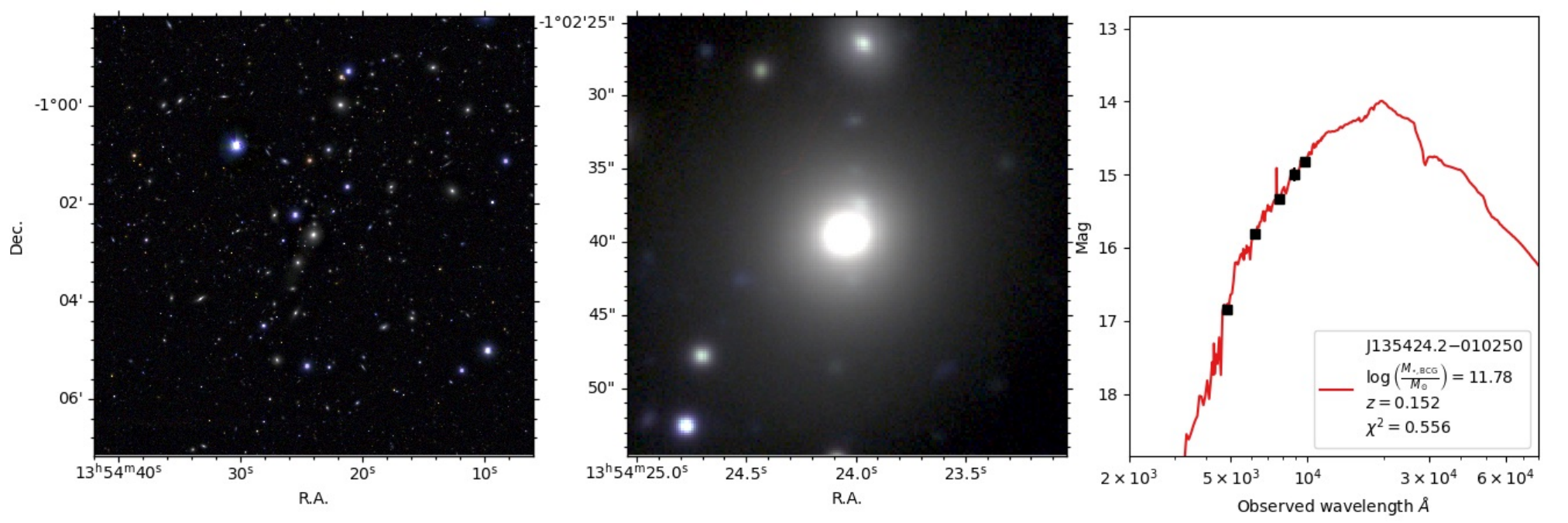}}\\
\resizebox{0.33\textwidth}{!}{\includegraphics[scale=1]{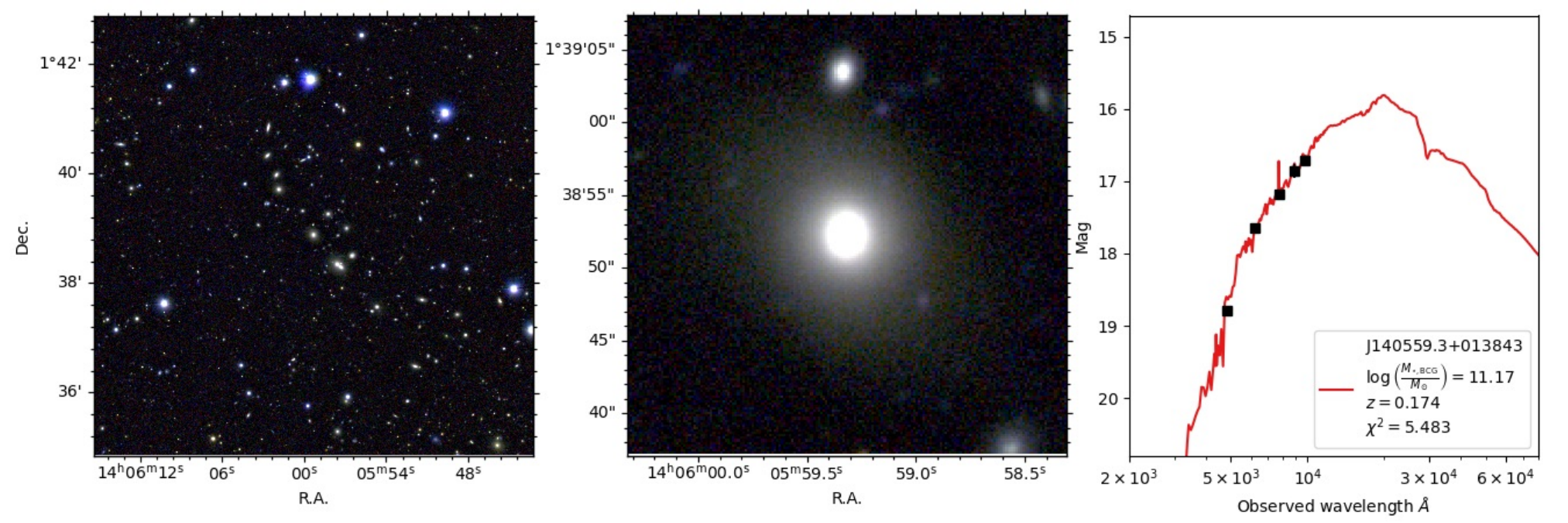}}
\resizebox{0.33\textwidth}{!}{\includegraphics[scale=1]{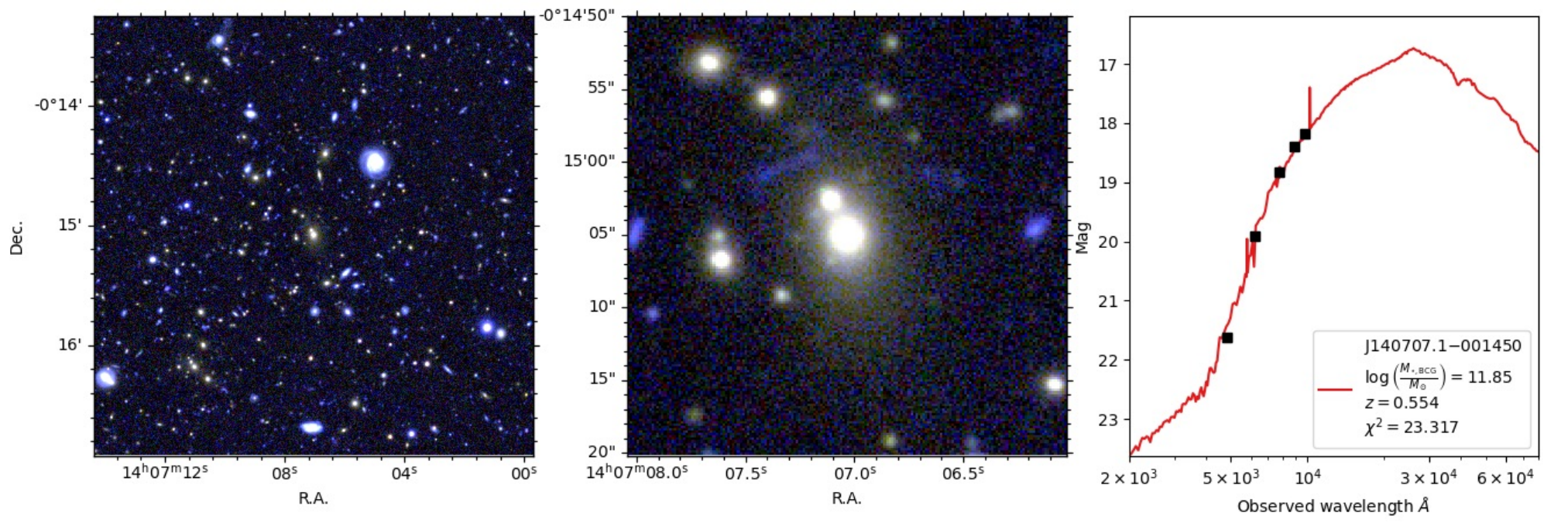}}
\resizebox{0.33\textwidth}{!}{\includegraphics[scale=1]{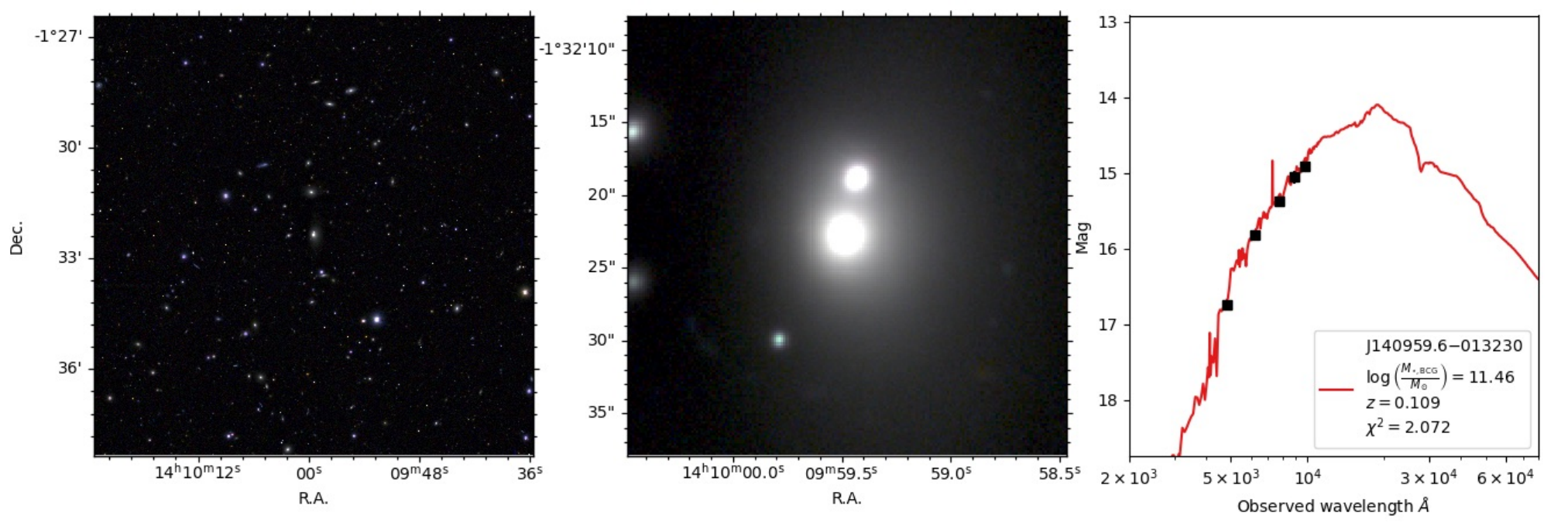}}\\
\resizebox{0.33\textwidth}{!}{\includegraphics[scale=1]{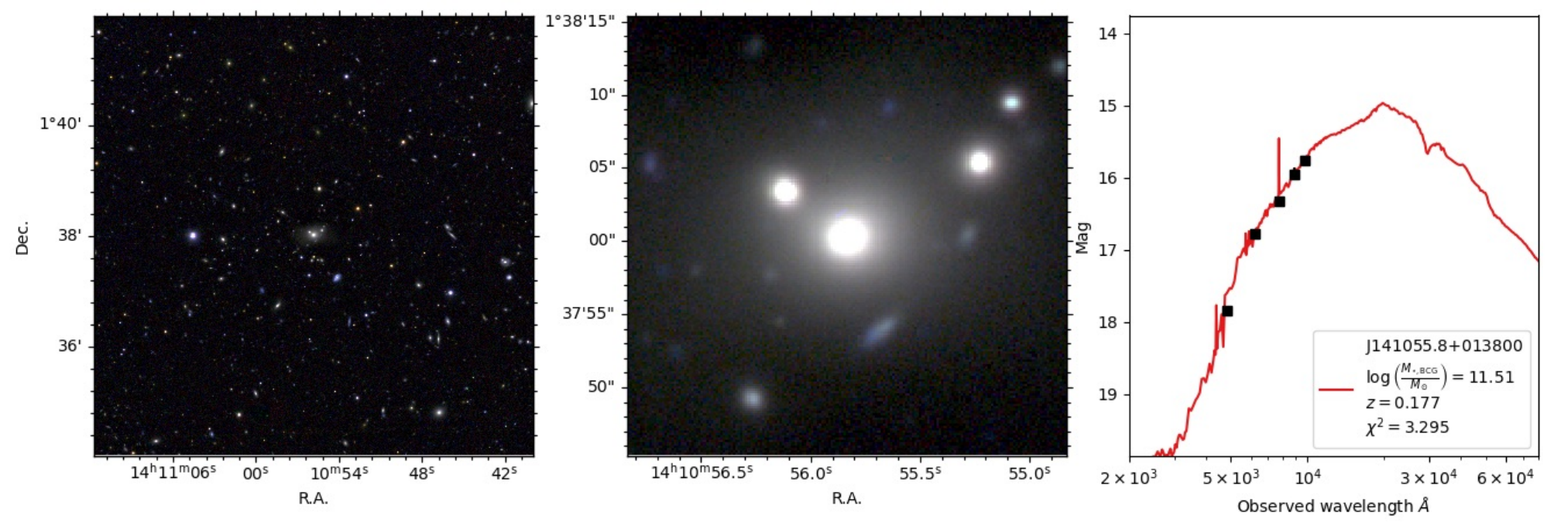}}
\resizebox{0.33\textwidth}{!}{\includegraphics[scale=1]{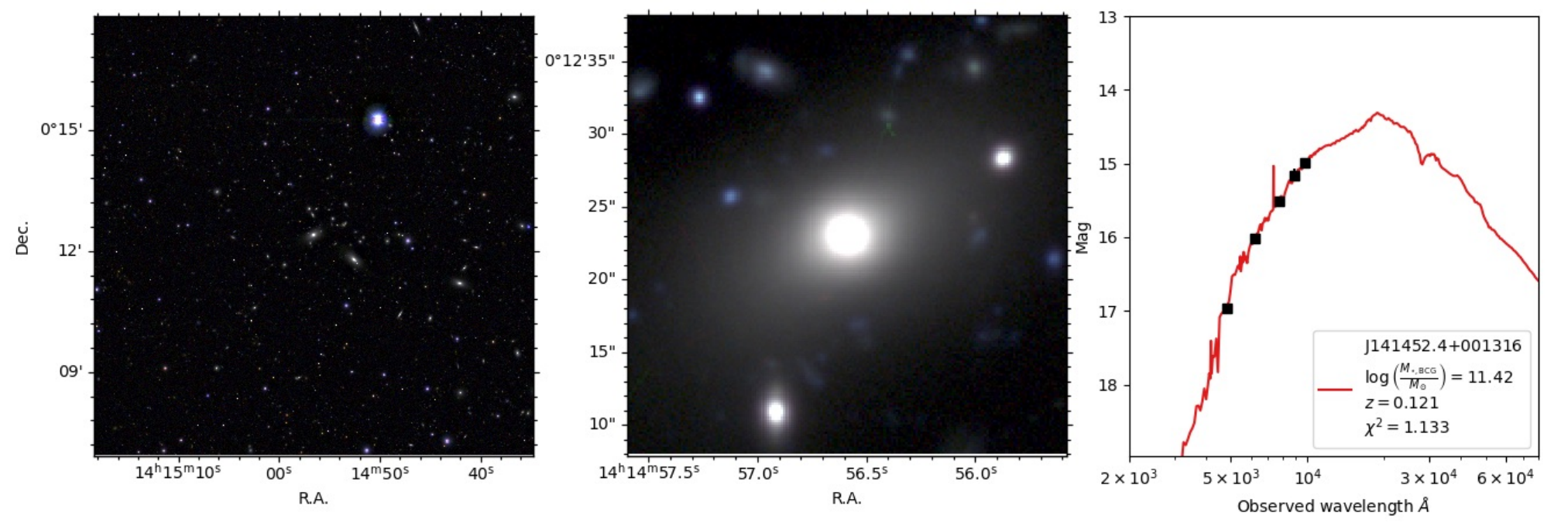}}
\resizebox{0.33\textwidth}{!}{\includegraphics[scale=1]{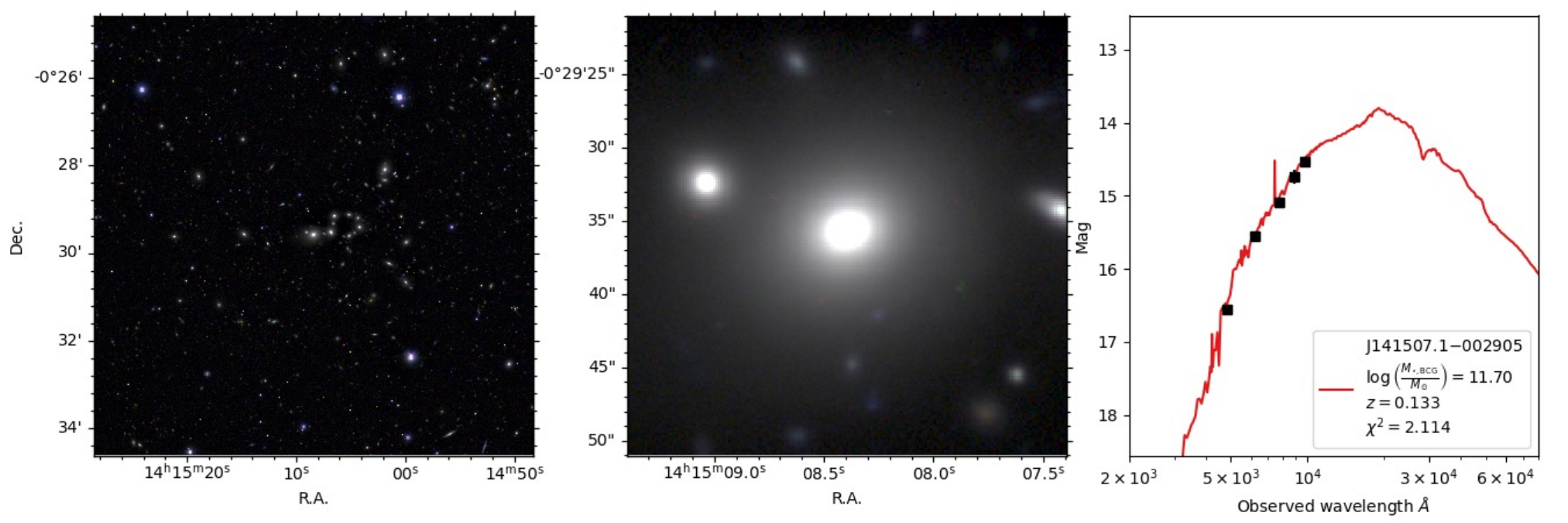}}\\
\resizebox{0.33\textwidth}{!}{\includegraphics[scale=1]{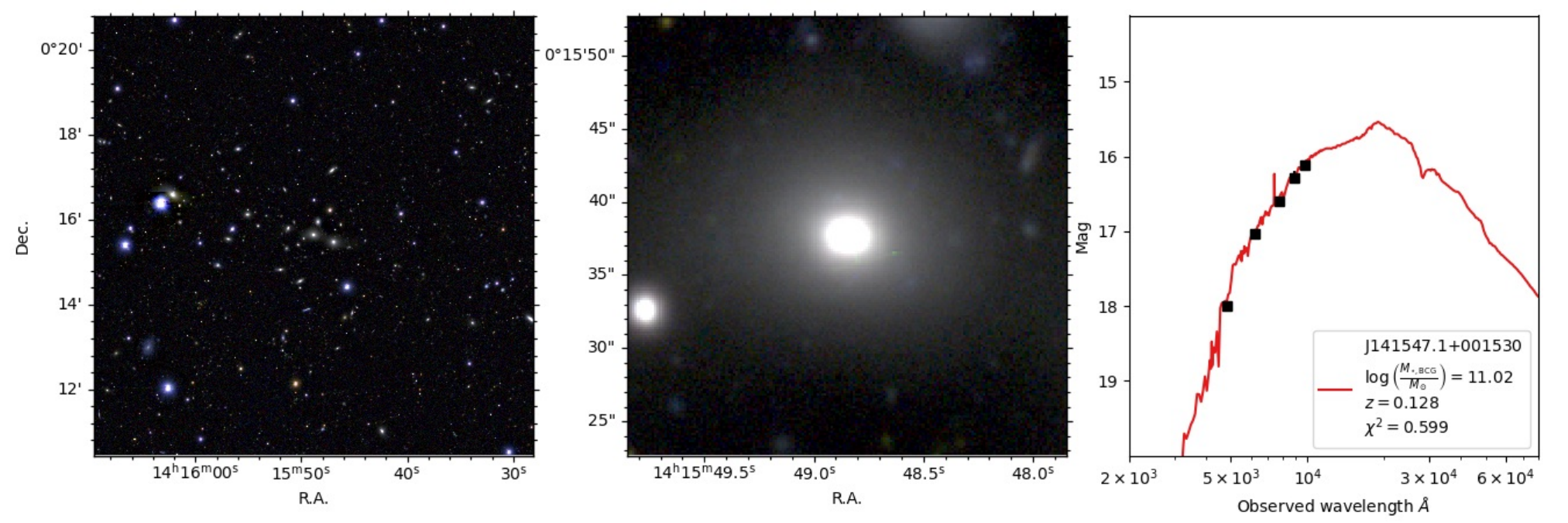}}
\resizebox{0.33\textwidth}{!}{\includegraphics[scale=1]{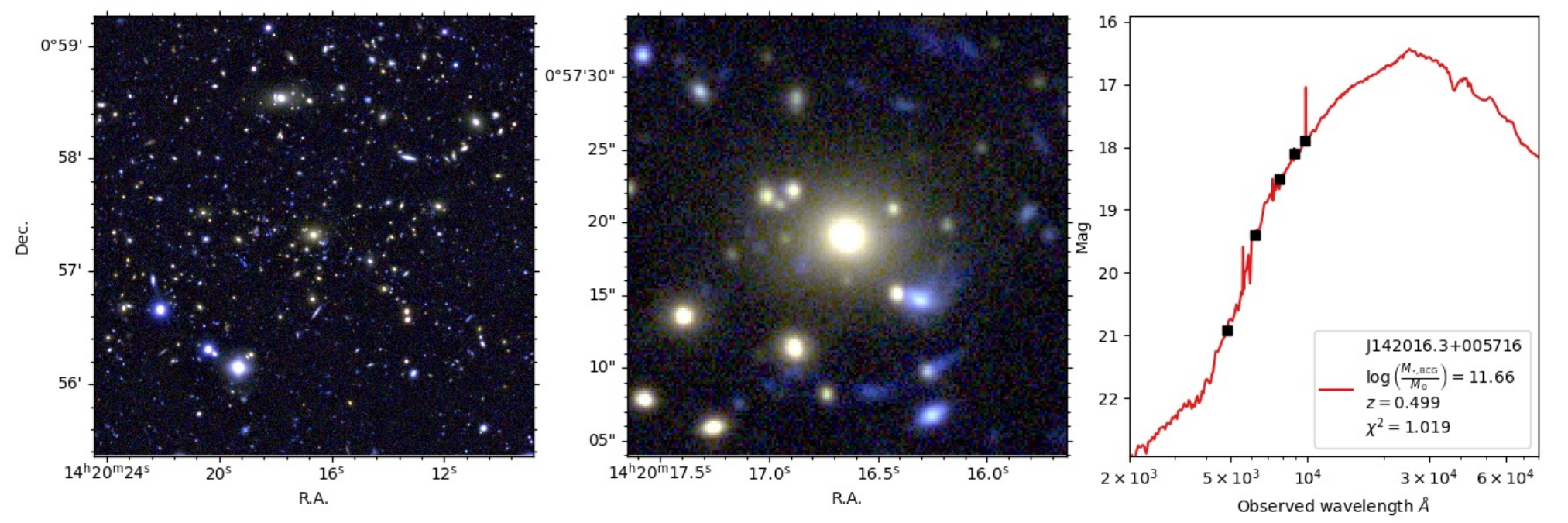}}
\resizebox{0.33\textwidth}{!}{\includegraphics[scale=1]{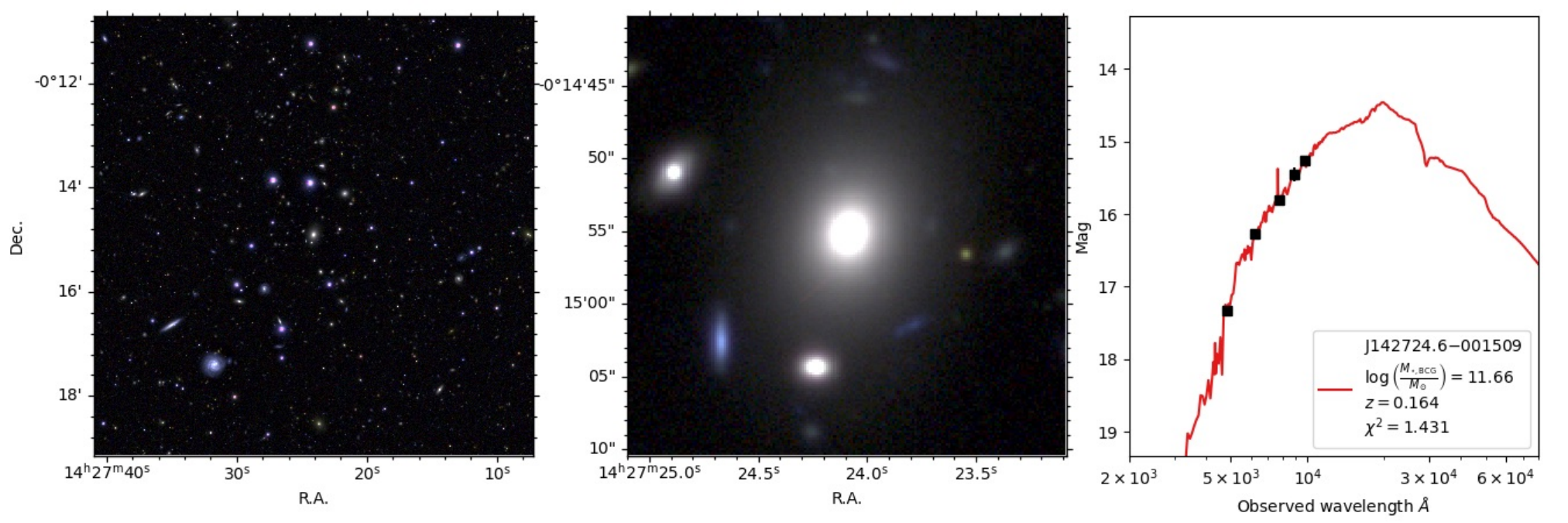}}\\
\resizebox{0.33\textwidth}{!}{\includegraphics[scale=1]{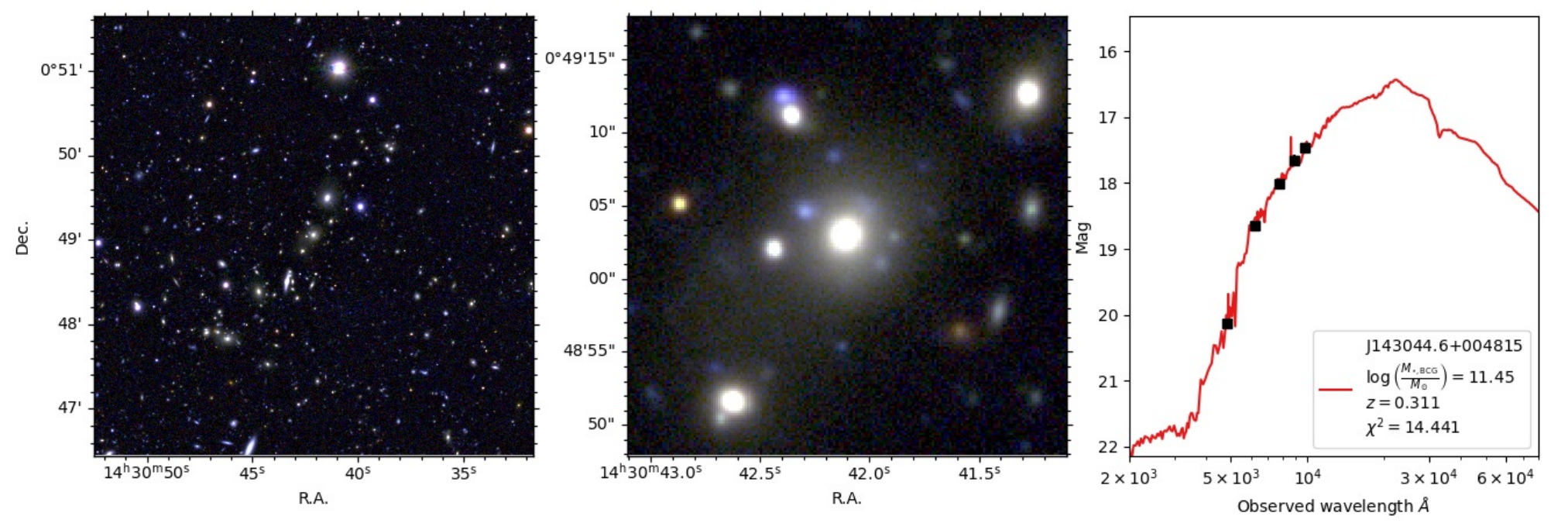}}
\resizebox{0.33\textwidth}{!}{\includegraphics[scale=1]{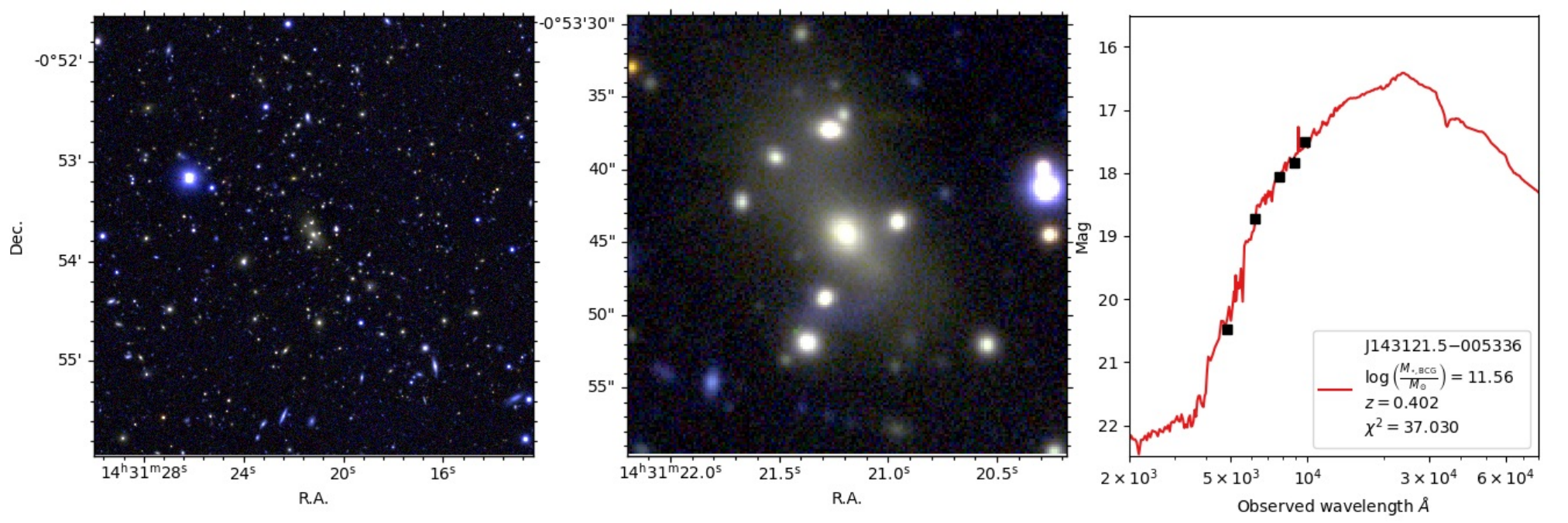}}
\resizebox{0.33\textwidth}{!}{\includegraphics[scale=1]{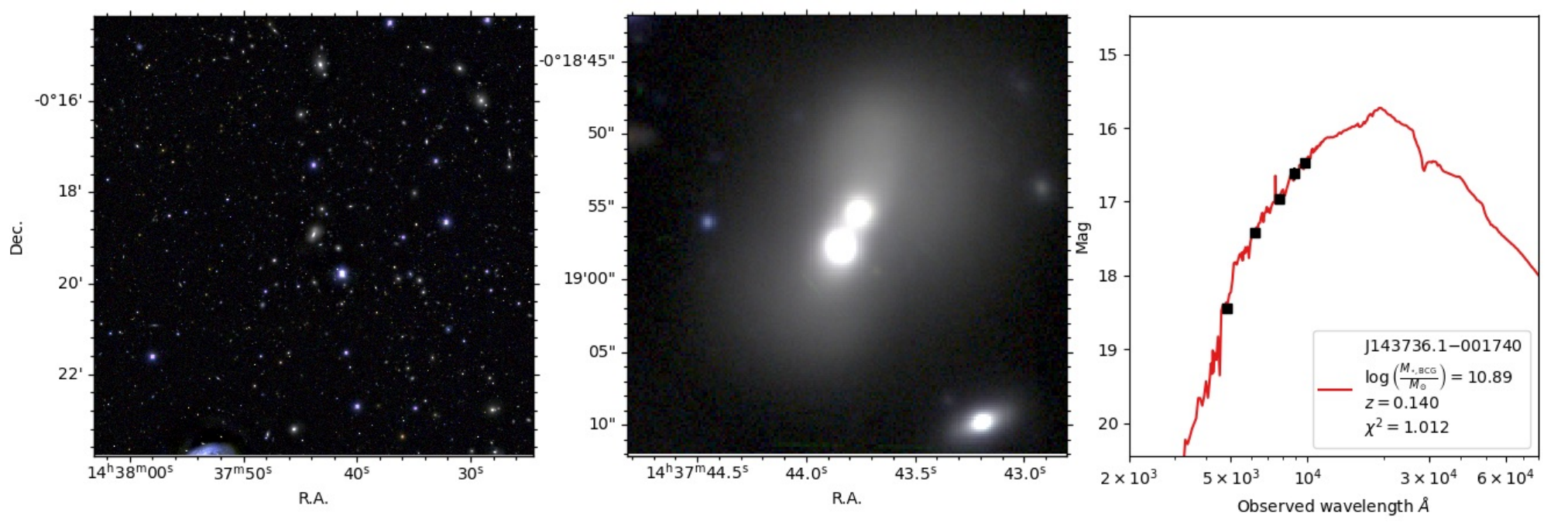}}\\
\resizebox{0.33\textwidth}{!}{\includegraphics[scale=1]{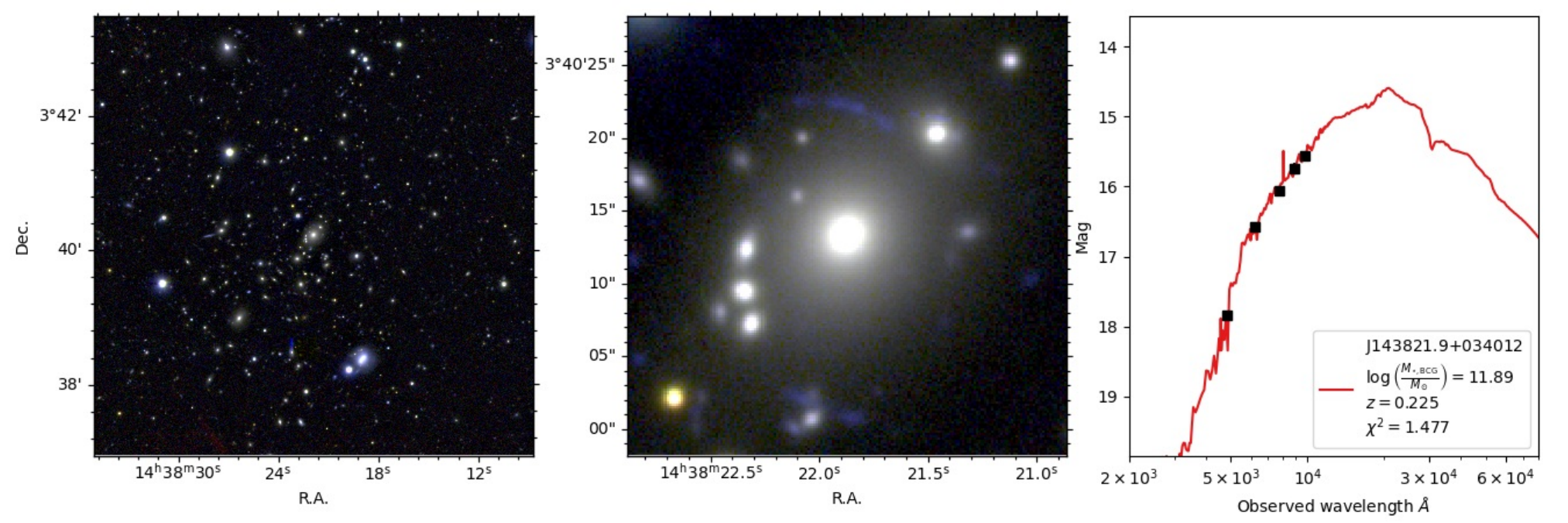}}
\resizebox{0.33\textwidth}{!}{\includegraphics[scale=1]{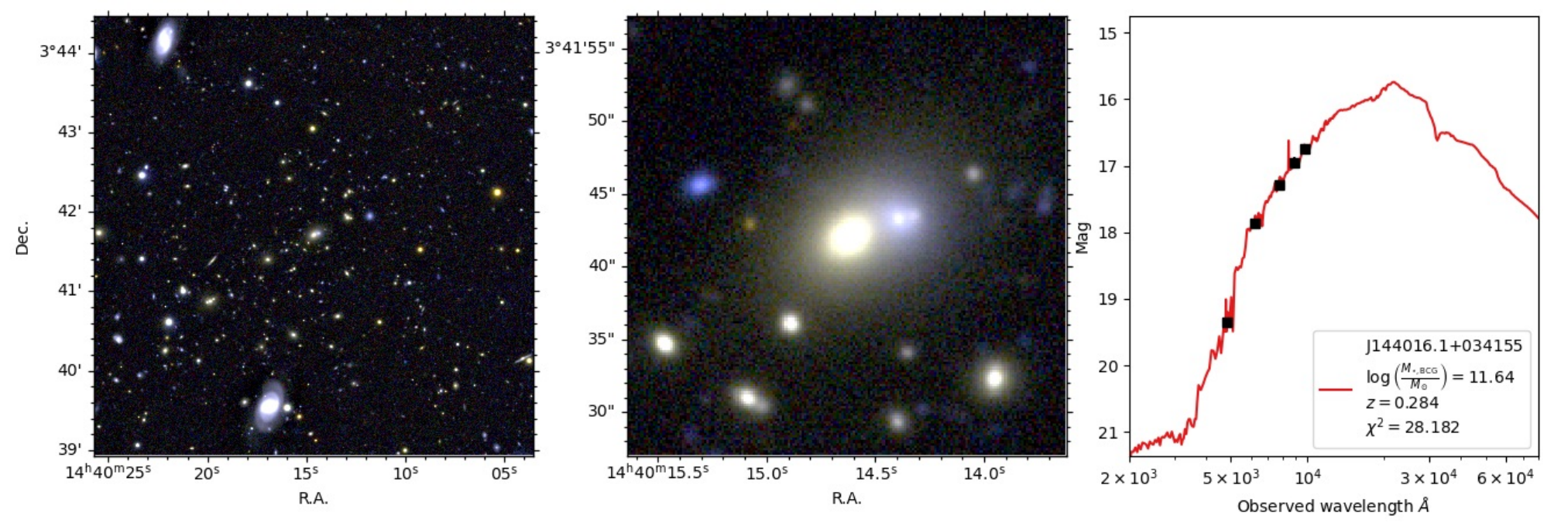}}
\resizebox{0.33\textwidth}{!}{\includegraphics[scale=1]{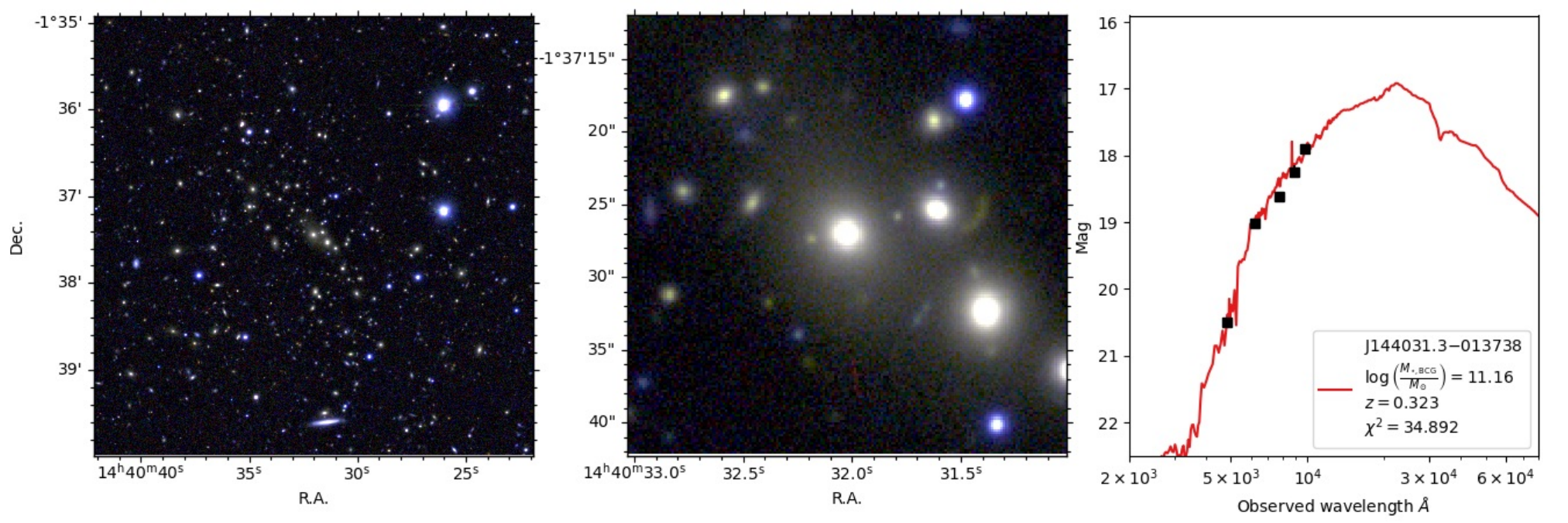}}\\
\resizebox{0.33\textwidth}{!}{\includegraphics[scale=1]{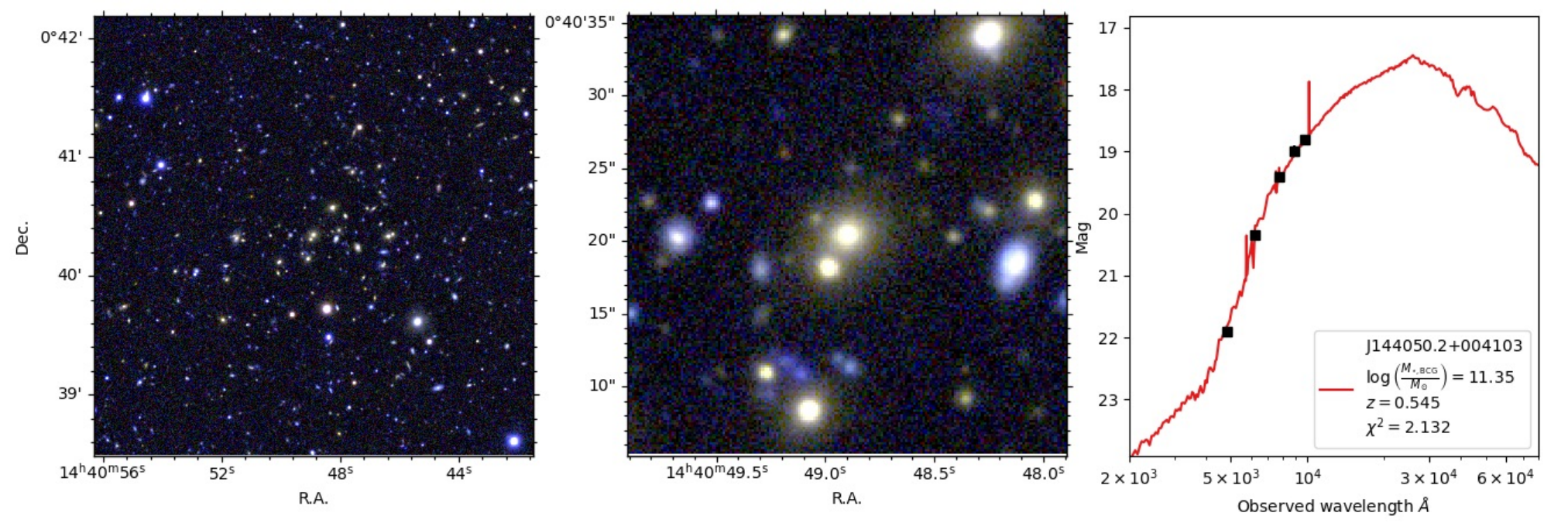}}
\resizebox{0.33\textwidth}{!}{\includegraphics[scale=1]{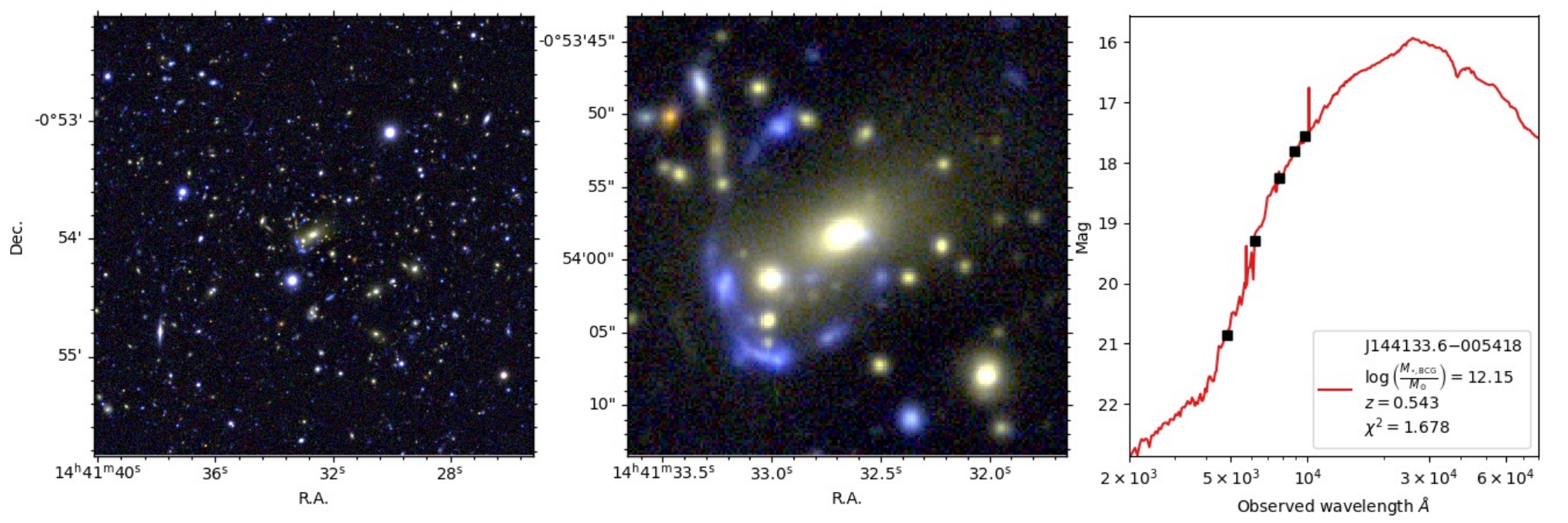}}
\resizebox{0.33\textwidth}{!}{\includegraphics[scale=1]{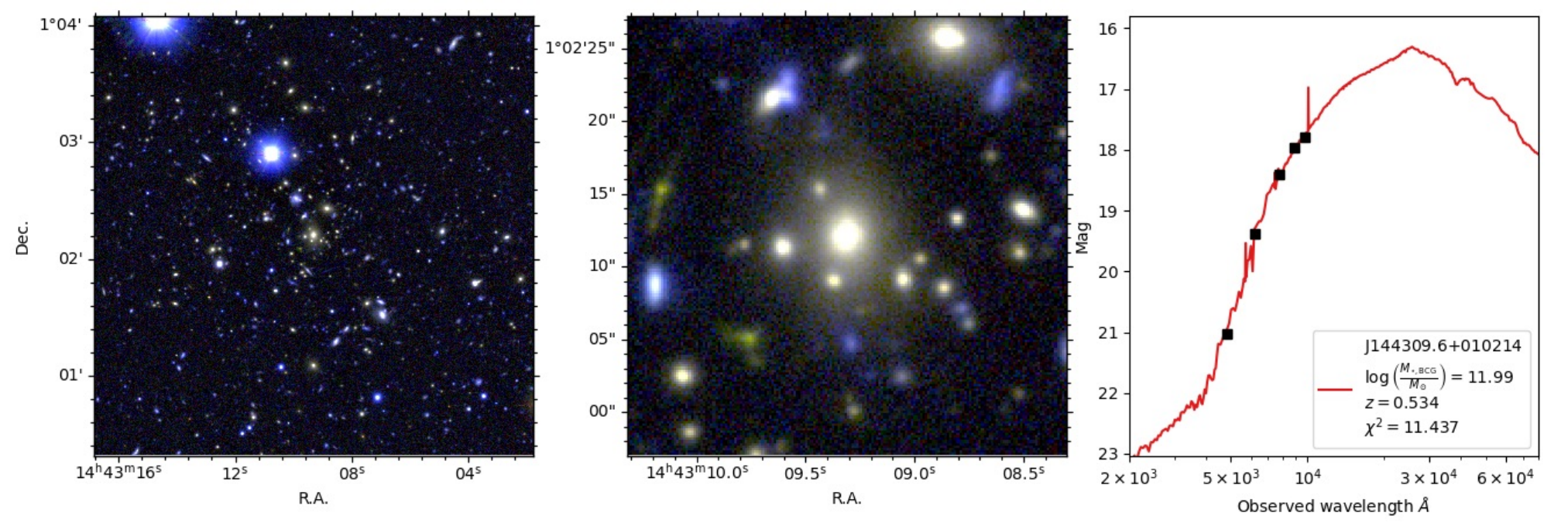}}\\
\resizebox{0.33\textwidth}{!}{\includegraphics[scale=1]{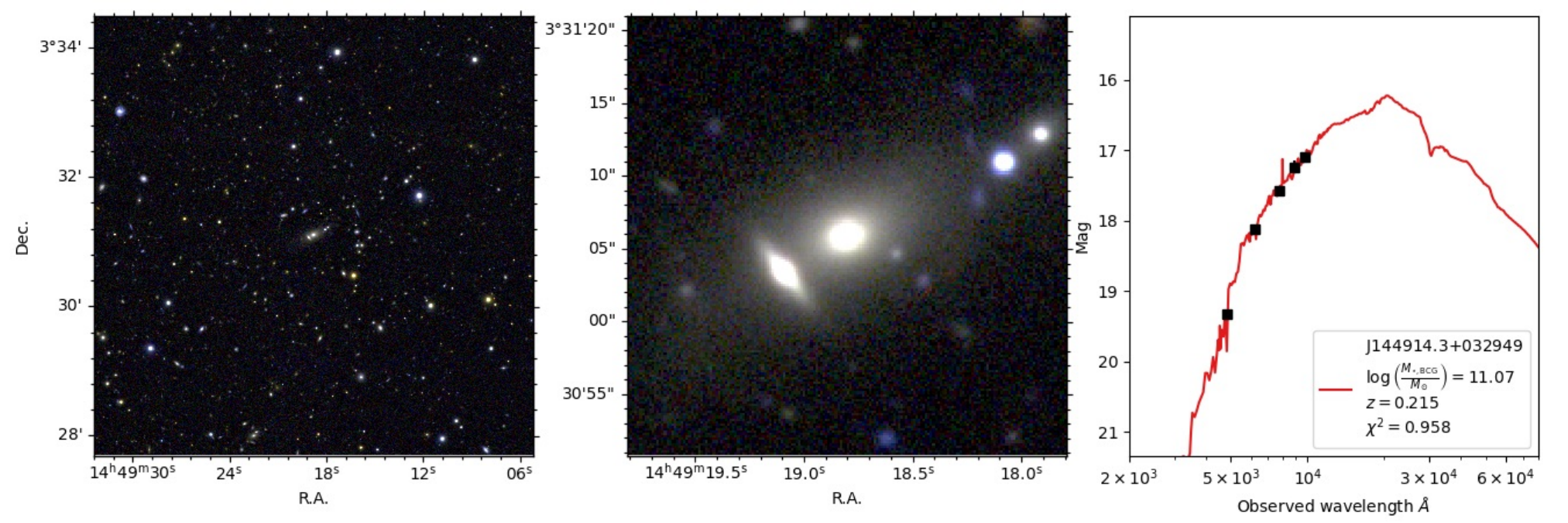}}
\resizebox{0.33\textwidth}{!}{\includegraphics[scale=1]{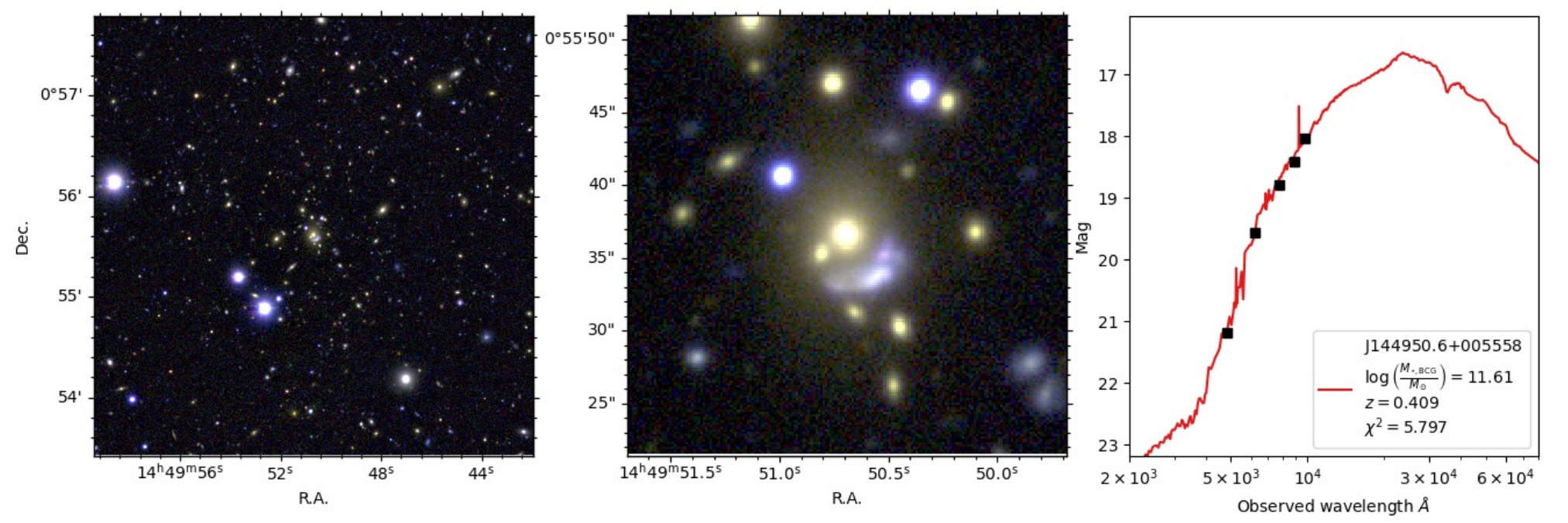}}
\resizebox{0.33\textwidth}{!}{\includegraphics[scale=1]{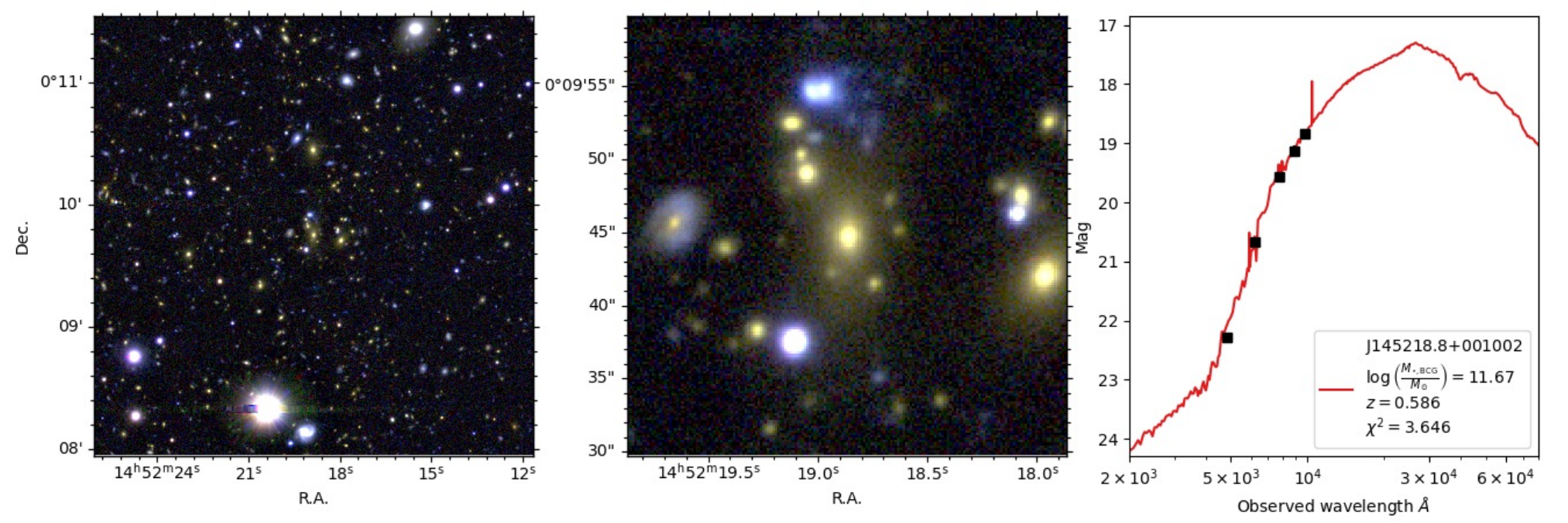}}\\
\resizebox{0.33\textwidth}{!}{\includegraphics[scale=1]{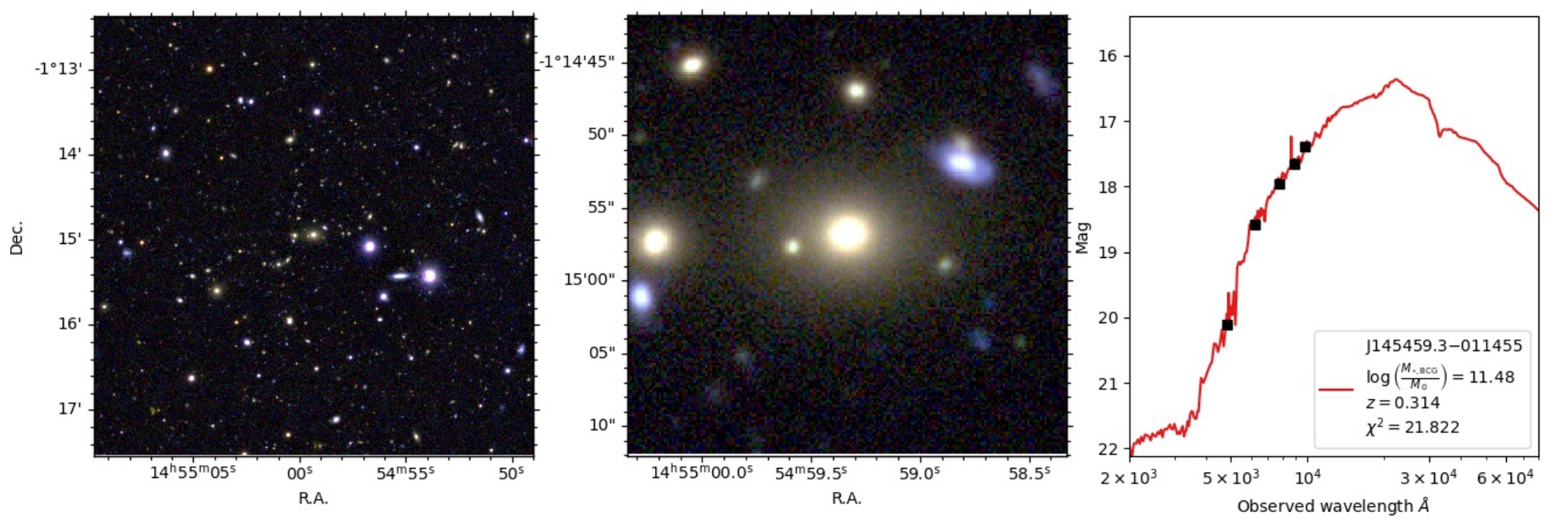}}
\resizebox{0.33\textwidth}{!}{\includegraphics[scale=1]{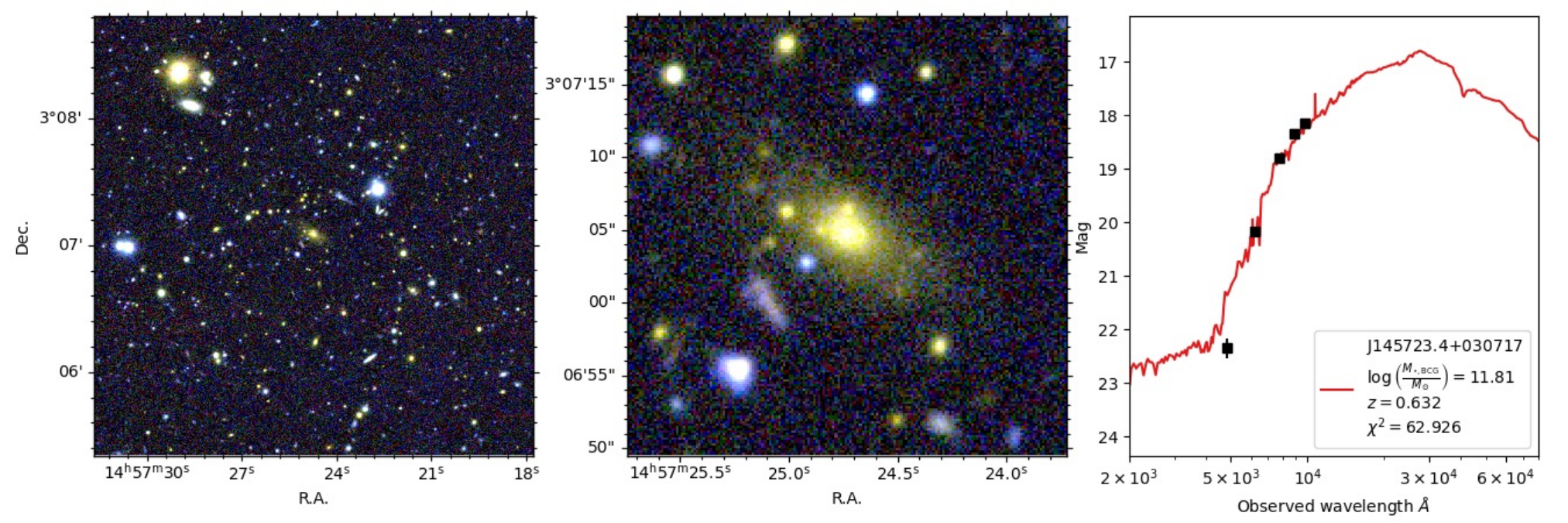}}
\caption{See Appendix~\ref{app:imaging} for details.}
\label{fig:cutout3}
\end{figure*}

We provide the cutout images of the $101$ \erass1 clusters and their BCGs in Figures~\ref{fig:cutout1}~to~\ref{fig:cutout3}.
Each figure contains multiple subplots, showing the results of individual clusters in three panels:
The cutout HSC images of individual \erass1 clusters ($0.5\Mpch\times0.5\Mpch$) and their BCGs ($15\arcsec\times15\arcsec$) are displayed in the left and middle panels, respectively.
Both cutout images center at the BCG.
The $grizY$ magnitudes (after applying all systematic corrections) and the errorbars are shown in the right panel with the best-fit template.
The cluster name, redshift, best-fit BCG stellar mass, and the $\chi^2$ of the best-fit template are indicated in the right panel.

In Table~\ref{tab:measurements}, we provide the mass measurements of individual clusters studied in this work.

{\small
\begin{longtable}{ccccccc}
\caption{
The measurements of \erass1 clusters.
}
\label{tab:measurements} \\
\hline
Name
& $\left(\alpha_{\mathrm{BCG}}, \delta_{\mathrm{BCG}}\right)$
& Data availability
& redshift 
& $\left\langle\log\left(\frac{\Mfiveoo}{\Msunh}\right)\right\rangle$ 
& Sampled \Mfiveoo
& $\log\left(\frac{\mbcg}{\Msun}\right)$  \\ 
\hline
 1eRASS~J083652.6$+$030000 & $( 129.21437, 3.00035 )$ & WL & $ 0.19 $ & $ 14.22 \pm 0.13 $ & $ 14.11 $& $ \cdots $ \\
 1eRASS~J083811.9$-$015938 & $( 129.54900, -1.99318 )$ & BCG & $ 0.56 $ & $ 14.66 \pm 0.18 $ & $ 14.44 $& $ 11.28 \pm 0.1113 $ \\
 1eRASS~J083934.2$-$014035 & $( 129.88909, -1.67908 )$ & BCG & $ 0.29 $ & $ 14.51 \pm 0.16 $ & $ 14.61 $& $ 11.57 \pm 0.0722 $ \\
 1eRASS~J084306.7$+$002834 & $( 130.78317, 0.48309 )$ & WL$+$BCG & $ 0.27 $ & $ 14.01 \pm 0.15 $ & $ 14.06 $& $ 11.50 \pm 0.0850 $ \\
 1eRASS~J084342.8$+$040323 & $( 130.92944, 4.05961 )$ & WL & $ 0.21 $ & $ 14.13 \pm 0.15 $ & $ 14.12 $& $ \cdots $ \\
 1eRASS~J084527.7$+$032736 & $( 131.36569, 3.46079 )$ & WL$+$BCG & $ 0.33 $ & $ 14.62 \pm 0.10 $ & $ 14.67 $& $ 11.75 \pm 0.0802 $ \\
 1eRASS~J085029.4$+$001453 & $( 132.61573, 0.25044 )$ & WL & $ 0.20 $ & $ 14.24 \pm 0.13 $ & $ 14.42 $& $ \cdots $ \\
 1eRASS~J085217.5$-$010126 & $( 133.06939, -1.02658 )$ & WL & $ 0.46 $ & $ 14.52 \pm 0.13 $ & $ 14.43 $& $ \cdots $ \\
 1eRASS~J085230.8$+$002509 & $( 133.12620, 0.41976 )$ & WL & $ 0.28 $ & $ 14.27 \pm 0.14 $ & $ 14.08 $& $ \cdots $ \\
 1eRASS~J085435.8$+$003858 & $( 133.65257, 0.64252 )$ & WL$+$BCG & $ 0.11 $ & $ 14.07 \pm 0.13 $ & $ 13.90 $& $ 11.46 \pm 0.0777 $ \\
 1eRASS~J085751.0$+$031014 & $( 134.47521, 3.17645 )$ & WL$+$BCG & $ 0.20 $ & $ 14.49 \pm 0.12 $ & $ 14.50 $& $ 11.02 \pm 0.0665 $ \\
 1eRASS~J085932.4$+$030832 & $( 134.88466, 3.14468 )$ & WL$+$BCG & $ 0.20 $ & $ 14.28 \pm 0.12 $ & $ 14.31 $& $ 11.76 \pm 0.0610 $ \\
 1eRASS~J090131.4$+$030055 & $( 135.37951, 3.01564 )$ & WL & $ 0.20 $ & $ 14.36 \pm 0.13 $ & $ 14.56 $& $ \cdots $ \\
 1eRASS~J090539.7$+$043433 & $( 136.41610, 4.57755 )$ & WL & $ 0.23 $ & $ 14.26 \pm 0.15 $ & $ 14.38 $& $ \cdots $ \\
 1eRASS~J091414.7$+$001922 & $( 138.56511, 0.32408 )$ & WL & $ 0.17 $ & $ 13.97 \pm 0.15 $ & $ 13.99 $& $ \cdots $ \\
 1eRASS~J091453.6$+$041611 & $( 138.71895, 4.26534 )$ & WL & $ 0.14 $ & $ 14.09 \pm 0.14 $ & $ 14.04 $& $ \cdots $ \\
 1eRASS~J091608.3$-$002355 & $( 139.03845, -0.40456 )$ & WL$+$BCG & $ 0.33 $ & $ 14.62 \pm 0.10 $ & $ 14.57 $& $ 11.70 \pm 0.0821 $ \\
 1eRASS~J092050.7$+$024512 & $( 140.20707, 2.75389 )$ & WL$+$BCG & $ 0.28 $ & $ 14.40 \pm 0.11 $ & $ 14.41 $& $ 11.38 \pm 0.0517 $ \\
 1eRASS~J092121.0$+$031735 & $( 140.33795, 3.28701 )$ & WL$+$BCG & $ 0.35 $ & $ 14.67 \pm 0.11 $ & $ 14.66 $& $ 11.89 \pm 0.0743 $ \\
 1eRASS~J092210.0$+$034626 & $( 140.53185, 3.76631 )$ & WL$+$BCG & $ 0.28 $ & $ 14.53 \pm 0.11 $ & $ 14.67 $& $ 11.62 \pm 0.0787 $ \\
 1eRASS~J092546.2$-$014321 & $( 141.43867, -1.72877 )$ & BCG & $ 0.23 $ & $ 14.10 \pm 0.20 $ & $ 14.11 $& $ 11.26 \pm 0.0735 $ \\
 1eRASS~J093025.3$+$021707 & $( 142.60267, 2.29030 )$ & WL$+$BCG & $ 0.54 $ & $ 14.61 \pm 0.13 $ & $ 14.68 $& $ 11.34 \pm 0.1110 $ \\
 1eRASS~J093150.9$-$002220 & $( 142.96262, -0.37049 )$ & WL$+$BCG & $ 0.35 $ & $ 14.39 \pm 0.12 $ & $ 14.50 $& $ 11.87 \pm 0.0884 $ \\
 1eRASS~J093459.9$+$005438 & $( 143.75282, 0.90407 )$ & WL$+$BCG & $ 0.38 $ & $ 14.52 \pm 0.11 $ & $ 14.43 $& $ 11.59 \pm 0.1221 $ \\
 1eRASS~J093512.7$+$004735 & $( 143.80125, 0.82562 )$ & WL$+$BCG & $ 0.36 $ & $ 14.78 \pm 0.09 $ & $ 14.73 $& $ 11.48 \pm 0.0487 $ \\
 1eRASS~J093521.3$+$023222 & $( 143.83756, 2.54326 )$ & WL$+$BCG & $ 0.50 $ & $ 14.71 \pm 0.14 $ & $ 14.82 $& $ 11.54 \pm 0.0727 $ \\
 1eRASS~J093831.3$-$012524 & $( 144.62560, -1.42579 )$ & WL & $ 0.41 $ & $ 14.25 \pm 0.16 $ & $ 13.77 $& $ \cdots $ \\
 1eRASS~J094023.3$+$022824 & $( 145.10245, 2.47763 )$ & WL$+$BCG & $ 0.15 $ & $ 14.48 \pm 0.10 $ & $ 14.60 $& $ 11.41 \pm 0.0700 $ \\
 1eRASS~J094611.9$+$022201 & $( 146.54988, 2.36884 )$ & WL & $ 0.12 $ & $ 14.10 \pm 0.13 $ & $ 14.11 $& $ \cdots $ \\
 1eRASS~J094844.3$+$020019 & $( 147.17811, 2.01425 )$ & WL & $ 0.49 $ & $ 14.54 \pm 0.13 $ & $ 14.50 $& $ \cdots $ \\
 1eRASS~J095341.6$+$014200 & $( 148.42250, 1.70030 )$ & WL & $ 0.11 $ & $ 14.04 \pm 0.14 $ & $ 14.15 $& $ \cdots $ \\
 1eRASS~J095728.1$+$033954 & $( 149.36589, 3.66553 )$ & BCG & $ 0.16 $ & $ 14.11 \pm 0.19 $ & $ 14.03 $& $ 11.32 \pm 0.0753 $ \\
 1eRASS~J095736.6$+$023430 & $( 149.40426, 2.57383 )$ & WL$+$BCG & $ 0.38 $ & $ 14.48 \pm 0.12 $ & $ 14.63 $& $ 12.03 \pm 0.0818 $ \\
 1eRASS~J095759.2$+$032732 & $( 149.49721, 3.45722 )$ & WL & $ 0.16 $ & $ 14.27 \pm 0.13 $ & $ 14.22 $& $ \cdots $ \\
 1eRASS~J095858.4$-$001323 & $( 149.74577, -0.20602 )$ & WL$+$BCG & $ 0.17 $ & $ 14.07 \pm 0.15 $ & $ 13.94 $& $ 11.70 \pm 0.0894 $ \\
 1eRASS~J100023.4$+$044406 & $( 150.09319, 4.73525 )$ & BCG & $ 0.36 $ & $ 14.61 \pm 0.16 $ & $ 14.63 $& $ 11.45 \pm 0.1125 $ \\
 1eRASS~J100502.4$+$045249 & $( 151.26224, 4.87969 )$ & BCG & $ 0.45 $ & $ 14.54 \pm 0.16 $ & $ 14.51 $& $ 11.48 \pm 0.1212 $ \\
 1eRASS~J101231.5$+$023933 & $( 153.13326, 2.65806 )$ & BCG & $ 0.24 $ & $ 14.43 \pm 0.17 $ & $ 14.36 $& $ 11.34 \pm 0.0497 $ \\
 1eRASS~J101536.3$+$023411 & $( 153.90268, 2.56672 )$ & BCG & $ 0.29 $ & $ 14.30 \pm 0.18 $ & $ 13.91 $& $ 11.70 \pm 0.0926 $ \\
 1eRASS~J101846.9$-$013110 & $( 154.69095, -1.51743 )$ & WL$+$BCG & $ 0.39 $ & $ 14.64 \pm 0.12 $ & $ 14.44 $& $ 11.82 \pm 0.0072 $ \\
 1eRASS~J101851.0$-$010105 & $( 154.71202, -1.01761 )$ & WL$+$BCG & $ 0.11 $ & $ 13.85 \pm 0.18 $ & $ 14.13 $& $ 11.28 \pm 0.0791 $ \\
 1eRASS~J102250.3$-$000309 & $( 155.70726, -0.05382 )$ & WL$+$BCG & $ 0.30 $ & $ 14.07 \pm 0.17 $ & $ 14.23 $& $ 10.97 \pm 0.0922 $ \\
 1eRASS~J102310.1$+$022157 & $( 155.79244, 2.36822 )$ & BCG & $ 0.19 $ & $ 14.32 \pm 0.18 $ & $ 14.52 $& $ 11.84 \pm 0.0492 $ \\
 1eRASS~J102339.7$+$041108 & $( 155.91508, 4.18628 )$ & BCG & $ 0.29 $ & $ 14.95 \pm 0.18 $ & $ 14.82 $& $ 11.44 \pm 0.0568 $ \\
 1eRASS~J103450.9$+$042431 & $( 158.70247, 4.40556 )$ & BCG & $ 0.16 $ & $ 14.17 \pm 0.19 $ & $ 14.27 $& $ 11.76 \pm 0.0499 $ \\
 1eRASS~J104003.4$+$015948 & $( 160.02298, 1.98560 )$ & BCG & $ 0.39 $ & $ 14.33 \pm 0.20 $ & $ 14.29 $& $ 11.48 \pm 0.0661 $ \\
 1eRASS~J104723.6$-$010449 & $( 161.84752, -1.07442 )$ & BCG & $ 0.47 $ & $ 14.65 \pm 0.16 $ & $ 14.65 $& $ 11.59 \pm 0.0929 $ \\
 1eRASS~J105039.5$+$001707 & $( 162.66629, 0.28534 )$ & WL$+$BCG & $ 0.60 $ & $ 14.77 \pm 0.12 $ & $ 14.70 $& $ 12.12 \pm 0.0320 $ \\
 1eRASS~J111111.5$+$004454 & $( 167.79674, 0.75210 )$ & WL$+$BCG & $ 0.19 $ & $ 14.35 \pm 0.11 $ & $ 14.19 $& $ 11.52 \pm 0.0738 $ \\
 1eRASS~J111552.0$+$012954 & $( 168.96630, 1.49861 )$ & BCG & $ 0.36 $ & $ 14.87 \pm 0.18 $ & $ 14.95 $& $ 11.37 \pm 0.1269 $ \\
 1eRASS~J112626.5$+$003625 & $( 171.61081, 0.60571 )$ & WL$+$BCG & $ 0.31 $ & $ 14.08 \pm 0.16 $ & $ 14.13 $& $ 11.59 \pm 0.0669 $ \\
 1eRASS~J112818.3$-$005859 & $( 172.07373, -0.98338 )$ & WL$+$BCG & $ 0.46 $ & $ 14.62 \pm 0.11 $ & $ 14.44 $& $ 11.73 \pm 0.3267 $ \\
 1eRASS~J112949.4$+$025546 & $( 172.46347, 2.93823 )$ & BCG & $ 0.24 $ & $ 14.43 \pm 0.17 $ & $ 14.47 $& $ 11.68 \pm 0.0650 $ \\
 1eRASS~J113655.7$+$000612 & $( 174.22743, 0.09299 )$ & WL$+$BCG & $ 0.59 $ & $ 14.74 \pm 0.12 $ & $ 14.85 $& $ 11.74 \pm 0.1017 $ \\
 1eRASS~J113843.1$+$031510 & $( 174.68360, 3.26052 )$ & WL$+$BCG & $ 0.14 $ & $ 13.91 \pm 0.18 $ & $ 13.87 $& $ 11.08 \pm 0.0731 $ \\
 1eRASS~J114341.5$-$014429 & $( 175.87352, -1.74169 )$ & WL & $ 0.12 $ & $ 14.15 \pm 0.17 $ & $ 14.28 $& $ \cdots $ \\
 1eRASS~J114441.9$+$004414 & $( 176.17379, 0.73734 )$ & WL$+$BCG & $ 0.33 $ & $ 14.35 \pm 0.12 $ & $ 14.23 $& $ 11.47 \pm 0.1185 $ \\
 1eRASS~J114647.4$-$012428 & $( 176.69839, -1.41111 )$ & WL & $ 0.33 $ & $ 14.08 \pm 0.17 $ & $ 13.99 $& $ \cdots $ \\
 1eRASS~J115019.2$-$003637 & $( 177.58504, -0.59318 )$ & WL$+$BCG & $ 0.14 $ & $ 14.06 \pm 0.13 $ & $ 14.02 $& $ 11.12 \pm 0.0651 $ \\
 1eRASS~J115208.5$-$004726 & $( 178.02708, -0.78046 )$ & WL$+$BCG & $ 0.26 $ & $ 13.41 \pm 0.29 $ & $ 13.33 $& $ 11.13 \pm 0.0274 $ \\
 1eRASS~J115214.5$+$003057 & $( 178.05915, 0.52395 )$ & WL$+$BCG & $ 0.47 $ & $ 14.45 \pm 0.13 $ & $ 14.54 $& $ 11.37 \pm 0.0492 $ \\
 1eRASS~J115235.7$+$035642 & $( 178.14592, 3.93986 )$ & WL$+$BCG & $ 0.68 $ & $ 14.53 \pm 0.16 $ & $ 14.59 $& $ 11.23 \pm 0.1730 $ \\
 1eRASS~J115417.0$+$022123 & $( 178.57200, 2.35657 )$ & WL$+$BCG & $ 0.74 $ & $ 14.64 \pm 0.16 $ & $ 14.64 $& $ 11.86 \pm 0.1339 $ \\
 1eRASS~J115620.0$-$001220 & $( 179.08350, -0.20553 )$ & WL$+$BCG & $ 0.11 $ & $ 14.03 \pm 0.13 $ & $ 13.98 $& $ 11.47 \pm 0.0750 $ \\
 1eRASS~J120024.4$+$032112 & $( 180.10549, 3.34696 )$ & WL$+$BCG & $ 0.14 $ & $ 14.71 \pm 0.10 $ & $ 14.89 $& $ 11.55 \pm 0.0848 $ \\
 1eRASS~J120143.7$-$001110 & $( 180.43196, -0.18456 )$ & WL$+$BCG & $ 0.17 $ & $ 14.41 \pm 0.11 $ & $ 14.46 $& $ 11.10 \pm 0.0721 $ \\
 1eRASS~J120358.5$+$012624 & $( 180.99168, 1.42579 )$ & WL$+$BCG & $ 0.39 $ & $ 14.64 \pm 0.12 $ & $ 14.55 $& $ 11.81 \pm 0.0750 $ \\
 1eRASS~J120417.3$+$025402 & $( 181.07295, 2.89815 )$ & WL & $ 0.15 $ & $ 14.06 \pm 0.14 $ & $ 14.08 $& $ \cdots $ \\
 1eRASS~J120954.5$-$003328 & $( 182.47791, -0.55770 )$ & WL$+$BCG & $ 0.18 $ & $ 14.09 \pm 0.14 $ & $ 14.17 $& $ 11.70 \pm 0.0629 $ \\
 1eRASS~J121016.3$+$022338 & $( 182.57381, 2.39546 )$ & WL$+$BCG & $ 0.38 $ & $ 14.56 \pm 0.13 $ & $ 14.62 $& $ 12.04 \pm 0.0051 $ \\
 1eRASS~J121336.5$+$025331 & $( 183.39372, 2.89887 )$ & WL & $ 0.39 $ & $ 14.32 \pm 0.13 $ & $ 14.14 $& $ \cdots $ \\
 1eRASS~J121503.9$-$012741 & $( 183.77197, -1.46098 )$ & WL$+$BCG & $ 0.20 $ & $ 14.22 \pm 0.15 $ & $ 14.23 $& $ 11.66 \pm 0.0405 $ \\
 1eRASS~J122226.1$-$012659 & $( 185.61026, -1.45326 )$ & WL$+$BCG & $ 0.28 $ & $ 14.36 \pm 0.13 $ & $ 14.58 $& $ 11.37 \pm 0.0836 $ \\
 1eRASS~J122242.4$-$013029 & $( 185.67767, -1.51167 )$ & WL$+$BCG & $ 0.29 $ & $ 14.36 \pm 0.14 $ & $ 14.53 $& $ 11.46 \pm 0.0829 $ \\
 1eRASS~J122420.6$+$021206 & $( 186.09158, 2.20303 )$ & WL$+$BCG & $ 0.45 $ & $ 14.62 \pm 0.12 $ & $ 14.62 $& $ 12.11 \pm 0.1678 $ \\
 1eRASS~J122528.9$+$004237 & $( 186.36363, 0.71004 )$ & WL$+$BCG & $ 0.24 $ & $ 14.08 \pm 0.16 $ & $ 14.13 $& $ 11.51 \pm 0.0758 $ \\
 1eRASS~J122644.5$-$003724 & $( 186.68702, -0.62162 )$ & WL$+$BCG & $ 0.16 $ & $ 13.98 \pm 0.15 $ & $ 13.81 $& $ 11.34 \pm 0.0754 $ \\
 1eRASS~J123055.0$+$024716 & $( 187.72970, 2.78710 )$ & BCG & $ 0.39 $ & $ 14.57 \pm 0.16 $ & $ 14.65 $& $ 11.49 \pm 0.1543 $ \\
 1eRASS~J123108.4$+$003653 & $( 187.78493, 0.61363 )$ & WL & $ 0.47 $ & $ 14.45 \pm 0.16 $ & $ 14.33 $& $ \cdots $ \\
 1eRASS~J123740.7$-$012119 & $( 189.42311, -1.36062 )$ & WL$+$BCG & $ 0.16 $ & $ 14.13 \pm 0.15 $ & $ 14.18 $& $ 11.88 \pm 0.0702 $ \\
 1eRASS~J123755.2$-$001611 & $( 189.48058, -0.27471 )$ & WL$+$BCG & $ 0.14 $ & $ 13.98 \pm 0.14 $ & $ 14.03 $& $ 11.20 \pm 0.0730 $ \\
 1eRASS~J124503.8$-$002823 & $( 191.28558, -0.46177 )$ & WL & $ 0.23 $ & $ 14.00 \pm 0.16 $ & $ 14.14 $& $ \cdots $ \\
 1eRASS~J124942.4$+$014447 & $( 192.43193, 1.74645 )$ & BCG & $ 0.20 $ & $ 14.09 \pm 0.20 $ & $ 14.15 $& $ 11.59 \pm 0.0818 $ \\
 1eRASS~J125035.8$+$003646 & $( 192.64934, 0.61281 )$ & WL$+$BCG & $ 0.63 $ & $ 14.54 \pm 0.14 $ & $ 14.23 $& $ 11.83 \pm 0.0398 $ \\
 1eRASS~J125728.4$+$011841 & $( 194.37388, 1.31586 )$ & BCG & $ 0.37 $ & $ 14.19 \pm 0.19 $ & $ 13.75 $& $ 11.35 \pm 0.0685 $ \\
 1eRASS~J130037.3$-$015635 & $( 195.15752, -1.94352 )$ & BCG & $ 0.44 $ & $ 14.39 \pm 0.17 $ & $ 14.27 $& $ 11.73 \pm 0.1211 $ \\
 1eRASS~J130709.0$+$011222 & $( 196.77408, 1.20512 )$ & WL$+$BCG & $ 0.19 $ & $ 13.95 \pm 0.18 $ & $ 14.13 $& $ 11.31 \pm 0.0666 $ \\
 1eRASS~J131129.4$-$012026 & $( 197.87297, -1.34109 )$ & WL$+$BCG & $ 0.18 $ & $ 14.98 \pm 0.11 $ & $ 15.01 $& $ 11.24 \pm 0.0561 $ \\
 1eRASS~J131419.1$-$012723 & $( 198.58215, -1.45715 )$ & WL$+$BCG & $ 0.18 $ & $ 14.13 \pm 0.15 $ & $ 13.81 $& $ 11.73 \pm 0.0762 $ \\
 1eRASS~J131814.0$+$014253 & $( 199.55881, 1.71313 )$ & BCG & $ 0.23 $ & $ 14.05 \pm 0.20 $ & $ 14.18 $& $ 11.36 \pm 0.0761 $ \\
 1eRASS~J131815.6$-$003818 & $( 199.55704, -0.62596 )$ & WL$+$BCG & $ 0.10 $ & $ 13.63 \pm 0.22 $ & $ 13.67 $& $ 11.52 \pm 0.0776 $ \\
 1eRASS~J131815.3$-$011004 & $( 199.56356, -1.16890 )$ & WL & $ 0.22 $ & $ 14.29 \pm 0.18 $ & $ 14.18 $& $ \cdots $ \\
 1eRASS~J131823.9$-$003135 & $( 199.59810, -0.52953 )$ & WL$+$BCG & $ 0.11 $ & $ 13.76 \pm 0.20 $ & $ 13.96 $& $ 11.60 \pm 0.0573 $ \\
 1eRASS~J134304.9$-$000053 & $( 205.76907, -0.01546 )$ & BCG & $ 0.68 $ & $ 14.50 \pm 0.17 $ & $ 14.56 $& $ 11.60 \pm 0.1429 $ \\
 1eRASS~J134525.0$+$001207 & $( 206.34472, 0.20605 )$ & WL$+$BCG & $ 0.25 $ & $ 14.14 \pm 0.15 $ & $ 14.41 $& $ 11.48 \pm 0.0760 $ \\
 1eRASS~J134745.7$+$012441 & $( 206.94083, 1.41249 )$ & BCG & $ 0.57 $ & $ 14.53 \pm 0.16 $ & $ 14.30 $& $ 11.86 \pm 0.1267 $ \\
 1eRASS~J134840.1$+$003907 & $( 207.17319, 0.65456 )$ & WL$+$BCG & $ 0.42 $ & $ 14.33 \pm 0.14 $ & $ 14.34 $& $ 12.07 \pm 0.0624 $ \\
 1eRASS~J135326.5$+$000255 & $( 208.35942, 0.04664 )$ & WL$+$BCG & $ 0.11 $ & $ 13.93 \pm 0.16 $ & $ 14.04 $& $ 11.19 \pm 0.0841 $ \\
 1eRASS~J135424.2$-$010250 & $( 208.60017, -1.04431 )$ & WL$+$BCG & $ 0.15 $ & $ 14.40 \pm 0.11 $ & $ 14.37 $& $ 11.78 \pm 0.0725 $ \\
 1eRASS~J140559.3$+$013843 & $( 211.49715, 1.64785 )$ & BCG & $ 0.17 $ & $ 13.93 \pm 0.20 $ & $ 13.15 $& $ 11.17 \pm 0.0618 $ \\
 1eRASS~J140707.1$-$001450 & $( 211.77931, -0.25140 )$ & WL$+$BCG & $ 0.55 $ & $ 14.64 \pm 0.13 $ & $ 14.57 $& $ 11.85 \pm 0.0320 $ \\
 1eRASS~J140959.6$-$013230 & $( 212.49782, -1.53965 )$ & WL$+$BCG & $ 0.11 $ & $ 13.81 \pm 0.19 $ & $ 14.05 $& $ 11.46 \pm 0.0743 $ \\
 1eRASS~J141055.8$+$013800 & $( 212.73264, 1.63343 )$ & BCG & $ 0.18 $ & $ 14.09 \pm 0.18 $ & $ 14.24 $& $ 11.51 \pm 0.0710 $ \\
 1eRASS~J141452.4$+$001316 & $( 213.73578, 0.20639 )$ & WL$+$BCG & $ 0.12 $ & $ 13.69 \pm 0.17 $ & $ 13.70 $& $ 11.42 \pm 0.0731 $ \\
 1eRASS~J141457.8$-$002050 & $( 213.74060, -0.34965 )$ & WL & $ 0.13 $ & $ 14.51 \pm 0.12 $ & $ 14.46 $& $ \cdots $ \\
 1eRASS~J141507.1$-$002905 & $( 213.78498, -0.49333 )$ & WL$+$BCG & $ 0.13 $ & $ 14.31 \pm 0.12 $ & $ 14.08 $& $ 11.70 \pm 0.0776 $ \\
 1eRASS~J141547.1$+$001530 & $( 213.95354, 0.26047 )$ & WL$+$BCG & $ 0.13 $ & $ 14.02 \pm 0.15 $ & $ 14.08 $& $ 11.02 \pm 0.0818 $ \\
 1eRASS~J142016.3$+$005716 & $( 215.06934, 0.95531 )$ & WL$+$BCG & $ 0.50 $ & $ 14.55 \pm 0.11 $ & $ 14.63 $& $ 11.66 \pm 0.0759 $ \\
 1eRASS~J142724.6$-$001509 & $( 216.85038, -0.24871 )$ & WL$+$BCG & $ 0.16 $ & $ 13.95 \pm 0.14 $ & $ 13.99 $& $ 11.66 \pm 0.0682 $ \\
 1eRASS~J143044.6$+$004815 & $( 217.67543, 0.81749 )$ & WL$+$BCG & $ 0.31 $ & $ 14.37 \pm 0.11 $ & $ 14.36 $& $ 11.45 \pm 0.0692 $ \\
 1eRASS~J143121.5$-$005336 & $( 217.83829, -0.89570 )$ & WL$+$BCG & $ 0.40 $ & $ 14.51 \pm 0.12 $ & $ 14.55 $& $ 11.56 \pm 0.1208 $ \\
 1eRASS~J143736.1$-$001740 & $( 219.43256, -0.31583 )$ & WL$+$BCG & $ 0.14 $ & $ 14.24 \pm 0.12 $ & $ 14.09 $& $ 10.89 \pm 0.0800 $ \\
 1eRASS~J143821.9$+$034012 & $( 219.59115, 3.67036 )$ & BCG & $ 0.22 $ & $ 14.53 \pm 0.17 $ & $ 14.65 $& $ 11.89 \pm 0.0828 $ \\
 1eRASS~J144016.1$+$034155 & $( 220.06095, 3.69504 )$ & BCG & $ 0.28 $ & $ 14.36 \pm 0.17 $ & $ 14.26 $& $ 11.64 \pm 0.0653 $ \\
 1eRASS~J144031.3$-$013738 & $( 220.13343, -1.62420 )$ & WL$+$BCG & $ 0.32 $ & $ 14.57 \pm 0.15 $ & $ 14.48 $& $ 11.16 \pm 0.0481 $ \\
 1eRASS~J144050.2$+$004103 & $( 220.20374, 0.67232 )$ & WL$+$BCG & $ 0.54 $ & $ 14.22 \pm 0.19 $ & $ 14.52 $& $ 11.35 \pm 0.1599 $ \\
 1eRASS~J144133.6$-$005418 & $( 220.38605, -0.89954 )$ & WL$+$BCG & $ 0.54 $ & $ 14.56 \pm 0.13 $ & $ 14.67 $& $ 12.15 \pm 0.0625 $ \\
 1eRASS~J144309.6$+$010214 & $( 220.78878, 1.03670 )$ & WL$+$BCG & $ 0.53 $ & $ 14.57 \pm 0.13 $ & $ 14.59 $& $ 11.99 \pm 0.0655 $ \\
 1eRASS~J144914.3$+$032949 & $( 222.32834, 3.51829 )$ & BCG & $ 0.21 $ & $ 13.89 \pm 0.28 $ & $ 13.91 $& $ 11.07 \pm 0.0782 $ \\
 1eRASS~J144950.6$+$005558 & $( 222.46120, 0.92683 )$ & WL$+$BCG & $ 0.41 $ & $ 14.36 \pm 0.13 $ & $ 14.30 $& $ 11.61 \pm 0.0643 $ \\
 1eRASS~J145004.7$+$004931 & $( 222.52202, 0.82471 )$ & WL & $ 0.38 $ & $ 14.33 \pm 0.13 $ & $ 14.29 $& $ \cdots $ \\
 1eRASS~J145218.8$+$001002 & $( 223.07861, 0.16244 )$ & WL$+$BCG & $ 0.59 $ & $ 14.73 \pm 0.13 $ & $ 15.01 $& $ 11.67 \pm 0.0568 $ \\
 1eRASS~J145459.3$-$011455 & $( 223.74723, -1.24914 )$ & WL$+$BCG & $ 0.31 $ & $ 14.22 \pm 0.14 $ & $ 14.13 $& $ 11.48 \pm 0.0750 $ \\
 1eRASS~J145723.4$+$030717 & $( 224.35304, 3.11794 )$ & BCG & $ 0.63 $ & $ 14.45 \pm 0.17 $ & $ 14.74 $& $ 11.81 \pm 0.0181 $ \\
\hline
\end{longtable}
\tablefoot{From the left to right, the columns are the cluster name, the location of the identified BCG, the availability of the data, the cluster redshift, the mean and standard deviation of the mass posterior $\left\langle\log\left(\frac{\Mfiveoo}{\Msunh}\right)\right\rangle$, the posterior-sampled halo mass, and the BCG stellar mass.}
}

\section{Auxiliary figures}
\label{app:auxiliary}

\begin{figure*}
\centering
\resizebox{0.8\textwidth}{!}{
\includegraphics[scale=1]{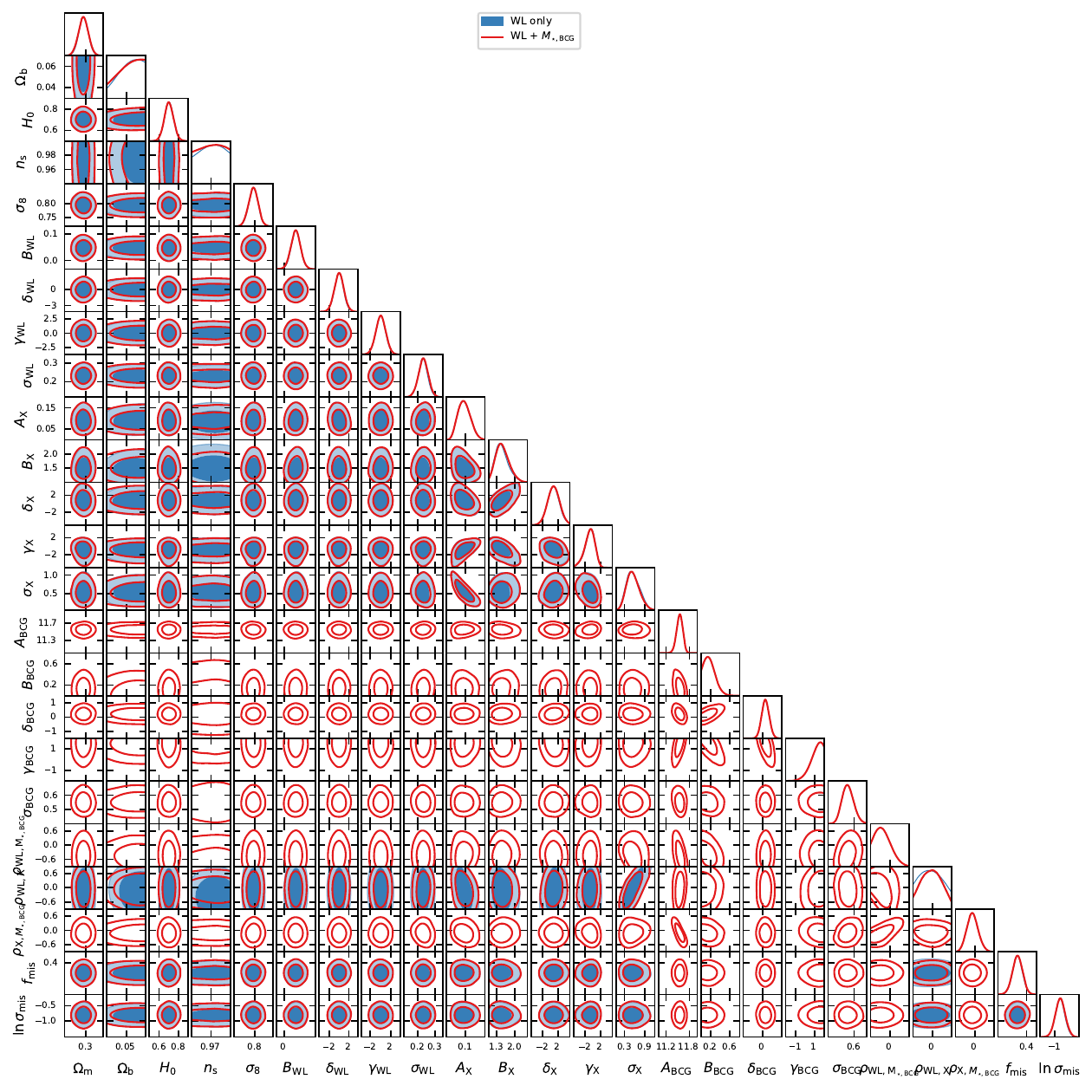}
}
\caption{
The same plot to Figure~\ref{fig:gtc_trunc} but for the full results of the parameter constraints in the WL-only modelling (blue contours) of the \rate--\mass--\redshift\ relation and the joint modelling (red lines) of the \rate--\mass--\redshift\ and \mbcg--\mass--\redshift\ relations.
}
\label{fig:gtc_allcomb}
\end{figure*}

\begin{figure*}
\centering
\resizebox{\textwidth}{!}{
\includegraphics[scale=1]{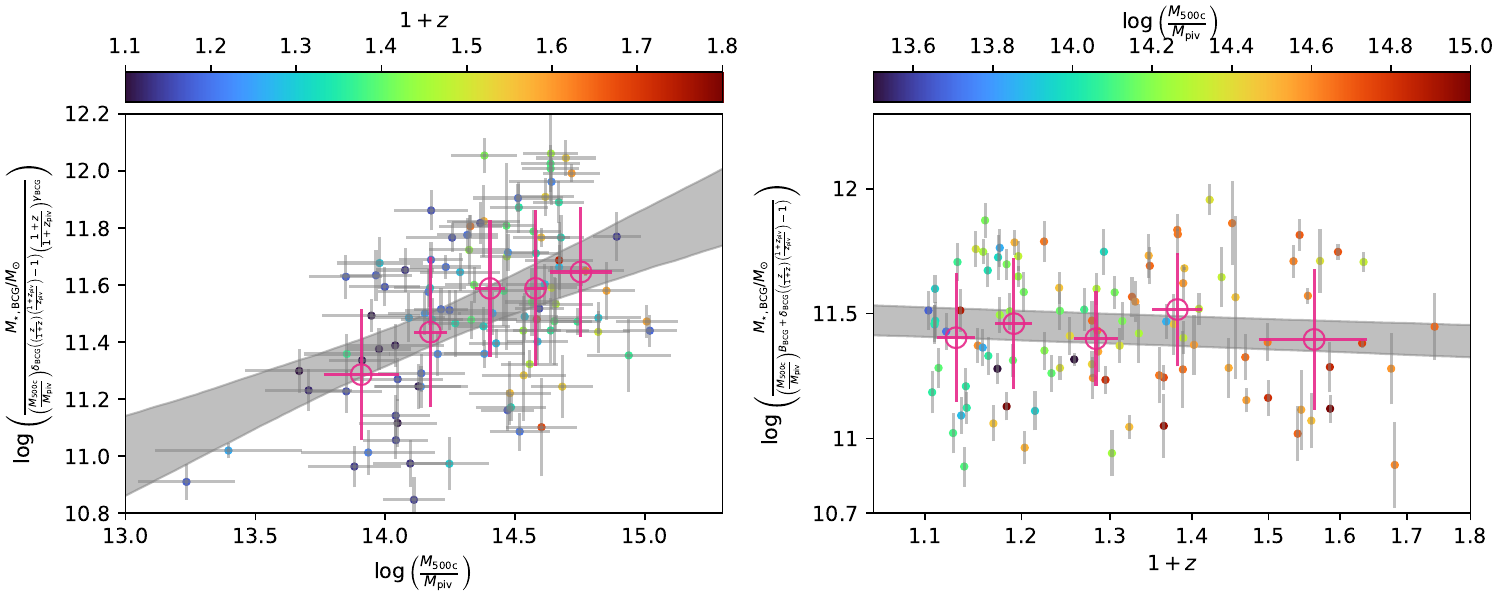}
}
\caption{
The sample plot as in Figure~\ref{fig:mbcg_m_z} but for the result of the \mbcg--\mass--\redshift\ relation (equation~(\ref{eq:bcg_relation_result_prioronz})) obtained with the Gaussian prior on \gammabcg.
}
\label{fig:mbcg_m_z_prioronz}
\end{figure*}

We provide the following materials in this appendix.
The full parameter constraints (marginalized posteriors and two-dimensional joint posteriors) of the WL-only and WL$+\mbcg$ modelling are contained in Figure~\ref{fig:gtc_allcomb}.
Figure~\ref{fig:mbcg_m_z_prioronz} shows the mass and redshift trends of the BCG stellar mass of the \erass1 clusters, using the modelling including the Gaussian prior on \gammabcg\ (see the text in Section~\ref{sec:bcg_relation_results}).


\section{The BCG stellar mass with the inclusion of the \wise\ photometry}
\label{app:wise_data}

The spectral energy distribution (SED) of the old stellar population in galaxies typically peaks at the wavelength range between $1\mu$m and $2\mu$m, while the optical emission is more subject to recent star-forming activities.
Owing to the cosmological redshifting, the mid-infrared wavelength provides an excellent window with negative $k$-correction to probe the stellar light of the old stellar population.
As BCGs are dominated by the old stellar population, the inclusion of mid-infrared photometry is expected to provide stronger constraints on the stellar mass estimates than using optical photometry alone.
Independently of the fiducial analysis, we include the infrared data from the \wise\ all-sky survey to estimate the BCG stellar mass and examine the associating systematic uncertainty to the final results.

We include the mid-infrared photometry at the wavelength of $3.4\mu$m and $4.6\mu$m, labelled as the filters \wone\ and \wtwo, respectively, from the all-sky surveys observed by the \textit{Wide-field Infrared Survey Explorer} \citep[\wise;][]{wright10} to estimate the BCG stellar mass \mbcg.
Specifically, we use the unWISE catalog \citep{lang16} with forced photometry extracted at the location of the SDSS-DR13 sources \citep{albareti17} on the unWISE coadds \citep{lang14}.
These coadds were reprocessed using an improved pipeline applied to the images collected in the ALLWISE \citep{wright10} survey without additional blurring.

To select galaxies from the unWISE catalog, we discard bright stars by applying the flag of $\mathtt{pointsource}!=1$ and removing objects with the \wone\ magnitude brighter than $14$~mag.
We then match the BCG catalog of the \erass1 clusters to the resulting unWISE catalog to obtain the \wone\wtwo\ photometry.
Among the $101$ clusters with the available HSC photometry for the BCGs, $96$ have the \wone\wtwo\ photometry from the unWISE catalog with a median matching separation of $\approx0.20$~arcsec.

We re-estimate the BCG stellar mass for those $96$ clusters using the seven-band photometry $grizY\wone\wtwo$, and re-run the subsequent analyses to derive the \mbcg--\mass--\redshift\ relation.
The resulting constraints are 
$\left(\Abcg, \Bbcg, \deltabcg, \gammabcg, \sigmabcg\right) = 
\left(
11.450^{+0.081}_{-0.074},
0.193^{+0.074}_{-0.16},
0.14 \pm 0.35,
0.61^{+0.92}_{-0.83}
0.471^{+0.052}_{-0.063},
\right)$, which are consistent with the fiducial results (without \wone\wtwo).
Therefore, we conclude that incorporating mid-infrared photometry does not have a significant impact on our final results. 

For homogeneous photometric measurements, we choose not to include \wise\ photometry in our fiducial analyses.

\end{document}